\documentclass[aps,a4paper,superscriptaddress,preprintnumbers,showpacs,amsmath,amssymb]{revtex4}
\usepackage{graphicx} \usepackage{dcolumn} \usepackage{color}
\usepackage{latexsym,amsfonts}

\usepackage{textcomp}

\usepackage{bm}

\usepackage{ulem}

\begin{document}

\title{Precision Theoretical Analysis\\ of Neutron Radiative Beta
  Decay to Order $O(\alpha^2/\pi^2)$}

\author{A. N. Ivanov}\email{ivanov@kph.tuwien.ac.at}
\affiliation{Atominstitut, Technische Universit\"at Wien, Stadionallee
  2, A-1020 Wien, Austria}
\author{R.~H\"ollwieser}\email{roman.hoellwieser@gmail.com}
\affiliation{Atominstitut, Technische Universit\"at Wien, Stadionallee
  2, A-1020 Wien, Austria}\affiliation{Department of Physics, New
  Mexico State University, Las Cruces, New Mexico 88003, USA}
\author{N. I. Troitskaya}\email{natroitskaya@yandex.ru}
\affiliation{Atominstitut, Technische Universit\"at Wien, Stadionallee
  2, A-1020 Wien, Austria}
\author{M. Wellenzohn}\email{max.wellenzohn@gmail.com}
\affiliation{Atominstitut, Technische Universit\"at Wien, Stadionallee
  2, A-1020 Wien, Austria} \affiliation{FH Campus Wien, University of
  Applied Sciences, Favoritenstra\ss e 226, 1100 Wien, Austria}
\author{Ya. A. Berdnikov}\email{berdnikov@spbstu.ru}
\affiliation{Peter the Great St. Petersburg Polytechnic University,
  Polytechnicheskaya 29, 195251, Russian Federation}

\date{\today}

\begin{abstract}
In the Standard Model (SM) we calculate the decay rate of the neutron
radiative $\beta^-$--decay to order $O(\alpha^2/\pi^2 \sim 10^{-5})$,
where $\alpha$ is the fine--structure constant, and radiative
corrections to order $O(\alpha/\pi \sim 10^{-3})$. The obtained
results together with the recent analysis of the neutron radiative
$\beta^-$--decay to next--to--leading order in the large proton--mass
expansion, performed by Ivanov {\it et al.}  Phys. Rev. D {\bf 95},
033007 (2017), describe recent experimental data by the RDK II
Collaboration (Bales {\it et al.}, Phys. Rev. Lett. {\bf 116}, 242501
(2016)) within $1.5$ standard deviations.
We argue a substantial influence of strong low--energy interactions of
hadrons coupled to photons on the properties of the amplitude of the
neutron radiative $\beta^-$--decay under gauge transformations of real
and virtual photons.
\end{abstract}
\maketitle

\section{Introduction}
\label{sec:introduction}

During a long period the radiative $\beta^-$--decay of a free neutron
$n \to p + e^- + \bar{\nu}_e + \gamma$ was used as an auxiliary
process in the analysis of the radiative corrections to the neutron
$\beta^-$--decay for the cancellation of infrared divergences, coming
from the virtual photon exchanges
\cite{Berman1958}--\cite{Shann1971}. Only starting from 1996 it has
been accepted as a physical process because of the work by Gaponov and
Khafizov \cite{Gaponov1996}, who made first calculation of the energy
spectrum and the decay rate. Then, the neutron radiative
$\beta^-$--decay was reinvestigated in \cite{Bernard2004} and
\cite{Ivanov2013,Ivanov2013a}. The first experimental data ${\rm
  BR}_{\beta \gamma} = 3.13(35)\times 10^{-3}$ and ${\rm BR}_{\beta
  \gamma} = 3.09(32)\times 10^{-3}$, measured by Nico {\it et al.}
\cite{Nico2006} and Cooper {\it et al.}  \cite{Cooper2010}, for the
photon--energy region $15\,{\rm keV} \le \omega \le 340\,{\rm keV}$,
were in agreement within one standard deviation with the theoretical
values ${\rm BR}_{\beta \gamma} = 2.87 \times 10^{-3}$
\cite{Ivanov2013} and ${\rm BR}_{\beta \gamma} = 2.85\times 10^{-3}$,
calculated by Gardner \cite{Nico2006} using the theoretical decay
rate, published in \cite{Bernard2004}.  Recently new precise
experimental values of the branching ratios of the radiative
$\beta^-$--decay of a free neutron have been reported by the RDK II
Collaboration Bales {\it et al.}  \cite{Bales2016}: ${\rm
  BR}^{(\exp)}_{\beta \gamma} = 3.35(16)\times 10^{-3}$ and ${\rm
  BR}^{(\exp)}_{\beta \gamma} = 5.82(66)\times 10^{-3}$, measured for
the photon--energy regions $14\,{\rm keV} \le \omega \le 782\,{\rm
  keV}$ and $0.4\,{\rm keV} \le \omega \le 14\,{\rm keV}$,
respectively.  Recently \cite{Ivanov2017} the rate of the neutron
radiative $\beta^-$--decay has been recalculated in the Standard Model
(SM) and in the tree--approximation to next--to--leading order in the
large proton mass expansion by taking into account the contributions
of the weak magnetism and proton recoil. As has been found the new
theoretical values of the branching ratios ${\rm BR}_{\beta \gamma} =
3.04 \times 10^{-3}$ and ${\rm BR}_{\beta \gamma} = 5.08 \times
10^{-3}$, calculated for experimental photon--energy regions $14\,{\rm
  keV} \le \omega \le 782\,{\rm keV}$ and $0.4\,{\rm keV} \le \omega
\le 14\,{\rm keV}$, respectively, agree with new experimental values
${\rm BR}^{(\exp)}_{\beta \gamma} = 3.35(16)\times 10^{-3}$ and ${\rm
  BR}^{(\exp)}_{\beta \gamma} = 5.82(66)\times 10^{-3}$ only within 2
and 1.2 standard deviations.  As has been shown in \cite{Ivanov2017}
the relative contributions of the weak magnetism and proton recoil to
the branching ratios of the neutron radiative $\beta^-$--decay are of
about $0.7\,\%$. Of course, these contributions are small compared to
the error bars of the experimental values but they are by a factor 4
larger than the contribution of the weak
magnetism and proton recoil $0.16\,\%$ to the rate of the neutron
$\beta^-$--decay \cite{Ivanov2013}. As has been pointed out in
\cite{Ivanov2017} the contributions to the rate of the neutron
radiative $\beta^-$--decay, calculated in the SM and in the
tree--approximation to next--to--leading order in the large baryon
mass expansion including the contributions of baryon resonances (see,
for example, Bernard {\it et al.}  \cite{Bernard2004}), cannot in
principle exceed $1.5\,\%$. So one may expect some tangible
contributions only beyond the tree--approximation, taking into
account, for example, one--virtual--photon exchanges to leading order
in the large proton mass expansion, i.e. the radiative corrections of
order $O(\alpha/\pi)$.  We would like to remind that radiative
corrections of order $O(\alpha/\pi)$ change the rate of the neutron
$\beta^-$--decay by about $3.75\,\%$
\cite{Ivanov2013,Ivanov2017}. Because of an enhancement of the
contributions of order $1/M$, where $2M = m_n + m_p$ is an averaged
nucleon mass \cite{Ivanov2013,Ivanov2017}, to the rate of the neutron
radiative $\beta^-$--decay, one may also expect an enhancement of the
relative contributions of the radiative corrections of order
$O(\alpha/\pi)$.

For the first time the radiative corrections of order $O(\alpha/ \pi)$
for the analysis of $T$--odd momentum correlations in the neutron
radiative $\beta^-$--decay to order $O(\alpha^2/\pi^2)$ have been
calculated by Gardner and He \cite{Gardner2012,Gardner2013}. In this
paper we give a complete analysis of the radiative corrections to
order $O(\alpha/\pi)$ to the rate of the neutron radiative
$\beta^-$--decay, caused by pure Quantum Electrodynamics (QED), where
photons couple to point--like proton and electron with a contribution
of strong low--energy interactions defined by the axial couping
constant $\lambda$ only.

\begin{figure}
\centering \includegraphics[height=0.12\textheight]{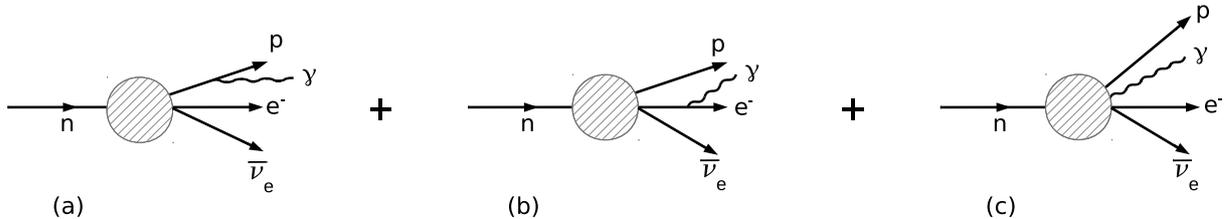}
  \caption{The Feynman diagrams, defining the contribution to the
    amplitude of the neutron radiative $\beta^-$--decay in the
    tree--approximation to order $e$.}
\label{fig:fig1}
\end{figure}
\begin{figure}
\centering \includegraphics[height=0.23\textheight]{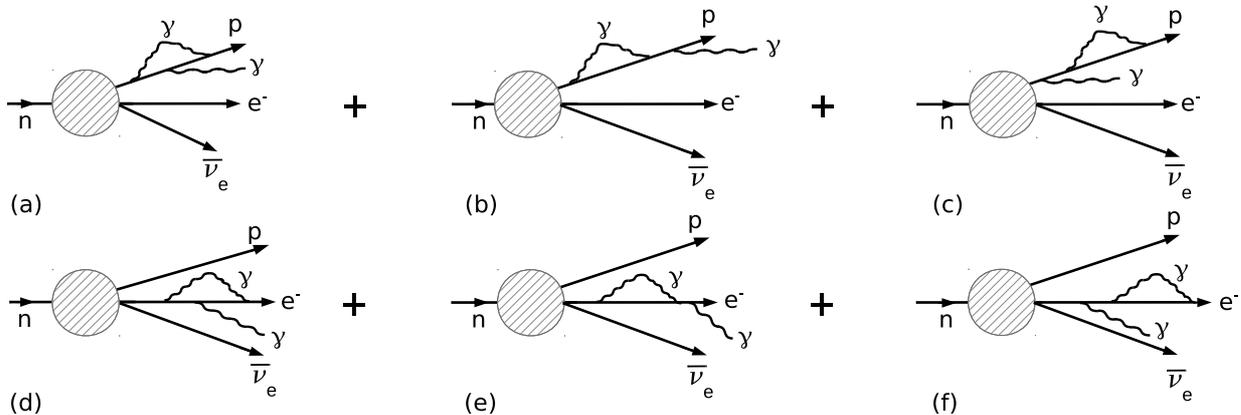}
  \caption{The Feynman diagrams, defining the contribution to the
    amplitude of the neutron radiative $\beta^-$--decay of order
    $e^3$, caused by pure QCD.}
\label{fig:fig2}
\end{figure}
\begin{figure}
\centering \includegraphics[height=0.23\textheight]{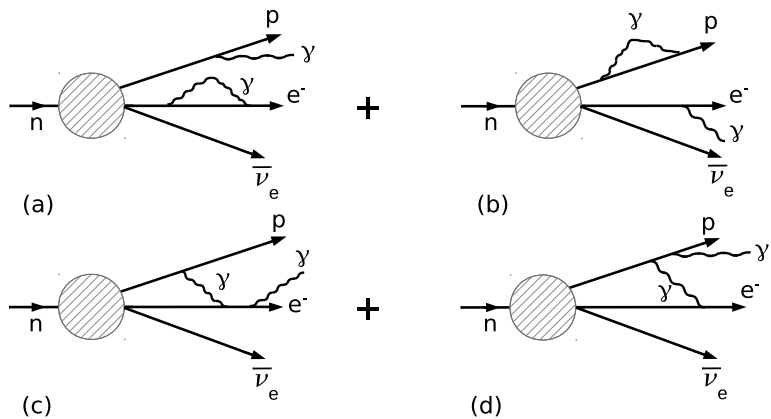}
  \caption{The Feynman diagrams, defining the main contribution to the
    amplitude of the neutron radiative $\beta^-$--decay of order
    $e^3$.}
\label{fig:fig3}
\end{figure}
\begin{figure}
\centering \includegraphics[height=0.35\textheight]{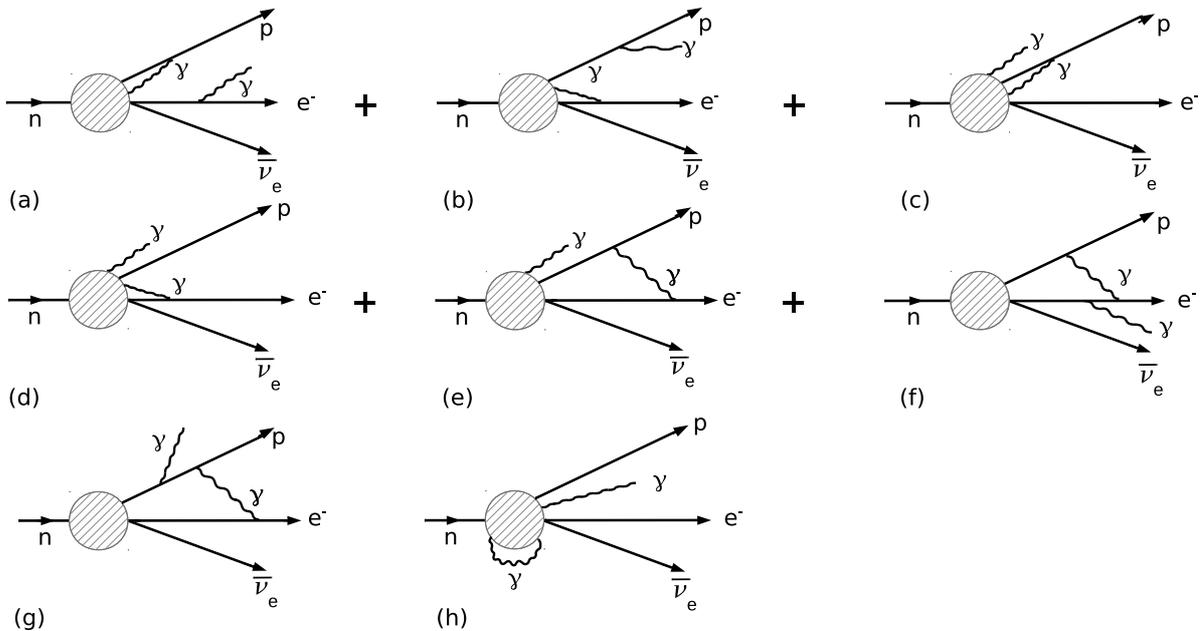}
  \caption{The Feynman diagrams, responsible for restoration of gauge
    invariance of the Feynman diagrams Fig.\,\ref{fig:fig3}c and
    Fig.\,\ref{fig:fig3}d.}
\label{fig:fig4}
\end{figure}
\begin{figure}
\centering \includegraphics[height=0.12\textheight]{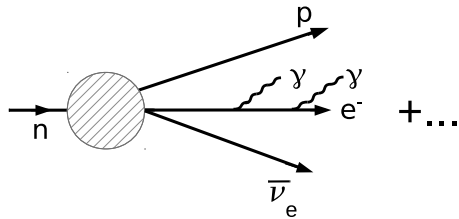}
  \caption{The Feynman diagrams, defining the main contribution of order
    $O(\alpha^2/\pi^2)$ to the rate of the neutron radiative
    $\beta^-$--decay $n \to p + e^- + \bar{\nu}_e + \gamma + \gamma$
    with one detected and one undetected photon.}
\label{fig:fig5}
\end{figure}
\begin{figure}
\centering \includegraphics[height=0.12\textheight]{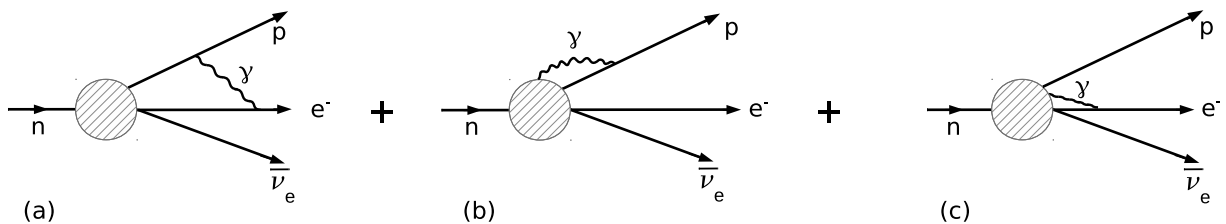}
  \caption{The Feynman diagrams, defining the main contribution of the
    radiative corrections of order $O(\alpha/\pi)$, caused by
    one--virtual photon exchanges, to the neutron $\beta^-$--decay (
    see Sirlin \cite{Sirlin1967}).}
\label{fig:fig6}
\end{figure}

A complete set of Feynman diagrams, describing the amplitude of the
neutron radiative $\beta^-$--decay in the tree and one--loop
approximation, are shown in Fig.\,\ref{fig:fig1},
Fig.\,\ref{fig:fig2}, Fig.\,\ref{fig:fig3}, Fig.\,\ref{fig:fig4} and
Fig.\,\ref{fig:fig5}, respectively.  In Fig.\,\ref{fig:fig1},
Fig.\,\ref{fig:fig2}, Fig.\,\ref{fig:fig3} and Fig.\,\ref{fig:fig4}
the states of real and virtual photons with 4--momenta $k$ and $q$,
respectively, are described by the polarization vector
$\varepsilon^*_{\lambda'}(k)$ with $\lambda' = 1,2$ and a Green
function $D_{\alpha\beta}(q) = (\eta_{\alpha\beta} - (1 -
\xi)\,q_{\alpha}q_{\beta}/q^2)/(q^2 + i0)$ \cite{Ivanov2013}, where
the polarization vector obeys the constraint
$\varepsilon^*_{\lambda'}(k) \cdot k = 0$ with $k^2 = 0$ and $\xi$ is
a gauge parameter. The Feynman diagrams in Fig.\,\ref{fig:fig5}
describe a neutron radiative $\beta^-$-decay with two real photons in
the final state. Integrating over degrees of freedom of one of the
photons one obtains the contribution of order $O(\alpha^2/\pi^2)$ to
the rate of the neutron radiative $\beta^-$--decay with one real
photon in the final state.  The contributions of strong low--energy
hadronic interactions in the Feynman diagrams Fig.\,\ref{fig:fig1},
Fig.\,\ref{fig:fig2}, Fig.\,\ref{fig:fig3}, Fig.\,\ref{fig:fig4} and
Fig.\,\ref{fig:fig5} (see also Fig.\,\ref{fig:fig6}) are denoted by
shaded regions.

The contributions of pure QED are given by the Feynman diagrams in
Fig.\,\ref{fig:fig1}a, Fig.\,\ref{fig:fig1}b, Fig.\,\ref{fig:fig2} and
Fig.\,\ref{fig:fig5}, where real and virtual photons couple to the
point--like proton and electron and strong low--energy hadronic and
electromagnetic interactions are factorized. The contribution of
strong low--energy interactions is described by the axial coupling
constant $\lambda$ only. In the diagram Fig\,\ref{fig:fig1}c a real
photon is emitted by a hadronic block. In spite of a possible
dependence of the contribution of this diagram on electron and photon
energies it has been neglected in the first calculations of the
neutron radiative $\beta^-$--decay by Gaponov and Khafizov
\cite{Gaponov1996} and in the subsequent calculations by Bernard {\it
  et al.}  \cite{Bernard2004} and Ivanov {\it et al.}
\cite{Ivanov2013,Ivanov2017}. In this paper we also accept such an
approximation. We neglect the contributions of all Feynman diagrams,
where even if one photon (real or virtual) is emitted or absorbed by a
hadronic block. In section \ref{sec:conclusion} we propose a
justification of the neglect of the contribution of the diagram in
Fig.\,\ref{fig:fig1}c. However, an analysis of contributions of strong
low--energy hadronic interactions in the diagrams in
Fig.\,\ref{fig:fig3} and Fig.\,\ref{fig:fig4} demands a special
consideration and goes beyond the scope of this paper.

It is well known that the amplitude of the neutron radiative
$\beta^-$--decay should be gauge invariant. This means that when
making a gauge transformation of a real photon wave function,
i.e. replacing the photon polarization vector
$\varepsilon^*_{\lambda'}(k)$ by $\varepsilon^*_{\lambda'}(k) \to
\varepsilon^*_{\lambda'}(k) + c\,k$, where $c$ is an arbitrary
constant, the contribution proportional to $c\,k$ should vanish
\cite{Itzykson1980} (see also \cite{Ivanov1973}). In Appendices A and
B of the Supplemental Material we investigate the properties of Feynman
diagrams in Fig.\,\ref{fig:fig1} and Fig.\,\ref{fig:fig2} with respect
to a gauge transformation $\varepsilon^*_{\lambda'}(k) \to
\varepsilon^*_{\lambda'}(k) + c\,k$. By means of a direct calculation
we show that in Fig.\,\ref{fig:fig1} the sum of the diagrams
Fig.\,\ref{fig:fig1}a and Fig.\,\ref{fig:fig1}b is gauge
invariant. This implies that the diagram Fig.\,\ref{fig:fig1}c should
be gauge invariant by itself. In turn, in Fig.\,\ref{fig:fig2} the
diagrams with photons coupled to the proton (Fig.\,\ref{fig:fig2}a,
Fig.\,\ref{fig:fig2}b and Fig.\,\ref{fig:fig2}c) and electron
(Fig.\,\ref{fig:fig2}d, Fig.\,\ref{fig:fig2}e and
Fig.\,\ref{fig:fig2}f) are invariant under a gauge transformation
$\varepsilon^*_{\lambda'}(k) \to \varepsilon^*_{\lambda'}(k) + c\,k$
separately.
We show that invariance of the diagrams in Fig.\,\ref{fig:fig2} with
respect to a gauge transformation $\varepsilon^*_{\lambda'}(k) \to
\varepsilon^*_{\lambda'}(k) + c\,k$ leads to Ward identities, which
impose well--known constraints on the renormalization parameters
\cite{Itzykson1980} and certain constraints on the structure functions
(see Appendix B of the Supplemental material). 
It is important to emphasize that to leading order in the large proton
mass expansion the contribution of the diagram in
Fig.\,\ref{fig:fig1}a is proportional to the time--component of the
photon polarization vector $\varepsilon^{0*}_{\lambda'}(k)$, which
vanishes in the physical gauge $\varepsilon^*_{\lambda'}(k) = (0,
\vec{\varepsilon}^{\,*}_{\lambda'}(\vec{k}\,))$, where the
polarization vector $\vec{\varepsilon}^{\,*}_{\lambda'}(\vec{k}\,)$
obeys the constraint $\vec{k}\cdot
\vec{\varepsilon}^{\,*}_{\lambda'}(\vec{k}\,) = 0$ \cite{BD1967} (see
also \cite{Ivanov2013,Ivanov2017,Ivanov2013b}). As has been shown in
\cite{Gaponov1996,Bernard2004,Ivanov2013,Ivanov2017} the contribution
of the diagrams in Fig.\,\ref{fig:fig1}, taken to leading order in the
large proton mass expansion with a real photon in the physical gauge
$\varepsilon^*_{\lambda'}(k) = (0,
\vec{\varepsilon}^{\,*}_{\lambda'}(\vec{k}\,))$, describes well the
main part of the branching ratio of the neutron radiative
$\beta^-$--decay (see Table I).

As regards the diagrams in Fig.\,\ref{fig:fig2}, to leading order in
the large proton mass expansion the contribution of the diagrams
Fig.\,\ref{fig:fig2}a, Fig.\,\ref{fig:fig2}b and Fig.\,\ref{fig:fig2}c
becomes proportional to $\varepsilon^{0*}_{\lambda'}(k)$ and vanishes
in the physical gauge $\varepsilon^*_{\lambda'}(k) = (0,
\vec{\varepsilon}^{\,*}_{\lambda'}(\vec{k}\,))$. As a result, only the
diagrams Fig.\,\ref{fig:fig2}d, Fig.\,\ref{fig:fig2}e and
Fig.\,\ref{fig:fig2}f give a contribution to the amplitude of the
neutron radiative $\beta^-$--decay, calculated to leading order in the
large proton mass expansion with a real photon in the physical gauge
$\varepsilon^*_{\lambda'}(k) = (0,
\vec{\varepsilon}^{\,*}_{\lambda'}(\vec{k}\,))$ (see Appendix B of the
Supplemental Material).

According to Sirlin \cite{Sirlin1967}, the contributions of the
Feynman diagrams with one--loop corrections, which are shown in
Fig.\,\ref{fig:fig2}, Fig.\,\ref{fig:fig3} and Fig.\,\ref{fig:fig4},
should be also invariant under a gauge transformation of a virtual
photon, which reduces to a redefinition of a longitudinal part of a
photon Green function $D_{\alpha\beta}(q) \to D_{\alpha\beta}(q) +
c(q^2)\,q_{\alpha} q_{\beta}$, where $c(q^2)$ is an arbitrary function
of $q^2$ \cite{Sirlin1967}. In Appendix B of the Supplemental Material
we show that the contributions of the diagrams in Fig.\,\ref{fig:fig2}
are invariant also under a gauge transformation $D_{\alpha\beta}(q)
\to D_{\alpha\beta}(q) + c(q^2)\,q_{\alpha} q_{\beta}$.

Unlike the Feynman diagrams in Fig.\,\ref{fig:fig2} the properties and
calculation of the set of Feynman diagrams in Fig.\,\ref{fig:fig3} and
Fig.\,\ref{fig:fig4} are not so simple and transparent. In Appendix C
of the Supplemental Material we show that the contributions of the
diagrams Fig.\,\ref{fig:fig3}a and Fig.\,\ref{fig:fig3}b, where strong
low--energy and electromagnetic interactions are factorized, vanish
after renormalization of masses and wave functions of the proton and
electron. In turn, the diagrams Fig.\,\ref{fig:fig3}c and
Fig.\,\ref{fig:fig3}d cannot be treated separately from the diagrams
in Fig.\,\ref{fig:fig4}, since by themselves they are not invariant
under gauge transformations $\varepsilon^*_{\lambda'}(k) \to
\varepsilon^*_{\lambda'}(k) + c\,k$ and $D_{\alpha\beta}(q) \to
D_{\alpha\beta}(q) + c(q^2)\,q_{\alpha} q_{\beta}$. Following Sirlin
\cite{Sirlin1967} we assume that required gauge invariance can be
fulfilled only for a sum of the Feynman diagrams
Fig.\,\ref{fig:fig3}c, Fig.\,\ref{fig:fig3}d and Fig.\,\ref{fig:fig4},
where strong low--energy hadronic and electromagnetic interactions are
overlapped and photons (real and virtual) are emitted or absorbed by a
hadronic block.

Such an assertion is not proved but based on the following
observation. After a removal of the lines of a real photon emission
the diagrams Fig.\,\ref{fig:fig3}c and Fig.\,\ref{fig:fig3}d reduce
themselves to the diagram Fig.\,\ref{fig:fig6}a, which, as has been
shown by Sirlin \cite{Sirlin1967}, gives the main contribution of the
radiative corrections of order $O(\alpha/\pi)$ to the rate of the
neutron $\beta^-$--decay. However, the diagram Fig.\,\ref{fig:fig6}a
by itself is not invariant under a gauge transformation
$D_{\alpha\beta}(q) \to D_{\alpha\beta}(q) + c(q^2)\,q_{\alpha}
q_{\beta}$. As has been pointed out by Sirlin \cite{Sirlin1967}, only
a sum of the diagrams in Fig.\,\ref{fig:fig6} should be gauge
invariant.
However, an exact calculation of the diagrams Fig.\,\ref{fig:fig6}b
and Fig.\,\ref{fig:fig6}c demands a certain model of strong
low--energy interactions of hadrons coupled to photons at low
energies.  Nevertheless, Sirlin, using the current algebra approach
\cite{Sirlin1967,Sirlin1978}, has succeeded in showing that the
contributions of the diagrams Fig.\,\ref{fig:fig6}b and
Fig.\,\ref{fig:fig6}c do not depend on the electron energy $E_e$. Such
a remarkable property of these diagrams has allowed Sirlin to
decompose the contribution of the diagram Fig.\,\ref{fig:fig6}a into
invariant and non--invariant parts with respect to a gauge
transformation $D_{\alpha\beta}(q) \to D_{\alpha\beta}(q) +
c(q^2)\,q_{\alpha} q_{\beta}$ in such a way that a gauge--non--ivariant
part does not depend on the electron energy. Then, a constant
gauge--non--invariant part has been merely absorbed by formal
renormalization of the Fermi weak coupling constant $G_F$ and the
axial coupling constant $\lambda$. We would like to emphasize that,
unfortunately, the diagrams Fig.\,\ref{fig:fig3}c and
Fig.\,\ref{fig:fig3}d do not possess such a remarkable property.
Nevertheless, it is obvious that different insertions of real photon
lines transform the diagrams in Fig.\,\ref{fig:fig6} into a set of
Feynman diagrams Fig.\,\ref{fig:fig3}c, Fig.\,\ref{fig:fig3}d and
Fig.\,\ref{fig:fig4} and should not destroy gauge properties of these
diagrams with respect to a gauge transformation $D_{\alpha\beta}(q)
\to D_{\alpha\beta}(q) + c(q^2)\,q_{\alpha}q_{\beta}$. As a result,
the analytical analysis of the diagrams Fig.\,\ref{fig:fig3}c,
Fig.\,\ref{fig:fig3}d and Fig.\,\ref{fig:fig4}, which is performed in
Appendices C and D of the Supplemental Material, runs as
follows. Firstly, we show that to leading order in the large proton
mass expansion the diagram Fig.\,\ref{fig:fig3}d, calculated with the
contribution of strong low--energy hadronic interactions given by the
axial coupling constant $\lambda$ only, vanishes in the physical gauge
of a real photon $\varepsilon^*_{\lambda'}(k) = (0,
\vec{\varepsilon}^{\,*}_{\lambda'}(\vec{k}\,))$. Secondly, we
calculate the diagram Fig.\,\ref{fig:fig3}c to leading order in the
large proton mass expansion and in the physical gauge of a real
photon.  After that we decompose the contribution of the diagram
Fig.\,\ref{fig:fig3}c into invariant and non--invariant part with
respect to a gauge transformation $D_{\alpha\beta}(q) \to
D_{\alpha\beta}(q) + c(q^2)\,q_{\alpha} q_{\beta}$. Keeping only the
part, that is invariant under a gauge transformation
$D_{\alpha\beta}(q) \to D_{\alpha\beta}(q) + c(q^2)\,q_{\alpha}
q_{\beta}$, and removing from it a part independent of the electron
$E_e$ and photon $\omega$ energy by renormalization of the Fermi weak
coupling and axial coupling constant, we obtain a contribution, which
can be accepted as a physical contribution of the diagram
Fig.\,\ref{fig:fig3}c to the amplitude and rate of the neutron
radiative $\beta^-$--decay to order $O(\alpha^2/\pi^2)$. What then is
the role of the Feynman diagrams in Fig.\,\ref{fig:fig4}\,?

As regards the diagrams in Fig.\,\ref{fig:fig4}, since the
contribution of them cannot be calculated in a model--independent way,
we follow Sirlin \cite{Sirlin1967} and assume that the diagrams in
Fig.\,\ref{fig:fig4} i) cancel a gauge--non--invariant part of the
diagram Fig.\,\ref{fig:fig3}c, determined relative to a gauge
transformation $D_{\alpha\beta}(q) \to D_{\alpha\beta}(q) +
c(q^2)\,q_{\alpha} q_{\beta}$, and the rest ii) either vanishes to
leading order in the proton mass expansion in the physical gauge of a
real photon $\varepsilon^*_{\lambda'}(k) = (0,
\vec{\varepsilon}^{\,*}_{\lambda'}(\vec{k}\,))$ (see Appendix D of the
Supplemental Material) or iii) is a constant, which can be absorbed by
renormalization of the Fermi coupling constant $G_F$ and the axial
coupling constant $\lambda$. This agrees also well with an assumption
that different insertions of real photons' lines into the diagrams in
Fig.\,\ref{fig:fig6} do not corrupt the properties of the Feynman
diagrams in Fig.\,\ref{fig:fig3}c, Fig.\,\ref{fig:fig3}d and
Fig.\,\ref{fig:fig4} under a gauge transformation of a photon Green
function $D_{\alpha\beta}(q) \to D_{\alpha\beta}(q) +
c(q^2)\,q_{\alpha} q_{\beta}$ even if to leading order in the large
proton mass expansion. In Appendix D of the Supplemental Material we
analyse the contributions of the diagrams Fig.\,\ref{fig:fig4}f and
Fig.\,\ref{fig:fig4}g, where strong low--energy interactions are given
by the axial coupling constant $\lambda$ only. We show that to leading
order in the large proton mass expansion the contributions of these
diagrams vanish.
Hence, an important contribution, which may cancel a
gauge--non--invariant part of the diagram Fig.\,\ref{fig:fig3}c, is
able to come only from the diagrams, where a real or virtual photon
couple to a hadronic block.

The diagram in Fig.\,\ref{fig:fig5} defines one of a set of Feynman
diagrams of the neutron radiative $\beta^-$--decay with emission of
two real photons.  Such a process with one undetected photon can
imitate a contribution of order $O(\alpha^2/\pi^2)$ to the rate of the
neutron radiative $\beta^-$--decay. All diagrams of the neutron
radiative $\beta^-$--decay with emission of one or two photons by the
proton, calculated to leading order in the large proton mass
expansion, do not contribute to the rate of the neutron radiative
$\beta^-$--decay in the physical gauge of real photons. Then, the
contributions of the diagrams with emission of photons from the
hadronic blocks are neglected (see a discussion in section
\ref{sec:conclusion}). Thus, in the accepted approximation the main
contribution to the rate of the neutron radiative $\beta^-$--decay is
defined by the Feynman diagrams in Fig.\,\ref{fig:fig5} with the
account for the contributions, caused by symmetry of the final state
with respect to symmetry properties of the two photons in the final
state of the decay. For the analytical calculation of the diagram in
Fig.\,\ref{fig:fig5} the contribution of strong low--energy
interactions is defined by the axial coupling only. The analytical
calculation of the diagrams in Fig.\,\ref{fig:fig5} is given in
Appendix E of the Supplemental Material.

The paper is organized as follows. In section \ref{sec:first} we give
a short description of the renormalization procedure of effective
low--energy electroweak interactions for the neutron radiative
$\beta^-$--decay. In section \ref{sec:rate} we adduce the
contributions of the Feynman diagrams in Fig.\,\ref{fig:fig1},
Fig.\,\ref{fig:fig2}, Fig.\,\ref{fig:fig3}, Fig.\,\ref{fig:fig4} and
Fig.\,\ref{fig:fig5} to the rate of the neutron radiative
$\beta^-$--decay . The numerical values of the branching ratio of the
neutron radiative $\beta^-$--decay for the three regions of photon
energies i)$ 15\,{\rm keV} \le \omega \le 350\,{\rm keV}$, ii)
$14\,{\rm keV} \le \omega \le 782\,{\rm keV}$ and iii) $0.4\,{\rm keV}
\le \omega \le 14\,{\rm keV}$, are given in Table I. In section
\ref{sec:conclusion} we discuss the obtained results. In the
Supplemental Material we give i) detailed analytical calculations and
analysis of the contributions of Feynman diagrams in
Fig.\,\ref{fig:fig1}, Fig.\,\ref{fig:fig2}, Fig.\,\ref{fig:fig3},
Fig.\,\ref{fig:fig4} and Fig.\,\ref{fig:fig5} to the amplitude and
rate of the neutron radiative $\beta^-$--decay.

Of course, we have to confess that the main problem of our analysis of
the radiative corrections to order $O(\alpha/\pi)$, defining
corrections to order $O(\alpha^2/\pi^2)$ to the rate of the neutron
radiative $\beta^-$--decay, concerns the contributions of diagrams
with real or virtual photons coupled to a hadronic block. A
justification of our assumption concerning the properties of these
diagrams within a certain model of strong low--energy interactions of
hadrons coupled to photons should be important for a confirmation of
the approximation accepted in this paper and the results obtained
therein. We would like to accentuate that unlike a passive role of
strong low--energy hadronic interactions in the radiative corrections
of order $O(\alpha/\pi)$ to the rate of the neutron $\beta^-$--decay,
strong low--energy interactions of hadrons coupled to real and virtual
photons in the diagrams in Fig.\,\ref{fig:fig4}, should play a more
important role, going beyond a formal renormalization of the Fermi
weak coupling and axial coupling constant, but give some
contributions, which depend on the electron and photon energies and
momenta, and should cancel a gauge--non--invariant part of the diagram
Fig.\,\ref{fig:fig3}c. The observed peculiarities of the Feynman
diagrams Fig.\,\ref{fig:fig3} and Fig.\,\ref{fig:fig4} agree well with
an important role of strong low--energy hadronic interactions in decay
processes that have been already pointed out by Weinberg
\cite{Weinberg1956}. Thus, the problem of strong low--energy hadronic
interactions in the neutron radiative $\beta^-$--decay to order
$O(\alpha^2/\pi^2)$ demands a special analysis and we are planning to
perform such a model--dependent analysis of the neutron radiative
$\beta^-$--decay to order $O(\alpha^2/\pi^2)$ in our forthcoming
publication.

\section{Renormalization procedure of effective low--energy electroweak
 interactions for the neutron radiative $\beta^-$--decay}
\label{sec:first}

In the Standard Model of electroweak interactions the neutron
radiative $\beta^-$--decay, defined in the one--loop approximation
with one--virtual--photon exchanges, is described by the following
interactions
\begin{eqnarray}\label{eq:1}
\hspace{-0.3in}{\cal L}_{\rm int}(x) = {\cal L}_{\rm W}(x) + {\cal
  L}_{\rm em}(x),
\end{eqnarray}
where ${\cal L}_{\rm W}(x)$ is the effective Lagrangian of low--energy
$V-A$ interactions with a real axial coupling constant $\lambda = -
1.2750(9)$ \cite{Abele2008} (see also \cite{Ivanov2013,Ivanov2017})
\begin{eqnarray}\label{eq:2}
\hspace{-0.3in}{\cal L}_{\rm W}(x) = -
\frac{G_F}{\sqrt{2}}\,V_{ud}\,[\bar{\psi}_p(x)\gamma_{\mu}(1 + \lambda
  \gamma^5)\psi_n(x)]\, [\bar{\psi}_e(x)\gamma^{\mu}(1 -
  \gamma^5)\psi_{\nu}(x)],
\end{eqnarray}
where $G_F = 1.1664 \times 10^{-11}\,{\rm MeV^{-2}}$ is the Fermi
coupling constant, and $|V_{ud}| = 0.97417(21)$ is the
Cabibbo--Kobayashi--Maskawa matrix element \cite{PDG2016}. Then,
$\psi_p(x)$, $\psi_n(x)$, $\psi_e(x)$ and $\psi_{\nu}(x)$ are the
field operators of the proton, neutron, electron and antineutrino,
respectively, and $\gamma^{\mu}$ and $\gamma^5$ are the Dirac matrices
\cite{Itzykson1980}. Since we calculate the radiative corrections of
order $O(\alpha/\pi)$ to the neutron radiative $\beta^-$--decay to
leading order in the large proton mass expansion, in the effective
Lagrangian ${\cal L}_{\rm W}(x)$ we do not take into account the
contribution of the weak magnetism proportional to $1/M$, where $2M =
m_n + m_p$ is an averaged nucleon mass \cite{Ivanov2017}.

For the calculation of the radiative corrections to order
$O(\alpha/\pi)$ the Lagrangian of the electromagnetic interaction
${\cal L}_{\rm em}(x)$ we take in the following form
\begin{eqnarray}\label{eq:3}
{\cal L}_{\rm em}(x) &=& - \frac{1}{4}\,F^{(0)}_{\mu\nu}(x)F^{(0)\mu\nu}(x) -
\frac{1}{2\xi_0}\,\Big(\partial_{\mu}A^{(0)\mu}(x)\Big)^2 \nonumber\\ && +
\bar{\psi}_{0e}(x)(i\gamma^{\mu}\partial_{\mu} - m_{0e})\psi_{0e}(x) - (-
e_0)\, \bar{\psi}_{0e}(x)\gamma^{\mu}\psi_{0e}(x)A^{(0)}_{\mu}(x)\nonumber\\ &&
+ \bar{\psi}_{0p}(x)(i\gamma^{\mu}\partial_{\mu} - m_{0p})\psi_{0p}(x) - (+ e_0)
\bar{\psi}_{0p}(x)\gamma^{\mu}\psi_{0p}(x) A^{(0)}_{\mu}(x),
\end{eqnarray}
where $F^{(0)}_{\mu\nu}(x) = \partial_{\mu}A^{(0)}_{\nu}(x) -
\partial_{\nu}A^{(0)}_{\mu}(x)$ is the electromagnetic field strength
tensor of the {\it bare} (unrenormalized) electromagnetic field
operator $A^{(0)}_{\mu}(x)$; $\psi_{0e}(x)$ and $\psi_{0p}(x)$ are
{\it bare} operators of the electron and proton fields with {\it bare}
masses $m_{0e}$ and $m_{0p}$, respectively; $- e_0$ and $+ e_0$ are
{\it bare} electric charges of the electron and proton, respectively.
Then, $\xi_0$ is a {\it bare} gauge parameter. After the calculation
of the one--loop corrections of order $O(\alpha/\pi)$ a transition to
the renormalized field operators, masses and electric charges is
defined by the Lagrangian
\begin{eqnarray}\label{eq:4}
{\cal L}_{\rm em}(x) &=& - \frac{1}{4}\,F_{\mu\nu}(x)F^{\mu\nu}(x) -
\frac{1}{2\xi}\,\Big(\partial_{\mu}A^{\mu}(x)\Big)^2\nonumber\\ && +
\bar{\psi}_e(x)(i\gamma^{\mu}\partial_{\mu} - m_{e})\psi_e(x) - (- e)\,
\bar{\psi}_e(x) \gamma^{\mu} \psi_e(x) A_{\mu}(x)\nonumber\\ && +
\bar{\psi}_p(x)(i\gamma^{\mu}\partial_{\mu} - m_p )\psi_p(x) - (+ e)\,
\bar{\psi}_p(x) \gamma^{\mu}\psi_p(x) A_{\mu}(x) + \delta {\cal L}_{\rm em}(x),
\end{eqnarray}
where $A_{\mu}(x)$, $\psi_e(x)$ and $\psi_p(x)$ are the renormalized
operators of the electromagnetic, electron and proton fields,
respectively; $m_e$ and $m_p$ are the renormalized masses of the
electron and proton; $e$ is the renormalized electric charge; and
$\xi$ is the renormalized gauge parameter. The Lagrangian $\delta
{\cal L}_{\rm em}(x)$ contains a complete set of the counterterms
\cite{Weinberg1995},
\begin{eqnarray}\label{eq:5}
\hspace{-0.3in}\delta {\cal L}_{\rm em}(x) &=& - \frac{1}{4}\,(Z_3 -
1)\,F_{\mu\nu}(x)F^{\mu\nu}(x) - \frac{Z_3 -
  1}{Z_{\xi}}\,\frac{1}{2\xi}\,\Big(\partial_{\mu}A^{\mu}(x)\Big)^2\nonumber\\\hspace{-0.3in}
&& + (Z^{(e)}_2 - 1)\,\bar{\psi}_e(x)(i\gamma^{\mu}\partial_{\mu} -
m_{e})\psi_e(x) - (Z^{(e)}_1 - 1)\,(- e)\,\bar{\psi}_e (x)\gamma^{\mu}
\psi_e(x) A_{\mu}(x) - Z^{(e)}_2 \delta m_e \bar{\psi}_e(x)\psi_e(x)
\nonumber\\ \hspace{-0.3in}&& + (Z^{(p)}_2 -
1)\,\bar{\psi}_p(x)(i\gamma^{\mu}\partial_{\mu} - m_p )\psi_p(x) -
(Z^{(p)}_1 - 1)\,( + e) \,\bar{\psi}_p(x) \gamma^{\mu}\psi_p(x)
A_{\mu}(x) - Z^{(p)}_2 \delta m_p \bar{\psi}_p(x) \psi_p(x),
\end{eqnarray}
where $Z_3$, $Z^{(e)}_2$, $Z^{(e)}_1$, $Z^{(p)}_2$, $Z^{(p)}_1$,
$\delta m_e$ and $\delta m_p$ are the counterterms. Here $Z_3$ is the
renormalization constant of the electromagnetic field operator
$A_{\mu}$, $Z^{(e)}_2$ and $Z^{(e)}_1$ are the renormalization
constants of the electron field operator $\psi_e$ and the
electron--electron--photon ($e^-e^-\gamma$) vertex, respectively;
$Z^{(p)}_2$ and $Z^{(p)}_1$ are the renormalization constants of the
proton field operator $\psi_p$ and the proton--proton--photon ($p p
\gamma$) vertex, respectively. Then, $(- e)$ and $(+ e)$, $m_e$ and
$m_p$ and $\delta m_e$ and $\delta m_p$ are the renormalized electric
charges and masses and the mass--counterterms of the electron and
proton, respectively. Rescaling the field operators
\cite{Weinberg1995,Bogoliubov1959}
\begin{eqnarray}\label{eq:6}
\sqrt{Z_3}\, A_{\mu}(x) = A^{(0)}_{\mu}(x)\quad,\quad
\sqrt{Z^{(e)}_2}\,\psi_e(x) = \psi_{0e}(x)\quad,\quad
\sqrt{Z^{(p)}_2}\,\psi_p(x) = \psi_{0p}(x)
\end{eqnarray}
and denoting $m_e + \delta m_e = m_{0e}$, $m_p + \delta m_p = m_{0p}$
and $Z_{\xi} \xi = \xi_0$ we arrive at the Lagrangian
\begin{eqnarray}\label{eq:7}
\hspace{-0.3in}{\cal L}_{\rm em}(x) &=& -
\frac{1}{4}\,F^{(0)}_{\mu\nu}(x)F^{(0)\mu\nu}(x) -
\frac{1}{2\xi_0}\,\Big(\partial_{\mu}A^{(0)\mu}(x)\Big)^2\nonumber\\ 
\hspace{-0.3in}&&+
  \bar{\psi}_{0e}(x)(i\gamma^{\mu}\partial_{\mu} - m_{0e})\psi_{0e}(x)
  - ( - e)\,Z^{(e)}_1 (Z^{(e)}_2)^{-1} Z^{-1/2}_3
  \bar{\psi}_{0e}(x)\gamma^{\mu}\psi_{0e}(x)A^{(0)}_{\mu}(x)\nonumber\\ 
\hspace{-0.3in}&& + \bar{\psi}_{0p}(x)(i\gamma^{\mu}\partial_{\mu} -
m_{0p})\psi_{0p}(x) - (+ e)\, Z^{(p)}_1(Z^{(p)}_2)^{-1} Z^{-1/2}_3
\bar{\psi}_{0p}(x)\gamma^{\mu}\psi_{0p}(x)A^{(0)}_{\mu}(x).
\end{eqnarray}
Because of the Ward identities $Z^{(e)}_1 = Z^{(e)}_2$ and $Z^{(p)}_1
= Z^{(p)}_2$ \cite{Itzykson1980,Weinberg1995,Bogoliubov1959}, we may
replace $(-e)\,Z^{-1/2}_3 = - e_0$ and $(+ e)\,Z^{-1/2}_3 = +
e_0$. This brings Eq.(\ref{eq:7}) to the form of Eq.(\ref{eq:3}). We
would like to emphasize that to order $O(\alpha/\pi)$ the
renormalization constant $Z_3$ is equal to unity, i.e., $Z_3 =
1$. This is because of the absence of closed fermion loops, giving
contributions of order $O(\alpha^2/\pi^2)$ to the amplitude of the
neutron radiative $\beta^-$--decay that goes beyond the accepted
approximation $O(\alpha/\pi)$ for the amplitude and
$O(\alpha^2/\pi^2)$ for the rate of the neutron radiative
$\beta^-$--decay. Hence, to order $O(\alpha/\pi)$ the {\it bare} $e_0$
and renormalized $e$ electric charges are equal, i.e. $e_0 = e$. Now
we may proceed to the discussion of the contributions of the radiative
corrections of order $O(\alpha/\pi)$, where $\alpha = e^2/4\pi =
1/137.036$ is the fine--structure constant \cite{PDG2016}, to the
amplitude and rate of the neutron radiative $\beta^-$--decay. The
detailed calculations and analysis of the Feynman diagrams in
Fig.\,\ref{fig:fig2}, Fig.\,\ref{fig:fig3}, Fig.\,\ref{fig:fig4} and
Fig.\,\ref{fig:fig5}, defining a complete set of radiative corrections
of order $O(\alpha/\pi)$, we give in the Supplemental Material. In
section \ref{sec:rate} we adduce the analytical expressions for the
contributions of the diagrams in Fig.\,\ref{fig:fig2},
Fig.\,\ref{fig:fig3}, Fig.\,\ref{fig:fig4} and Fig.\,\ref{fig:fig5} to
the rate of the neutron radiative $\beta^-$--decay. The numerical
values are collected in Table I. For completeness we take into account
the tree--level contribution, given by the Feynman diagrams in
Fig.\,\ref{fig:fig1} and calculated in \cite{Ivanov2017} to order
$1/M$, including corrections of the weak magnetism and proton recoil.

\section{Rate of neutron radiative $\beta^-$--decay with one detected 
photon}
\label{sec:rate}

The rate of the neutron radiative $\beta^-$--decay with a photon,
detected in the photon energy region $\omega_{\rm min} \le \omega \le
\omega_{\rm max}$, is given by
\begin{eqnarray}\label{eq:8}
\hspace{-0.3in}\lambda_{\beta\gamma}(\omega_{\rm max}, \omega_{\rm
  min}) = \sum^5_{j = 1}\lambda^{(\rm Fig\,j)}_{\beta\gamma}(\omega_{\rm
  max}, \omega_{\rm min}),
\end{eqnarray}
where $\lambda^{(\rm Fig\,j)}_{\beta\gamma}(\omega_{\rm max},
\omega_{\rm min})$ are the rates, caused by the contributions of the
diagrams in Fig.\,j for $j = 1,2,\ldots,5$.  They are calculated in
the Supplemental Material. To leading order in the large proton mass
expansion the contribution of the diagrams in Fig.\,\ref{fig:fig1} is
equal to \cite{Ivanov2013}
\begin{eqnarray}\label{eq:9}
\lambda^{(\rm Fig.\,\ref{fig:fig1})}_{\beta\gamma}(\omega_{\rm
  max},\omega_{\rm min}) &=& (1 + 3
\lambda^2)\,\frac{\alpha}{\pi}\,\frac{G^2_F|V_{ud}|^2}{2\pi^3}\int^{\omega_{\rm
    max}}_{\omega_{\rm min}}\frac{d\omega}{\omega}\int^{E_0 -
  \omega}_{m_e}dE_e \,\sqrt{E^2_e - m^2_e}\,E_e\,F(E_e, Z = 1)\,(E_0 -
E_e - \omega)^2\nonumber\\ &&\times\Big\{\Big(1 + \frac{\omega}{E_e} +
\frac{1}{2}\frac{\omega^2}{E^2_e}\Big)\,\Big[\frac{1}{\beta}\,{\ell
    n}\Big(\frac{1 + \beta}{1 - \beta}\Big) - 2\Big] +
\frac{\omega^2}{E^2_e}\Big\},
\end{eqnarray}
where $E_0 = (m^2_n - m^2_p + m^2_e)/2 m_n$ is the end--point energy
of the electron--energy spectrum of the neutron $\beta^-$--decay
\cite{Ivanov2013}; $\omega$ is a photon energy; $\beta = k_e/E_e =
\sqrt{E^2_e - m^2_e}/E_e$ is a velocity of the electron with a
momentum $k_e$; and $F(E_e, Z = 1)$ is the relativistic Fermi function,
describing the Coulomb proton--electron interaction in the final state
of the decay. It is equal to
\begin{eqnarray}\label{eq:10}
\hspace{-0.3in}F(E_e, Z = 1 ) =  \Big(1 +
\frac{1}{2}\gamma\Big)\,\frac{4(2 r_pm_e\beta)^{2\gamma}}{\Gamma^2(3 +
  2\gamma)}\,\frac{\displaystyle e^{\,\pi
 \alpha/\beta}}{(1 - \beta^2)^{\gamma}}\,\Big|\Gamma\Big(1 + \gamma +
 i\,\frac{\alpha }{\beta}\Big)\Big|^2,
\end{eqnarray}
where $\gamma = \sqrt{1 - \alpha^2} - 1$, $r_p$ is the electric radius
of the proton and $\alpha = 1/137.036$ is the fine--structure
constant.  In numerical calculations we shall use $r_p = 0.841\,{\rm
  fm}$ \cite{Pohl2010}. The rate of the neutron radiative
$\beta^-$--decay, calculated to next--to--leading order in the large
proton mass expansion, taking into account the contributions of the
weak magnetism and proton recoil to order $1/M$, where $2M = m_n +
m_p$ is the averaged nucleon mass, has been calculated in
\cite{Ivanov2017}. The result is
\begin{eqnarray}\label{eq:11}
\hspace{-0.3in}\lambda^{(\rm Fig.\,\ref{fig:fig1})}_{\beta
  \gamma}(\omega_{\rm max},\omega_{\rm min}) &=& (1 + 3
\lambda^2)\frac{\alpha}{\pi} \frac{G^2_F |V_{ud}|^2}{2\pi^3}
\int^{\omega_{\rm max}}_{\omega_{\rm
    min}}\frac{d\omega}{\omega}\int^{E_0 - \omega}_{m_e}
dE_e\,E_e\sqrt{E^2_e - m^2_e}\, (E_0 - E_e - \omega)^2\nonumber\\
\hspace{-0.3in}&&\times \,F(E_e, Z = 1)\,\rho^{(\rm
  Fig.\,\ref{fig:fig1})}_{\beta\gamma}(E_e,\omega).
\end{eqnarray}
The function $\rho^{(\rm Fig.\,\ref{fig:fig1})}_{\beta
  \gamma}(E_e,\omega)$ is given by the integral \cite{Ivanov2017}
\begin{eqnarray}\label{eq:12}
\hspace{-0.3in}&&\rho^{(\rm
  Fig.\,\ref{fig:fig1})}_{\beta\gamma}(E_e,\omega) =
\int\frac{d\Omega_{e\gamma}}{4\pi}\,\Big\{\Big[1 +
  2\,\frac{\omega}{M}\,\frac{E_e - \vec{k}_e\cdot
    \vec{n}_{\vec{k}}}{E_0 - E_e - \omega} + \frac{3}{M}\,\Big(E_e +
  \omega - \frac{1}{3}\,E_0\Big) + \frac{\lambda^2 - 2(\kappa +
    1)\lambda + 1}{1 + 3\lambda^2}\,\frac{E_0 - E_e -
    \omega}{M}\Big]\nonumber\\
\hspace{-0.3in}&&\times\,\Big[\Big(1 +
  \frac{\omega}{E_e}\Big)\,\frac{k^2_e - (\vec{k}_e\cdot
    \vec{n}_{\vec{k}})^2}{(E_e - \vec{k}_e\cdot \vec{n}_{\vec{k}})^2}
  + \frac{\omega^2}{E_e}\,\frac{1}{E_e - \vec{k}_e\cdot
    \vec{n}_{\vec{k}}}\Big] + \frac{3\lambda^2 - 1}{1 + 3
  \lambda^2}\,\frac{1}{M}\,\Big(\frac{k^2_e + \omega \vec{k}_e\cdot
  \vec{n}_{\vec{k}}}{E_e}\,\Big[\frac{k^2_e - (\vec{k}_e\cdot
  \vec{n}_{\vec{k}})^2}{(E_e - \vec{k}_e\cdot \vec{n}_{\vec{k}})^2} +
\frac{\omega}{E_e - \vec{k}_e\cdot \vec{n}_{\vec{k}}}\Big]\nonumber\\
\hspace{-0.3in}&& + (\omega + \vec{k}_e\cdot
\vec{n}_{\vec{k}})\Big[\Big(1 +
  \frac{\omega}{E_e}\Big)\frac{\omega}{E_e - \vec{k}_e\cdot
    \vec{n}_{\vec{k}}} - \frac{m^2_e}{E_e}\,\frac{\omega}{(E_e -
    \vec{k}_e\cdot \vec{n}_{\vec{k}})^2}\Big]\Big) - \frac{\lambda^2 + 2
  (\kappa + 1)\lambda - 1}{1 +
  3\lambda^2}\,\frac{1}{M}\,\Big[\frac{k^2_e + \omega^2 + 2\omega
    \vec{k}_e\cdot \vec{n}_{\vec{k}}}{E_e}\nonumber\\
\hspace{-0.3in}&&\times\,\frac{k^2_e - (\vec{k}_e\cdot
  \vec{n}_{\vec{k}})^2}{(E_e - \vec{k}_e\cdot \vec{n}_{\vec{k}})^2} +
\frac{\omega}{E_e}\,\frac{k^2_e - (\vec{k}_e\cdot
  \vec{n}_{\vec{k}})^2}{E_e - \vec{k}_e\cdot \vec{n}_{\vec{k}}} +
\frac{\omega^2}{E_e}\,\frac{\omega + \vec{k}_e\cdot
  \vec{n}_{\vec{k}}}{E_e - \vec{k}_e\cdot \vec{n}_{\vec{k}}}\Big] -
\frac{\lambda(\lambda - 1)}{1 +
  3\lambda^2}\,\frac{1}{M}\,\Big[\frac{\omega}{E_e}\, \frac{k^2_e -
    (\vec{k}_e\cdot \vec{n}_{\vec{k}})^2}{E_e - \vec{k}_e\cdot
    \vec{n}_{\vec{k}}} + 3\,\frac{\omega^2}{E_e}\Big]\Big\},
\end{eqnarray}
where $\kappa = \kappa_p - \kappa_n = 3.70589$ is the isovector
anomalous magnetic moment of the nucleon \cite{Ivanov2013,Ivanov2017},
$d\Omega_{e\gamma}$ is an infinitesimal solid angle of the
electron--photon momentum correlations $\vec{k}_e\cdot
\vec{n}_{\vec{k}} = k_e \cos\theta_{e\gamma}$ and $\vec{n}_{\vec{k}} =
\vec{k}/\omega$ is a unit vector along the photon 3--momentum
\cite{Ivanov2013,Ivanov2013a,Ivanov2017}. The contribution of the
diagrams in Fig.\,\ref{fig:fig2} is equal to (see Appendix B of
the Supplemental Material)
\begin{eqnarray}\label{eq:13}
\hspace{-0.3in}&&\lambda^{(\rm
  Fig.\,\ref{fig:fig2})}_{\beta\gamma}(\omega_{\rm max},\omega_{\rm
  min}) = (1 +
3\lambda^2)\,\frac{\alpha^2}{\pi^2}\,\frac{G^2_FV_{ud}|^2}{4\pi^3}
\int^{\omega_{\rm max}}_{\omega_{\rm min}} d\omega\int^{E_0 -
  \omega}_{m_e} dE_e\,(E_0 - E_e - \omega)^2\,\sqrt{E^2_e -
  m^2_e}\nonumber\\
\hspace{-0.3in}&&\times \,F(E_e, Z = 1) \,
\int\frac{d\Omega_{e\gamma}}{4\pi}\,\Big\{\frac{k^2_e - (\vec{k}_e
  \cdot \vec{n}_{\vec{k}})^2}{(E_e - \vec{k}_e\cdot
  \vec{n}_{\vec{k}})^2}\,{\rm Re}\,F_4 + \frac{\omega}{E_e -
  \vec{k}_e\cdot \vec{n}_{\vec{k}}}\, {\rm Re}\,(2 F_2 - F_3 -
2F_4)\Big\},
\end{eqnarray}
where $F_2$, $F_3$ and $F_4$ are given in Eq.(B-71) of the
Supplemental Material as functions of $k_e\cdot k = \omega\,(E_e -
\vec{k}_e\cdot \vec{n}_{\vec{k}})$. The contribution of the diagrams
in Fig.\,\ref{fig:fig3} and Fig.\,\ref{fig:fig4} we define as (see
Appendix C of the Supplemental Material)
\begin{eqnarray}\label{eq:14}
\hspace{-0.3in}&&\lambda^{(\rm
  Fig.\,\ref{fig:fig3})}_{\beta\gamma}(\omega_{\rm max},\omega) = (1 +
3\lambda^2)\,\frac{\alpha^2}{\pi^2}\,\frac{G^2_F|V_{ud}|^2}{4\pi^3}
\int^{\omega_{\rm max}}_{\omega_{\rm min}}
\frac{d\omega}{\omega}\int^{E_0 - \omega}_{m_e} dE_e\,F(E_e, Z =
1)\,(E_0 - E_e - \omega)^2\,\sqrt{E^2_e - m^2_e}\nonumber\\
\hspace{-0.3in}&&\times \,
\int\frac{d\Omega_{e\gamma}}{4\pi}\,\Big\{f_1(E_e, \vec{k}_e, \omega,
\vec{k}\,)\, \Big[(E_e + \omega)\, \frac{k^2_e - (\vec{k}_e \cdot
    \vec{n}_{\vec{k}})^2}{(E_e - \vec{k}_e \cdot \vec{n}_{\vec{k}})^2}
  + \frac{\omega^2}{E_e - \vec{k}_e \cdot \vec{n}_{\vec{k}}}\Big] +
f_2(E_e, \vec{k}_e, \omega,
\vec{k}\,)\, \Big[\Big(2\, (E_e + \omega)^2 - m^2_e\nonumber\\
\hspace{-0.3in}&& - \omega \,(E_e - \vec{k}_e \cdot
\vec{n}_{\vec{k}})\Big)\, \frac{k^2_e - (\vec{k}_e \cdot
  \vec{n}_{\vec{k}})^2}{(E_e - \vec{k}_e \cdot \vec{n}_{\vec{k}})^2} +
2\, (E_e + \omega)\, \frac{\omega^2}{E_e - \vec{k}_e \cdot
  \vec{n}_{\vec{k}}} - \omega^2\Big]\Big\},
\end{eqnarray}
where the functions $f_1(E_e, \vec{k}_e, \omega, \vec{k}\,)$ and
$f_2(E_e, \vec{k}_e, \omega, \vec{k}\,)$ are given in
Eq.(\ref{eq:C.33}) of the Supplemental material.  They are defined by
the contribution of the diagram Fig.\,\ref{fig:fig3}c, since to
leading order in the large proton mass expansion and in the physical
gauge of a real photon the contribution of the diagram in
Fig.\,\ref{fig:fig3}d vanishes. Then, the rate $\lambda^{(\rm
  Fig.\,\ref{fig:fig3})}_{\beta\gamma}(\omega_{\rm max},\omega_{\rm
  min})$ is defined by a part of the diagram Fig.\,\ref{fig:fig3}c,
which is invariant under a gauge transformation $D_{\alpha\beta}(q)
\to D_{\alpha\beta}(q) + c(q^2)\,q_{\alpha}q_{\beta}$. A
non--invariant part of the diagram Fig.\,\ref{fig:fig3}c is absorbed
by the diagrams in Fig.\,\ref{fig:fig4}. We assume that the
contribution of the diagrams in Fig.\,\ref{fig:fig4}, calculated to
leading order in the large proton mass expansion and in the physical
gauge of a real photon, contains only i) an electron--photon--energy
dependent part, cancelling a part of the diagram Fig.\,\ref{fig:fig3}c
that is non--invariant under the gauge transformation
$D_{\alpha\beta}(q) \to D_{\alpha\beta}(q) +
c(q^2)\,q_{\alpha}q_{\beta}$, and ii) a constant, which can be
absorbed by renormalization of the Fermi weak coupling constant $G_F$
and the axial coupling constant $\lambda$ similar to Sirlin's analysis
of the radiative corrections to the rate of the neutron
$\beta^-$--decay \cite{Sirlin1967}. Of course, our assumption is much
stronger than Sirlin's one. Nevertheless, we believe that it is
correct and it might be confirmed by a model--dependent way within a
model of strong interactions of hadrons coupled to photons at low
energies (see a discussion in section \ref{sec:conclusion}).

The contribution of the diagrams in Fig.\,\ref{fig:fig5} of the
neutron radiative $\beta^-$--decay with two real photons and only one
detected photon is equal to (see Appendix E of the Supplemental
Material)
\begin{eqnarray}\label{eq:15}
\hspace{-0.15in}\lambda^{(\rm Fig.\,\ref{fig:fig5})}_{\beta
  \gamma}(\omega_{\rm max}, \omega_{\rm min}) &=& (1 +
3\lambda^2)\,\frac{\alpha^2}{\pi^2}\,\frac{G^2_F|V_{ud}|^2}{16\pi^3}
\int^{\omega_{\rm max}}_{\omega_{\rm min}} \!\!\!d\omega\int^{E_0 -
  \omega}_{m_e} \!\!\!dE_e \,\sqrt{E^2_e - m^2_e}\int^{E_0 - E_e - \omega}_0\!\!\!
dq_0\,(E_0 - E_e - \omega - q_0)^2\,\nonumber\\
\hspace{-0.15in}&&\times F(E_e, Z = 1)
\int\frac{d\Omega_{e\gamma}}{4\pi}\int\frac{d\Omega_{e\gamma'}}{4\pi}\,\Big(\rho^{(1)}_{e\gamma\gamma'}(E_e,\vec{k}_e,
\omega, \vec{n}_{\vec{k}}, q_0, \vec{n}_{\vec{q}}) +
\rho^{(2)}_{e\gamma\gamma'}(E_e,\vec{k}_e, \omega, \vec{n}_{\vec{k}},
q_0, \vec{n}_{\vec{q}})\nonumber\\
\hspace{-0.15in}&&+ \rho^{(2)}_{e\gamma\gamma'}(E_e,\vec{k}_e, q_0,
\vec{n}_{\vec{q}}, \omega, \vec{n}_{\vec{k}})\Big),
\end{eqnarray}
where $q_0$ is the energy of an undetected photon and
$\vec{n}_{\vec{q}} = \vec{q}/q_0$ is a unit vector along its
3--momentum $\vec{q}$. The functions
$\rho^{(1)}_{e\gamma\gamma'}(E_e,\vec{k}_e, \omega, \vec{n}_{\vec{k}},
q_0, \vec{n}_{\vec{q}})$, $\rho^{(2)}_{e\gamma\gamma'}(E_e,\vec{k}_e,
\omega, \vec{n}_{\vec{k}}, q_0, \vec{n}_{\vec{q}})$ and
$\rho^{(2)}_{e\gamma\gamma'}(E_e,\vec{k}_e, q_0, \vec{n}_{\vec{q}},
\omega, \vec{n}_{\vec{k}})$ are given by Eq.(\ref{eq:E.14}),
Eq.(\ref{eq:E.15}) and Eq.(\ref{eq:E.16}) in the Supplemental
Material.

The numerical values of the branching ratios ${\rm BR}^{(\rm
  Fig.j)}_{\beta\gamma} = \tau_n\,\lambda_{\beta\gamma}(\omega_{\rm
  max}, \omega_{\rm min})_{\rm Fig\,j}$ for $j = 1,2,\ldots,5$ and
their total contribution are given in Table I for the three
photon--energy regions i) $15\,{\rm keV} \le \omega \le 340\,{\rm
  keV}$, ii) $14\,{\rm keV} \le \omega \le 782\,{\rm keV}$ and iii)
$0.4\,{\rm keV} \le \omega \le 14\,{\rm keV}$. The branching ratios
${\rm BR}^{(\rm Fig.j)}_{\beta\gamma}$ are obtained relative to the
neutron lifetime $\tau_n = 879.6(1.1)\,{\rm s}$, calculated in
\cite{Ivanov2013} and agreeing well with the world--averaged value
$\tau_n = 880.2(1.0)\,{\rm s}$ \cite{PDG2016}.

\begin{table}[h]
\begin{tabular}{|c|c|c|c|c|c|c|}
\hline $\omega\, [\rm keV]$ & ${\rm BR}_{\beta\gamma} (\rm
Experiment)$ & ${\rm BR}^{(\rm Fig.\ref{fig:fig1})}_{\beta\gamma} $ &
${\rm BR}^{(\rm Fig.\ref{fig:fig2})}_{\beta\gamma}$ & $ {\rm BR}^{(\rm
  Fig.\ref{fig:fig3})}_{\beta\gamma}$ & $ {\rm BR}^{(\rm
  Fig.\ref{fig:fig5})}_{\beta\gamma}$ & $ {\rm BR}_{\beta\gamma}(\rm
Theory)$\\ \hline $15 \le \omega \le 340$ & $~~~~~~~~~~~~~~(3.09 \pm
0.32)\times 10^{-3}$ \quad~~~~~~\cite{Cooper2010} & $2.89 \times
10^{-3}$ & $0.95\times 10^{-7}$ & $0.65\times 10^{-4}$ & $0.52\times
10^{-5}$ & $2.960\times 10^{-3}$\\\hline $14 \le \omega \le 782$ &
$(3.35 \pm 0.05\,[\rm stat] \pm 0.15\,[syst])\times
10^{-3}$\;\cite{Bales2016} & $ 3.04 \times 10^{-3}$ & $1.23 \times
10^{-7}$ & $0.68\times 10^{-4}$ & $0.55\times 10^{-5}$ & $3.114\times
10^{-3}$\\ \hline $0.4 \le \omega \le 14$ & $(5.82 \pm 0.23\,[\rm
  stat]\pm 0.62\,[syst])\times 10^{-3}$\;\cite{Bales2016} & $5.08
\times 10^{-3}$ & $0.03 \times 10^{-7}$ & $1.54\times 10^{-4}$ & $3.23
\times 10^{-5}$ & $5.266\times 10^{-3}$\\\hline
\end{tabular} 
\caption{Branching ratios of the neutron radiative $\beta^-$--decay
  for three photon--energy regions, calculated for the lifetime of the
  neutron $\tau_n = 879.6(1.1)\,{\rm s}$ \cite{Ivanov2013}. The
  branching ratio ${\rm BR}^{(\rm Fig.\ref{fig:fig1})}_{\beta\gamma} $
  takes into account the contributions of the weak magnetism and
  proton recoil, calculated in \cite{Ivanov2017} to next--to--leading
  order in the large proton mass expansion.}
\end{table}

\section{Conclusion}
\label{sec:conclusion}

We have proposed a precision analysis of the rate of the neutron
radiative $\beta^-$--decay $n \to p + e^- + \bar{\nu}_e + \gamma$ to
order $O(\alpha^2/\pi^2)$, defined by the $1/M$ corrections, caused by
the weak magnetism and proton recoil \cite{Ivanov2017}, and radiative
corrections of order $O(\alpha/\pi)$ in the one--virtual--photon
approximation, and the contribution of the neutron radiative
$\beta^-$--decay with two real photons $n \to p + e^- + \bar{\nu}_e +
\gamma + \gamma$. Integrating over degrees of freedom of one of two
photons one arrives at the contribution of order $O(\alpha^2/\pi^2)$
to the rate of the neutron radiative $\beta^-$--decay $n \to p + e^- +
\bar{\nu}_e + \gamma$. The contributions of the one--virtual--photon
exchanges we have classified by the Feynman diagrams in
Fig.\,\ref{fig:fig2}, Fig.\,\ref{fig:fig3} and
Fig.\,\ref{fig:fig4}. In the diagrams in Fig.\,\ref{fig:fig2} the
contributions of strong low--energy and electromagnetic interactions
are factorized, and both the real and virtual photons couple to the
point--like proton and electron. The contribution of strong
low--energy interactions of hadrons is given by the axial coupling
constant $\lambda$ only. All divergences, caused by virtual photon
exchanges, are absorbed by renormalization of masses and wave
functions of the proton and electron, the proton--proton--photon
$(pp\gamma)$ and electron--electron--photon $(e^-e^-\gamma)$
vertices. Therewith, the counterterms of renormalization of the wave
functions and vertices obey standard Ward identities
\cite{Itzykson1980,Weinberg1995,Bogoliubov1959}. The diagrams in
Fig.\,\ref{fig:fig2} are invariant under gauge transformations
$\varepsilon^*_{\lambda'}(k) \to \varepsilon^*_{\lambda'}(k) + c k$ of
a real photon wave function and $D_{\alpha\beta}(q) \to
D_{\alpha\beta}(q) + c(q^2)\,q_{\alpha}q_{\beta}$ of a photon Green
function, respectively. The structure functions, defining the
renormalized contribution of the Feynman diagrams in
Fig.\,\ref{fig:fig2} to the amplitude of the neutron radiative
$\beta^-$--decay of order $O(\alpha/\pi)$, obey Ward identities. The
contribution of the Feynman diagrams in Fig.\,\ref{fig:fig2} to the
branching ratio is of order $10^{-7}$ (see Table I).

The dominant but most problematic contribution comes from the Feynman
diagrams in Fig.\,\ref{fig:fig3} and Fig.\,\ref{fig:fig4}. For the
calculation of the contribution of these diagrams we follow Sirlin's
assumption for the calculation of the radiative corrections of order
$O(\alpha/\pi)$ to the rate of the neutron $\beta^-$--decay
\cite{Sirlin1967}. This means that we assume that the contribution of
the diagrams in Fig.\,\ref{fig:fig4}, which survives to leading order
in the large proton mass expansion in the physical gauge of a real
photon, contains i) a part of the diagram Fig.\,\ref{fig:fig3}c, which
is not invariant under gauge transformations of a real photon wave
function $\varepsilon^*_{\lambda'}(k) \to \varepsilon^*_{\lambda'}(k)
+ c k$ and of a photon Green function $D_{\alpha\beta}(q) \to
D_{\alpha\beta}(q) + c(q^2)\,q_{\alpha}q_{\beta}$, respectively, and
ii) a part independent of the electron and photon energies, which can
be absorbed by renormalization of the Fermi weak coupling constant
$G_F$ and the axial coupling constant $\lambda$. This is, of course,
an extended interpretation of Sirlin's assumption, since in the
neutron $\beta^-$--decay the Feynman diagrams similar to the diagrams
in Fig.\,\ref{fig:fig4} (see Fig.\,\ref{fig:fig6}b and
Fig.\,\ref{fig:fig6}c) have been found independent of the electron
energy, the contribution of which has been absorbed by renormalization
of the Fermi weak and axial coupling constants. A confirmation of our
assumption, concerning the properties of the Feynman diagrams in
Fig.\,\ref{fig:fig3} and Fig.\,\ref{fig:fig4} might be supported by
the fact that all possible insertions of real photon external lines
transform the Feynman diagrams in Fig.\,\ref{fig:fig6} to the Feynman
diagrams Fig.\,\ref{fig:fig3}c, Fig.\,\ref{fig:fig3}d and
Fig.\,\ref{fig:fig4}. It is obvious that all possible insertions of
real photon external lines should not change the properties of the
diagrams with respect to a gauge transformation $D_{\alpha\beta}(q)
\to D_{\alpha\beta} + c(q^2)\,q_{\alpha}q_{\beta}$. Hence, all
diagrams in Fig.\,\ref{fig:fig4} should play an auxiliary role for the
diagram Fig.\,\ref{fig:fig3}c to leading order in the large proton
mass expansion. Thus, such an extended Sirlin's assumption, applied to
the calculation of the Feynman diagrams in Fig.\,\ref{fig:fig3}c,
Fig.\,\ref{fig:fig3}d and Fig.\,\ref{fig:fig4}, we have realized as
follows. Firstly, we have shown that in Fig.\,\ref{fig:fig3} only the
diagram Fig.\,\ref{fig:fig3}c survives to leading order in the large
proton mass expansion in the physical gauge of a real photon.
Secondly, we have decomposed the contribution of the diagram
Fig.\,\ref{fig:fig3}c into invariant and non--invariant parts with
respect to a gauge transformation of a photon Green function
$D_{\alpha\beta}(q) \to D_{\alpha\beta}(q) +
c(q^2)\,q_{\alpha}q_{\beta}$. Finally, we have omitted a
gauge--non--invariant part and the contributions, independent of the
electron and photon energies, we have removed by renormalization of
the Fermi weak coupling constant $G_F$ and the axial coupling constant
$\lambda$, respectively. The contribution of the diagrams in
Fig.\,\ref{fig:fig3} and Fig.\,\ref{fig:fig4} to the branching ratio
of the neutron radiative $\beta^-$--decay, obtained in such a way, is
of order $10^{-4}$ (see Table I).

It is important to emphasize that the renormalized contribution of the
diagram Fig.\,\ref{fig:fig3}c, which we have defined in terms of the
functions $f_1(E_e, \vec{k}_e, \omega, \vec{k}\,)$ and $f_2(E_e,
\vec{k}_e, \omega, \vec{k}\,)$ (see Eq.(\ref{eq:C.33}) of the
Supplemental Material), does not depend on the infrared cut--of $\mu$,
which is introduced as a photon mass \cite{Sirlin1967}. This is unlike
the contribution of the diagram Fig.\,\ref{fig:fig6}a to the rate of
the neutron $\beta^-$--decay, which has been found as a function of
the infrared cut--off $\mu$ \cite{Sirlin1967}. A $\mu$--dependence of
the radiative corrections, caused by the diagram
Fig.\,\ref{fig:fig6}a, has been cancelled only by the diagram
Fig.\,\ref{fig:fig1}b (see \cite{Sirlin1967}).

We would like to accentuate that the contribution of the diagrams in
Fig.\,\ref{fig:fig5}, describing a neutron radiative $\beta^-$--decay
with two real photons in the final state, is also infrared
stable. Having integrated over the momentum and energy of one of two
photons we have obtained the contribution of order $O(\alpha^2/\pi^2)$
to the rate of the neutron radiative $\beta^-$--decay with a photon,
detected in the energy region $\omega_{\rm min} \le \omega \le
\omega_{\rm max}$. The contribution of the neutron radiative
$\beta^-$--decay with two real photons in the final state, described
by the diagrams in Fig.\,\ref{fig:fig5}, is of order $10^{-5}$ (see
Table I).

Total contributions of the radiative corrections of order
$O(\alpha/\pi)$ to the rate of the neutron radiative $\beta^-$--decay
are about $2.42\,\%$, $2.44\,\%$ and $3.66\,\%$ for three
photon--energy regions $15\,{\rm keV} \le \omega \le 340\,{\rm keV}$,
$14\,{\rm keV} \le \omega \le 782\,{\rm keV}$ and $0.4\,{\rm keV} \le
\omega \le 14\,{\rm keV}$, respectively. They are commensurable with
the radiative correction $3.75\,\%$ to the rate of the neutron
$\beta^-$--decay \cite{Sirlin1967,Ivanov2013}. However, they are not
enhanced with respect to the contribution of the radiative corrections
to the rate of the neutron $\beta^-$--decay as we have expected
because of an enhancement of the corrections of order $1/M$, caused by
the weak magnetism and proton recoil \cite{Ivanov2017}. The
theoretical values of the branching ratios (see Table I) do not
contradict the experimental data within the experimental error
bars. Nevertheless, deviations of about $4.21\,\%$, $7.05\,\%$ and
$9.52\,\%$ of the mean values of the experimental data from the
theoretical values for three photon--energy regions $15\,{\rm keV} \le
\omega \le 340\,{\rm keV}$, $14\,{\rm keV} \le \omega \le 782\,{\rm
  keV}$ and $0.4\,{\rm keV} \le \omega \le 14\,{\rm keV}$,
respectively, might only imply that such a distinction cannot be
covered by the contributions of interactions beyond the Standard
Model. Therefore, apart from the experimental error bars one may
expect a better agreement between theory and experiment only from the
contributions of strong low--energy interactions of hadrons beyond the
axial coupling constant $\lambda$. One may expect that they might be
caused by the contributions of diagrams in Fig.\,\ref{fig:fig4}, where
real and virtual photon couple to a hadronic block.

In this connection we may confess that there are two problems of our
precision analysis of the rate of the neutron radiative
$\beta^-$--decay to order $O(\alpha^2/\pi^2)$. They are i) a
justification of a neglect of the diagrams with photons coupled to
hadronic blocks such as the diagram Fig.\,\ref{fig:fig1}c and so on
and ii) a justification of Sirlin's assumption for an extraction of a
physical contribution from the Feynman diagrams in
Fig.\,\ref{fig:fig3} and Fig\,\ref{fig:fig4}. As we have mentioned
above both of these problems can be investigated only by a
model--dependent way within certain models of strong low--energy
interactions of hadrons coupled to photons.

However, very likely that the contribution of the diagram
Fig.\,\ref{fig:fig1}c is really not important. One may show this at
the tree--level using the following effective low--energy
electromagnetic interactions of the neutron and proton
\cite{Itzykson1980}
\begin{eqnarray}\label{eq:16} 
\hspace{-0.15in}&& \delta {\cal L}_{\rm em}(x) = \frac{\kappa_n e}{4
  M}\,\bar{\psi}_n(x)\sigma_{\mu\nu}\psi_n(x)\,F^{\mu\nu}(x) +
\frac{\kappa_p e}{4
  M}\,\bar{\psi}_p(x)\sigma_{\mu\nu}\psi_p(x)\,F^{\mu\nu}(x), 
\end{eqnarray}
where $\kappa_n = - 1.91304$ and $\kappa_p = 1.79285$ are anomalous
magnetic moments of the neutron and proton \cite{PDG2016},
respectively, and $\sigma_{\mu\nu} = (i/2)(\gamma_{\mu}\gamma_{\nu} -
\gamma_{\nu}\gamma_{\mu})$ are Dirac matrices \cite{Itzykson1980}. The
contribution of the diagram Fig.\,\ref{fig:fig1}c to the amplitude of
the neutron radiative $\beta^-$--decay is equal to
\begin{eqnarray}\label{eq:17} 
\hspace{-0.3in}&&{\cal M}_{\rm Fig.\,\ref{fig:fig1}c}(n \to p e^- \bar{\nu}_e
  \gamma)_{\lambda'} =\nonumber\\ 
\hspace{-0.3in}&&= -
  \frac{\kappa_n}{2M}\,\Big[\bar{u}_p(\vec{k}_p,
    \sigma_p)\,\gamma^{\mu}(1 + \lambda \gamma^5)\,\frac{1}{m_n -
      \hat{k}_n + \hat{k} -
      i0}\,i\,\sigma_{\alpha\beta}\,k^{\alpha}\varepsilon^{\beta
      *}_{\lambda'}u_n(\vec{k}_n,
    \sigma_n)\Big]\,\Big[\bar{u}_e(\vec{k}_e,
    \sigma_e)\,\gamma_{\mu}(1 - \gamma^5)\,v_{\nu}(\vec{k}_{\nu}, +
    \frac{1}{2})\Big]\nonumber\\ 
\hspace{-0.3in}&&~~ +
  \frac{\kappa_p}{2M}\,\Big[\bar{u}_p(\vec{k}_p,
    \sigma_p)\,i\,\sigma_{\alpha\beta}\,k^{\alpha}\varepsilon^{\beta
      *}_{\lambda'}\,\gamma^{\mu}(1 + \lambda \gamma^5)\,\frac{1}{m_n
      - \hat{k}_p - \hat{k} - i0}u_n(\vec{k}_n,
    \sigma_n)\Big]\,\Big[\bar{u}_e(\vec{k}_e,
    \sigma_e)\,\gamma_{\mu}(1 - \gamma^5)\,v_{\nu}(\vec{k}_{\nu}, +
    \frac{1}{2})\Big].
\end{eqnarray}
One may see that the amplitude Eq.(\ref{eq:17}) is invariant under a
gauge transformation $\varepsilon^*_{\lambda'}(k) \to
\varepsilon^*_{\lambda'}(k) + c k$.  The contribution of the diagram
Fig.\,\ref{fig:fig1}c to the branching ratio is given by
\begin{eqnarray}\label{eq:18} 
B^{(\rm Fig.\,\ref{fig:fig1}c)}_{\beta\gamma}(\omega_{\rm max},
\omega_{\rm min}) &=&
\frac{\alpha}{\pi}\,\frac{G^2_F|V_{ud}|^2}{2\pi^3 M}\,\Big(2\lambda^2
  (\kappa_p + \kappa_n) - \lambda\,(\kappa_p -
  \kappa_n)\Big)\nonumber\\ &&\times \int^{\omega_{\rm
    max}}_{\omega_{\rm min}}d\omega\,\omega\int^{E_0 -
  \omega}_{m_e}dE_e\,\sqrt{E^2_e - m^2_e}\,(E_0 - E_e -
\omega)^2\,F(E_e, Z = 1).
\end{eqnarray} 
For the three photon energy regions (see Table I) the branching ratio
is equal to $B^{(\rm Fig.\,\ref{fig:fig1}c)}_{\beta\gamma} =
0.97\times 10^{-10}$, $B^{(\rm Fig.\,\ref{fig:fig1}c)}_{\beta\gamma} =
1.25\times 10^{-10}$ and $B^{(\rm
  Fig.\,\ref{fig:fig1}c)}_{\beta\gamma} = 4.90\times 10^{-13}$,
respectively. This may testify that the diagram Fig.\,\ref{fig:fig1}c
can actually be neglected. Such a neglect does not violate invariance
of the diagrams Fig.\,\ref{fig:fig1}a and Fig.\,\ref{fig:fig1}b with
respect to a gauge transformation $\varepsilon^*_{\lambda'}(k) \to
\varepsilon^*_{\lambda'}(k) + c k$. Our justification of a possible
neglect of the contribution of the diagram Fig.\,\ref{fig:fig1}c
confirms also a neglect of all diagrams with emission of a real photon
by a hadronic block in the radiative neutron $\beta^-$--decay with two
real photons in the final state, given by the diagram in
Fig.\,\ref{fig:fig5}.

Hence, the main contribution of strong low--energy interactions we may
expect only from the diagrams in Fig.\,\ref{fig:fig3} and
Fig.\,\ref{fig:fig4}. As a first step on the way of the analysis of
these diagrams we are planning to use the Standard Model of
electroweak interactions supplemented by the linear $\sigma$--model of
strong low--energy nucleon--pion interactions by Gell--Mann and Levy
\cite{GellMann1958} (see also \cite{DeAlfaro1973}). It is well--known
that a linear $\sigma$--model is a renormalizable one
\cite{GellMann1960,Kim1973,Johnson1998}. Renormalization of an
extended version of a linear $\sigma$--model has been investigated in
\cite{Pisarski1984,Grahl2013}. The observed peculiar properties of
strong low--energy hadronic interactions in the neutron radiative
$\beta^-$--decay to order of $O(\alpha^2/\pi^2)$ agree well with
assertion, pointed out by Weinberg \cite{Weinberg1956}, about the
important role of strong low--energy hadronic interactions in decay
processes.

We would like to emphasize that analysis of the rate of the neutron
radiative $\beta^-$--decay to order $O(\alpha^2/\pi^2)$ is a first
step toward the analysis of the neutron $\beta^-$--decay to order
$O(\alpha^2/\pi^2)$. One of the most intriguing theoretical features
of this analysis, which we anticipate, is a cancellation of the
infrared dependences in the sum of the contributions of the diagrams
with only virtual photon exchanges and the diagrams of the neutron
radiative $\beta^-$--decay with one and two photons in the final
state. For the analytical investigation of this problem the results,
obtained in this paper, are of great deal of importance. The
calculation of the neutron $\beta^-$--decay to order $\alpha^2/\pi^2
\sim 10^{-5}$ together with the contributions of order $(\alpha/\pi)
(E_e/M) \sim 3\times 10^{-6}$ and $E^2_e/M^2 \sim 10^{-6}$ should give
a new level of theoretical precision for the experimental search of
interactions beyond the Standard Model \cite{Ivanov2013}.

It is well known that in the limit $m_{\sigma} \to \infty$, where
$m_{\sigma}$ is a scalar $\sigma$--meson mass, a linear
$\sigma$--model is equivalent to current algebra
\cite{Weinberg1967,Gasiorowicz1969}.  This means that the results,
obtained in a linear $\sigma$--model and taken in the limit
$m_{\sigma} \to \infty$, should reproduce the results, obtained in
current algebra \cite{Weinberg1967,Gasiorowicz1969}, i.e. in a
model--independent approach. This bridges between the results, which
we are planning to obtain for the contributions of strong low--energy
interactions to the radiative corrections of order $O(\alpha^2/\pi^2)$
for the neutron radiative and neutron $\beta^-$--decays, and the
results, obtained by Sirlin \cite{Sirlin1967,Sirlin1978} for the
contributions of strong low--energy interactions to the radiative
corrections of order $O(\alpha/\pi)$ for the neutron $\beta^-$--decay.

\section{Acknowledgements}

The work of A. N. Ivanov was supported by the Austrian ``Fonds zur
F\"orderung der Wissenschaftlichen Forschung'' (FWF) under contracts
I689-N16, I862-N20, P26781-N20 and P26636-N20,  ``Deutsche
F\"orderungsgemeinschaft'' (DFG) AB 128/5-2 and by the \"OAW within
the New Frontiers Groups Programme, NFP 2013/09. The work of
R. H\"ollwieser was supported by the Erwin Schr\"odinger Fellowship
program of the Austrian Science Fund FWF (``Fonds zur F\"orderung der
wissenschaftlichen Forschung'') under Contract No. J3425-N27. The work
of M. Wellenzohn was supported by the MA 23 (FH-Call 16) under the
project ``Photonik - Stiftungsprofessur f\"ur Lehre''.

\newpage

\section{Supplemental material}

\section*{Appendix A: The amplitude and rate of the neutron 
radiative $\beta^-$--decay in the tree--approximation, described by
the Feynman diagram in Fig.\,\ref{fig:fig1}}
\renewcommand{\theequation}{A-\arabic{equation}}
\setcounter{equation}{0}

The amplitude of the neutron radiative $\beta^-$--decay, represented
by the diagrams in Fig.\,\ref{fig:fig1}, Fig.\,\ref{fig:fig2},
Fig.\,\ref{fig:fig3}, Fig.\,\ref{fig:fig4} and Fig.\,\ref{fig:fig5}, we
define as follows
\begin{eqnarray}\label{eq:A.1}
\hspace{-0.3in}M(n \to p e^- \bar{\nu}_e \gamma)_{\lambda'} =
e\,\frac{G_F}{\sqrt{2}}\,V_{ud}\sum^5_{j = 1}\,{\cal M}_{\rm
  Fig.\,j}(n \to p e^- \bar{\nu}_e \gamma)_{\lambda'},
\end{eqnarray}
where ${\cal M}_{\rm Fig.\,j}(n \to p e^- \bar{\nu}_e
\gamma)_{\lambda'}$ is the contribution of the diagram in Fig.\,j for
$j = 1,2,\ldots,5$. The amplitude ${\cal M}_{\rm Fig.\ref{fig:fig1}}(n
\to p e^- \bar{\nu}_e \gamma)_{\lambda'}$ of the neutron radiative
$\beta^-$--decay, defined by the diagrams in Fig.\ref{fig:fig1}, is
\cite{Ivanov2013,Ivanov2017}
\begin{eqnarray}\label{eq:A.2}
\hspace{-0.3in}{\cal M}_{\rm Fig.\ref{fig:fig1}}(n \to p e^-
\bar{\nu}_e \gamma)_{\lambda'} &=& \Big[\bar{u}_p(\vec{k}_p, \sigma_p)
\gamma^{\mu}(1 + \lambda \gamma^5)  u_n(\vec{k}_n, \sigma_n)\Big]
\Big[\bar{u}_e(\vec{k}_e,\sigma_e)\,\frac{1}{2k_e\cdot
    k}\,Q_e\,\gamma_{\mu} (1 - \gamma^5) v_{\nu}(\vec{k}_{\nu}, +
  \frac{1}{2})\Big]\nonumber\\
\hspace{-0.3in}&-& \Big[\bar{u}_p(\vec{k}_p, \sigma_p)\,Q_p
  \,\frac{1}{2k_p \cdot k}\,\gamma^{\mu}(1 + \lambda \gamma^5)
  u_n(\vec{k}_n, \sigma_n)\Big]\Big[\bar{u}_e(\vec{k}_e,\sigma_e)
  \gamma^{\mu} (1 - \gamma^5) v_{\nu}(\vec{k}_{\nu}, +
  \frac{1}{2})\Big],
\end{eqnarray}
where $Q_e$ and $Q_p$ are given by
\begin{eqnarray}\label{eq:A.3}
\hspace{-0.3in}Q_e = 2 \varepsilon^{*}_{\lambda'}\cdot k_e +
\hat{\varepsilon}^*_{\lambda'}\hat{k}\;,\; Q_p = 2
\varepsilon^{*}_{\lambda'}\cdot k_p +
\hat{\varepsilon}^*_{\lambda'}\hat{k}.
\end{eqnarray}
For the derivation of Eq.(\ref{eq:A.2}) we have used the Dirac
equations for the free proton and electron. Replacing
$\varepsilon^{*}_{\lambda'} \to k$ and using $k^2 = 0$ we get ${\cal
  M}(n \to p e^- \bar{\nu}_e \gamma)|_{\varepsilon^{*}_{\lambda'} \to
  k} = 0$. This confirms invariance of the amplitude Eq.(\ref{eq:A.2})
with respect to a gauge transformation $\varepsilon^*_{\lambda'}(k)
\to \varepsilon^*_{\lambda'}(k) + c\,k$, where $c$ is an arbitrary
constant.

To leading order in the large proton mass expansion the amplitude of
the neutron radiative $\beta^-$--decay, calculated in the
tree--approximation, is equal to \cite{Ivanov2013,Ivanov2017}
\begin{eqnarray}\label{eq:A.4}
\hspace{-0.3in}&&{\cal M}_{\rm Fig.\ref{fig:fig1}}(n \to p e^-
\bar{\nu}_e \gamma)_{\lambda'} = 2m_n\nonumber\\
\hspace{-0.3in}&&\times\,\Big\{[\varphi^{\dagger}_p\varphi_n]
\Big[\bar{u}_e(\vec{k}_e,\sigma_e)\,\frac{1}{2k_e\cdot
    k}\,Q_e\,\gamma^0 (1 - \gamma^5) v_{\nu}(\vec{k}, +
  \frac{1}{2})\Big] - \lambda\,[\varphi^{\dagger}_p\,\vec{\sigma}\,
  \varphi_n] \cdot
\Big[\bar{u}_e(\vec{k}_e,\sigma_e)\,\frac{1}{2k_e\cdot
    k}\,Q_e\,\vec{\gamma}\,(1 - \gamma^5) v_{\nu}(\vec{k}, +
  \frac{1}{2})\Big]\nonumber\\
\hspace{-0.3in}&& -
\frac{\varepsilon^{0*}_{\lambda'}}{\omega}[\varphi^{\dagger}_p
  \varphi_n] \Big[\bar{u}_e(\vec{k}_e,\sigma_e)\,\gamma^0 (1 -
  \gamma^5) v_{\nu}(\vec{k}, + \frac{1}{2})\Big] +
\frac{\varepsilon^{0*}_{\lambda'}}{\omega}\,\lambda\,[\varphi^{\dagger}_p\,
  \vec{\sigma}\,\varphi_n] \cdot
\Big[\bar{u}_e(\vec{k}_e,\sigma_e)\,\vec{\gamma}\,(1 - \gamma^5)
  v_{\nu}(\vec{k}, + \frac{1}{2})\Big]\Big\}.
\end{eqnarray}
The hermitian conjugate amplitude is 
\begin{eqnarray}\label{eq:A.5}
\hspace{-0.3in}&&{\cal M}^{\dagger}_{\rm Fig.\ref{fig:fig1}}(n \to p e^-
\bar{\nu}_e \gamma)_{\lambda'} = 2m_n\nonumber\\
\hspace{-0.3in}&& \times\,\Big\{[\varphi^{\dagger}_n \varphi_p\Big]
\Big[\bar{v}_{\nu}(\vec{k}, + \frac{1}{2})\,\frac{1}{2k_e\cdot
    k}\,\gamma^0\, \bar{Q}_e (1 - \gamma^5)
  u_e(\vec{k}_e,\sigma_e)\Big] -
\lambda\,[\varphi^{\dagger}_n\,\vec{\sigma}\, \varphi_p] \cdot
\Big[\bar{v}_{\nu}(\vec{k}, + \frac{1}{2})\,\frac{1}{2k_e\cdot
    k}\,\vec{\gamma}\,\bar{Q}_e\,(1 - \gamma^5)
  u_e(\vec{k}_e,\sigma_e)\Big]\nonumber\\
\hspace{-0.3in}&& -
\frac{\varepsilon^{0}_{\lambda'}}{\omega}[\varphi^{\dagger}_n
  \varphi_p]\, \Big[\bar{v}_{\nu}(\vec{k}, + \frac{1}{2})\,\gamma^0 (1
  - \gamma^5) u_e(\vec{k}_e,\sigma_e)\Big] +
\frac{\varepsilon^{0}_{\lambda'}}{\omega}\,\lambda\,[\varphi^{\dagger}_n\,
  \vec{\sigma}\varphi_p] \cdot
\Big[\bar{v}_{\nu}(\vec{k}, + \frac{1}{2})\,\vec{\gamma}\,(1 -
  \gamma^5) u_e(\vec{k}_e,\sigma_e)\Big]\Big\},
\end{eqnarray}
where $\bar{Q}_e = 2 k_e\cdot \varepsilon_{\lambda'} +
\hat{k}\,\hat{\varepsilon}_{\lambda'}$ \cite{Ivanov2013}. The
amplitudes Eq.(\ref{eq:A.4}) and Eq.(\ref{eq:A.5}) are invariant under
a gauge transformation $\varepsilon^*_{\lambda'}(k) \to
\varepsilon^*_{\lambda'}(k) + c\,k$ or $\varepsilon^{0*}_{\lambda'}(k)
\to \varepsilon^{0*}_{\lambda'}(k) + c\,\omega$ and
$\vec{\varepsilon}^{\,*}_{\lambda'}(k) \to
\vec{\varepsilon}^{\,*}_{\lambda'}(k) + c\,\vec{k}$.

As has been shown in \cite{Ivanov2013,Ivanov2017,Ivanov2013b}, for the
calculation of the rate of the neutron radiative $\beta^-$--decay one
may set $\varepsilon^{0*}_{\lambda'} = 0$ and deal with only physical
degrees of freedom of a photon \cite{BD1967} such as
$\varepsilon^*_{\lambda'} = (0, \vec{\varepsilon}^{\,*}_{\lambda'})$,
which obey the relations \cite{BD1967} (see also
\cite{Ivanov2013,Ivanov2013b,Ivanov2017})
\begin{eqnarray}\label{eq:A.6}
\hspace{-0.3in}\vec{k}\cdot \vec{\varepsilon}^{\,*}_{\lambda'} &=&
\vec{k}\cdot \vec{\varepsilon}_{\lambda'} =
0\;,\;\vec{\varepsilon}^{\,*}_{\lambda'}\cdot
\vec{\varepsilon}_{\lambda''} = \delta_{\lambda'\lambda''},\nonumber\\
\hspace{-0.3in}\sum_{\lambda' = 1,2}\vec{\varepsilon}^{\,i
  *}_{\lambda'}\vec{\varepsilon}^{\,j}_{\lambda'} &=& \delta^{ij} -
\frac{\vec{k}^{\,i} \vec{k}^{\,j}}{\omega^2} = \delta^{ij} -
\vec{n}^{\,i}_{\vec{k}} \vec{n}^{\,j}_{\vec{k}},
\end{eqnarray}
where $\vec{n}_{\vec{k}} = \vec{k}/\omega$ is a unit vector along the
photon 3--momentum \cite{Ivanov2013,Ivanov2013a}. The rate of the
neutron radiative $\beta^-$--decay with a photon, emitted in the
energy region $\omega_{\rm min} \le \omega \le \omega_{\rm max}$ and
calculated to leading order in the large proton mass expansion, is
\cite{Ivanov2013}
\begin{eqnarray*}
\lambda^{(\rm Fig.\,\ref{fig:fig1})}_{\beta\gamma}(\omega_{\rm
  max},\omega_{\rm min}) &=& \frac{\alpha}{\pi}\,(1 + 3
\lambda^2)\,\frac{G^2_F|V_{ud}|^2}{2\pi^3}\int^{\omega_{\rm
    max}}_{\omega_{\rm min}}\frac{d\omega}{\omega}\int^{E_0 -
  \omega}_{m_e}dE_e \,\sqrt{E^2_e - m^2_e}\,E_e\,F(E_e, Z = 1)\,(E_0 -
E_e - \omega)^2\nonumber\\ 
\end{eqnarray*}
\begin{eqnarray}\label{eq:A.7}
&&\times\Big\{\Big(1 + \frac{\omega}{E_e} +
\frac{1}{2}\frac{\omega^2}{E^2_e}\Big)\,\Big[\frac{1}{\beta}\,{\ell
    n}\Big(\frac{1 + \beta}{1 - \beta}\Big) - 2\Big] +
\frac{\omega^2}{E^2_e}\Big\},
\end{eqnarray}
where $\beta = k_e/E_e = \sqrt{E^2_e - m^2_e}/E_e$ is a velocity of
the electron with a 3--momentum $k_e$, the Fermi function $F(E_e, Z =
1)$ describes the proton--electron Coulomb interaction in the final
state of the decay \cite{Ivanov2013,Ivanov2017} and $E_0 = (m^2_n -
m^2_p + m^2_e)/2m_n = 1.2927\,{\rm MeV}$ is the end--point energy of
the electron--energy spectrum of the neutron $\beta^-$--decay
\cite{Ivanov2013}. The rate of the neutron radiative $\beta^-$--decay,
taken to next--to--leading order in the large proton mass expansion
and taking into account the contributions of the weak magnetism and
proton recoil, has been calculated in \cite{Ivanov2017}. It is given
by \cite{Ivanov2017}
\begin{eqnarray}\label{eq:A.8}
\hspace{-0.3in}\lambda^{(\rm Fig\,\ref{fig:fig1})}_{\beta
  \gamma}(\omega_{\rm max},\omega_{\rm min}) &=& (1 + 3
\lambda^2)\frac{\alpha}{\pi} \frac{G^2_F |V_{ud}|^2}{2\pi^3}
\int^{\omega_{\rm max}}_{\omega_{\rm
    min}}\frac{d\omega}{\omega}\int^{E_0 - \omega}_{m_e}
dE_e\,E_e\sqrt{E^2_e - m^2_e}\, (E_0 - E_e - \omega)^2\nonumber\\
\hspace{-0.3in}&&\times \,F(E_e, Z =
1)\,\rho^{(\rm Fig\,\ref{fig:fig1})}_{\beta\gamma}(E_e,\omega).
\end{eqnarray}
The function $\rho^{(\rm Fig\,\ref{fig:fig1})}_{\beta \gamma}(E_e,\omega)$ is
given by the integral \cite{Ivanov2017}
\begin{eqnarray}\label{eq:A.9}
\hspace{-0.3in}&&\rho^{(\rm
  Fig\,\ref{fig:fig1})}_{\beta\gamma}(E_e,\omega) =
\int\frac{d\Omega_{e\gamma}}{4\pi}\,\Big\{\Big[1 +
  2\,\frac{\omega}{M}\,\frac{E_e - \vec{k}_e\cdot
    \vec{n}_{\vec{k}}}{E_0 - E_e - \omega} + \frac{3}{M}\,\Big(E_e +
  \omega - \frac{1}{3}\,E_0\Big) + \frac{\lambda^2 - 2(\kappa +
    1)\lambda + 1}{1 + 3\lambda^2}\,\frac{E_0 - E_e -
    \omega}{M}\Big]\nonumber\\
\hspace{-0.3in}&&\times\,\Big[\Big(1 +
  \frac{\omega}{E_e}\Big)\,\frac{k^2_e - (\vec{k}_e\cdot
    \vec{n}_{\vec{k}})^2}{(E_e - \vec{k}_e\cdot \vec{n}_{\vec{k}})^2}
  + \frac{\omega^2}{E_e}\,\frac{1}{E_e - \vec{k}_e\cdot
    \vec{n}_{\vec{k}}}\Big] + \frac{3\lambda^2 - 1}{1 + 3
  \lambda^2}\,\frac{1}{M}\,\Big(\frac{k^2_e + \omega \vec{k}_e\cdot
  \vec{n}_{\vec{k}}}{E_e}\,\Big[\frac{k^2_e - (\vec{k}_e\cdot
    \vec{n}_{\vec{k}})^2}{(E_e - \vec{k}_e\cdot \vec{n}_{\vec{k}})^2}
  + \frac{\omega}{E_e - \vec{k}_e\cdot
    \vec{n}_{\vec{k}}}\Big]\nonumber\\
\hspace{-0.3in}&& + (\omega + \vec{k}_e\cdot
\vec{n}_{\vec{k}})\Big[\Big(1 +
  \frac{\omega}{E_e}\Big)\frac{\omega}{E_e - \vec{k}_e\cdot
    \vec{n}_{\vec{k}}} - \frac{m^2_e}{E_e}\,\frac{\omega}{(E_e -
    \vec{k}_e\cdot \vec{n}_{\vec{k}})^2}\Big]\Big) - \frac{\lambda^2 +
  2 (\kappa + 1)\lambda - 1}{1 +
  3\lambda^2}\,\frac{1}{M}\,\Big[\frac{k^2_e + \omega^2 + 2\omega
    \vec{k}_e\cdot \vec{n}_{\vec{k}}}{E_e}\nonumber\\
\hspace{-0.3in}&&\times\,\frac{k^2_e - (\vec{k}_e\cdot
  \vec{n}_{\vec{k}})^2}{(E_e - \vec{k}_e\cdot \vec{n}_{\vec{k}})^2} +
\frac{\omega}{E_e}\,\frac{k^2_e - (\vec{k}_e\cdot
  \vec{n}_{\vec{k}})^2}{E_e - \vec{k}_e\cdot \vec{n}_{\vec{k}}} +
\frac{\omega^2}{E_e}\,\frac{\omega + \vec{k}_e\cdot
  \vec{n}_{\vec{k}}}{E_e - \vec{k}_e\cdot \vec{n}_{\vec{k}}}\Big] -
\frac{\lambda(\lambda - 1)}{1 +
  3\lambda^2}\,\frac{1}{M}\,\Big[\frac{\omega}{E_e}\, \frac{k^2_e -
    (\vec{k}_e\cdot \vec{n}_{\vec{k}})^2}{E_e - \vec{k}_e\cdot
    \vec{n}_{\vec{k}}} + 3\,\frac{\omega^2}{E_e}\Big]\Big\},
\end{eqnarray}
where $d\Omega_{e\gamma}$ is an infinitesimal solid angle of the
electron--photon momentum correlations $\vec{k}_e\cdot
\vec{n}_{\vec{k}} = k_e \cos\theta_{e\gamma}$. The numerical value of
the rate Eq.(\ref{eq:A.7}) is given in Table I. It has been calculated
in \cite{Ivanov2017}.

\section*{Appendix B: Contributions of Feynman diagrams
in Fig.\,\ref{fig:fig2} to the amplitude and rate of the neutron
radiative $\beta^-$--decay}
\renewcommand{\theequation}{B-\arabic{equation}}
\setcounter{equation}{0}

The contribution of the Feynman diagrams in Fig.\,\ref{fig:fig2} to
the amplitude of the neutron radiative $\beta^-$--decay we define as
follows
\begin{eqnarray}\label{eq:B.1}
\hspace{-0.3in}&&{\cal M}_{\rm Fig.\,\ref{fig:fig2}}(n \to p e^-
\bar{\nu}_e \gamma)_{\lambda'} = \sum_{j = a,b,c}{\cal M}^{(p)}_{\rm
  Fig.\,\ref{fig:fig2}j}(n \to p e^- \bar{\nu}_e \gamma)_{\lambda'} +
\sum_{j = d,e,f}{\cal M}^{(e)}_{\rm Fig.\,\ref{fig:fig2}j}(n \to p e^-
\bar{\nu}_e \gamma)_{\lambda'},
\end{eqnarray}
where the amplitudes ${\cal M}^{(p)}_{\rm Fig.\,\ref{fig:fig2}j}(n \to
p e^- \bar{\nu}_e \gamma)_{\lambda'}$ and ${\cal M}^{(e)}_{\rm
  Fig.\,\ref{fig:fig2}j}(n \to p e^- \bar{\nu}_e \gamma)_{\lambda'}$
are given by the following analytical expressions
\begin{eqnarray}\label{eq:B.2}
{\cal M}^{(p)}_{\rm Fig.\,\ref{fig:fig2}a}(n \to p e^- \bar{\nu}_e
\gamma)_{\lambda'} &=& \Big[\bar{u}_p(\vec{k}_p, \sigma_p)
  \varepsilon^*_{\lambda'}\cdot \Lambda_p(k_p, k)\,\frac{1}{m_p -
    \hat{k}_p - \hat{k} - i0}\,\gamma^{\mu}(1 + \lambda \gamma^5)\,
  u_n(\vec{k}_n,
  \sigma_n)\Big]\nonumber\\ &&\times\,\Big[\bar{u}_e(\vec{k}_e,\sigma_e)\,\gamma_{\mu}
  (1 - \gamma^5)\, v_{\nu}(\vec{k}_{\nu}, + \frac{1}{2})\Big],
\end{eqnarray}
where $\Lambda^{\alpha}_p(k_p, k)$ is the proton--proton--photon
$(pp\gamma)$ vertex function
\begin{eqnarray}\label{eq:B.3}
\Lambda^{\alpha}_p(k_p, k) = (Z^{(p)}_1 - 1)\,\gamma^{\alpha} +
e^2\int \frac{d^4q}{(2\pi)^4i}\,\gamma^{\sigma}\,\frac{1}{m_p -
  \hat{k}_p - \hat{q} - i0}\,\gamma^{\alpha}\,\frac{1}{m_p - \hat{k}_p
  - \hat{q} - \hat{k} - i0}\,\gamma^{\beta}\,D_{\sigma\beta}(q)
\end{eqnarray}
 and
\begin{eqnarray}\label{eq:B.4} 
{\cal M}^{(p)}_{\rm Fig.\,\ref{fig:fig2}b}(n \to p e^- \bar{\nu}_e
\gamma)_{\lambda'} &=& \Big[\bar{u}_p(\vec{k}_p, \sigma_p)
  \,\hat{\varepsilon}^*_{\lambda'}\,\frac{1}{m_p - \hat{k}_p - \hat{k}
    - i0}\,\Sigma^{(p)}(k_p,k)\,\frac{1}{m_p - \hat{k}_p - \hat{k} -
    i0}\,\gamma^{\mu}(1 + \lambda \gamma^5)\, u_n(\vec{k}_n,
  \sigma_n)\Big]\nonumber\\ &&\times\,\Big[\bar{u}_e(\vec{k}_e,\sigma_e)\,\gamma_{\mu}
  (1 - \gamma^5)\, v_{\nu}(\vec{k}_{\nu}, + \frac{1}{2})\Big],
\end{eqnarray}
where $\Sigma^{(p)}(k_p, k)$ is the proton self--energy correction
\begin{eqnarray}\label{eq:B.5}
\Sigma^{(p)}(k_p, k) = - \delta m_p - (Z^{(p)}_2 - 1)\,(m_p -
\hat{k}_p - \hat{k}) + e^2\int
\frac{d^4q}{(2\pi)^4i}\,\gamma^{\sigma}\frac{1}{m_p - \hat{k}_p -
  \hat{q} - \hat{k} - i 0}\gamma^{\beta}\,D_{\sigma\beta}(q),
\end{eqnarray}
and 
\begin{eqnarray}\label{eq:B.6} 
{\cal M}^{(p)}_{\rm Fig.\,\ref{fig:fig2}c}(n \to p e^- \bar{\nu}_e
\gamma)_{\lambda'} &=& \Big[\bar{u}_p(\vec{k}_p, \sigma_p)
  \,\Sigma^{(p)}(k_p)\,\frac{1}{m_p - \hat{k}_p -
    i0}\,\hat{\varepsilon}^*_{\lambda'}\,\frac{1}{m_p - \hat{k}_p -
    \hat{k} - i0}\,\gamma^{\mu}(1 + \lambda \gamma^5)\, u_n(\vec{k}_n,
  \sigma_n)\Big]\nonumber\\ &&\times\,\Big[\bar{u}_e(\vec{k}_e,\sigma_e)
  \,\gamma_{\mu} (1 - \gamma^5)\, v_{\nu}(\vec{k}_{\nu}, +
  \frac{1}{2})\Big],
\end{eqnarray}
where $\Sigma^{(p)}(k_p)$ is the proton self--energy correction
\begin{eqnarray}\label{eq:B.7}
\Sigma^{(p)}(k_p) = - \delta m_p - (Z^{(p)}_2 - 1)\,(m_p - \hat{k}_p )
+ e^2\int \frac{d^4q}{(2\pi)^4i}\,\gamma^{\sigma}\frac{1}{m_p -
  \hat{k}_p - \hat{q} - i 0}\gamma^{\beta}\,D_{\sigma\beta}(q),
\end{eqnarray}
and 
\begin{eqnarray}\label{eq:B.8}
{\cal M}^{(e)}_{\rm Fig.\,\ref{fig:fig2}d}(n \to p e^- \bar{\nu}_e
\gamma)_{\lambda'} &=& - \Big[\bar{u}_p(\vec{k}_p, \sigma_p)
  \,\gamma^{\mu}(1 + \lambda \gamma^5)\,
  u_n(\vec{k}_n,
  \sigma_n)\Big]\nonumber\\ &&\times\,\Big[\bar{u}_e(\vec{k}_e,\sigma_e)\,\varepsilon^*_{\lambda'}\cdot \Lambda_e(k_e, k)\,\frac{1}{m_e -
    \hat{k}_e - \hat{k} - i0}\,\gamma_{\mu}
  (1 - \gamma^5)\, v_{\nu}(\vec{k}_{\nu}, + \frac{1}{2})\Big],
\end{eqnarray}
where $\Lambda^{\alpha}_e(k_e, k)$ is the electron--electron--photon
$(e^-e^-\gamma)$ vertex function
\begin{eqnarray}\label{eq:B.9}
\Lambda^{\alpha}_e(k_e, k) = (Z^{(e)}_1 - 1)\,\gamma^{\alpha} +
e^2\int \frac{d^4q}{(2\pi)^4i}\,\gamma^{\sigma}\,\frac{1}{m_e -
  \hat{k}_e - \hat{q} - i0}\,\gamma^{\alpha}\,\frac{1}{m_e - \hat{k}_e
  - \hat{q} - \hat{k} - i0}\,\gamma^{\beta}\,D_{\sigma\beta}(q),
\end{eqnarray}
and 
\begin{eqnarray}\label{eq:B.10} 
&&{\cal M}^{(e)}_{\rm Fig.\,\ref{fig:fig2}e}(n \to p e^- \bar{\nu}_e
  \gamma)_{\lambda'} = - \Big[\bar{u}_p(\vec{k}_p, \sigma_p)
    \,\gamma^{\mu}(1 + \lambda \gamma^5)\, u_n(\vec{k}_n,
    \sigma_n)\Big]\nonumber\\ &&\times\,
  \Big[\bar{u}_e(\vec{k}_e,\sigma_e)\,\hat{\varepsilon}^*_{\lambda'}\,
    \frac{1}{m_e - \hat{k}_e - \hat{k} -
      i0}\,\Sigma^{(e)}(k_e,k)\,\frac{1}{m_e - \hat{k}_e - \hat{k} -
      i0}\,\gamma_{\mu} (1 - \gamma^5)\, v_{\nu}(\vec{k}_{\nu}, +
    \frac{1}{2})\Big],
\end{eqnarray}
where $\Sigma^{(e)}(k_e, k)$ is the electron self--energy correction
\begin{eqnarray}\label{eq:B.11}
\Sigma^{(e)}(k_e, k) = - \delta m_e - (Z^{(e)}_2 - 1)\,(m_e -
\hat{k}_e - \hat{k}) + e^2\int
\frac{d^4q}{(2\pi)^4i}\,\gamma^{\sigma}\frac{1}{m_e - \hat{k}_e -
  \hat{q} - \hat{k} - i 0}\gamma^{\beta}\,D_{\sigma\beta}(q),
\end{eqnarray}
and 
\begin{eqnarray}\label{eq:B.12} 
&&{\cal M}^{(e)}_{\rm Fig.\,\ref{fig:fig2}f}(n \to p e^- \bar{\nu}_e
  \gamma)_{\lambda'} = -\Big[\bar{u}_p(\vec{k}_p, \sigma_p)
    \,\gamma^{\mu}(1 + \lambda \gamma^5)\, u_n(\vec{k}_n,
    \sigma_n)\Big]\nonumber\\ &&\times\,
  \Big[\bar{u}_e(\vec{k}_e,\sigma_e)\,\Sigma^{(e)}(k_e)\,\frac{1}{m_e
      - \hat{k}_e - i0}\,\hat{\varepsilon}^*_{\lambda'}\,\frac{1}{m_e
      - \hat{k}_e - \hat{k} - i0}\,\gamma_{\mu} (1 -
    \gamma^5)\,v_{\nu}(\vec{k}_{\nu}, + \frac{1}{2})\Big],
\end{eqnarray}
where $\Sigma^{(e)}(k_e)$ is the electron self--energy correction
\begin{eqnarray}\label{eq:B.13}
\Sigma^{(e)}(k_e) = - \delta m_e - (Z^{(e)}_2 - 1)\,(m_p - \hat{k}_e)
+ e^2\int \frac{d^4q}{(2\pi)^4i}\,\gamma^{\beta}\frac{1}{m_p -
  \hat{k}_e - \hat{q} - i 0}\gamma^{\beta}\,D_{\sigma\beta}(q),
\end{eqnarray}
where $D_{\sigma\beta}(q) = (\eta_{\sigma\beta} -(1 - \xi)\,
q_{\sigma}q_{\beta}/q^2)/(q^2 + i 0)$ is a photon Green function and
$\xi$ is a gauge parameter.

Now we may analyse the properties of the Feynman diagrams in
Fig.\,\ref{fig:fig2} with respect to gauge transformations
$\varepsilon^*_{\lambda'}(k) \to \varepsilon^*_{\lambda'}(k) + c\,k$
and $D_{\sigma\beta}(q) \to D_{\sigma\beta}(q) +
c(q^2)\,q_{\sigma}q_{\beta}$. Making a gauge transformation
$\varepsilon^*_{\lambda'}(k) \to \varepsilon^*_{\lambda'}(k) + c\,k$
for the scalar product $k\cdot \Lambda_p(k_p,k)$ we obtain the
following expression
\begin{eqnarray}\label{eq:B.14}
\hspace{-0.3in}&&k\cdot \Lambda_p(k_p,k) = (Z^{(p)}_1 - 1)\Big((m_p -
\hat{k}_p) - (m_p - \hat{k}_p - \hat{k})\Big) + e^2\int
\frac{d^4q}{(2\pi)^4i}\,\gamma^{\sigma}\frac{1}{m_p - \hat{k}_p -
  \hat{q} - \hat{k} - i
  0}\gamma^{\beta}\,D_{\sigma\beta}(q)\nonumber\\ 
\hspace{-0.3in}&&- e^2\int
\frac{d^4q}{(2\pi)^4i}\,\gamma^{\sigma}\frac{1}{m_p - \hat{k}_p -
  \hat{q} - i 0}\gamma^{\beta}\,D_{\sigma\beta}(q) = \Big[ - \delta
  m_p - (Z^{(p)}_1 - 1) (m_p - \hat{k}_p - \hat{k})\nonumber\\ 
\hspace{-0.3in}&& + e^2\int
\frac{d^4q}{(2\pi)^4i}\,\gamma^{\sigma}\frac{1}{m_p - \hat{k}_p -
  \hat{q} - \hat{k} - i 0}\gamma^{\beta}\,D_{\sigma\beta}(q)\Big] -
\Big[ - \delta m_p - (Z^{(p)}_1 - 1) (m_p - \hat{k}_p)\nonumber\\ 
\hspace{-0.3in}&& + e^2\int
\frac{d^4q}{(2\pi)^4i}\,\gamma^{\sigma}\frac{1}{m_p - \hat{k}_p -
  \hat{q}- i 0}\gamma^{\beta}\,D_{\sigma\beta}(q)\Big].
\end{eqnarray}
Since $Z^{(p)}_1 = Z^{(p)}_2$ \cite{Bogoliubov1959,Itzykson1980},
Eq.(\ref{eq:B.14}) can be transcribed into the standard form of the
Ward identity \cite{Bogoliubov1959,Itzykson1980}
\begin{eqnarray}\label{eq:B.15}
\hspace{-0.3in}k\cdot \Lambda_p(k_p,k) = \Sigma^{(p)}(k_p, k) -
\Sigma^{(p)}(k_p),
\end{eqnarray}
where $\Sigma^{(p)}(k_p, k)$ and $\Sigma^{(p)}(k_p)$ are given by
Eq.(\ref{eq:B.5}) and Eq.(\ref{eq:B.7}), respectively. Then, because
of the relation $Z^{(e)}_1 = Z^{(e)}_2$
\cite{Bogoliubov1959,Itzykson1980}, we get the Ward identity
\cite{Bogoliubov1959,Itzykson1980}
\begin{eqnarray}\label{eq:B.16}
\hspace{-0.3in}k\cdot \Lambda_e(k_e,k) = \Sigma^{(e)}(k_e, k) -
\Sigma^{(e)}(k_e),
\end{eqnarray}
where $\Sigma^{(e)}(k_e, k)$ and $\Sigma^{(e)}(k_e)$ are given by
Eq.(\ref{eq:B.9}) and Eq.(\ref{eq:B.11}), respectively.

Making a gauge transformation of a photon Green function
$D_{\sigma\beta}(q) \to D_{\sigma\beta}(q) +
c(q^2)\,q_{\sigma}q_{\beta}$ we obtain the following correction to the
$(pp\gamma)$ vertex diagram
\begin{eqnarray}\label{eq:B.17}
\hspace{-0.3in}\varepsilon^*_{\lambda'}\cdot \delta \Lambda_p(k_p,k) =
e^2 \int\frac{d^4q}{(2\pi)^4i}\,c(q^2)\,
\hat{\varepsilon}^*_{\lambda'}\,\frac{1}{m_p - \hat{k}_p - \hat{k} -
  i0} - e^2 \int\frac{d^4q}{(2\pi)^4i}\,c(q^2)\,
\hat{\varepsilon}^*_{\lambda'}\,\frac{1}{m_p - \hat{k}_p - \hat{q} -
  \hat{k} - i0},
\end{eqnarray}
where we have used the Dirac equation for a free proton. The
self--energy diagrams acquire the corrections
\begin{eqnarray}\label{eq:B.18}
\hspace{-0.3in}&&\hat{\varepsilon}^*_{\lambda'}\,\frac{1}{m_p -
  \hat{k}_p - \hat{k} - i0}\,\delta \Sigma^{(p)}(k_p,k)\,\frac{1}{m_p
  - \hat{k}_p - \hat{k} - i0} = - e^2
\int\frac{d^4q}{(2\pi)^4i}\,c(q^2)\,
\hat{\varepsilon}^*_{\lambda'}\,\frac{1}{m_p - \hat{k}_p - \hat{k} -
  i0}\,\hat{q}\,\frac{1}{m_p - \hat{k}_p - \hat{k} -
  i0}\nonumber\\ \hspace{-0.3in}&& - e^2
\int\frac{d^4q}{(2\pi)^4i}\,c(q^2)\,
\hat{\varepsilon}^*_{\lambda'}\,\frac{1}{m_p - \hat{k}_p - \hat{k} -
  i0} + e^2 \int\frac{d^4q}{(2\pi)^4i}\,c(q^2)\,
\hat{\varepsilon}^*_{\lambda'}\,\frac{1}{m_p - \hat{k}_p - \hat{q} -
  \hat{k} - i0}.
\end{eqnarray}
Since the first term in Eq.(\ref{eq:B.18}) is equal to zero, we get
\begin{eqnarray}\label{eq:B.19}
\hspace{-0.3in}&&\hat{\varepsilon}^*_{\lambda'}\,\frac{1}{m_p -
  \hat{k}_p - \hat{k} - i0}\,\delta \Sigma^{(p)}(k_p,k)\,\frac{1}{m_p
  - \hat{k}_p - \hat{k} - i0} = - e^2
\int\frac{d^4q}{(2\pi)^4i}\,c(q^2)\,
\hat{\varepsilon}^*_{\lambda'}\,\frac{1}{m_p - \hat{k}_p - \hat{k} -
  i0}\nonumber\\ \hspace{-0.3in}&& + e^2
\int\frac{d^4q}{(2\pi)^4i}\,c(q^2)\,
\hat{\varepsilon}^*_{\lambda'}\,\frac{1}{m_p - \hat{k}_p - \hat{q} -
  \hat{k} - i0}.
\end{eqnarray}
Then, the self--energy correction $\Sigma^{(p)}(k_p)$ is invariant under the gauge transformation $D_{\sigma\beta}(q) \to D_{\sigma\beta}(q) +
c(q^2)\,q_{\sigma}q_{\beta}$. We get
\begin{eqnarray}\label{eq:B.20}
\hspace{-0.3in}&&\delta \Sigma^{(p)}(k_p) = - e^2
\int\frac{d^4q}{(2\pi)^4i}\,c(q^2)\,\hat{q} = 0,
\end{eqnarray}
where we have used the Dirac equation for a free proton. Plugging
Eq.(\ref{eq:B.17}) and Eq.(\ref{eq:B.19}) into Eq.(\ref{eq:B.2}) and
Eq.(\ref{eq:B.4}) one may see that the acquired corrections cancel
each other in the first term of Eq.(\ref{eq:B.1}). This confirms
invariance of the Feynman diagrams in Fig.\,\ref{fig:fig2}a,
Fig.\,\ref{fig:fig2}b and Fig.\,\ref{fig:fig2}c under a gauge
transformation $D_{\sigma\beta}(q) \to D_{\sigma\beta}(q) +
c(q^2)\,q_{\sigma}q_{\beta}$. It is obvious that the Feynman diagrams
in Fig.\,\ref{fig:fig2}d, Fig.\,\ref{fig:fig2}e and
Fig.\,\ref{fig:fig2}f are also invariant under a transformation
$D_{\sigma\beta}(q) \to D_{\sigma\beta}(q) +
c(q^2)\,q_{\sigma}q_{\beta}$ of a photon Green function. The observed
gauge invariance allows to make calculations of the Feynman diagrams
in Fig.\,\ref{fig:fig2} in the Feynman gauge $\xi = 1$
\cite{Sirlin1967}.

For the calculation of $\Lambda^{\alpha}_p(k_p, q)$ we rewrite the
right-hand-side (r.h.s.) of Eq.(\ref{eq:B.3}) as follows
\begin{eqnarray}\label{eq:B.21}
\Lambda^{\alpha}_p(k_p, k) = (Z^{(p)}_1 - 1)\,\gamma^{\alpha} +
e^2\int \frac{d^4q}{(2\pi)^4i} \frac{\gamma^{\beta}(m_p
  + \hat{k}_p + \hat{q})\gamma^{\alpha}(m_p + \hat{k}_p + \hat{q} +
  \hat{k})\gamma_{\beta}}{[m^2_p - (k_p + q)^2 - i0][m^2_p - (k_p + q
    + k)^2 - i0]}\,\frac{1}{q^2 + i0}.
\end{eqnarray}
Since the integral over $q$ diverges, we have to regularize it.  For
this aim we use the Pauli--Villars regularization and make in
Eq.(\ref{eq:B.21}) a replacement \cite{Ivanov2013} (see also
\cite{Bogoliubov1959})
\begin{eqnarray}\label{eq:B.22}
\frac{1}{q^2 + i 0} \to \frac{1}{\Lambda^2 - q^2 - i0} -
\frac{1}{\mu^2 - q^2 - i0},
\end{eqnarray}
where $\Lambda$ and $\mu$ are the ultraviolet and infrared cut--off,
respectively, which should be finally taken in the limit $\Lambda \to
\infty$ and $\mu \to 0$.  The next step is to merge the
denominators. Merging the denominators of the proton propagators we get
\begin{eqnarray}\label{eq:B.23}
\hspace{-0.3in}\frac{1}{[m^2_p - (k_p + q)^2 - i0]}\frac{1}{[m^2_p -
    (k_p + q + k)^2 - i0]} = \int^1_0\frac{dx}{[m^2_p- k^2 x(1 - x) -
    (q + k_p + k x)^2 - i0]^2}.
\end{eqnarray}
Then, we have to take into account the contribution of the regularized
photon propagator. Using the formula \cite{Ivanov2013} 
\begin{eqnarray}\label{eq:B.24}
\frac{1}{A^2 B} 
= \int^1_0\frac{2y\,dy}{[A\,y + B\,(1 -y)]^3}
\end{eqnarray}
we obtain
\begin{eqnarray}\label{eq:B.25}
\hspace{-0.3in}&&\frac{1}{[m^2_p - (k_p + q)^2 - i0]} \frac{1}{[m^2_p
    - (k_p + q + k)^2 - i0]} \Big( \frac{1}{\Lambda^2 - q^2 - i0} -
\frac{1}{\mu^2 - q^2 - i0}\Big) =\nonumber\\
\hspace{-0.3in}&&= \int^1_0\!\!\!\int^1_0
\frac{2y\,dx\,dy}{[\Lambda^2(1 - y) + m^2_p y^2 - 2 k_p\cdot k x y (1
    - y) - k^2 x y (1 - x y) - (q - (k_p + k x)y)^2 -
    i0]^3}\nonumber\\
\hspace{-0.3in}&&- \int^1_0\!\!\!\int^1_0 \frac{2y\,dx\,dy}{[\mu^2(1 -
    y) + m^2_py^2 - 2 k_p\cdot k x y (1 - y) - k^2 x y (1 - x y) - (q
    - (k_p + k x)y)^2 - i0]^3}.
\end{eqnarray}
Now we take into account that $k^2 = 0$
\begin{eqnarray}\label{eq:B.26}
\hspace{-0.3in}&&\frac{1}{[m^2_p - (k_p + q)^2 - i0]} \frac{1}{[m^2_p
    - (k_p + q + k)^2 - i0]} \Big( \frac{1}{\Lambda^2 - q^2 - i0} -
\frac{1}{\mu^2 - q^2 - i0}\Big) =\nonumber\\
\hspace{-0.3in}&&= \int^1_0\!\!\!\int^1_0
\frac{2y\,dx\,dy}{[\Lambda^2(1 - y) + m^2_p y^2 - 2 k_p\cdot k x y (1
    - y) - (q + (k_p + k x)y)^2 - i0]^3}\nonumber\\
\hspace{-0.3in}&&- \int^1_0\!\!\!\int^1_0 \frac{2y\,dx\,dy}{[\mu^2(1 -
    y) + m^2_py^2 - 2 k_p\cdot k x y (1 - y) - (q + (k_p + k x)y)^2 -
    i0]^3}.
\end{eqnarray}
For the numerator of the integrand of Eq.(\ref{eq:B.21}) we get the
expression
\begin{eqnarray}\label{eq:B.27}
\gamma^{\beta}(m_p + \hat{k}_p + \hat{q})\gamma^{\alpha}(m_p +
\hat{k}_p + \hat{q} + \hat{k})\gamma_{\beta} &=& - 2m ^2_p
\gamma^{\alpha} + 4 m_p (2 k_p + 2 q + k)^{\alpha} + 2 (k_p +
q)^2\,\gamma^{\alpha} + 2\,\hat{k}\,(\hat{k}_p +
\hat{q})\,\gamma^{\alpha}\nonumber\\ && - 4\,(\hat{k}_p + \hat{q} +
\hat{k})\,(k_p + q)^{\alpha}.
\end{eqnarray}
Then, making a change of variables $q + (k_p + k x)y \to q$ and
integrating over the solid angle in the 4--dimensional $q$--space we
arrive at the expression
\begin{eqnarray}\label{eq:B.28}
\hspace{-0.3in}&&\gamma^{\beta}(m_p + \hat{k}_p +
\hat{q})\gamma^{\alpha}(m_p + \hat{k}_p + \hat{q} +
\hat{k})\gamma_{\beta} \to - 2m ^2_p \gamma^{\alpha} + 4 m_p (2 k_p +
2 q - 2(k_p + k x)y + k)^{\alpha} \nonumber\\ \hspace{-0.3in}&& + 2
(k_p + q - (k_p + k x)y)^2\,\gamma^{\alpha} + 2\,\hat{k}\,(\hat{k}_p +
\hat{q} - (\hat{k}_p + \hat{k} x)y)\,\gamma^{\alpha} - 4\,(\hat{k}_p +
\hat{q} - (\hat{k}_p + \hat{k} x) y +
\hat{k})\nonumber\\ \hspace{-0.3in}&&\times\,(k_p + q - (k_p + k
x)y)^{\alpha}.
\end{eqnarray}
Having integrated over the directions of the 4--vector $q$ we get
\begin{eqnarray}\label{eq:B.29}
\hspace{-0.3in}&&\gamma^{\beta}(m_p + \hat{k}_p +
\hat{q})\gamma^{\alpha}(m_p + \hat{k}_p + \hat{q} +
\hat{k})\gamma_{\beta} \to \Big[q^2 - 2m^2_py(2 - y) + 4k_p\cdot k (1
  - xy) (1 - y) - 2 m_p \hat{k} (1 - y)\nonumber\\ \hspace{-0.3in}&& -
  2 k^2 x y (1 - xy)\Big]\,\gamma^{\alpha} + \Big[4 m_p k^{\alpha}_p (1 - y^2)
  + 4 m_p k^{\alpha} (1 - x y(1 + y)) - 4 k^{\alpha}_p \hat{k} (1 -
  xy)(1 - y) + 4 \hat{k} k^{\alpha} x y(1 - x y)\Big].~~~
\end{eqnarray}
Now we may remove the terms, which vanish because of the relations
$\varepsilon^*_{\lambda'}\cdot k = 0$ and $k^2 = 0$. This gives the
following representation for the $(pp\gamma)$ vertex
\begin{eqnarray}\label{eq:B.30}
\hspace{-0.3in}&&\Lambda^{\alpha}_p(k_p, k) = (Z^{(p)}_1 -
1)\,\gamma^{\alpha} + e^2\int^1_0dx\int^1_0dy\,2y\,\int
\frac{d^4q}{(2\pi)^4i} \Big( \frac{1}{[\Lambda^2(1 - y) + m^2_p y^2 -
    2 k_p\cdot k x y (1 - y) - q^2 - i0]^3}\nonumber\\
\hspace{-0.3in}&&- \frac{1}{[\mu^2(1 - y) + m^2_py^2 - 2 k_p\cdot k x
    y (1 - y) - q^2 - i0]^3}\Big)\Big\{\Big[q^2 - 2m^2_py(2 - y) +
  4k_p\cdot k (1 - xy) (1 - y) - 2 m_p \hat{k} (1 -
  y)\Big]\,\gamma^{\alpha}\nonumber\\ \hspace{-0.3in}&& + \Big[4 m_p
  k^{\alpha}_p (1 - y^2) - 4 k^{\alpha}_p \hat{k} (1 - xy)(1 -
  y)\Big]\Big\}.
\end{eqnarray}
Making a Wick rotation \cite{Bogoliubov1959} we obtain the expression
\begin{eqnarray}\label{eq:B.31}
\hspace{-0.3in}&&\Lambda^{\alpha}_p(k_p, k) = (Z^{(p)}_1 -
1)\,\gamma^{\alpha} + e^2\int^1_0dx\int^1_0dy\,2y\,\int
\frac{d^4q}{(2\pi)^4} \Big( \frac{1}{[\Lambda^2(1 - y) + m^2_p y^2 -
    2 k_p\cdot k x y (1 - y) + q^2]^3}\nonumber\\
\hspace{-0.3in}&&- \frac{1}{[\mu^2(1 - y) + m^2_py^2 - 2 k_p\cdot k x
    y (1 - y) + q^2]^3}\Big)\Big\{\Big[- q^2 - 2m^2_py(2 - y) +
  4k_p\cdot k (1 - xy) (1 - y) - 2 m_p \hat{k} (1 -
  y)\Big]\,\gamma^{\alpha}\nonumber\\ \hspace{-0.3in}&& + \Big[4 m_p
  k^{\alpha}_p (1 - y^2) - 4 k^{\alpha}_p \hat{k} (1 - xy)(1 -
  y)\Big]\Big\},
\end{eqnarray}
which we transcribe into the form
\begin{eqnarray}\label{eq:B.32}
\hspace{-0.3in}&&\Lambda^{\alpha}_p(k_p, k) = (Z^{(p)}_1 -
1)\,\gamma^{\alpha} - e^2\,\gamma^{\alpha}
\int^1_0dx\int^1_0dy\,2y\nonumber\\
\hspace{-0.3in}&&\times \int \frac{d^4q}{(2\pi)^4} \Big(
\frac{1}{[\Lambda^2(1 - y) + m^2_p y^2 - 2 k_p\cdot k x y (1 - y) +
    q^2]^2} - \frac{1}{[\mu^2(1 - y) + m^2_py^2 - 2 k_p\cdot k x y (1
    - y) + q^2]^2}\Big)\nonumber\\
\hspace{-0.3in}&& + e^2\int^1_0dx\int^1_0dy\,2y\,\int
\frac{d^4q}{(2\pi)^4}\,\frac{1}{[\Lambda^2(1 - y) + m^2_p y^2 -
    2 k_p\cdot k x y (1 - y) + q^2]^3}\,\Big\{\Big[\Lambda^2(1 - y) -
  m^2_py(4 - 3y)\nonumber\\ \hspace{-0.3in}&& + 2k_p\cdot k (2 - 3xy)
  (1 - y) - 2 m_p \hat{k} (1 - y)\Big]\,\gamma^{\alpha} + \Big[4 m_p
  k^{\alpha}_p (1 - y^2) - 4 k^{\alpha}_p \hat{k} (1 - xy)(1 -
  y)\Big]\Big\}\nonumber\\
\hspace{-0.3in}&& - e^2\int^1_0dx\int^1_0dy\,2y\,\int
\frac{d^4q}{(2\pi)^4}\,\frac{1}{[\mu^2(1 - y) + m^2_p y^2 - 2 k_p\cdot
    k x y (1 - y) + q^2]^3}\,\Big\{\Big[\mu^2(1 - y) - m^2_py(4 -
  3y)\nonumber\\ \hspace{-0.3in}&& + 2k_p\cdot k (2 - 3xy) (1 - y) - 2
  m_p \hat{k} (1 - y)\Big]\,\gamma^{\alpha} + \Big[4 m_p k^{\alpha}_p
  (1 - y^2) - 4 k^{\alpha}_p \hat{k} (1 - xy)(1 - y)\Big]\Big\}
\end{eqnarray}
Having integrated over $q$ we get
\begin{eqnarray}\label{eq:B.33}
\hspace{-0.3in}&&\Lambda^{\alpha}_p(k_p, k) = (Z^{(p)}_1 -
1)\,\gamma^{\alpha} + \gamma^{\alpha}\,\frac{e^2}{16\pi^2}
\int^1_0dx\int^1_0dy\,2y\,{\ell n}\Big(\frac{\Lambda^2(1 - y) + m^2_p
  y^2 - 2 k_p\cdot k x y (1 - y)}{\mu^2(1 - y) + m^2_py^2 - 2 k_p\cdot
  k x y (1 - y)}\Big)\nonumber\\
\hspace{-0.3in}&& +
\frac{e^2}{32\pi^2}\int^1_0dx\int^1_0dy\,2y\,\frac{1}{\Lambda^2(1 - y)
  + m^2_p y^2 - 2 k_p\cdot k x y (1 - y)}\,\Big\{\Big[\Lambda^2(1 - y)
  - m^2_py(4 - 3y)\nonumber\\ \hspace{-0.3in}&& + 2k_p\cdot k (2 -
  3xy) (1 - y) - 2 m_p \hat{k} (1 - y)\Big]\,\gamma^{\alpha} + \Big[4
  m_p k^{\alpha}_p (1 - y^2) - 4 k^{\alpha}_p \hat{k} (1 - xy)(1 -
  y)\Big]\Big\}\nonumber\\
\hspace{-0.3in}&& -
\frac{e^2}{32\pi^2}\int^1_0dx\int^1_0dy\,2y\,\frac{1}{\mu^2(1 - y) +
    m^2_p y^2 - 2 k_p\cdot k x y (1 - y)}\,\Big\{\Big[\mu^2(1 - y) -
    m^2_py(4 - 3y)\nonumber\\ \hspace{-0.3in}&& + 2k_p\cdot k (2 -
    3xy) (1 - y) - 2 m_p \hat{k} (1 - y)\Big]\,\gamma^{\alpha} +
  \Big[4 m_p k^{\alpha}_p (1 - y^2) - 4 k^{\alpha}_p \hat{k} (1 -
    xy)(1 - y)\Big]\Big\}.
\end{eqnarray}
Then, we make a change of variables $xy \to x$. This gives
\begin{eqnarray}\label{eq:B.34}
\hspace{-0.3in}&&\Lambda^{\alpha}_p(k_p, k) = (Z^{(p)}_1 -
1)\,\gamma^{\alpha} + \gamma^{\alpha}\,\frac{e^2}{8\pi^2}
\int^1_0dy\int^y_0dx{\ell n}\Big(\frac{\Lambda^2(1 - y) + m^2_p
  y^2 - 2 k_p\cdot k x(1 - y)}{\mu^2(1 - y) + m^2_py^2 - 2 k_p\cdot
  k x (1 - y)}\Big)\nonumber\\
\hspace{-0.3in}&& +
\frac{e^2}{16\pi^2}\int^1_0dy\int^y_0dx\,\frac{1}{\Lambda^2(1 - y) +
  m^2_p y^2 - 2 k_p\cdot k x (1 - y)}\,\Big\{\Big[\Lambda^2(1 - y) -
  m^2_py(4 - 3y)\nonumber\\ \hspace{-0.3in}&& + 2k_p\cdot k (2 - 3 x)
  (1 - y) - 2 m_p \hat{k} (1 - y)\Big]\,\gamma^{\alpha} + \Big[4 m_p
  k^{\alpha}_p (1 - y^2) - 4 k^{\alpha}_p \hat{k} (1 - x)(1 -
  y)\Big]\Big\}\nonumber\\
\hspace{-0.3in}&& -
\frac{e^2}{16\pi^2}\int^1_0dy\int^y_0dx\,\frac{1}{\mu^2(1 - y) + m^2_p
  y^2 - 2 k_p\cdot k x (1 - y)}\,\Big\{\Big[\mu^2(1 - y) - m^2_py(4 -
  3y)\nonumber\\ \hspace{-0.3in}&& + 2k_p\cdot k (2 - 3 x) (1 - y) - 2
  m_p \hat{k} (1 - y)\Big]\,\gamma^{\alpha} + \Big[4 m_p k^{\alpha}_p
  (1 - y^2) - 4 k^{\alpha}_p \hat{k} (1 - x)(1 - y)\Big]\Big\}.
\end{eqnarray}
Taking into account the limit $\Lambda \gg m_p$ we get
\begin{eqnarray}\label{eq:B.35} 
\hspace{-0.3in}&&\Lambda^{\alpha}_p(k_p, k) = (Z^{(p)}_1 -
1)\,\gamma^{\alpha} + \gamma^{\alpha}\,\frac{e^2}{8\pi^2}
\int^1_0dy\int^y_0dx\,{\ell n}\Big(\frac{\Lambda^2(1 - y) }{\mu^2(1 -
  y) + m^2_py^2 - 2 k_p\cdot k x (1 -
  y)}\Big)\nonumber\\ \hspace{-0.3in}&&+
\frac{e^2}{8\pi^2}\int^1_0dy\int^y_0dx\,\frac{1}{ \mu^2(1 - y) + m^2_p
  y^2 - 2 k_p\cdot k x (1 - y)}\,\Big\{\Big[ m^2_py(2 - y) - 2
  k_p\cdot k (1 - x) (1 - y)\nonumber\\ \hspace{-0.3in}&& + m_p
  \hat{k} (1 - y)\Big]\,\gamma^{\alpha} + \Big[- 2 m_p k^{\alpha}_p (1
  - y^2) + 2 k^{\alpha}_p \hat{k} (1 - x)(1 - y)\Big]\Big\}.
\end{eqnarray}
To leading order in the large proton mass expansion we obtain the
following result
\begin{eqnarray}\label{eq:B.36}
\hspace{-0.3in}&&\Lambda^{\alpha}_p(k_p, q) = (Z^{(p)}_1 -
1)\,\gamma^{\alpha} + \frac{e^2}{8\pi^2}\,\Big({\ell
  n}\Big(\frac{\Lambda}{m_p}\Big) + \frac{5}{4}\Big)\,\gamma^{\alpha}
+ \frac{e^2}{8\pi^2}\,\Big(2 {\ell n}\Big(- \frac{2 k_p\cdot
  k}{m^2_p}\Big) - 1\Big)\,\frac{k^{\alpha}_p}{m_p} 
\end{eqnarray}
(see Eq.(\ref{eq:B.66}) with the replacement $m_e \to m_p$ and $k_e
\to k_p$, where we have dropped the terms of order $O(1/m_p)$). The
calculation of the proton self--energy corrections in
Eq.(\ref{eq:B.5}) and Eq.(\ref{eq:B.7}) runs as follows. For the
calculation of the integrals over $q$ we regularize the photon
propagator
\begin{eqnarray}\label{eq:B.37}
\frac{1}{q^2 + i 0} \to \frac{1}{\Lambda^2 - q^2 - i0} -
\frac{1}{\mu^2 - q^2 - i0}.
\end{eqnarray}
Then, we rewrite Eq.(\ref{eq:B.5}) and Eq.(\ref{eq:B.7}) as follows
\begin{eqnarray}\label{eq:B.38}
\Sigma^{(p)}(k_p, k) &=& - \delta m_p - (Z^{(p)}_2 - 1)\,(m_p -
\hat{k}_p - \hat{k}) + e^2\int
\frac{d^4q}{(2\pi)^4i}\,\frac{\gamma^{\beta}(m_p + \hat{k}_p + \hat{k}
  + \hat{q})\gamma_{\beta}}{[m^2_p - (k_p + k + q)^2 - i
    0]}\nonumber\\ &&\times\,\Big( \frac{1}{\Lambda^2 - q^2 - i0} -
\frac{1}{\mu^2 - q^2 - i0}\Big)
\end{eqnarray}
and 
\begin{eqnarray}\label{eq:B.39}
\Sigma^{(p)}(k_p) &=& - \delta m_p - (Z^{(p)}_2 - 1)\,(m_p - \hat{k}_p)
+ e^2\int \frac{d^4q}{(2\pi)^4i}\,\frac{\gamma^{\beta}(m_p + \hat{k}_p
  + \hat{q})\gamma_{\beta}}{[m^2_p - (k_p + q)^2 - i
    0]}\nonumber\\ &&\times\,\Big( \frac{1}{\Lambda^2 - q^2 - i0} -
\frac{1}{\mu^2 - q^2 - i0}\Big).
\end{eqnarray}
Using the algebra of the Dirac matrices we get
\begin{eqnarray}\label{eq:B.40}
\Sigma^{(p)}(k_p, k) &=& - \delta m_p - (Z^{(p)}_2 - 1)\,(m_p -
\hat{k}_p - \hat{k}) + e^2\int \frac{d^4q}{(2\pi)^4i}\,\frac{4m_p -
  2(\hat{k}_p + \hat{k} + \hat{q})}{[m^2_p - (k_p + k + q)^2 - i
    0]}\nonumber\\ &&\times\,\Big( \frac{1}{\Lambda^2 - q^2 - i0} -
\frac{1}{\mu^2 - q^2 - i0}\Big)
\end{eqnarray}
and 
\begin{eqnarray}\label{eq:B.41}
\Sigma^{(p)}(k_p) &=& - \delta m_p - (Z^{(p)}_2 - 1)\,(m_p -
\hat{k}_p) + e^2\int \frac{d^4q}{(2\pi)^4i}\,\frac{4 m_p - 2(\hat{k}_p
  + \hat{q})}{[m^2_p - (k_p + q)^2 - i 0]}\nonumber\\ &&\times\,\Big(
\frac{1}{\Lambda^2 - q^2 - i0} - \frac{1}{\mu^2 - q^2 - i0}\Big).
\end{eqnarray}
Merging the proton and photon propagators we obtain
\begin{eqnarray}\label{eq:B.42}
\hspace{-0.3in}&&\frac{1}{[m^2_p - (k_p+ k + q)^2 - i 0]}\Big(
\frac{1}{\Lambda^2 - q^2 - i0} - \frac{1}{\mu^2 - q^2 - i0}\Big) =
\nonumber\\
\hspace{-0.3in}&& = \int^1_0 dx\,\Big(\frac{1}{[\Lambda^2 (1 - x ) +
    m^2_p x - (k_p + k)^2 x (1 - x) - (q + (k_p + k)x)^2 - i0]^2}
\nonumber\\
\hspace{-0.3in}&& - \frac{1}{[\mu^2 (1 - x ) + m^2_p x - (k_p + k)^2 x
    (1 - x) - (q + (k_p + k)x)^2 - i0]^2}
\end{eqnarray}
and 
\begin{eqnarray}\label{eq:B.43}
\hspace{-0.3in}&&\frac{1}{[m^2_p - (k_p + q)^2 - i 0]}\Big(
\frac{1}{\Lambda^2 - q^2 - i0} - \frac{1}{\mu^2 - q^2 - i0}\Big) =
\nonumber\\
\hspace{-0.3in}&& = \int^1_0 dx\,\Big(\frac{1}{[\Lambda^2 (1 - x ) +
    m^2_p x - k^2_p x (1 - x) - (q + k_p x)^2 - i0]^2} \nonumber\\
\hspace{-0.3in}&& - \frac{1}{[\mu^2 (1 - x ) + m^2_p x - k^2_p x (1 -
    x) - (q + k_p x)^2 - i0]^2}
\end{eqnarray}
Plugging Eq.(\ref{eq:B.42}) and Eq.(\ref{eq:B.43}) into
Eq.(\ref{eq:B.40}) and Eq.(\ref{eq:B.41}), respectively, making a
shift of variables $q + (k_p + k)x \to q$ and $q + k_p x \to q$ and
integrating over the 4--dimensional solid angle in the $q$--space we
arrive at the expressions
\begin{eqnarray}\label{eq:B.44}
\hspace{-0.3in}\Sigma^{(p)}(k_p, k) &=& - \delta m_p - (Z^{(p)}_2 -
1)\,(m_p - \hat{k}_p - \hat{k})\nonumber\\
\hspace{-0.3in}&& + \frac{e^2}{8\pi^2}\int^1_0dx\int
\frac{d^4q}{\pi^2i}\,\Big\{\frac{m_p(1 + x) + (m_p - \hat{k}_p -
  \hat{k})(1 - x) }{[\Lambda^2 (1 - x ) + m^2_p x - (k_p + k)^2 x (1 -
    x) - q^2 - i0]^2}\nonumber\\
\hspace{-0.3in}&& - \frac{m_p(1 + x) + (m_p - \hat{k}_p - \hat{k})(1 -
  x) }{[\mu^2 (1 - x ) + m^2_p x - (k_p + k)^2 x (1 - x) - q^2 -
    i0]^2}\Big\}
\end{eqnarray}
and 
\begin{eqnarray}\label{eq:B.45}
\hspace{-0.3in}&&\Sigma^{(p)}(k_p) = - \delta m_p - (Z^{(p)}_2 - 1)\,
(m_p - \hat{k}_p)\nonumber\\
\hspace{-0.3in}&& + \frac{e^2}{8\pi^2}\int^1_0dx\int
\frac{d^4k}{\pi^2i}\,\Big\{\frac{m_p(1 + x) + (m_p - \hat{k}_p)(1 - x)
}{[\Lambda^2 (1 - x ) + m^2_p x - k^2_p x (1 - x) - q^2 -
    i0]^2}\nonumber\\
\hspace{-0.3in}&& - \frac{m_p(1 + x) + (m_p - \hat{k}_p )(1 - x)
}{[\mu^2 (1 - x ) + m^2_p x - k^2_p x (1 - x) - q^2 - i0]^2}\Big\}.
\end{eqnarray}
Making the Wick rotation and integrating over $q^2$ we get
\begin{eqnarray}\label{eq:B.46}
\hspace{-0.3in}\Sigma^{(p)}(k_p, k) &=& - \delta m_p - (Z^{(p)}_2 - 1)
\,(m_p - \hat{k}_p - \hat{k}) - \frac{e^2}{8\pi^2}\int^1_0dx\,
\Big(m_p(1 + x) + (m_p - \hat{k}_p - \hat{k})(1 - x)\Big)\nonumber\\
\hspace{-0.3in}&&\times\,{\ell n}\Big(\frac{\Lambda^2 (1 - x ) + m^2_p
  x - (k_p + k)^2 x (1 - x)}{\mu^2 (1 - x ) + m^2_p x - (k_p + k)^2 x
  (1 - x)}\Big)
\end{eqnarray}
and 
\begin{eqnarray}\label{eq:B.47}
\hspace{-0.3in}\Sigma^{(p)}(k_p) &=& - \delta m_p - (Z^{(p)}_2 -
1)\,(m_p - \hat{k}_p) - \frac{e^2}{8\pi^2}\int^1_0dx\Big(m_p(1 + x) + (m_p
- \hat{k}_p)(1 - x)\Big)\nonumber\\
\hspace{-0.3in}&&\times\,{\ell n}\Big(\frac{\Lambda^2 (1 - x ) + m^2_p
  x - k^2_p x (1 - x)}{\mu^2 (1 - x ) + m^2_p x - k^2_p x (1 - x)
}\Big).
\end{eqnarray}
Since $\Lambda \gg m_p$, we may reduces the integrands to the form
\begin{eqnarray}\label{eq:B.48}
\hspace{-0.3in}\Sigma^{(p)}(k_p, k) &=& - \delta m_p - (Z^{(p)}_2 -
1)\, (m_p - \hat{k}_p - \hat{k}) -
\frac{e^2}{8\pi^2}\int^1_0dx\,\Big(m_p(1 + x) + (m_p - \hat{k}_p -
\hat{k})(1 - x)\Big)\nonumber\\
\hspace{-0.3in}&& \times\,\Big\{{\ell n}\Big(\frac{\Lambda^2 }{\mu^2
  (1 - x ) + m^2_p x - (k_p + k)^2 x (1 - x)}\Big) + {\ell n}(1 -
x)\Big\}
\end{eqnarray}
and 
\begin{eqnarray}\label{eq:B.49}
\hspace{-0.3in}\Sigma^{(p)}(k_p) &=& - \delta m_p - (Z^{(p)}_2 -
1)\,(m_p - \hat{k}_p) - \frac{e^2}{8\pi^2}\int^1_0dx\Big(m_p(1 + x) +
(m_p - \hat{k}_p)(1 - x)\Big)\nonumber\\
\hspace{-0.3in}&&\times\,\Big\{{\ell n}\Big(\frac{\Lambda^2}{\mu^2 (1
  - x ) + m^2_p x - k^2_p x (1 - x) }\Big) + {\ell n}(1 - x)\Big\}.
\end{eqnarray}
The next step is to rewrite Eq.(\ref{eq:B.48}) and Eq.(\ref{eq:B.49})
in the following form
\begin{eqnarray}\label{eq:B.50}
\hspace{-0.3in}\Sigma^{(p)}(k_p, k) &=& - \delta m_p - (Z^{(p)}_2 -
1)\, (m_p - \hat{k}_p - \hat{k}) - \frac{e^2}{8\pi^2}\int^1_0dx\,\Big(m_p(1 + x) +
(m_p - \hat{k}_p - \hat{k})(1 - x)\Big)\nonumber\\
\hspace{-0.3in}&& \times\,\Big\{{\ell n}\Big(\frac{\Lambda^2 }{m^2_p
  x^2 + \mu^2 (1 - x )} \Big) + {\ell n}(1 - x) - {\ell n}\Big(1 +
\frac{(m^2_p - (k_p + k)^2)x(1 - x)}{m^2_p x^2 + \mu^2 (1 - x )}\Big)
\Big\}
\end{eqnarray}
and 
\begin{eqnarray}\label{eq:B.51}
\hspace{-0.3in}\Sigma^{(p)}(k_p) &=& - \delta m_p - (Z^{(p)}_2 -
1)\,(m_p - \hat{k}_p) - \frac{e^2}{8\pi^2}\int^1_0dx\Big(m_p(1 + x) +
(m_p - \hat{k}_p)(1 - x)\Big)\nonumber\\
\hspace{-0.3in}&&\times\,\Big\{{\ell n}\Big(\frac{\Lambda^2}{m^2_p x^2
  + \mu^2 (1 - x )}\Big) + {\ell n}(1 - x) - {\ell n}\Big(1 +
\frac{(m^2_p - k^2_p)x(1 - x)}{m^2_p x^2 + \mu^2 (1 - x )}\Big)\Big\}.
\end{eqnarray}
For the proton on--mass shell $k^2_p = m^2_p$ we transcribe
Eq.(\ref{eq:B.50}) and Eq.(\ref{eq:B.51}) into the form
\begin{eqnarray}\label{eq:B.52}
\hspace{-0.3in}&&\Sigma^{(p)}(k_p, k) = - \delta m_p - (Z^{(p)}_2 -
1)\, (m_p - \hat{k}_p - \hat{k}) - m_p\frac{e^2}{8\pi^2}\int^1_0dx\,
(1 + x)\Big\{{\ell n}\Big(\frac{\Lambda^2 }{m^2_p x^2 + \mu^2 (1 -
  x )} \Big) + {\ell n}(1 - x)\Big\}\nonumber\\
\hspace{-0.3in}&& - (m_p - \hat{k}_p -
\hat{k})\,\frac{e^2}{8\pi^2}\int^1_0dx\,(1 - x)\,\Big\{{\ell
  n}\Big(\frac{\Lambda^2 }{m^2_p x^2 + \mu^2 (1 - x )} \Big) + {\ell
  n}(1 - x) - {\ell n}\Big(1 - \frac{2 k_p\cdot k\,x(1 - x)}{m^2_p x^2
  + \mu^2 (1 - x )}\Big) \Big\}
\end{eqnarray}
and
\begin{eqnarray}\label{eq:B.53}
\hspace{-0.3in}\Sigma^{(p)}(k_p) &=& - \delta m_p - (Z^{(p)}_2 -
1)\,(m_p - \hat{k}_p) - m_p\frac{e^2}{8\pi^2}\int^1_0dx\, (1 +
x)\Big\{{\ell n}\Big(\frac{\Lambda^2 }{m^2_p x^2 + \mu^2 (1 - x )}
\Big) + {\ell n}(1 - x)\Big\}\nonumber\\
\hspace{-0.3in}&& - (m_p - \hat{k}_p
)\,\frac{e^2}{8\pi^2}\int^1_0dx\,(1 - x)\,\Big\{{\ell
  n}\Big(\frac{\Lambda^2 }{m^2_p x^2 + \mu^2 (1 - x )} \Big) +{\ell
  n}(1 - x)\Big\}.
\end{eqnarray}
Keeping only the leading order contributions in the large proton mass
expansion we get
\begin{eqnarray}\label{eq:B.54}
\hspace{-0.3in}\Sigma^{(p)}(k_p, k) &=& - \delta m_p - (Z^{(p)}_2 -
1)\, (m_p - \hat{k}_p - \hat{k}) - 3 m_p\, \frac{e^2}{8\pi^2}\,\Big({\ell
  n}\Big(\frac{\Lambda}{m_p}\Big) + \frac{1}{4}\Big)\nonumber\\
\hspace{-0.3in}&& - (m_p - \hat{k}_p -
\hat{k})\,\frac{e^2}{8\pi^2}\,\Big({\ell
  n}\Big(\frac{\Lambda}{m_p}\Big) + \frac{5}{4}\Big) +
\frac{e^2}{8\pi^2}\,\frac{k_p\cdot k}{m_p}\,\Big(2\,{\ell n}\Big(-
\frac{2 k_p\cdot k}{m_p}\Big) - 1\Big)
\end{eqnarray}
(see Eq.(\ref{eq:B.67}) with the replacement $m_e \to m_p$ and $k_e
\to k_p$, where we have dropped the terms of order $O(1/m_p)$) and
\begin{eqnarray}\label{eq:B.55}
\hspace{-0.3in}\Sigma^{(p)}(k_p) &=& - \delta m_p - (Z^{(p)}_2 -
1)\,(m_p - \hat{k}_p) - 3 m_p\frac{e^2}{8\pi^2}\,\Big({\ell
  n}\Big(\frac{\Lambda}{m_p}\Big) + \frac{1}{4}\Big)\nonumber\\
\hspace{-0.3in}&& - (m_p - \hat{k}_p )\,\frac{e^2}{8\pi^2}\,\Big({\ell
  n}\Big(\frac{\Lambda}{m_p}\Big) + \frac{5}{4}\Big).
\end{eqnarray}
As a result, the renormalization parameters $Z^{(p)}_1$, $\delta m_p$
and $Z^{(p)}_2$ are given by
\begin{eqnarray}\label{eq:B.56}
\hspace{-0.3in}\delta m_p &=& - 3 m_p\frac{e^2}{8\pi^2}\,\Big({\ell
  n}\Big(\frac{\Lambda}{m_p}\Big) + \frac{1}{4}\Big)\nonumber\\
\hspace{-0.3in}Z^{(p)}_1 &=& Z^{(p)}_2 = 1 -
\frac{e^2}{8\pi^2}\,\Big({\ell n}\Big(\frac{\Lambda}{m_p}\Big) +
\frac{5}{4}\Big).
\end{eqnarray}
They are calculated in agreement with the Ward identity $Z^{(p)}_1 =
Z^{(p)}_2$, required by gauge invariance.  Thus, the renormalized
$(pp\gamma)$ vertex $ \bar{\Lambda}^{\alpha}_p(k_p, k)$ and proton
self--energy corrections $\bar{\Sigma}^{(p)}(k_p,k)$ and
$\bar{\Sigma}^{(p)}(k_p)$ are equal to
\begin{eqnarray}\label{eq:B.57}
\bar{\Lambda}^{\alpha}_p(k_p,k) &=& \frac{e^2}{8\pi^2}\,\Big(2{\ell
  n}\Big(- \frac{2 k_p \cdot k}{m^2_p}\Big) -
1\Big)\,\frac{k^{\alpha}_p}{m_p},\nonumber\\ \bar{\Sigma}^{(p)}(k_p,k)
&=& \frac{e^2}{8\pi^2}\,\frac{k_p \cdot k}{m_p}\,\Big(2\,{\ell
  n}\Big(- \frac{2 k_p \cdot k}{m^2_p}\Big) -
1\Big),\nonumber\\ \bar{\Sigma}^{(p)}(k_p) &=& 0,
\end{eqnarray}
The sum of the renormalized amplitudes in Fig.\,\ref{fig:fig2}a,
\ref{fig:fig2}b and \ref{fig:fig2}c is given by
\begin{eqnarray}\label{eq:B.58}
\hspace{-0.3in}&&\sum_{j =a,b,c}{\cal M}_{\rm Fig.\,\ref{fig:fig2}j}(n
\to p e^- \bar{\nu}_e \gamma)_{\lambda'} = \Big[\bar{u}_p(\vec{k}_p,
  \sigma_p)\, \varepsilon^*_{\lambda'}\cdot \bar{\Lambda}_p(k_p,
  k)\,\frac{1}{m_p - \hat{k}_p - \hat{k} - i0}\,\gamma^{\mu}(1 +
  \lambda \gamma^5)\, u_n(\vec{k}_n,
  \sigma_n)\Big]\nonumber\\ &&\times\,\Big[\bar{u}_e(\vec{k}_e,\sigma_e)
  \,\gamma_{\mu} (1 - \gamma^5)\, v_{\nu}(\vec{k}_{\nu}, +
  \frac{1}{2})\Big]\nonumber\\ && + \Big[\bar{u}_p(\vec{k}_p,
  \sigma_p) \, \hat{\varepsilon}^*_{\lambda'}\,\frac{1}{m_p -
    \hat{k}_p - \hat{k} - i0}\,
  \bar{\Sigma}^{(p)}(k_p,k)\,\frac{1}{m_p - \hat{k}_p - \hat{k} -
    i0}\,\gamma^{\mu}(1 + \lambda \gamma^5)\, u_n(\vec{k}_n,
  \sigma_n)\Big]\nonumber\\ &&\times\,\Big[\bar{u}_e(\vec{k}_e,\sigma_e)
  \,\gamma_{\mu} (1 - \gamma^5)\, v_{\nu}(\vec{k}_{\nu}, +
  \frac{1}{2})\Big],
\end{eqnarray}
where we have taken into account that $\bar{\Sigma}^{(p)}(k_p) =
0$. Since the renormalized $(pp\gamma)$ vertex
$\bar{\Lambda}^{\alpha}_p(k_p,k)$ and proton self--energy correction
$\bar{\Sigma}^{(p)}(k_p,k)$ obey the Ward identity \cite{Itzykson1980}
\begin{eqnarray}\label{eq:B.59}
k \cdot \bar{\Lambda}_p(k_p,k) = \bar{\Sigma}^{(p)}(k_p,k),
\end{eqnarray}
the amplitude Eq.(\ref{eq:B.58}) is invariant under the gauge
transformation $\varepsilon^*_{\lambda'}(k) \to
\varepsilon^*_{\lambda'}(k) + c\,k$. This confirms the correctness of
the calculation of the diagrams in Fig.\,\ref{fig:fig2}a,
\ref{fig:fig2}b and \ref{fig:fig2}c.

Now we may proceed to the calculation of the Feynman diagrams in
Fig.\,\ref{fig:fig2}d, \ref{fig:fig2}e and \ref{fig:fig2}f.  Skipping
intermediate calculations, which are similar to those we have
performed for the diagrams in Fig.\,\ref{fig:fig2}a, \ref{fig:fig2}b
and \ref{fig:fig2}c, we arrive at the following expressions for the
$(e^-e^-\gamma)$ vertex function and electron self--energy corrections
\begin{eqnarray}\label{eq:B.60} 
\hspace{-0.3in}&&\Lambda^{\alpha}_e(k_e, k) = (Z^{(e)}_1 -
1)\,\gamma^{\alpha} + \gamma^{\alpha}\,\frac{e^2}{8\pi^2}
\int^1_0dy\int^y_0dx\,{\ell n}\Big(\frac{\Lambda^2(1 - y) }{\mu^2(1 -
  y) + m^2_ey^2 - 2 k_e\cdot k x (1 -
  y)}\Big)\nonumber\\ \hspace{-0.3in}&&+
\frac{e^2}{8\pi^2}\int^1_0dy\int^y_0dx\,\frac{1}{ \mu^2(1 - y) +
  m^2_e y^2 - 2 k_e\cdot k x (1 - y)}\,\Big\{\Big[m^2_ey(2 - y) -
  2 k_e\cdot k (1 - x) (1 - y)\nonumber\\ \hspace{-0.3in}&& + m_e
  \hat{k} (1 - y)\Big]\,\gamma^{\alpha} + \Big[- 2 m_e k^{\alpha}_e (1 -
  y^2) + 2 k^{\alpha}_e \hat{k} (1 - x)(1 - y)\Big]\Big\}
\end{eqnarray}
and 
\begin{eqnarray}\label{eq:B.61}
\hspace{-0.3in}\Sigma^{(e)}(k_e, k) &=& - \delta m_e - (Z^{(e)}_2 -
1)\, (m_e - \hat{k}_e - \hat{k}) -
\frac{e^2}{8\pi^2}\int^1_0dx\,\Big(m_e(1 + x) + (m_e - \hat{k}_e -
\hat{k})(1 - x)\Big)\nonumber\\
\hspace{-0.3in}&& \times\,\Big\{{\ell n}\Big(\frac{\Lambda^2 }{m^2_e
  x^2 + \mu^2 (1 - x )} \Big) + {\ell n}(1 - x) - {\ell n}\Big(1 -
\frac{2k_e\cdot k\,x(1 - x)}{m^2_e x^2 + \mu^2 (1 - x )}\Big) \Big\}
\end{eqnarray}
and 
\begin{eqnarray}\label{eq:B.62}
\hspace{-0.3in}\Sigma^{(e)}(k_e) = - \delta m_e - (Z^{(e)}_2 -
1)\,(m_e - \hat{k}_e) - 3 m_e\frac{e^2}{8\pi^2}\,\Big({\ell
  n}\Big(\frac{\Lambda}{m_e}\Big) + \frac{1}{4}\Big) - (m_e -
\hat{k}_e )\,\frac{e^2}{8\pi^2}\,\Big({\ell
  n}\Big(\frac{\Lambda}{m_e}\Big) + \frac{5}{4}\Big),
\end{eqnarray}
where we have kept the electron on--mass shell $k^2_e = m^2_e$. For
the calculation of $\Lambda^{\alpha}_e(k_e,k)$ we propose to
transcribe it into the form
\begin{eqnarray}\label{eq:B.63}
\hspace{-0.3in}&&\Lambda^{\alpha}_e(k_e, k) = (Z^{(e)}_1 -
1)\,\gamma^{\alpha} + \frac{e^2}{8\pi^2}\,\gamma^{\alpha}
\int^1_0dy\int^y_0 dx\,\Big\{{\ell
  n}\Big(\frac{\Lambda^2}{m^2_e y^2 + \mu^2(1 - y)}\Big) + {\ell n}(1
- y)\Big\}\nonumber\\
\hspace{-0.3in}&&+ \frac{e^2}{8\pi^2}\,\gamma^{\alpha}
\int^1_0dy\int^y_0 dx\,{\ell n}\Big(\frac{m^2_e y^2 + \mu^2(1 -
  y)}{m^2_e y^2 + \mu^2(1 - y) - 2 k_e\cdot k\, x (1 -
  y)}\Big)\nonumber\\
\hspace{-0.3in}&&+ \frac{e^2}{8\pi^2}\int^1_0dy\int^y_0dx\,
\frac{1}{m^2_e y^2 + \mu^2(1 - y) - 2 k_e\cdot k\, x (1 -
  y)}\,\Big\{\Big[m^2_e\,y\,(2 - y) - 2 k_e\cdot k\, (1 - x) (1 - y) 
+  m_e
  \hat{k}\, (1 - y)\Big]\,\gamma^{\alpha}\nonumber\\
\hspace{-0.3in}&& + \Big[ - 2 m_e k^{\alpha}_e (1 - y^2) + 2
  k^{\alpha}_e \hat{k}\, (1 - x)(1 - y)\Big]\Big\}.
\end{eqnarray}
The result of the calculation of the first integral in
Eq.(\ref{eq:B.63}) is equal to
\begin{eqnarray}\label{eq:B.64}
\hspace{-0.3in}&& \gamma^{\alpha}\int^1_0dy\int^y_0 dx\,\Big\{{\ell
  n}\Big(\frac{\Lambda^2}{m^2_e y^2 + \mu^2(1 - y)}\Big) + {\ell n}(1
- y)\Big\} = \gamma^{\alpha}\,\Big({\ell
  n}\Big(\frac{\Lambda}{m_e}\Big) + \frac{5}{4}\Big).
\end{eqnarray}
For the last two integrals in Eq.(\ref{eq:B.63}) we obtain the
following result
\begin{eqnarray}\label{eq:B.65}
\hspace{-0.3in}&& \gamma^{\alpha} \int^1_0dy\int^y_0 dx\,{\ell
  n}\Big(\frac{m^2_e y^2 + \mu^2(1 - y)}{m^2_e y^2 + \mu^2(1 - y) - 2
  k_e\cdot k\, x (1 - y)}\Big) + \int^1_0dy\int^y_0dx\, \frac{1}{m^2_e
  y^2 + \mu^2(1 - y) - 2 k_e\cdot k\, x (1 - y)}\nonumber\\
\hspace{-0.3in}&&\times\,\Big\{\Big[m^2_e\, y\, (2 - y) - 2 k_e\cdot
  k\, (1 - x) (1 - y) + m_e \hat{k}\, (1 - y)\Big]\,\gamma^{\alpha} +
\Big[ - 2 m_e k^{\alpha}_e (1 - y^2) + 2 k^{\alpha}_e \hat{k} (1 -
  x)(1 - y)\Big]\Big\}=\nonumber\\
\hspace{-0.3in}&& = \gamma^{\alpha}\Big\{ \frac{1}{2} + {\ell n}\Big(-
\frac{2 k_e\cdot k}{m^2_e}\Big)\,\Big[ \frac{m^2_e + k_e\cdot k}{m^2_e
    + 2 k_e\cdot k} - \frac{m^2_e}{2 k_e\cdot k}\,{\ell n}\Big(1 +
  \frac{2k_e\cdot k}{m^2_e}\Big)\Big] - \frac{m^2_e}{2 k_e\cdot
  k}\,{\rm Li}_2\Big(- \frac{2k_e\cdot k}{m^2_e}\Big) + \frac{m_e
  \hat{k}}{m^2_e + 2k_e\cdot k}\nonumber\\
\hspace{-0.3in}&&\times\,{\ell n}\Big(- \frac{2 k_e\cdot
  k}{m^2_e}\Big)\Big\} + \frac{k^{\alpha}_e}{m_e}\Big\{- \frac{
  m^2_e}{m^2_e + 2 k_e\cdot k} + 2\,\frac{m^2_e(m^2_e + 3 k_e\cdot
  k)}{(m^2_e + 2 k_e\cdot k)^2}\,{\ell n}\Big(- \frac{2 k_e\cdot
  k}{m^2_e}\Big)\Big\} + \frac{k^{\alpha}_e\hat{k}}{k_e\cdot
  k}\,\Big\{\frac{m^2_e + k_e\cdot k}{m^2_e + 2 k_e\cdot k}\nonumber\\
\hspace{-0.3in}&& + {\ell n}\Big(- \frac{2 k_e\cdot
  k}{m^2_e}\Big)\,\Big[- \frac{(m^2_e + k_e\cdot k)(m^2_e + 4 k_e\cdot
    k)}{(m^2_e + 2 k_e\cdot k)^2} + \frac{m^2_e}{2 k_e\cdot k}\,{\ell
    n}\Big(1 + \frac{2k_e\cdot k}{m^2_e}\Big)\Big] + \frac{m^2_e}{2
  k_e\cdot k}\,{\rm Li}_2\Big(- \frac{2k_e\cdot k}{m^2_e}\Big)\Big\},
\end{eqnarray}
where we have used a relation $\hat{k}\gamma^{\alpha} = -
\gamma^{\alpha}\hat{k} + 2k^{\alpha}$ and omitted the term
$2k^{\alpha}$.  Then, ${\rm Li}_2(- 2k_e \cdot k/m^2_e)$ is the Spence
function. Summing up the contributions, for the vertex
$\Lambda^{\alpha}_e(k_e,k)$ we obtain the following analytical
expression
\begin{eqnarray}\label{eq:B.66}
\hspace{-0.3in}&&\Lambda^{\alpha}_e(k_e, k) = (Z^{(e)}_1 -
1)\,\gamma^{\alpha} + \frac{e^2}{8\pi^2}\,\gamma^{\alpha}\,\Big({\ell
  n}\Big(\frac{\Lambda}{m_e}\Big) + \frac{5}{4}\Big)\nonumber\\
\hspace{-0.3in}&& + \frac{e^2}{8\pi^2}\,\gamma^{\alpha}\Big\{ -1 +
       {\ell n}\Big(- \frac{2 k_e\cdot k}{m^2_e}\Big)\,\Big[
         \frac{m^2_e + k_e\cdot k}{m^2_e + 2 k_e\cdot k} -
         \frac{m^2_e}{2 k_e\cdot k}\,{\ell n}\Big(1 + \frac{2k_e\cdot
           k}{m^2_e}\Big)\Big] - \frac{m^2_e}{2 k_e\cdot k}\,{\rm
         Li}_2\Big(- \frac{2k_e\cdot k}{m^2_e}\Big)\nonumber\\
\hspace{-0.3in}&& + \frac{m_e \hat{k}}{m^2_e + 2k_e\cdot k}\,{\ell
  n}\Big(- \frac{2 k_e\cdot k}{m^2_e}\Big)\Big\} +
\frac{e^2}{8\pi^2}\,\frac{k^{\alpha}_e}{m_e}\Big\{- \frac{m^2_e}{m^2_e
  + 2 k_e\cdot k} + 2\,\frac{m^2_e(m^2_e + 3 k_e\cdot k)}{(m^2_e + 2
  k_e\cdot k)^2}\,{\ell n}\Big(- \frac{2 k_e\cdot
  k}{m^2_e}\Big)\Big\}\nonumber\\
\hspace{-0.3in}&& +
\frac{e^2}{8\pi^2}\,\frac{k^{\alpha}_e\hat{k}}{k_e\cdot
  k}\,\Big\{\frac{m^2_e + k_e\cdot k}{m^2_e + 2 k_e\cdot k} - {\ell
  n}\Big(- \frac{2 k_e\cdot k}{m^2_e}\Big)\,\Big[\frac{(m^2_e +
    k_e\cdot k)(m^2_e + 4 k_e\cdot k)}{(m^2_e + 2 k_e\cdot k)^2} -
  \frac{m^2_e}{2 k_e\cdot k}\,{\ell n}\Big(1 + \frac{2k_e\cdot
    k}{m^2_e}\Big)\Big]\nonumber\\
\hspace{-0.3in}&& + \frac{m^2_e}{2 k_e\cdot k}\,{\rm Li}_2\Big(-
\frac{2k_e\cdot k}{m^2_e}\Big)\Big\}.
\end{eqnarray}
Now we may proceed to the calculation of the electron self--energy
correction $\Sigma^{(e)}(k_e,k)$, defined by Eq.(\ref{eq:B.61}). The
result is
\begin{eqnarray}\label{eq:B.67}
\hspace{-0.3in}&&\Sigma^{(e)}(k_e, k) = - \delta m_e - (Z^{(e)}_2 -
1)\, (m_e - \hat{k}_e - \hat{k}) - 3
m_e\,\frac{e^2}{8\pi^2}\,\Big({\ell n}\Big(\frac{\Lambda}{m_e}\Big) +
\frac{1}{4}\Big) - (m_e - \hat{k}_e -
\hat{k})\,\frac{e^2}{8\pi^2}\,\Big({\ell
  n}\Big(\frac{\Lambda}{m_e}\Big) + \frac{5}{4}\Big) \nonumber\\
\hspace{-0.3in}&& + (m_e - \hat{k}_e -
\hat{k})\,\frac{e^2}{8\pi^2}\,\Big[\frac{k_e\cdot k}{m^2_e + 2
    k_e\cdot k} + 2\,\frac{(k_e\cdot k)(m^2_e + k_e\cdot k)}{(m^2_e +
    2 k_e\cdot k)^2}\,{\ell n}\Big(- \frac{2 k_e\cdot
    k}{m^2_e}\Big)\Big] + m_e\,\frac{e^2}{8\pi^2}\Big[- \frac{
    k_e\cdot k}{m^2_e + 2 k_e \cdot k}\nonumber\\
\hspace{-0.3in}&& + 2\,\frac{(k_e\cdot k)(m^2_e + 3 k_e\cdot
  k)}{(m^2_e + 2 k_e \cdot k)^2}\,{\ell n}\Big(- \frac{2 k_e\cdot
  k}{m^2_e}\Big)\Big].
\end{eqnarray}
The renormalization parameters $Z^{(e)}_1$, $\delta m_e$ and
$Z^{(e)}_2$ are given by
\begin{eqnarray}\label{eq:B.68}
\hspace{-0.3in}\delta m_e &=& - 3 m_e\frac{e^2}{8\pi^2}\,\Big({\ell
  n}\Big(\frac{\Lambda}{m_e}\Big) + \frac{1}{4}\Big)\nonumber\\
\hspace{-0.3in}Z^{(e)}_1 &=& Z^{(e)}_2 = 1 -
\frac{e^2}{8\pi^2}\,\Big({\ell n}\Big(\frac{\Lambda}{m_e}\Big) +
\frac{5}{4}\Big),
\end{eqnarray}
where the counterterms $Z^{(e)}_1$ and $Z^{(e)}_2$ obey the Ward
identity $Z^{(e)}_1 = Z^{(e)}_2 $ \cite{Itzykson1980}. Thus, for the
renormalized $(e^-e^-\gamma)$ vertex $\bar{\Lambda}^{\alpha}_e(k_e,
k)$ and electron self--energy corrections $\bar{\Sigma}^{(e)}(k_e,k)$
and $\bar{\Sigma}^{(e)}(k_e)$ we obtain the following expressions
\begin{eqnarray}\label{eq:B.69}
\hspace{-0.3in}&& \varepsilon^*_{\lambda'}\cdot \bar{\Lambda}_e(k_e,
k) =
\frac{e^2}{8\pi^2}\,\Big[\hat{\varepsilon}^*_{\lambda'}\,F_1(k_e\cdot
  k) + \frac{\hat{\varepsilon}^*_{\lambda'}\hat{k}}{m_e}\,F_2(k_e\cdot
  k) + \frac{\varepsilon^*_{\lambda'}\cdot k_e}{m_e}\,F_3(k_e\cdot k)
  + \frac{\varepsilon^*_{\lambda'}\cdot k_e}{k_e\cdot
    k}\,\hat{k}\,F_4(k_e\cdot k)\Big],\nonumber\\
\hspace{-0.3in}&&\bar{\Sigma}^{(e)}(k_e,k) =
m_e\,\frac{e^2}{8\pi^2}\,F_5(k_e\cdot k) + (m_e - \hat{k}_e -
\hat{k})\,\frac{e^2}{8\pi^2}\,F_6(k_e\cdot k),\nonumber\\
\hspace{-0.3in}&&\bar{\Sigma}^{(e)}(k_e) = 0,
\end{eqnarray}
where the functions $F_j(k_e\cdot k)$ for $j = 1,2,\ldots,6$ are equal
to
\begin{eqnarray}\label{eq:B.70}
\hspace{-0.3in}&&F_1(k_e\cdot k) = -1 + {\ell n}\Big(- \frac{2
  k_e\cdot k}{m^2_e}\Big)\,\Big[ \frac{m^2_e + k_e\cdot k}{m^2_e + 2
    k_e\cdot k} - \frac{m^2_e}{2 k_e\cdot k}\,{\ell n}\Big(1 +
  \frac{2k_e\cdot k}{m^2_e}\Big)\Big] - \frac{m^2_e}{k_e\cdot k}\,{\rm
  Li}_2\Big(- \frac{2k_e\cdot k}{m^2_e}\Big),\nonumber\\
\hspace{-0.3in}&&F_2(k_e\cdot k) = \frac{ m^2_e}{m^2_e + 2k_e\cdot
  k}\,{\ell n}\Big(- \frac{2k_e\cdot k}{m^2_e}\Big),\nonumber\\
\hspace{-0.3in}&&F_3(k_e\cdot k) = - \frac{m^2_e}{m^2_e + 2 k_e\cdot
  k} + 2\,\frac{m^2_e(m^2_e + 3 k_e\cdot k)}{(m^2_e + 2 k_e\cdot
  k)^2}\,{\ell n}\Big(- \frac{2k_e\cdot k}{m^2_e}\Big),\nonumber\\
\hspace{-0.3in}&&F_4(k_e\cdot k) = \frac{m^2_e + k_e\cdot k}{m^2_e + 2
  k_e\cdot k} - {\ell n}\Big(- \frac{2 k_e\cdot
  k}{m^2_e}\Big)\,\Big[\frac{(m^2_e + k_e\cdot k)(m^2_e + 4 k_e\cdot
    k)}{(m^2_e + 2 k_e\cdot k)^2} - \frac{m^2_e}{2 k_e\cdot k}\,{\ell
    n}\Big(1 + \frac{2k_e\cdot k}{m^2_e}\Big)\Big]\nonumber\\
\hspace{-0.3in}&& + \frac{m^2_e}{2 k_e\cdot k}\,{\rm Li}_2\Big(-
\frac{2k_e\cdot k}{m^2_e}\Big),\nonumber\\
\hspace{-0.3in}&&F_5(k_e\cdot k) = - \frac{k_e\cdot k}{m^2_e + 2 k_e
  \cdot k} + 2\,\frac{(k_e\cdot k)(m^2_e + 3 k_e\cdot k)}{(m^2_e + 2
  k_e \cdot k)^2}\,{\ell n}\Big(- \frac{2 k_e\cdot
  k}{m^2_e}\Big),\nonumber\\
\hspace{-0.3in}&&F_6(k_e\cdot k) = \frac{k_e\cdot k}{m^2_e + 2
  k_e\cdot k} + 2\,\frac{(k_e\cdot k)(m^2_e + k_e\cdot k)}{(m^2_e + 2
  k_e\cdot k)^2}\,{\ell n}\Big(- \frac{2 k_e\cdot k}{m^2_e}\Big).
\end{eqnarray}
After renormalization the sum of the diagrams in
Fig.\,\ref{fig:fig2}d, Fig.\,\ref{fig:fig2}e and Fig.\,\ref{fig:fig2}f
is equal to
\begin{eqnarray}\label{eq:B.71}
\hspace{-0.3in}&&\sum_{j =d,e,f}{\cal M}_{\rm Fig.\,\ref{fig:fig2}j}(n
\to p e^- \bar{\nu}_e \gamma)_{\lambda'} = - \Big[\bar{u}_p(\vec{k}_p,
  \sigma_p) \,\gamma^{\mu}(1 + \lambda \gamma^5)\, u_n(\vec{k}_n,
  \sigma_n)\Big]\,\Big[\bar{u}_e(\vec{k}_e,\sigma_e)\,
  \varepsilon^*_{\lambda'}\cdot \bar{\Lambda}_e(k_e, k)\,\frac{1}{m_e
    - \hat{k}_e - \hat{k} - i0} \nonumber\\
\hspace{-0.3in}&&\times\,\gamma_{\mu} (1 - \gamma^5)\,
v_{\nu}(\vec{k}_{\nu}, + \frac{1}{2})\Big] - \Big[\bar{u}_p(\vec{k}_p,
  \sigma_p) \,\gamma^{\mu}(1 + \lambda \gamma^5)\, u_n(\vec{k}_n,
  \sigma_n)\Big]\,\Big[\bar{u}_e(\vec{k}_e,\sigma_e)\,
  \hat{\varepsilon}^*_{\lambda'}\,\frac{1}{m_e - \hat{k}_e - \hat{k} -
    i0}\,\bar{\Sigma}^{(e)}(k_e,k)\nonumber\\
\hspace{-0.3in}&&\times\,\frac{1}{m_e - \hat{k}_e - \hat{k} -
  i0}\,\gamma_{\mu} (1 - \gamma^5)\, v_{\nu}(\vec{k}_{\nu}, +
\frac{1}{2})\Big],
\end{eqnarray}
where we have taken into account that $\bar{\Sigma}^{(e)}(k_e) = 0$.
The correctness of the calculation of the renormalized
$(e^-e^-\gamma)$ vertex function and electron self--energy corrections
we verify by using the Ward identity \cite{Itzykson1980} (see also
\cite{Ivanov1973}). Indeed, one may show that the amplitude
Eq.(\ref{eq:B.71}) is invariant under the gauge transformation
$\varepsilon^*_{\lambda'}(k) \to \varepsilon^*_{\lambda'}(k) + c\,k$
if $\bar{\Lambda}^{\alpha}_e(k_e,k)$ and $\bar{\Sigma}^{(e)}(k_e, k)$
obey the Wald identity
\begin{eqnarray}\label{eq:B.72}
k\cdot \bar{\Lambda}_e(k_e,k) =
\bar{\Sigma}^{(e)}(k_e,k),
\end{eqnarray}
multiplied by $\bar{u}_e(\vec{k}_e, \sigma_e)$ or at $\hat{k}_e = m_e$
\cite{Itzykson1980}, The Ward identity Eq.(\ref{eq:B.72}) imposes the
following relations between the functions $F_j(k_e\cdot k)$:
\begin{eqnarray}\label{eq:B.73}
F_1(k_e,k) + F_4(k_e\cdot k) &=& - F_6(k_e\cdot k),\nonumber\\
\frac{k_e\cdot k}{m_e}\,F_3(k_e\cdot k) &=& m_e\,F_5(k_e\cdot k).
\end{eqnarray}
The functions $F_j(k_e\cdot k)$ with $j = 1,3,4,5,6$ in
Eq.(\ref{eq:B.70}) fulfil the constraints Eq.(\ref{eq:B.73}).  

Since to leading order in the large proton mass expansion and in the
physical gauge of a photon the diagrams Fig.\,\ref{fig:fig2}a,
Fig.\,\ref{fig:fig2}b and Fig.\,\ref{fig:fig2}c vanish, the
contribution of the diagrams in Fig.\,\ref{fig:fig2} to the amplitude
of the neutron radiative $\beta^-$--decay is defined fully by the
diagrams Fig.\,\ref{fig:fig2}d, Fig.\,\ref{fig:fig2}e and
Fig.\,\ref{fig:fig2}f, respectively.  In the non--relativistic limit
the contribution of the diagrams Fig.\,\ref{fig:fig2} is given by
\begin{eqnarray}\label{eq:B.74}
\hspace{-0.3in}&&{\cal M}_{\rm Fig.\,\ref{fig:fig2}}(n \to p e^-
\bar{\nu}_e \gamma)_{\lambda'} = 2 m_n\,
\frac{e^2}{8\pi^2}\,\frac{1}{2k_e\cdot
  k}\,\Big\{[\varphi^{\dagger}_p\varphi_n]\,\Big\{
\bar{u}_e(\vec{k}_e,\sigma_e)\,\Big[2\,\varepsilon_{\lambda'}\cdot k_e
  \,\Big(F_1 + F_3 + F_4 - \frac{m^2_e}{k_e\cdot k}\,F_5 +
  F_6\Big)\nonumber\\
\hspace{-0.3in}&& + \hat{\varepsilon}^*_{\lambda'}\,\hat{k}\,\Big(F_1
+ 2F_2 - \frac{m^2_e}{k_e\cdot k}\,F_5 + F_6\Big) + \frac{2k_e\cdot
  k}{m_e}\, \hat{\varepsilon}^*_{\lambda'}\,\Big(F_2 - \frac{m^2_e}{2
  k_e\cdot k}\,F_5\Big) + \frac{\varepsilon^*_{\lambda'}\cdot
  k_e}{m_e}\,\hat{k}\,\Big(- 2 F_2 + F_3\Big)\Big]\,\gamma^0
\nonumber\\
\hspace{-0.3in}&&\times\, \,(1 - \gamma^5)\,
v_{\nu}(\vec{k}_{\nu}, + \frac{1}{2})\Big\} - \lambda
       [\varphi^{\dagger}_p\vec{\sigma}\,\varphi_n] \cdot
       \Big\{\bar{u}_e(\vec{k}_e,\sigma_e)\,\Big[2\,
         \varepsilon_{\lambda'}\cdot k_e \,\Big(F_1 + F_3 + F_4 -
         \frac{m^2_e}{k_e\cdot k}\,F_5 + F_6\Big)\nonumber\\
\hspace{-0.3in}&& + \hat{\varepsilon}^*_{\lambda'}\,\hat{k}\,\Big(F_1
+ 2F_2 - \frac{m^2_e}{k_e\cdot k}\,F_5 + F_6\Big) + \frac{2k_e\cdot
  k}{m_e}\, \hat{\varepsilon}^*_{\lambda'}\,\Big(F_2 - \frac{m^2_e}{2
  k_e\cdot k}\,F_5\Big) + \frac{\varepsilon^*_{\lambda'}\cdot
  k_e}{m_e}\,\hat{k}\,\Big(- 2 F_2 +
F_3\Big)\Big]\,\vec{\gamma}\nonumber\\
\hspace{-0.3in}&&\times \, (1 - \gamma^5)\,
v_{\nu}(\vec{k}_{\nu}, + \frac{1}{2})\Big]\Big\}\Big\}.
\end{eqnarray}
Using the relations Eq.(\ref{eq:B.73}), which are imposed by gauge
invariance, we transcribe Eq.(\ref{eq:B.74}) into the form
\begin{eqnarray}\label{eq:B.75}
\hspace{-0.3in}&&{\cal M}_{\rm Fig.\,\ref{fig:fig2}}(n
\to p e^- \bar{\nu}_e \gamma)_{\lambda'} =
2 m_n\,\frac{e^2}{8\pi^2}\,\frac{1}{2k_e\cdot
  k}\,\Big\{[\varphi^{\dagger}_p\varphi_n]\,\Big\{
\bar{u}_e(\vec{k}_e,\sigma_e)\,\Big[
  \hat{\varepsilon}^*_{\lambda'}\,\hat{k}\,\Big(2 F_2 - F_3 - F_4\Big)
  + \frac{k_e\cdot k}{m_e}\, \hat{\varepsilon}^*_{\lambda'}\,\Big(2
  F_2 - F_3\Big)\nonumber\\
\hspace{-0.3in}&& + \frac{\varepsilon^*_{\lambda'}\cdot
  k_e}{m_e}\,\hat{k}\,\Big(- 2 F_2 + F_3\Big)\Big]\,\gamma^0 \,(1 -
\gamma^5)\, v_{\nu}(\vec{k}_{\nu}, + \frac{1}{2})\Big\} - \lambda
      [\varphi^{\dagger}_p\vec{\sigma}\,\varphi_n] \cdot
      \Big\{\bar{u}_e(\vec{k}_e,\sigma_e)\,\Big[
        \hat{\varepsilon}^*_{\lambda'}\,\hat{k}\,\Big(2 F_2 - F_3 -
        F_4\Big)\nonumber\\
\hspace{-0.3in}&& + \frac{k_e\cdot k}{m_e}\,
\hat{\varepsilon}^*_{\lambda'}\,\Big(2 F_2 - F_3\Big) +
\frac{\varepsilon^*_{\lambda'}\cdot k_e}{m_e}\,\hat{k}\,\Big(- 2 F_2 +
F_3\Big)\Big] \,\vec{\gamma}\, (1 - \gamma^5)\, v_{\nu}(\vec{k}_{\nu},
      + \frac{1}{2})\Big\}\Big\}.
\end{eqnarray}
The amplitude Eq.(\ref{eq:B.75}) is gauge invariant. It vanishes after
the replacement $\varepsilon^*_{\lambda'} \to k$ for a photon on--mass
shell $k^2 = 0$. The contribution of the diagrams in
Fig.\,\ref{fig:fig2} to the rate of the neutron radiative
$\beta^-$--decay with photon energies from the interval $\omega_{\rm
  min} \le \omega \le \omega_{\rm max}$ is given by
\begin{eqnarray}\label{eq:B.76}
\hspace{-0.3in}&&\lambda^{(\rm
  Fig\,\ref{fig:fig2})}_{\beta\gamma}(\omega_{\rm max},\omega_{\rm
  min}) = (1 +
3\lambda^2)\,\frac{\alpha^2}{\pi^2}\,\frac{G^2_FV_{ud}|^2}{32\pi^3}
\int^{\omega_{\rm max}}_{\omega_{\rm min}} d\omega\int^{E_0 -
  \omega}_{m_e} dE_e\,F(E_e, Z = 1)\,(E_0 - E_e - \omega)\,\sqrt{E^2_e
  - m^2_e}\;\omega\nonumber\\
\hspace{-0.3in}&&\times \,
\int\frac{d\Omega_{e\gamma}}{4\pi}\int\frac{d\Omega_{\nu}}{4\pi}\,
\Big[\frac{1}{2}\sum_{\rm pol, \lambda'}\frac{1}{1 +
    3\lambda^2}\Big({\cal M}^{\dagger}_{\rm Fig.\,\ref{fig:fig1}}(n
  \to p e^- \bar{\nu}_e \gamma)_{\lambda'} \tilde{\cal M}_{\rm
    Fig.\,\ref{fig:fig2}}(n \to p e^- \bar{\nu}_e
  \gamma)_{\lambda'} + {\rm h.c.}\Big)\Big]\Big|_{E_{\nu} = E_0 - E_e - \omega},
\end{eqnarray}
where the abbreviation ${\rm h.c.}$ means ``hermitian
conjugate''. Then, $\tilde{\cal M}_{\rm Fig.\,\ref{fig:fig2}} =
(8\pi^2/2m_n e^2)\,{\cal M}_{\rm Fig.\,\ref{fig:fig2}}$ and
$d\Omega_{e\gamma}$ and $d\Omega_{\nu}$ are infinitesimal solid angle
elements of the electron--photon momentum correlations and
antineutrino, respectively. The sum over polarizations of interacting
particles is defined by the following traces over Dirac matrices
\begin{eqnarray}\label{eq:B.77}
\hspace{-0.3in}&&\sum_{\rm pol, \lambda'}\frac{1}{1 +
  3\lambda^2}\Big({\cal M}^{\dagger}_{\rm Fig.\,\ref{fig:fig1}}(n \to
p e^- \bar{\nu}_e \gamma)_{\lambda'} \tilde{\cal M}_{\rm
  Fig.\,\ref{fig:fig2}}(n \to p e^- \bar{\nu}_e \gamma)_{\lambda'} +
{\rm h. c.}\Big)= \frac{1}{1 + 3\lambda^2}\, \frac{2}{(2k_e\cdot
  k)^2}\nonumber\\
\hspace{-0.3in}&&\times\,\sum_{ \lambda'}\Big\{{\rm tr}\{(m_e +
\hat{k}_e)\,\Big[ \hat{\varepsilon}^*_{\lambda'}\,\hat{k}\,\Big(2 F_2
  - F_3 - F_4\Big) + \frac{k_e\cdot k}{m_e}\,
  \hat{\varepsilon}^*_{\lambda'}\,\Big(2 F_2 - F_3\Big) +
  \frac{\varepsilon^*_{\lambda'}\cdot k_e}{m_e}\,\hat{k}\,\Big(- 2 F_2
  + F_3\Big)\Big]\,\gamma^0 \,(1 - \gamma^5)\nonumber\\
\hspace{-0.3in}&&\times\,\hat{k}_{\nu} \gamma^0 \Big(2 k_e\cdot
\varepsilon_{\lambda'} + \hat{k} \hat{\varepsilon}_{\lambda'}\Big)\,(1
- \gamma^5)\} + \lambda\,\delta^{ij}\, {\rm tr}\{(m_e +
\hat{k}_e)\,\Big[ \hat{\varepsilon}^*_{\lambda'}\,\hat{k}\,\Big(2 F_2
  - F_3 - F_4\Big) + \frac{k_e\cdot k}{m_e}\,
  \hat{\varepsilon}^*_{\lambda'}\,\Big(2 F_2 - F_3\Big)\nonumber\\
\hspace{-0.3in}&& + \frac{\varepsilon^*_{\lambda'}\cdot
  k_e}{m_e}\,\hat{k}\,\Big(- 2 F_2 + F_3\Big)\Big]\,\vec{\gamma}^{\,i}
\,(1 - \gamma^5)\,\hat{k}_{\nu} \vec{\gamma}^{\,j} \Big(2 k_e\cdot
\varepsilon_{\lambda'} + \hat{k} \hat{\varepsilon}_{\lambda'}\Big)\,(1
- \gamma^5)\} + {\rm h.c.}\Big\}.
\end{eqnarray}
Having integrated over directions of the antineutrino momentum
$\vec{k}_{\nu}$ we get
\begin{eqnarray}\label{eq:B.78}
\hspace{-0.3in}&&\int \frac{d\Omega_{\nu}}{4\pi}\,\sum_{\rm pol, \lambda'}
\frac{1}{1 + 3\lambda^2}\Big({\cal M}^{\dagger}_{\rm
  Fig.\,\ref{fig:fig1}}(n \to p e^- \bar{\nu}_e \gamma)_{\lambda'}
\tilde{\cal M}_{\rm Fig.\,\ref{fig:fig2}}(n \to p e^- \bar{\nu}_e
\gamma)_{\lambda'} + {\rm h. c.}\Big)= \frac{E_{\nu}}{(k_e\cdot
  k)^2}\nonumber\\
\hspace{-0.3in}&&\times\,\sum_{ \lambda'}\Big\{{\rm tr}\{(m_e +
\hat{k}_e)\,\Big[ \hat{\varepsilon}^*_{\lambda'}\,\hat{k}\,\Big(2 F_2
  - F_3 - F_4\Big) + \frac{k_e\cdot k}{m_e}\,
  \hat{\varepsilon}^*_{\lambda'}\,\Big(2 F_2 - F_3\Big) +
  \frac{\varepsilon^*_{\lambda'}\cdot k_e}{m_e}\,\hat{k}\,\Big(- 2 F_2
  + F_3\Big)\Big]\nonumber\\
\hspace{-0.3in}&&\times\,\gamma^0\,\Big(2 k_e\cdot
\varepsilon_{\lambda'} + \hat{k} \hat{\varepsilon}_{\lambda'}\Big)\,(1
- \gamma^5)\} + {\rm h.c.}\Big\}.
\end{eqnarray}
The traces over the Dirac matrices are equal to 
\begin{eqnarray}\label{eq:B.79}
\sum_{ \lambda'}{\rm tr}\{(m_e +
\hat{k}_e)\,\hat{\varepsilon}^*_{\lambda'}\,\hat{k}\,\gamma^0\,\Big(2
k_e\cdot \varepsilon_{\lambda'} + \hat{k}
\hat{\varepsilon}_{\lambda'}\Big)\,(1 - \gamma^5)\} &=&
8\,\omega\,\Big(k^2_e - (\vec{k}_e\cdot \vec{n}_{\vec{k}})^2\Big) +
16\,\omega^2\,\Big(E_e - \vec{k}_e\cdot
\vec{n}_{\vec{k}}\Big),\nonumber\\ \sum_{ \lambda'}{\rm tr}\{(m_e +
\hat{k}_e)\,\hat{\varepsilon}^*_{\lambda'}\,\gamma^0\,\Big(2 k_e\cdot
\varepsilon_{\lambda'} + \hat{k} \hat{\varepsilon}_{\lambda'}\Big)\,(1
- \gamma^5)\} &=& - 8\,m_e \,\omega,\nonumber\\ \sum_{
  \lambda'}(k_e\cdot \varepsilon^*_{\lambda'})\,{\rm tr}\{(m_e +
\hat{k}_e)\,\hat{k}\,\gamma^0\,\Big(2 k_e\cdot \varepsilon_{\lambda'}
+ \hat{k} \hat{\varepsilon}_{\lambda'}\Big)\,(1 - \gamma^5)\} &=& 
8\,m_e \,\omega \,\Big(k^2_e - (\vec{k}_e\cdot \vec{n}_{\vec{k}})^2\Big).
\end{eqnarray}
Thus, we get
\begin{eqnarray}\label{eq:B.80}
\hspace{-0.3in}&&\int
\frac{d\Omega_{\nu}}{4\pi}\,\frac{1}{2}\,\sum_{\rm pol, \lambda'}
\frac{1}{1 + 3\lambda^2}\Big({\cal M}^{\dagger}_{\rm
  Fig.\,\ref{fig:fig1}}(n \to p e^- \bar{\nu}_e \gamma)_{\lambda'}
\tilde{\cal M}_{\rm Fig.\,\ref{fig:fig2}}(n \to p e^- \bar{\nu}_e
\gamma)_{\lambda'} + {\rm h. c.}\Big) = \frac{E_{\nu}}{(k_e\cdot
  k)^2}\nonumber\\
\hspace{-0.3in}&&\times\,\Big\{8\,\omega\,{\rm Re}\,F_4\,\Big(k^2_e -
(\vec{k}_e\cdot \vec{n}_{\vec{k}})^2\Big) + 8\,\omega^2\, {\rm Re}\,(2
F_2 - F_3 - 2F_4)\,\Big(E_e - \vec{k}_e \cdot
\vec{n}_{\vec{k}}\Big)\Big\},
\end{eqnarray}
where ${\rm Re}F_4$ and ${\rm Re}(2 F_2 - F_3 - 2F_4)$ are the real
parts of the functions $F_2$, $F_3$ and $F_4$, i.e. ${\rm Re}\,F_j =
(F_j + F^*_j)/2$ for $j = 2,3,4$. Plugging Eq.(\ref{eq:B.80}) into
Eq.(\ref{eq:B.76}) we obtain
\begin{eqnarray}\label{eq:B.81}
\hspace{-0.3in}\lambda^{(\rm
  Fig\,\ref{fig:fig2})}_{\beta\gamma}(\omega_{\rm max},\omega_{\rm
  min}) &=& (1 +
3\lambda^2)\,\frac{\alpha^2}{\pi^2}\,\frac{G^2_FV_{ud}|^2}{4\pi^3}
\int^{\omega_{\rm max}}_{\omega_{\rm min}} d\omega\int^{E_0 -
  \omega}_{m_e} dE_e\,F(E_e, Z = 1)\,(E_0 - E_e -
\omega)^2\,\sqrt{E^2_e - m^2_e}\nonumber\\
\hspace{-0.3in}&&\times \,
\int\frac{d\Omega_{e\gamma}}{4\pi}\,\Big\{\frac{k^2_e - (\vec{k}_e
  \cdot \vec{n}_{\vec{k}})^2}{(E_e - \vec{k}_e\cdot
  \vec{n}_{\vec{k}})^2}\,{\rm Re}\,F_4 + \frac{\omega}{E_e - \vec{k}_e\cdot
  \vec{n}_{\vec{k}}}\, {\rm Re}\,(2 F_2 - F_3 - 2F_4)\Big\},
\end{eqnarray}
where $F_2$, $F_3$ and $F_4$ are given in Eq.(\ref{eq:B.70}) as
functions of $k_e\cdot k = \omega\,(E_e - \vec{k}_e\cdot
\vec{n}_{\vec{k}})$. It is important to emphasize that the
contribution of the diagrams in Fig.\,\ref{fig:fig2} to the rate of
the neutron radiative $\beta^-$--decay is not infrared divergent.

\section*{Appendix C: The amplitude of the neutron 
radiative $\beta^-$--decay, described by Feynman diagrams in
Fig.\,\ref{fig:fig3}} \renewcommand{\theequation}{C-\arabic{equation}}
\setcounter{equation}{0}

Since, according to our calculations in Appendix B, the renormalized
self--energy corrections $\Sigma^{(p)}(k_p)$ and $\Sigma^{(e)}(k_e)$
of the proton and electron vanish, the contributions of the Feynman
diagrams Fig.\,\ref{fig:fig3}a and Fig.\,\ref{fig:fig3}b to the
amplitude of the neutron radiative $\beta^-$--decay vanish. At first
glimpse, non--trivial contributions of the diagrams in
Fig.\,\ref{fig:fig3} are given by the diagrams in
Fig.\,\ref{fig:fig3}c and Fig.\,\ref{fig:fig3}d. The analytical
expression for the diagrams Fig.\,\ref{fig:fig3}c and
Fig.\,\ref{fig:fig3}d are
\begin{eqnarray}\label{eq:C.1}
&&{\cal M}_{\rm Fig.\,\ref{fig:fig3}c}(n \to p e^-
  \bar{\nu}_e\gamma)_{\lambda'} =\nonumber\\ &&= 
  e^2\int\frac{d^4q}{(2\pi)^4i}\,\Big[\bar{u}_p(\vec{k}_p,\sigma_p)\,
    \gamma^{\alpha}\frac{1}{m_p - \hat{k}_p + \hat{q}
      -i0}\,J^{\mu}(k_p, k_p + k)\,
    u_n(\vec{k}_n,\sigma_n)\Big]\nonumber\\ &&\times
  \,\Big[\bar{u}_e(\vec{k}_e,\sigma_e)\,\hat{\varepsilon}^*_{\lambda'}\,
    \frac{1}{m_e - \hat{k}_e - \hat{k} - i
      0}\,\gamma^{\beta}\frac{1}{m_e - \hat{k}_e - \hat{k}- \hat{q} -
      i0}\gamma_{\mu}(1 - \gamma^5) v_{\nu}(\vec{k}_{\nu}, +
    \frac{1}{2}\,)\Big]\,D_{\alpha\beta}(q)
\end{eqnarray}
and
\begin{eqnarray}\label{eq:C.2}
&&{\cal M}_{\rm Fig.\,\ref{fig:fig3}d}(n \to p e^-
  \bar{\nu}_e\gamma)_{\lambda'} =\nonumber\\ &&= -
  e^2\int\frac{d^4q}{(2\pi)^4i}\,\Big[\bar{u}_p(\vec{k}_p,\sigma_p)\,
    \hat{\varepsilon}^*_{\lambda'}\,\frac{1}{m_p - \hat{k}_p - \hat{k}
      - i0} \gamma^{\alpha}\frac{1}{m_p - \hat{k}_p + \hat{q} - \hat{k}
        -i0}\,J^{\mu}(k_p, k_p + k)\,
      u_n(\vec{k}_n,\sigma_n)\Big]\nonumber\\ &&\times
    \,\Big[\bar{u}_e(\vec{k}_e,\sigma_e)\,\gamma^{\beta}\frac{1}{m_e -
        \hat{k}_e - \hat{q} - i0}\gamma_{\mu}(1 - \gamma^5)
      v_{\nu}(\vec{k}_{\nu}, +
      \frac{1}{2}\,)\Big]\,D_{\alpha\beta}(q),
\end{eqnarray}
where $J^{\mu}(k_p, k_p + k)$ is a hadronic current, described by the
shaded region of the diagrams. In case of strong interactions, defined
by only the axial coupling constant $\lambda$, the hadronic current
$J^{\mu}(k_p, k_p + k)$ is equal to $J^{\mu}(k_p, k_p + k) =
\gamma^{\mu}(1 + \lambda\,\gamma^5)$. Let us check invariance of the
Feynman diagrams Fig.\,\ref{fig:fig3}c and Fig.\,\ref{fig:fig3}d under
gauge transformations $\varepsilon^*_{\lambda'}(k) \to
\varepsilon^*_{\lambda'}(k) + ck$ and $D_{\alpha\beta}(q) \to
D_{\alpha\beta}(q) + c(q^2)\,q_{\alpha}q_{\beta}$. Making, first, a
gauge transformation $\varepsilon^*_{\lambda'}(k) \to
\varepsilon^*_{\lambda'}(k) + ck$ for the contributions of the term
$ck$ we get the following expressions
\begin{eqnarray}\label{eq:C.3}
&&{\cal M}_{\rm Fig.\,\ref{fig:fig3}c}(n \to p e^-
  \bar{\nu}_e\gamma)_{\lambda'}\Big|_{\varepsilon^*_{\lambda'}(k)\to
    k} =\nonumber\\ &&= -
  e^2\int\frac{d^4q}{(2\pi)^4i}\,\Big[\bar{u}_p(\vec{k}_p,\sigma_p)\,
    \gamma^{\alpha}\frac{1}{m_p - \hat{k}_p + \hat{q}
      -i0}\,J^{\mu}(k_p, k_p + k)\,
    u_n(\vec{k}_n,\sigma_n)\Big]\nonumber\\ &&\times
  \,\Big[\bar{u}_e(\vec{k}_e,\sigma_e)\,\gamma^{\beta}\frac{1}{m_e -
      \hat{k}_e - \hat{k}- \hat{q} - i0}\gamma_{\mu}(1 - \gamma^5)
    v_{\nu}(\vec{k}_{\nu}, + \frac{1}{2}\,)\Big]\,D_{\alpha\beta}(q)
\end{eqnarray}
and 
\begin{eqnarray}\label{eq:C.4}
&&{\cal M}_{\rm Fig.\,\ref{fig:fig3}d}(n \to p e^-
  \bar{\nu}_e\gamma)_{\lambda'}\Big|_{\varepsilon^*_{\lambda'} \to k}
  =\nonumber\\ &&= +
  e^2\int\frac{d^4q}{(2\pi)^4i}\,\Big[\bar{u}_p(\vec{k}_p,\sigma_p)\,
    \gamma^{\alpha}\frac{1}{m_p - \hat{k}_p + \hat{q} - \hat{k}
      -i0}\,J^{\mu}(k_p, k_p + k)\,
    u_n(\vec{k}_n,\sigma_n)\Big]\nonumber\\ &&\times
  \,\Big[\bar{u}_e(\vec{k}_e,\sigma_e)\,\gamma^{\beta}\frac{1}{m_e -
      \hat{k}_e - \hat{q} - i0}\gamma_{\mu}(1 - \gamma^5)
    v_{\nu}(\vec{k}_{\nu}, + \frac{1}{2}\,)\Big]\,D_{\alpha\beta}(q),
\end{eqnarray}
where we have used the Dirac equations for the free proton and
electron. One may see that the sum of the diagrams
Fig.\,\ref{fig:fig3}c and Fig.\,\ref{fig:fig3}d is not invariant under
a gauge transformation $\varepsilon^*_{\lambda'}(k) \to
\varepsilon^*_{\lambda'}(k) + ck$. It is obvious that the sum of the
diagrams Fig.\,\ref{fig:fig3}c and Fig.\,\ref{fig:fig3}d is not also
invariant under a gauge transformation $D_{\alpha\beta}(q) \to
D_{\alpha\beta}(q) + c(q^2)\,q_{\alpha} q_{\beta}$.

Now we may proceed to the calculation of the Feynman diagrams
Fig.\,\ref{fig:fig3}c and Fig.\,\ref{fig:fig3}d. For this aim we
replace $J^{\mu}(k_p, k_p + k)$ by $J^{\mu}(k_p, k_p + k) \to
\gamma^{\mu}(1 + \lambda\,\gamma^5)$. Then, merging denominators and
skipping standard intermediate calculations we transcribe the
r.h.s. of Eq.(\ref{eq:C.1}) and Eq.(\ref{eq:C.2}) into the form
\begin{eqnarray*}
\hspace{-0.3in}&&{\cal M}_{\rm Fig.\,\ref{fig:fig3}c}(n \to p e^-
\bar{\nu}_e\gamma)_{\lambda'} = e^2 \int^1_0dx\int^1_0 dy\,2 y
\int\frac{d^4q}{(2\pi)^4i}\,\Big[\bar{u}_p(\vec{k}_p,\sigma_p)\,
  \gamma^{\alpha}(m_p + \hat{k}_p - \hat{k}_p(x) y) \,\gamma^{\mu}(1 +
  \lambda \gamma^5) u_n(\vec{k}_n,\sigma_n)\Big]\nonumber\\
\hspace{-0.3in}&&\times
\,\Big[\bar{u}_e(\vec{k}_e,\sigma_e)\,\hat{\varepsilon}^*_{\lambda'}\,
  \frac{1}{m_e - \hat{k}_e - \hat{k} - i 0}\,\gamma_{\alpha}(m_e +
  \hat{k}_e + \hat{k} + \hat{k}_p(x)y)\,\gamma_{\mu}(1 - \gamma^5)
  v_{\nu}(\vec{k}_{\nu}, + \frac{1}{2}\,)\Big]\nonumber\\
\hspace{-0.3in}&&\times\,\Big(\frac{1}{[q^2 - k^2_p(x)y^2 + 2(k_e\cdot
    k)(1 - x)y - \mu^2 (1 - y) + i 0]^3} - \frac{1}{[q^2 - k^2_p(x)y^2 +
    2(k_e\cdot k)(1 - x)y - \Lambda^2 (1 - y) + i 0]^3}\Big)\nonumber\\
\hspace{-0.3in}&& - e^2\,\frac{1}{4}\,\Big[\bar{u}_p(\vec{k}_p,\sigma_p)\,
  \gamma^{\alpha}\,\gamma^{\beta}\,\gamma^{\mu}(1 + \lambda \gamma^5)
  u_n(\vec{k}_n,\sigma_n)\Big]
\,\Big[\bar{u}_e(\vec{k}_e,\sigma_e)\,\hat{\varepsilon}^*_{\lambda'}\,
  \frac{1}{m_e - \hat{k}_e - \hat{k} - i
    0}\,\gamma_{\alpha}\,\gamma_{\beta}\,\gamma_{\mu}(1 - \gamma^5)
  v_{\nu}(\vec{k}_{\nu}, + \frac{1}{2}\,)\Big]\nonumber\\ 
\hspace{-0.3in}&&\times \int^1_0dx\int^1_0 dy\,2 y
\int\frac{d^4q}{(2\pi)^4i}\,\Big(\frac{q^2}{[q^2 - k^2_p(x)y^2 +
    2(k_e\cdot k)(1 - x)y - \mu^2 (1 - y) + i 0]^3}\nonumber\\
\end{eqnarray*}
\begin{eqnarray}\label{eq:C.5}
\hspace{-0.3in}&& - \frac{q^2}{[q^2
    - k^2_p(x)y^2 + 2(k_e\cdot k)(1 - x)y - \Lambda^2 (1 - y) + i
    0]^3}\Big),
\end{eqnarray}
and
\begin{eqnarray}\label{eq:C.6}
\hspace{-0.3in}&&{\cal M}_{\rm Fig.\,\ref{fig:fig3}d}(n \to p e^-
  \bar{\nu}_e\gamma)_{\lambda'} =\nonumber\\ 
\hspace{-0.3in}&&= -
  e^2\int^1_0dx\int^1_0 dy\,2 y
  \int\frac{d^4q}{(2\pi)^4i}\,\Big[\bar{u}_p(\vec{k}_p,\sigma_p)\,
    \hat{\varepsilon}^*_{\lambda'}\,\frac{1}{m_p - \hat{k}_p - \hat{k}
      - i0} \gamma^{\alpha}(m_p + \hat{k}_p + \hat{k}+ \hat{k}_e(x)
    y)\,\gamma^{\mu}(1 + \lambda\,\gamma^5)\,
    u_n(\vec{k}_n,\sigma_n)\Big]\nonumber\\ 
\hspace{-0.3in}&&\times
  \,\Big[\bar{u}_e(\vec{k}_e,\sigma_e)\,\gamma_{\alpha}(m_e +
    \hat{k}_e - \hat{k}_e(x)y)\,\gamma_{\mu}(1 - \gamma^5)
    v_{\nu}(\vec{k}_{\nu}, + \frac{1}{2}\,)\Big]\nonumber\\ 
\hspace{-0.3in}&&
  \times\,\Big(\frac{1}{[q^2 - k^2_e(x)y^2 + 2(k_p\cdot k)(1 - x) y -
      \mu^2 (1 - y) + i 0]^3} - \frac{1}{[q^2 - k^2_e(x)y^2 +
      2(k_p\cdot k)(1 - x) y - \Lambda^2 (1 - y) + i
      0]^3}\Big)\nonumber\\ 
\hspace{-0.3in}&& +
  e^2\,\frac{1}{4}\,\Big[\bar{u}_p(\vec{k}_p,\sigma_p)\,\hat{\varepsilon}^*_{\lambda'}\,
    \frac{1}{m_e - \hat{k}_p - \hat{k} - i 0}\,
    \gamma^{\alpha}\,\gamma^{\beta}\,\gamma^{\mu}(1 + \lambda
    \gamma^5) u_n(\vec{k}_n,\sigma_n)\Big]
  \,\Big[\bar{u}_e(\vec{k}_e,\sigma_e)\,\gamma_{\alpha}\,\gamma_{\beta}\,\gamma_{\mu}(1
    - \gamma^5) v_{\nu}(\vec{k}_{\nu}, +
    \frac{1}{2}\,)\Big]\nonumber\\
\hspace{-0.3in}&&\times \int^1_0dx\int^1_0 dy\,2 y
\int\frac{d^4q}{(2\pi)^4i}\,\Big(\frac{q^2}{[q^2 - k^2_e(x)y^2 +
    2(k_p\cdot k)(1 - x)y - \mu^2 (1 - y) + i 0]^3}\nonumber\\
\hspace{-0.3in}&& - \frac{q^2}{[q^2 - k^2_e(x)y^2 + 2(k_p\cdot k)(1 -
    x)y - \Lambda^2 (1 - y) + i 0]^3}\Big),
\end{eqnarray}
where we have denoted $k_p(x) = k_p x - (k_e + k)\,(1-x)$ and $k_e(x)
= k_e x - (k_p + k)\,(1 - x)$, respectively. Making the Wick rotation
and integrating over $q^2$ we arrive at the expressions
\begin{eqnarray}\label{eq:C.7}
\hspace{-0.3in}&&{\cal M}_{\rm Fig.\,\ref{fig:fig3}c}(n \to p e^-
\bar{\nu}_e\gamma)_{\lambda'} = -\,\frac{e^2}{32\pi^2}
\int^1_0dx\int^1_0 dy\,2 y \,\Big[\bar{u}_p(\vec{k}_p,\sigma_p)\,
  \gamma^{\alpha}(m_p + \hat{k}_p - \hat{k}_p(x) y) \,\gamma^{\mu}(1 +
  \lambda \gamma^5) u_n(\vec{k}_n,\sigma_n)\Big]\nonumber\\
\hspace{-0.3in}&&\times
\,\Big[\bar{u}_e(\vec{k}_e,\sigma_e)\,\hat{\varepsilon}^*_{\lambda'}\,
  \frac{1}{m_e - \hat{k}_e - \hat{k} - i 0}\,\gamma_{\alpha}(m_e +
  \hat{k}_e + \hat{k} + \hat{k}_p(x)y)\,\gamma_{\mu}(1 - \gamma^5)
  v_{\nu}(\vec{k}_{\nu}, + \frac{1}{2}\,)\Big]\nonumber\\
\hspace{-0.3in}&&\times\,\Big(\frac{1}{k^2_p(x)y^2 - 2(k_e\cdot k)(1 -
  x)y + \mu^2 (1 - y)} - \frac{1}{k^2_p(x)y^2 - 2(k_e\cdot k)(1 - x)y
  + \Lambda^2 (1 - y)}\Big)\nonumber\\
\hspace{-0.3in}&& - \frac{e^2}{64\pi^2}\,\Big[\bar{u}_p(\vec{k}_p,\sigma_p)\,
  \gamma^{\alpha}\,\gamma^{\beta}\,\gamma^{\mu}(1 + \lambda \gamma^5)
  u_n(\vec{k}_n,\sigma_n)\Big]
\,\Big[\bar{u}_e(\vec{k}_e,\sigma_e)\,\hat{\varepsilon}^*_{\lambda'}\,
  \frac{1}{m_e - \hat{k}_e - \hat{k} - i
    0}\,\gamma_{\alpha}\,\gamma_{\beta}\,\gamma_{\mu}(1 - \gamma^5)
  v_{\nu}(\vec{k}_{\nu}, + \frac{1}{2}\,)\Big]\nonumber\\ 
\hspace{-0.3in}&&\times \int^1_0dx\int^1_0 dy\,2 y\,{\ell
  n}\Big(\frac{k^2_p(x)y^2 - 2(k_e\cdot k)(1 - x)y + \Lambda^2 (1 - y)}{k^2_p(x)y^2 - 2(k_e\cdot k)(1 - x)y + \mu^2 (1 - y)}\Big)
\end{eqnarray}
and
\begin{eqnarray}\label{eq:C.8}
\hspace{-0.3in}&&{\cal M}_{\rm Fig.\,\ref{fig:fig3}d}(n \to p e^-
\bar{\nu}_e\gamma)_{\lambda'} =\nonumber\\ 
\hspace{-0.3in}&&= +
e^2\int^1_0dx\int^1_0 dy\,2 y\,\Big[\bar{u}_p(\vec{k}_p,\sigma_p)\,
  \hat{\varepsilon}^*_{\lambda'}\,\frac{1}{m_p - \hat{k}_p - \hat{k} -
    i0} \gamma^{\alpha}(m_p + \hat{k}_p + \hat{k}+ \hat{k}_e(x)
  y)\,\gamma^{\mu}(1 + \lambda\,\gamma^5)\,
  u_n(\vec{k}_n,\sigma_n)\Big]\nonumber\\
\hspace{-0.3in}&&\times
  \,\Big[\bar{u}_e(\vec{k}_e,\sigma_e)\,\gamma_{\alpha}(m_e +
    \hat{k}_e - \hat{k}_e(x)y)\,\gamma_{\mu}(1 - \gamma^5)
    v_{\nu}(\vec{k}_{\nu}, + \frac{1}{2}\,)\Big]\nonumber\\ &&
  \times\,\Big(\frac{1}{k^2_e(x)y^2 - 2(k_p\cdot k)(1 - x) y + \mu^2
    (1 - y)} - \frac{1}{k^2_e(x)y^2 - 2(k_p\cdot k)(1 - x) y +
    \Lambda^2 (1 - y)}\Big)\nonumber\\ 
\hspace{-0.3in}&& +
\frac{e^2}{64\pi^2}\,\Big[\bar{u}_p(\vec{k}_p,\sigma_p)\,
  \hat{\varepsilon}^*_{\lambda'}\, \frac{1}{m_e - \hat{k}_p - \hat{k}
    - i 0}\, \gamma^{\alpha}\,\gamma^{\beta}\,\gamma^{\mu}(1 + \lambda
  \gamma^5) u_n(\vec{k}_n,\sigma_n)\Big]
\,\Big[\bar{u}_e(\vec{k}_e,\sigma_e)\,\gamma_{\alpha}\,\gamma_{\beta}\,
  \gamma_{\mu}(1 - \gamma^5) v_{\nu}(\vec{k}_{\nu}, +
  \frac{1}{2}\,)\Big]\nonumber\\
\hspace{-0.3in}&&\times \int^1_0dx\int^1_0 dy\,2 y\,{\ell
  n}\Big(\frac{k^2_e(x)y^2 - 2(k_p\cdot k)(1 - x)y + \Lambda^2 (1 -
  y)}{k^2_e(x)y^2 - 2(k_p\cdot k)(1 - x)y + \mu^2 (1 - y)}\Big).
\end{eqnarray}
For $\Lambda \gg m_p$ we get
\begin{eqnarray*}
\hspace{-0.3in}&&{\cal M}_{\rm Fig.\,\ref{fig:fig3}c}(n \to p e^-
\bar{\nu}_e\gamma)_{\lambda'} = -\,\frac{e^2}{32\pi^2}
\int^1_0dx\int^1_0 dy\,2 y \,\Big[\bar{u}_p(\vec{k}_p,\sigma_p)\,
  \gamma^{\alpha}(m_p + \hat{k}_p - \hat{k}_p(x) y) \,\gamma^{\mu}(1 +
  \lambda \gamma^5) u_n(\vec{k}_n,\sigma_n)\Big]\nonumber\\
\hspace{-0.3in}&&\times
\,\Big[\bar{u}_e(\vec{k}_e,\sigma_e)\,\hat{\varepsilon}^*_{\lambda'}\,
  \frac{1}{m_e - \hat{k}_e - \hat{k} - i 0}\,\gamma_{\alpha}(m_e +
  \hat{k}_e + \hat{k} + \hat{k}_p(x)y)\,\gamma_{\mu}(1 - \gamma^5)
  v_{\nu}(\vec{k}_{\nu}, + \frac{1}{2}\,)\Big]\nonumber\\
\end{eqnarray*}
\begin{eqnarray}\label{eq:C.9}
\hspace{-0.3in}&&\times\,\frac{1}{k^2_p(x)y^2 - 2(k_e\cdot k)(1 -
  x)y + \mu^2 (1 - y)}\nonumber\\
\hspace{-0.3in}&& - \frac{e^2}{64\pi^2}\,\Big[\bar{u}_p(\vec{k}_p,\sigma_p)\,
  \gamma^{\alpha}\,\gamma^{\beta}\,\gamma^{\mu}(1 + \lambda \gamma^5)
  u_n(\vec{k}_n,\sigma_n)\Big]
\,\Big[\bar{u}_e(\vec{k}_e,\sigma_e)\,\hat{\varepsilon}^*_{\lambda'}\,
  \frac{1}{m_e - \hat{k}_e - \hat{k} - i
    0}\,\gamma_{\alpha}\,\gamma_{\beta}\,\gamma_{\mu}(1 - \gamma^5)
  v_{\nu}(\vec{k}_{\nu}, + \frac{1}{2}\,)\Big]\nonumber\\ 
\hspace{-0.3in}&&\times \int^1_0dx\int^1_0 dy\,2 y\,{\ell
  n}\Big(\frac{\Lambda^2 (1 - y)}{k^2_p(x)y^2 - 2(k_e\cdot k)(1 - x)y
  + \mu^2 (1 - y)}\Big)
\end{eqnarray}
and
\begin{eqnarray}\label{eq:C.10}
\hspace{-0.3in}&&{\cal M}_{\rm Fig.\,\ref{fig:fig3}d}(n \to p e^-
  \bar{\nu}_e\gamma)_{\lambda'} =\nonumber\\ \hspace{-0.3in}&&= +
  \frac{e^2}{32\pi^2}\int^1_0dx\int^1_0 dy\,2
  y\,\Big[\bar{u}_p(\vec{k}_p,\sigma_p)\,
    \hat{\varepsilon}^*_{\lambda'}\,\frac{1}{m_p - \hat{k}_p - \hat{k}
      - i0} \gamma^{\alpha}(m_p + \hat{k}_p + \hat{k}+ \hat{k}_e(x)
    y)\,\gamma^{\mu}(1 + \lambda\,\gamma^5)\,
    u_n(\vec{k}_n,\sigma_n)\Big]\nonumber\\ &&\times
  \,\Big[\bar{u}_e(\vec{k}_e,\sigma_e)\,\gamma_{\alpha}(m_e +
    \hat{k}_e - \hat{k}_e(x)y)\,\gamma_{\mu}(1 - \gamma^5)
    v_{\nu}(\vec{k}_{\nu}, + \frac{1}{2}\,)\Big]\nonumber\\ &&
  \times\,\frac{1}{k^2_e(x)y^2 - 2(k_p\cdot k)(1 - x) y + \mu^2 (1 -
    y)}\nonumber\\ && +
  \frac{e^2}{64\pi^2}\,\Big[\bar{u}_p(\vec{k}_p,\sigma_p)\,
    \hat{\varepsilon}^*_{\lambda'}\, \frac{1}{m_e - \hat{k}_p -
      \hat{k} - i 0}\, \gamma^{\alpha}\,\gamma^{\beta}\,\gamma^{\mu}(1
    + \lambda \gamma^5) u_n(\vec{k}_n,\sigma_n)\Big]
  \,\Big[\bar{u}_e(\vec{k}_e,\sigma_e)\,\gamma_{\alpha}\,\gamma_{\beta}\,
    \gamma_{\mu}(1 - \gamma^5) v_{\nu}(\vec{k}_{\nu}, +
    \frac{1}{2}\,)\Big]\nonumber\\
\hspace{-0.3in}&&\times \int^1_0dx\int^1_0 dy\,2 y\,{\ell
  n}\Big(\frac{\Lambda^2 (1 - y)}{k^2_e(x)y^2 - 2(k_p\cdot k)(1 - x)y
  + \mu^2 (1 - y)}\Big).
\end{eqnarray}
First, in the last terms of Eq.(\ref{eq:C.9}) and Eq.(\ref{eq:C.10})
we integrate over $x$ and $y$. Keeping only the leading terms in the
large proton mass expansion we get
\begin{eqnarray}\label{eq:C.11}
\hspace{-0.3in}&&{\cal M}_{\rm Fig.\,\ref{fig:fig3}c}(n \to p e^-
\bar{\nu}_e\gamma)_{\lambda'} = -\,\frac{e^2}{32\pi^2}
\int^1_0dx\int^1_0 dy\,2 y \,\Big[\bar{u}_p(\vec{k}_p,\sigma_p)\,
  \gamma^{\alpha}(m_p + \hat{k}_p - \hat{k}_p(x) y) \,\gamma^{\mu}(1 +
  \lambda \gamma^5) u_n(\vec{k}_n,\sigma_n)\Big]\nonumber\\
\hspace{-0.3in}&&\times
\,\Big[\bar{u}_e(\vec{k}_e,\sigma_e)\,\hat{\varepsilon}^*_{\lambda'}\,
  \frac{1}{m_e - \hat{k}_e - \hat{k} - i 0}\,\gamma_{\alpha}(m_e +
  \hat{k}_e + \hat{k} + \hat{k}_p(x)y)\,\gamma_{\mu}(1 - \gamma^5)
  v_{\nu}(\vec{k}_{\nu}, + \frac{1}{2}\,)\Big]\nonumber\\
\hspace{-0.3in}&&\times\,\frac{1}{k^2_p(x)y^2 - 2(k_e\cdot k)(1 -
  x)y + \mu^2 (1 - y)} - \frac{e^2}{32\pi^2}\,\Big({\ell
  n}\Big(\frac{\Lambda}{m_p}\Big) + \frac{3}{4}\Big)\, \Big[\bar{u}_p(\vec{k}_p,\sigma_p)\,
  \gamma^{\alpha}\,\gamma^{\beta}\,\gamma^{\mu}(1 + \lambda \gamma^5)
  u_n(\vec{k}_n,\sigma_n)\Big]\nonumber\\
\hspace{-0.3in}&&\times
\,\Big[\bar{u}_e(\vec{k}_e,\sigma_e)\,\hat{\varepsilon}^*_{\lambda'}\,
  \frac{1}{m_e - \hat{k}_e - \hat{k} - i
    0}\,\gamma_{\alpha}\,\gamma_{\beta}\,\gamma_{\mu}(1 - \gamma^5)
  v_{\nu}(\vec{k}_{\nu}, + \frac{1}{2}\,)\Big]
\end{eqnarray}
and
\begin{eqnarray}\label{eq:C.12}
\hspace{-0.3in}&&{\cal M}_{\rm Fig.\,\ref{fig:fig3}d}(n \to p e^-
\bar{\nu}_e\gamma)_{\lambda'} =\nonumber\\ \hspace{-0.3in}&&= +
\frac{e^2}{32\pi^2}\int^1_0dx\int^1_0 dy\,2
y\,\Big[\bar{u}_p(\vec{k}_p,\sigma_p)\,
  \hat{\varepsilon}^*_{\lambda'}\,\frac{1}{m_p - \hat{k}_p - \hat{k} -
    i0} \gamma^{\alpha}(m_p + \hat{k}_p + \hat{k}+ \hat{k}_e(x)
  y)\,\gamma^{\mu}(1 + \lambda\,\gamma^5)\,
  u_n(\vec{k}_n,\sigma_n)\Big]\nonumber\\ \hspace{-0.3in}&&\times
\,\Big[\bar{u}_e(\vec{k}_e,\sigma_e)\,\gamma_{\alpha}(m_e + \hat{k}_e
  - \hat{k}_e(x)y)\,\gamma_{\mu}(1 - \gamma^5) v_{\nu}(\vec{k}_{\nu},
  + \frac{1}{2}\,)\Big]\,\frac{1}{k^2_e(x)y^2 - 2(k_p\cdot k)(1 - x) y
  + \mu^2 (1 - y)}\nonumber\\ \hspace{-0.3in}&& +
\frac{e^2}{32\pi^2}\,\Big({\ell n}\Big(\frac{\Lambda}{m_p}\Big) +
\frac{3}{4}\Big)\,\Big[\bar{u}_p(\vec{k}_p,\sigma_p)\,
  \hat{\varepsilon}^*_{\lambda'}\, \frac{1}{m_e - \hat{k}_p - \hat{k}
    - i 0}\, \gamma^{\alpha}\,\gamma^{\beta}\,\gamma^{\mu}(1 + \lambda
  \gamma^5)
  u_n(\vec{k}_n,\sigma_n)\Big]\nonumber\\ \hspace{-0.3in}&&\times
\,\Big[\bar{u}_e(\vec{k}_e,\sigma_e)\,\gamma_{\alpha}\,\gamma_{\beta}\,
  \gamma_{\mu}(1 - \gamma^5) v_{\nu}(\vec{k}_{\nu}, +
  \frac{1}{2}\,)\Big].
\end{eqnarray}
One may see that the terms dependent on the ultra--violet cut--off
$\Lambda$ are invariant under a gauge transformation
$\varepsilon^*_{\lambda'}(k) \to \varepsilon^*_{\lambda'}(k) + c
k$. For the integration over $x$ and $y$ in the first terms of
Eq.(\ref{eq:C.11}) and Eq.(\ref{eq:C.12}) we transcribe them as
follows
\begin{eqnarray*}
\hspace{-0.3in}&&{\cal M}_{\rm Fig.\,\ref{fig:fig3}c}(n \to p e^-
\bar{\nu}_e\gamma)_{\lambda'} = -\,\frac{e^2}{32\pi^2}
\,\Big[\bar{u}_p(\vec{k}_p,\sigma_p)\, \gamma^{\alpha}(m_p +
  \hat{k}_p) \,\gamma^{\mu}(1 + \lambda \gamma^5)
  u_n(\vec{k}_n,\sigma_n)\Big]\nonumber\\
\hspace{-0.3in}&&\times
\,\Big[\bar{u}_e(\vec{k}_e,\sigma_e)\,\hat{\varepsilon}^*_{\lambda'}\,
  \frac{1}{m_e - \hat{k}_e - \hat{k} - i 0}\,\gamma_{\alpha}(m_e +
  \hat{k}_e + \hat{k})\,\gamma_{\mu}(1 - \gamma^5)
  v_{\nu}(\vec{k}_{\nu}, + \frac{1}{2}\,)\Big]\nonumber\\
\hspace{-0.3in}&&\times\int^1_0dx\int^1_0 dy\,2 y
\,\frac{1}{k^2_p(x)y^2 - 2(k_e\cdot k)(1 - x)y + \mu^2 (1 - y)}
\nonumber\\
\end{eqnarray*}
\begin{eqnarray}\label{eq:C.13}
\hspace{-0.3in}&&+\,\frac{e^2}{32\pi^2}
\int^1_0dx\int^1_0 dy\,2 y^2 \,\Big[\bar{u}_p(\vec{k}_p,\sigma_p)\,
  \gamma^{\alpha}\hat{k}_p(x)\,\gamma^{\mu}(1 +
  \lambda \gamma^5) u_n(\vec{k}_n,\sigma_n)\Big]\nonumber\\
\hspace{-0.3in}&&\times
\,\Big[\bar{u}_e(\vec{k}_e,\sigma_e)\,\hat{\varepsilon}^*_{\lambda'}\,
  \frac{1}{m_e - \hat{k}_e - \hat{k} - i 0}\,\gamma_{\alpha}(m_e +
  \hat{k}_e + \hat{k})\,\gamma_{\mu}(1 - \gamma^5)
  v_{\nu}(\vec{k}_{\nu}, + \frac{1}{2}\,)\Big]\nonumber\\
\hspace{-0.3in}&&\times\,\frac{1}{k^2_p(x)y^2 - 2(k_e\cdot k)(1 -
  x)y + \mu^2 (1 - y)}\nonumber\\
\hspace{-0.3in}&&-\,\frac{e^2}{32\pi^2}
\int^1_0dx\int^1_0 dy\,2 y^2 \,\Big[\bar{u}_p(\vec{k}_p,\sigma_p)\,
  \gamma^{\alpha}(m_p + \hat{k}_p) \,\gamma^{\mu}(1 +
  \lambda \gamma^5) u_n(\vec{k}_n,\sigma_n)\Big]\nonumber\\
\hspace{-0.3in}&&\times
\,\Big[\bar{u}_e(\vec{k}_e,\sigma_e)\,\hat{\varepsilon}^*_{\lambda'}\,
  \frac{1}{m_e - \hat{k}_e - \hat{k} - i 0}\,\gamma_{\alpha} \hat{k}_p(x)\,\gamma_{\mu}(1 - \gamma^5)
  v_{\nu}(\vec{k}_{\nu}, + \frac{1}{2}\,)\Big]\nonumber\\
\hspace{-0.3in}&&\times\,\frac{1}{k^2_p(x)y^2 - 2(k_e\cdot k)(1 - x)y
  + \mu^2 (1 - y)}\nonumber\\
\hspace{-0.3in}&&+\,\frac{e^2}{32\pi^2}
\int^1_0dx\int^1_0 dy\,2 y^3 \,\Big[\bar{u}_p(\vec{k}_p,\sigma_p)\,
  \gamma^{\alpha}\hat{k}_p(x)\,\gamma^{\mu}(1 +
  \lambda \gamma^5) u_n(\vec{k}_n,\sigma_n)\Big]\nonumber\\
\hspace{-0.3in}&&\times
\,\Big[\bar{u}_e(\vec{k}_e,\sigma_e)\,\hat{\varepsilon}^*_{\lambda'}\,
  \frac{1}{m_e - \hat{k}_e - \hat{k} - i
    0}\,\gamma_{\alpha}\hat{k}_p(x)\,\gamma_{\mu}(1 - \gamma^5)
  v_{\nu}(\vec{k}_{\nu}, + \frac{1}{2}\,)\Big]\nonumber\\
\hspace{-0.3in}&&\times\,\frac{1}{k^2_p(x)y^2 - 2(k_e\cdot k)(1 -
  x)y + \mu^2 (1 - y)}\nonumber\\
\hspace{-0.3in}&& -
\frac{e^2}{32\pi^2}\,\Big({\ell n}\Big(\frac{\Lambda}{m_p}\Big) +
\frac{3}{4}\Big)\, \Big[\bar{u}_p(\vec{k}_p,\sigma_p)\,
  \gamma^{\alpha}\,\gamma^{\beta}\,\gamma^{\mu}(1 + \lambda \gamma^5)
  u_n(\vec{k}_n,\sigma_n)\Big]\nonumber\\
\hspace{-0.3in}&&\times
\,\Big[\bar{u}_e(\vec{k}_e,\sigma_e)\,\hat{\varepsilon}^*_{\lambda'}\,
  \frac{1}{m_e - \hat{k}_e - \hat{k} - i
    0}\,\gamma_{\alpha}\,\gamma_{\beta}\,\gamma_{\mu}(1 - \gamma^5)
  v_{\nu}(\vec{k}_{\nu}, + \frac{1}{2}\,)\Big]
\end{eqnarray}
and 
\begin{eqnarray}\label{eq:C.14}
\hspace{-0.3in}&&{\cal M}_{\rm Fig.\,\ref{fig:fig3}d}(n \to p e^-
\bar{\nu}_e\gamma)_{\lambda'} = +
\frac{e^2}{32\pi^2}\Big[\bar{u}_p(\vec{k}_p,\sigma_p)\,
  \hat{\varepsilon}^*_{\lambda'}\,\frac{1}{m_p - \hat{k}_p - \hat{k} -
    i0} \gamma^{\alpha}(m_p + \hat{k}_p + \hat{k})\,\gamma^{\mu}(1 +
  \lambda\,\gamma^5)\,
  u_n(\vec{k}_n,\sigma_n)\Big]\nonumber\\ \hspace{-0.3in}&&\times
\,\Big[\bar{u}_e(\vec{k}_e,\sigma_e)\,\gamma_{\alpha}(m_e +
  \hat{k}_e)\,\gamma_{\mu}(1 - \gamma^5) v_{\nu}(\vec{k}_{\nu}, +
  \frac{1}{2}\,)\Big]\,\int^1_0dx\int^1_0 dy\,2
y\,\frac{1}{k^2_e(x)y^2 - 2(k_p\cdot k)(1 - x) y + \mu^2 (1 -
  y)}\nonumber\\
\hspace{-0.3in}&&+\frac{e^2}{32\pi^2}\int^1_0dx\int^1_0 dy\,2
y^2\,\Big[\bar{u}_p(\vec{k}_p,\sigma_p)\,
  \hat{\varepsilon}^*_{\lambda'}\,\frac{1}{m_p - \hat{k}_p - \hat{k} -
    i0} \gamma^{\alpha}\hat{k}_e(x)\,\gamma^{\mu}(1 +
  \lambda\,\gamma^5)\,
  u_n(\vec{k}_n,\sigma_n)\Big]\nonumber\\ \hspace{-0.3in}&&\times
\,\Big[\bar{u}_e(\vec{k}_e,\sigma_e)\,\gamma_{\alpha}(m_e +
  \hat{k}_e)\,\gamma_{\mu}(1 - \gamma^5) v_{\nu}(\vec{k}_{\nu}, +
  \frac{1}{2}\,)\Big]\,\frac{1}{k^2_e(x)y^2 - 2(k_p\cdot k)(1 - x) y +
  \mu^2 (1 - y)}\nonumber\\
\hspace{-0.3in}&&-\frac{e^2}{32\pi^2}\int^1_0dx\int^1_0 dy\,2
y^2\,\Big[\bar{u}_p(\vec{k}_p,\sigma_p)\,
  \hat{\varepsilon}^*_{\lambda'}\,\frac{1}{m_p - \hat{k}_p - \hat{k} -
    i0} \gamma^{\alpha}(m_p + \hat{k}_p + \hat{k})\,\gamma^{\mu}(1 +
  \lambda\,\gamma^5)\,
  u_n(\vec{k}_n,\sigma_n)\Big]\nonumber\\ \hspace{-0.3in}&&\times
\,\Big[\bar{u}_e(\vec{k}_e,\sigma_e)\,\gamma_{\alpha}
  \hat{k}_e(x)\,\gamma_{\mu}(1 - \gamma^5) v_{\nu}(\vec{k}_{\nu}, +
  \frac{1}{2}\,)\Big]\,\frac{1}{k^2_e(x)y^2 - 2(k_p\cdot k)(1 - x) y +
  \mu^2 (1 - y)}\nonumber\\
\hspace{-0.3in}&&-\frac{e^2}{32\pi^2}\int^1_0dx\int^1_0 dy\,2
y^3\,\Big[\bar{u}_p(\vec{k}_p,\sigma_p)\,
  \hat{\varepsilon}^*_{\lambda'}\,\frac{1}{m_p - \hat{k}_p - \hat{k} -
    i0} \gamma^{\alpha}\hat{k}_e(x)\,\gamma^{\mu}(1 +
  \lambda\,\gamma^5)\,
  u_n(\vec{k}_n,\sigma_n)\Big]\nonumber\\ \hspace{-0.3in}&&\times
\,\Big[\bar{u}_e(\vec{k}_e,\sigma_e)\,\gamma_{\alpha}\hat{k}_e(x)\,
  \gamma_{\mu}(1 - \gamma^5) v_{\nu}(\vec{k}_{\nu}, +
  \frac{1}{2}\,)\Big]\, \frac{1}{k^2_e(x)y^2 - 2(k_p\cdot k)(1 - x) y
  + \mu^2 (1 - y)}\nonumber\\
\hspace{-0.3in}&& +
\frac{e^2}{32\pi^2}\,\Big({\ell n}\Big(\frac{\Lambda}{m_p}\Big) +
\frac{3}{4}\Big)\,\Big[\bar{u}_p(\vec{k}_p,\sigma_p)\,
  \hat{\varepsilon}^*_{\lambda'}\, \frac{1}{m_e - \hat{k}_p - \hat{k}
    - i 0}\, \gamma^{\alpha}\,\gamma^{\beta}\,\gamma^{\mu}(1 + \lambda
  \gamma^5)
  u_n(\vec{k}_n,\sigma_n)\Big]\nonumber\\ \hspace{-0.3in}&&\times
\,\Big[\bar{u}_e(\vec{k}_e,\sigma_e)\,\gamma_{\alpha}\,\gamma_{\beta}\,
  \gamma_{\mu}(1 - \gamma^5) v_{\nu}(\vec{k}_{\nu}, +
  \frac{1}{2}\,)\Big].
\end{eqnarray}
Using the Dirac equations for the free proton and electron we get
\begin{eqnarray*}
\hspace{-0.3in}&&{\cal M}_{\rm Fig.\,\ref{fig:fig3}c}(n \to p e^-
\bar{\nu}_e\gamma)_{\lambda'} = -\,\frac{e^2}{16\pi^2}
\,\Big[\bar{u}_p(\vec{k}_p,\sigma_p)\, \gamma^{\mu}(1 + \lambda
  \gamma^5) u_n(\vec{k}_n,\sigma_n)\Big]\nonumber\\
\end{eqnarray*}
\begin{eqnarray}\label{eq:C.15}
\hspace{-0.3in}&&\times
\,\Big[\bar{u}_e(\vec{k}_e,\sigma_e)\,\hat{\varepsilon}^*_{\lambda'}\,
  \frac{1}{m_e - \hat{k}_e - \hat{k} - i 0}\,\hat{k}_p(m_e + \hat{k}_e
  + \hat{k})\,\gamma_{\mu}(1 - \gamma^5) v_{\nu}(\vec{k}_{\nu}, +
  \frac{1}{2}\,)\Big]\nonumber\\
\hspace{-0.3in}&&\times\int^1_0dx\int^1_0 dy\,2 y
\,\frac{1}{k^2_p(x)y^2 - 2(k_e\cdot k)(1 - x)y + \mu^2 (1 - y)}
\nonumber\\
\hspace{-0.3in}&&+\,\frac{e^2}{32\pi^2}
\int^1_0dx\int^1_0 dy\,2 y^2 \,\Big[\bar{u}_p(\vec{k}_p,\sigma_p)\,
  \gamma^{\alpha}\hat{k}_p(x)\,\gamma^{\mu}(1 +
  \lambda \gamma^5) u_n(\vec{k}_n,\sigma_n)\Big]\nonumber\\
\hspace{-0.3in}&&\times
\,\Big[\bar{u}_e(\vec{k}_e,\sigma_e)\,\hat{\varepsilon}^*_{\lambda'}\,
  \frac{1}{m_e - \hat{k}_e - \hat{k} - i 0}\,\gamma_{\alpha}(m_e +
  \hat{k}_e + \hat{k})\,\gamma_{\mu}(1 - \gamma^5)
  v_{\nu}(\vec{k}_{\nu}, + \frac{1}{2}\,)\Big]\nonumber\\
\hspace{-0.3in}&&\times\,\frac{1}{k^2_p(x)y^2 - 2(k_e\cdot k)(1 -
  x)y + \mu^2 (1 - y)}\nonumber\\
\hspace{-0.3in}&&-\,\frac{e^2}{16\pi^2} \int^1_0dx\int^1_0 dy\,2 y^2
\,\Big[\bar{u}_p(\vec{k}_p,\sigma_p)\, \gamma^{\mu}(1 + \lambda
  \gamma^5) u_n(\vec{k}_n,\sigma_n)\Big]\nonumber\\
\hspace{-0.3in}&&\times
\,\Big[\bar{u}_e(\vec{k}_e,\sigma_e)\,\hat{\varepsilon}^*_{\lambda'}\,
  \frac{1}{m_e - \hat{k}_e - \hat{k} - i
    0}\,\hat{k}_p\hat{k}_p(x)\,\gamma_{\mu}(1 - \gamma^5)
  v_{\nu}(\vec{k}_{\nu}, + \frac{1}{2}\,)\Big]\nonumber\\
\hspace{-0.3in}&&\times\,\frac{1}{k^2_p(x)y^2 - 2(k_e\cdot k)(1 - x)y
  + \mu^2 (1 - y)}\nonumber\\
\hspace{-0.3in}&&+\,\frac{e^2}{32\pi^2}
\int^1_0dx\int^1_0 dy\,2 y^3 \,\Big[\bar{u}_p(\vec{k}_p,\sigma_p)\,
  \gamma^{\alpha}\hat{k}_p(x)\,\gamma^{\mu}(1 +
  \lambda \gamma^5) u_n(\vec{k}_n,\sigma_n)\Big]\nonumber\\
\hspace{-0.3in}&&\times
\,\Big[\bar{u}_e(\vec{k}_e,\sigma_e)\,\hat{\varepsilon}^*_{\lambda'}\,
  \frac{1}{m_e - \hat{k}_e - \hat{k} - i
    0}\,\gamma_{\alpha}\hat{k}_p(x)\,\gamma_{\mu}(1 - \gamma^5)
  v_{\nu}(\vec{k}_{\nu}, + \frac{1}{2}\,)\Big]\nonumber\\
\hspace{-0.3in}&&\times\,\frac{1}{k^2_p(x)y^2 - 2(k_e\cdot k)(1 -
  x)y + \mu^2 (1 - y)}\nonumber\\
\hspace{-0.3in}&& -
\frac{e^2}{32\pi^2}\,\Big({\ell n}\Big(\frac{\Lambda}{m_p}\Big) +
\frac{3}{4}\Big)\, \Big[\bar{u}_p(\vec{k}_p,\sigma_p)\,
  \gamma^{\alpha}\,\gamma^{\beta}\,\gamma^{\mu}(1 + \lambda \gamma^5)
  u_n(\vec{k}_n,\sigma_n)\Big]\nonumber\\
\hspace{-0.3in}&&\times
\,\Big[\bar{u}_e(\vec{k}_e,\sigma_e)\,\hat{\varepsilon}^*_{\lambda'}\,
  \frac{1}{m_e - \hat{k}_e - \hat{k} - i
    0}\,\gamma_{\alpha}\,\gamma_{\beta}\,\gamma_{\mu}(1 - \gamma^5)
  v_{\nu}(\vec{k}_{\nu}, + \frac{1}{2}\,)\Big]
\end{eqnarray}
and 
\begin{eqnarray}\label{eq:C.16}
\hspace{-0.3in}&&{\cal M}_{\rm Fig.\,\ref{fig:fig3}d}(n \to p e^-
\bar{\nu}_e\gamma)_{\lambda'} = +
\frac{e^2}{16\pi^2}\Big[\bar{u}_p(\vec{k}_p,\sigma_p)\,
  \hat{\varepsilon}^*_{\lambda'}\,\frac{1}{m_p - \hat{k}_p - \hat{k} -
    i0}\,\hat{k}_e (m_p + \hat{k}_p + \hat{k})\,\gamma^{\mu}(1 +
  \lambda\,\gamma^5)\,
  u_n(\vec{k}_n,\sigma_n)\Big]\nonumber\\ \hspace{-0.3in}&&\times
\,\Big[\bar{u}_e(\vec{k}_e,\sigma_e)\,\gamma_{\mu}(1 - \gamma^5)
  v_{\nu}(\vec{k}_{\nu}, + \frac{1}{2}\,)\Big]\,\int^1_0dx\int^1_0
dy\,2 y\,\frac{1}{k^2_e(x)y^2 - 2(k_p\cdot k)(1 - x) y + \mu^2 (1 -
  y)}\nonumber\\
\hspace{-0.3in}&&+\frac{e^2}{16\pi^2}\int^1_0dx\int^1_0 dy\,2
y^2\,\Big[\bar{u}_p(\vec{k}_p,\sigma_p)\,
  \hat{\varepsilon}^*_{\lambda'}\,\frac{1}{m_p - \hat{k}_p - \hat{k} -
    i0}\,\hat{k}_e \hat{k}_e(x)\,\gamma^{\mu}(1 + \lambda\,\gamma^5)\,
  u_n(\vec{k}_n,\sigma_n)\Big]\nonumber\\ \hspace{-0.3in}&&\times
\,\Big[\bar{u}_e(\vec{k}_e,\sigma_e)\,\gamma_{\mu}(1 - \gamma^5)
  v_{\nu}(\vec{k}_{\nu}, + \frac{1}{2}\,)\Big]\,\frac{1}{k^2_e(x)y^2 -
  2(k_p\cdot k)(1 - x) y + \mu^2 (1 - y)}\nonumber\\
\hspace{-0.3in}&&-\frac{e^2}{32\pi^2}\int^1_0dx\int^1_0 dy\,2
y^2\,\Big[\bar{u}_p(\vec{k}_p,\sigma_p)\,
  \hat{\varepsilon}^*_{\lambda'}\,\frac{1}{m_p - \hat{k}_p - \hat{k} -
    i0} \gamma^{\alpha}(m_p + \hat{k}_p + \hat{k})\,\gamma^{\mu}(1 +
  \lambda\,\gamma^5)\,
  u_n(\vec{k}_n,\sigma_n)\Big]\nonumber\\ \hspace{-0.3in}&&\times
\,\Big[\bar{u}_e(\vec{k}_e,\sigma_e)\,\gamma_{\alpha}
  \hat{k}_e(x)\,\gamma_{\mu}(1 - \gamma^5) v_{\nu}(\vec{k}_{\nu}, +
  \frac{1}{2}\,)\Big]\,\frac{1}{k^2_e(x)y^2 - 2(k_p\cdot k)(1 - x) y +
  \mu^2 (1 - y)}\nonumber\\
\hspace{-0.3in}&&-\frac{e^2}{32\pi^2}\int^1_0dx\int^1_0 dy\,2
y^3\,\Big[\bar{u}_p(\vec{k}_p,\sigma_p)\,
  \hat{\varepsilon}^*_{\lambda'}\,\frac{1}{m_p - \hat{k}_p - \hat{k} -
    i0} \gamma^{\alpha}\hat{k}_e(x)\,\gamma^{\mu}(1 +
  \lambda\,\gamma^5)\,
  u_n(\vec{k}_n,\sigma_n)\Big]\nonumber\\ \hspace{-0.3in}&&\times
\,\Big[\bar{u}_e(\vec{k}_e,\sigma_e)\,\gamma_{\alpha}\hat{k}_e(x)\,
  \gamma_{\mu}(1 - \gamma^5) v_{\nu}(\vec{k}_{\nu}, +
  \frac{1}{2}\,)\Big]\, \frac{1}{k^2_e(x)y^2 - 2(k_p\cdot k)(1 - x) y
  + \mu^2 (1 - y)}\nonumber\\
\hspace{-0.3in}&& +
\frac{e^2}{32\pi^2}\,\Big({\ell n}\Big(\frac{\Lambda}{m_p}\Big) +
\frac{3}{4}\Big)\,\Big[\bar{u}_p(\vec{k}_p,\sigma_p)\,
  \hat{\varepsilon}^*_{\lambda'}\, \frac{1}{m_e - \hat{k}_p - \hat{k}
    - i 0}\, \gamma^{\alpha}\,\gamma^{\beta}\,\gamma^{\mu}(1 + \lambda
  \gamma^5)
  u_n(\vec{k}_n,\sigma_n)\Big]\nonumber\\ \hspace{-0.3in}&&\times
\,\Big[\bar{u}_e(\vec{k}_e,\sigma_e)\,\gamma_{\alpha}\,\gamma_{\beta}\,
  \gamma_{\mu}(1 - \gamma^5) v_{\nu}(\vec{k}_{\nu}, +
  \frac{1}{2}\,)\Big].
\end{eqnarray}
It is convenient to rewrite Eq.(\ref{eq:C.15}) and Eq.(\ref{eq:C.16})
as follows
\begin{eqnarray}\label{eq:C.17}
\hspace{-0.3in}&&{\cal M}_{\rm Fig.\,\ref{fig:fig3}c}(n \to p e^-
\bar{\nu}_e\gamma)_{\lambda'} = -\,\frac{e^2}{16\pi^2}
\,\Big[\bar{u}_p(\vec{k}_p,\sigma_p)\, \gamma^{\mu}(1 + \lambda
  \gamma^5) u_n(\vec{k}_n,\sigma_n)\Big]
\,\Big\{\Big[\bar{u}_e(\vec{k}_e,\sigma_e)\,\hat{\varepsilon}^*_{\lambda'}\,
  \hat{k}_p\,\gamma_{\mu}(1 - \gamma^5) v_{\nu}(\vec{k}_{\nu}, +
  \frac{1}{2}\,)\Big]\nonumber\\
\hspace{-0.3in}&& + 2 k_p\cdot (k_e +
k)\,\Big[\bar{u}_e(\vec{k}_e,\sigma_e)\,\hat{\varepsilon}^*_{\lambda'}\frac{1}{m_e
    - \hat{k}_e - \hat{k} - i 0}\,\gamma_{\mu}(1 - \gamma^5)
  v_{\nu}(\vec{k}_{\nu}, + \frac{1}{2}\,)\Big]\Big\}\nonumber\\
\hspace{-0.3in}&&\times\int^1_0dx\int^1_0 dy\,2 y
\,\frac{1}{k^2_p(x)y^2 - 2(k_e\cdot k)(1 - x)y + \mu^2 (1 - y)}
\nonumber\\
\hspace{-0.3in}&&+\,\frac{e^2}{32\pi^2}
\int^1_0dx\int^1_0 dy\,2 y^2 \,\Big[\bar{u}_p(\vec{k}_p,\sigma_p)\,
  \gamma^{\alpha}\hat{k}_p(x)\,\gamma^{\mu}(1 +
  \lambda \gamma^5) u_n(\vec{k}_n,\sigma_n)\Big]\,\Big\{\Big[\bar{u}_e(\vec{k}_e,\sigma_e)\,\hat{\varepsilon}^*_{\lambda'}\,
  \gamma_{\alpha}\,\gamma_{\mu}(1 - \gamma^5) v_{\nu}(\vec{k}_{\nu}, +
  \frac{1}{2}\,)\Big]\nonumber\\
\hspace{-0.3in}&& + 2(k_e +
k)_{\alpha}\Big[\bar{u}_e(\vec{k}_e,\sigma_e)\,\hat{\varepsilon}^*_{\lambda'}\,
  \frac{1}{m_e - \hat{k}_e - \hat{k} - i 0}\,\gamma_{\mu}(1 -
  \gamma^5) v_{\nu}(\vec{k}_{\nu}, +
  \frac{1}{2}\,)\Big]\Big\}\nonumber\\
\hspace{-0.3in}&&\times\,\frac{1}{k^2_p(x)y^2 - 2(k_e\cdot k)(1 -
  x)y + \mu^2 (1 - y)}\nonumber\\
\hspace{-0.3in}&&-\,\frac{e^2}{16\pi^2} \int^1_0dx\int^1_0 dy\,2 y^2
\,\Big[\bar{u}_p(\vec{k}_p,\sigma_p)\, \gamma^{\mu}(1 + \lambda
  \gamma^5) u_n(\vec{k}_n,\sigma_n)\Big]\nonumber\\
\hspace{-0.3in}&&\times
\,\Big[\bar{u}_e(\vec{k}_e,\sigma_e)\,\hat{\varepsilon}^*_{\lambda'}\,
  \frac{1}{m_e - \hat{k}_e - \hat{k} - i
    0}\,\hat{k}_p\hat{k}_p(x)\,\gamma_{\mu}(1 - \gamma^5)
  v_{\nu}(\vec{k}_{\nu}, + \frac{1}{2}\,)\Big]\nonumber\\
\hspace{-0.3in}&&\times\,\frac{1}{k^2_p(x)y^2 - 2(k_e\cdot k)(1 - x)y
  + \mu^2 (1 - y)}\nonumber\\
\hspace{-0.3in}&&+\,\frac{e^2}{32\pi^2}
\int^1_0dx\int^1_0 dy\,2 y^3 \,\Big[\bar{u}_p(\vec{k}_p,\sigma_p)\,
  \gamma^{\alpha}\hat{k}_p(x)\,\gamma^{\mu}(1 +
  \lambda \gamma^5) u_n(\vec{k}_n,\sigma_n)\Big]\nonumber\\
\hspace{-0.3in}&&\times
\,\Big[\bar{u}_e(\vec{k}_e,\sigma_e)\,\hat{\varepsilon}^*_{\lambda'}\,
  \frac{1}{m_e - \hat{k}_e - \hat{k} - i
    0}\,\gamma_{\alpha}\hat{k}_p(x)\,\gamma_{\mu}(1 - \gamma^5)
  v_{\nu}(\vec{k}_{\nu}, + \frac{1}{2}\,)\Big]\nonumber\\
\hspace{-0.3in}&&\times\,\frac{1}{k^2_p(x)y^2 - 2(k_e\cdot k)(1 -
  x)y + \mu^2 (1 - y)}\nonumber\\
\hspace{-0.3in}&& -
\frac{e^2}{32\pi^2}\,\Big({\ell n}\Big(\frac{\Lambda}{m_p}\Big) +
\frac{3}{4}\Big)\, \Big[\bar{u}_p(\vec{k}_p,\sigma_p)\,
  \gamma^{\alpha}\,\gamma^{\beta}\,\gamma^{\mu}(1 + \lambda \gamma^5)
  u_n(\vec{k}_n,\sigma_n)\Big]\nonumber\\
\hspace{-0.3in}&&\times
\,\Big[\bar{u}_e(\vec{k}_e,\sigma_e)\,\hat{\varepsilon}^*_{\lambda'}\,
  \frac{1}{m_e - \hat{k}_e - \hat{k} - i
    0}\,\gamma_{\alpha}\,\gamma_{\beta}\,\gamma_{\mu}(1 - \gamma^5)
  v_{\nu}(\vec{k}_{\nu}, + \frac{1}{2}\,)\Big]
\end{eqnarray}
and 
\begin{eqnarray*}
\hspace{-0.3in}&&{\cal M}_{\rm Fig.\,\ref{fig:fig3}d}(n \to p e^-
\bar{\nu}_e\gamma)_{\lambda'} = +
\frac{e^2}{16\pi^2}\Big\{\Big[\bar{u}_p(\vec{k}_p,\sigma_p)\,
  \hat{\varepsilon}^*_{\lambda'}\,\hat{k}_e\,\gamma^{\mu}(1 +
  \lambda\,\gamma^5)\,
  u_n(\vec{k}_n,\sigma_n)\Big]\nonumber\\
\hspace{-0.3in}&& + 2k_e\cdot (k_p +
k)\,\Big[\bar{u}_p(\vec{k}_p,\sigma_p)\,
  \hat{\varepsilon}^*_{\lambda'}\, \frac{1}{m_p - \hat{k}_p - \hat{k}
    - i0}\,\gamma^{\mu}(1 + \lambda\,\gamma^5)\,
  u_n(\vec{k}_n,\sigma_n)\Big]\Big\}\nonumber\\ 
\hspace{-0.3in}&&\times
\,\Big[\bar{u}_e(\vec{k}_e,\sigma_e)\,\gamma_{\mu}(1 - \gamma^5)
  v_{\nu}(\vec{k}_{\nu}, + \frac{1}{2}\,)\Big]\,\int^1_0dx\int^1_0
dy\,2 y\,\frac{1}{k^2_e(x)y^2 - 2(k_p\cdot k)(1 - x) y + \mu^2 (1 -
  y)}\nonumber\\
\hspace{-0.3in}&&+\frac{e^2}{16\pi^2}\int^1_0dx\int^1_0 dy\,2
y^2\,\Big[\bar{u}_p(\vec{k}_p,\sigma_p)\,
  \hat{\varepsilon}^*_{\lambda'}\,\frac{1}{m_p - \hat{k}_p - \hat{k} -
    i0}\,\hat{k}_e \hat{k}_e(x)\,\gamma^{\mu}(1 + \lambda\,\gamma^5)\,
  u_n(\vec{k}_n,\sigma_n)\Big]\nonumber\\ \hspace{-0.3in}&&\times
\,\Big[\bar{u}_e(\vec{k}_e,\sigma_e)\,\gamma_{\mu}(1 - \gamma^5)
  v_{\nu}(\vec{k}_{\nu}, + \frac{1}{2}\,)\Big]\,\frac{1}{k^2_e(x)y^2 -
  2(k_p\cdot k)(1 - x) y + \mu^2 (1 - y)}\nonumber\\
\hspace{-0.3in}&&-\frac{e^2}{32\pi^2}\int^1_0dx\int^1_0 dy\,2
y^2\,\Big\{\Big[\bar{u}_p(\vec{k}_p,\sigma_p)\,\hat{\varepsilon}^*_{\lambda'}\,\gamma^{\alpha}\,\gamma^{\mu}(1
  + \lambda\,\gamma^5)\,
  u_n(\vec{k}_n,\sigma_n)\Big]\nonumber\\ 
\hspace{-0.3in}&& + 2(k_p +
k)_{\alpha}
\Big[\bar{u}_p(\vec{k}_p,\sigma_p)\,\hat{\varepsilon}^*_{\lambda'}\,\frac{1}{m_p
    - \hat{k}_p - \hat{k} - i0}\,\gamma^{\mu}(1 + \lambda\,\gamma^5)\,
  u_n(\vec{k}_n,\sigma_n)\Big]\Big\}\nonumber\\ \hspace{-0.3in}&&\times
\,\Big[\bar{u}_e(\vec{k}_e,\sigma_e)\,\gamma_{\alpha}
  \hat{k}_e(x)\,\gamma_{\mu}(1 - \gamma^5) v_{\nu}(\vec{k}_{\nu}, +
  \frac{1}{2}\,)\Big]\,\frac{1}{k^2_e(x)y^2 - 2(k_p\cdot k)(1 - x) y +
  \mu^2 (1 - y)}\nonumber\\
\end{eqnarray*}
\begin{eqnarray}\label{eq:C.18}
\hspace{-0.3in}&&-\frac{e^2}{32\pi^2}\int^1_0dx\int^1_0 dy\,2
y^3\,\Big[\bar{u}_p(\vec{k}_p,\sigma_p)\,
  \hat{\varepsilon}^*_{\lambda'}\,\frac{1}{m_p - \hat{k}_p - \hat{k} -
    i0} \gamma^{\alpha}\hat{k}_e(x)\,\gamma^{\mu}(1 +
  \lambda\,\gamma^5)\,
  u_n(\vec{k}_n,\sigma_n)\Big]\nonumber\\ \hspace{-0.3in}&&\times
\,\Big[\bar{u}_e(\vec{k}_e,\sigma_e)\,\gamma_{\alpha}\hat{k}_e(x)\,
  \gamma_{\mu}(1 - \gamma^5) v_{\nu}(\vec{k}_{\nu}, +
  \frac{1}{2}\,)\Big]\, \frac{1}{k^2_e(x)y^2 - 2(k_p\cdot k)(1 - x) y
  + \mu^2 (1 - y)}\nonumber\\
\hspace{-0.3in}&& +
\frac{e^2}{32\pi^2}\,\Big({\ell n}\Big(\frac{\Lambda}{m_p}\Big) +
\frac{3}{4}\Big)\,\Big[\bar{u}_p(\vec{k}_p,\sigma_p)\,
  \hat{\varepsilon}^*_{\lambda'}\, \frac{1}{m_e - \hat{k}_p - \hat{k}
    - i 0}\, \gamma^{\alpha}\,\gamma^{\beta}\,\gamma^{\mu}(1 + \lambda
  \gamma^5)
  u_n(\vec{k}_n,\sigma_n)\Big]\nonumber\\ 
\hspace{-0.3in}&&\times
\,\Big[\bar{u}_e(\vec{k}_e,\sigma_e)\,\gamma_{\alpha}\,\gamma_{\beta}\,
  \gamma_{\mu}(1 - \gamma^5) v_{\nu}(\vec{k}_{\nu}, +
  \frac{1}{2}\,)\Big].
\end{eqnarray}
The calculation of the integrals over $x$ and $y$ to leading order in
the large proton mass expansion runs as follows
\begin{eqnarray}\label{eq:C.19}
\hspace{-0.3in}&& \int^1_0dx\int^1_0 dy\,2 y
\,\frac{k^{\alpha}_p}{k^2_p(x)y^2 - 2(k_e\cdot k)(1 - x)y + \mu^2 (1 -
  y)} = 2k^{\alpha}_p\int^1_0\frac{dx}{k^2_p(x)}\,{\ell n}\Big(-
\frac{k^2_p(x)}{2k_e\cdot k}\Big)=\nonumber\\
\hspace{-0.3in}&& = \frac{k^{\alpha}_p}{m_p}\,\frac{1}{|\vec{k}_e +
  \vec{k}\,|}\Big\{\frac{\pi^2}{3} - {\ell n}\Big(\frac{2|\vec{k}_e + \vec{k}\,|^2}{-
  k_e\cdot k}\Big)\,{\ell n}\Big(\frac{E_e + \omega + |\vec{k}_e +
  \vec{k}\,|}{E_e + \omega - |\vec{k}_e + \vec{k}\,|}\Big) -
\frac{1}{2}\,{\ell n}^2\Big(\frac{E_e + \omega + |\vec{k}_e +
  \vec{k}\,|}{E_e + \omega - |\vec{k}_e + \vec{k}\,|}\Big)\nonumber\\
\hspace{-0.3in}&& -2 {\rm Li}_2\Big(\frac{E_e + \omega + |\vec{k}_e +
  \vec{k}\,|}{E_e + \omega - |\vec{k}_e +
  \vec{k}\,|}\Big)\Big\},\nonumber\\
\hspace{-0.3in}&&\int^1_0dx\int^1_0 dy\,2 y^2
\,\frac{\hat{k}_p}{k^2_p(x)y^2 - 2(k_e\cdot k)(1 - x)y + \mu^2 (1 - y)} =
2\int^1_0\frac{dx}{k^2_p(x)}\hat{k}_p =
\frac{1}{m_p}\,\Big\{\frac{\hat{k}_p}{m_p}\,\Big[{\ell
    n}\Big(\frac{m^2_p}{(k_e + k)^2}\Big)\nonumber\\ 
\hspace{-0.3in}&& - \frac{E_e +
    \omega}{|\vec{k}_e + \vec{k}\,|}\,{\ell n}\Big(\frac{E_e + \omega
    + |\vec{k}_e + \vec{k}\,|}{E_e + \omega - |\vec{k}_e +
    \vec{k}\,|}\Big)\Big] + \frac{\hat{k}_e + \hat{k}}{|\vec{k}_e +
  \vec{k}\,|}\,{\ell n}\Big(\frac{E_e + \omega + |\vec{k}_e +
  \vec{k}\,|}{E_e + \omega - |\vec{k}_e + \vec{k}\,|}\Big)\Big\},
\nonumber\\
\hspace{-0.3in}&&\int^1_0dx\int^1_0 dy\,2 y^2
\,\frac{\hat{k}_p\hat{k}_p(x)}{k^2_p(x)y^2 - 2(k_e\cdot k)(1 - x)y +
  \mu^2 (1 - y)} = 2\int^1_0\frac{
  dx}{k^2_p(x)}\,\hat{k}_p\hat{k}_p(x) = \Big[{\ell
    n}\Big(\frac{m^2_p}{(k_e + k)^2}\Big)\nonumber\\ 
\hspace{-0.3in}&& - \frac{E_e + \omega}{|\vec{k}_e +
  \vec{k}\,|}\,{\ell n}\Big(\frac{E_e + \omega + |\vec{k}_e +
  \vec{k}\,|}{E_e + \omega - |\vec{k}_e + \vec{k}\,|}\Big)\Big] +
\frac{\hat{k}_p(\hat{k}_e + \hat{k})}{m_p |\vec{k}_e +
  \vec{k}\,|}\,{\ell n}\Big(\frac{E_e + \omega + |\vec{k}_e +
  \vec{k}\,|}{E_e + \omega - |\vec{k}_e + \vec{k}\,|}\Big),\nonumber\\
\hspace{-0.3in}&&\int^1_0dx\int^1_0 dy\,2 y^3 \,\frac{k^{\alpha}_p(x)
  k^{\beta}_p(x)}{k^2_p(x)y^2 - 2(k_e\cdot k)(1 - x)y + \mu^2 (1 - y)}
= \int^1_0\frac{dx}{k^2_p(x)}\,k^{\alpha}_p(x) k^{\beta}_p(x) =
\eta^{0\alpha}\eta^{0\beta}.
\end{eqnarray}
One may see that in Eq.(\ref{eq:C.19}) the second integral from above,
calculated to leading order in the large proton mass expansion, does
not contribute to the amplitude Eq.(\ref{eq:C.17}). Then, to leading
order in the large proton mass expansion the contribution of the
diagram Fig.\,\ref{fig:fig3}d is proportional to
$\varepsilon^{0*}_{\lambda'}$ and vanishes in the physical gauge
$\varepsilon^*_{\lambda'} = (0, \vec{\varepsilon}^{\,*}_{\lambda'})$
(see Appendix A and \cite{Ivanov2013,Ivanov2013b,Ivanov2017}). Thus,
below we may discuss the diagram Fig.\,\ref{fig:fig3}c only.
For the extraction of a physical contribution of the diagram
Fig.\,\ref{fig:fig3}c we have to investigate the property of this
diagram with respect to the gauge transformation $D_{\alpha\beta}(q)
\to D_{\alpha\beta}(q) + c(q^2)\,q_{\alpha}q_{\beta}$. For this aim we
rewrite Eq.(\ref{eq:C.2}) as follows \cite{Ivanov2013}
\begin{eqnarray}\label{eq:C.20}
\hspace{-0.3in}&&{\cal M}_{\rm Fig.\,\ref{fig:fig3}c}(n \to p e^-
\bar{\nu}_e\gamma)_{\lambda'} = -
\,e^2\int\frac{d^4q}{(2\pi)^4i}\,D_{\alpha\beta}(q)\,
\frac{[\bar{u}_p(\vec{k}_p,\sigma_p)\,(2 k^{\alpha}_p - q^{\alpha} +
    i\,\sigma^{\alpha\rho}q_{\rho})\,\gamma^{\mu}(1 + \lambda
    \gamma^5) u_n(\vec{k}_n,\sigma_n)]}{q^2 - 2 k_p\cdot q +
  i0}\nonumber\\
\hspace{-0.3in}&&\times
  \,\Big[\bar{u}_e(\vec{k}_e,\sigma_e)\,\hat{\varepsilon}^*_{\lambda'}\,
    \frac{1}{m_e - \hat{k}_e - \hat{k} - i
      0}\,\gamma^{\beta}\frac{1}{m_e - \hat{k}_e - \hat{k}- \hat{q} -
      i0}\,\gamma_{\mu}(1 - \gamma^5) v_{\nu}(\vec{k}_{\nu}, +
    \frac{1}{2}\,)\Big],
\end{eqnarray}
where $\sigma^{\alpha\rho} = \frac{i}{2}(\gamma^{\alpha}\gamma^{\rho}
- \gamma^{\rho} \gamma^{\alpha})$ are the Dirac matrices
\cite{Itzykson1980} and the amplitude
\begin{eqnarray}\label{eq:C.21}
\hspace{-0.3in}&&{\cal M}^{(1)}_{\rm Fig.\,\ref{fig:fig3}c}(n \to p
e^- \bar{\nu}_e\gamma)_{\lambda'} = -
\,e^2\int\frac{d^4q}{(2\pi)^4i}\,D_{\alpha\beta}(q)\,
\frac{[\bar{u}_p(\vec{k}_p,\sigma_p)\,
    i\,\sigma^{\alpha\rho}q_{\rho}\,\gamma^{\mu}(1 + \lambda \gamma^5)
    u_n(\vec{k}_n,\sigma_n)]}{q^2 - 2 k_p\cdot q + i0}\nonumber\\
\hspace{-0.3in}&&\times
  \,\Big[\bar{u}_e(\vec{k}_e,\sigma_e)\,\hat{\varepsilon}^*_{\lambda'}\,
    \frac{1}{m_e - \hat{k}_e - \hat{k} - i
      0}\,\gamma^{\beta}\frac{1}{m_e - \hat{k}_e - \hat{k}- \hat{q} -
      i0}\gamma_{\mu}(1 - \gamma^5) v_{\nu}(\vec{k}_{\nu}, +
    \frac{1}{2}\,)\Big]
\end{eqnarray}
is invariant under a gauge transformation $D_{\alpha\beta}(q) \to
D_{\alpha\beta}(q) + c(q^2)\,q_{\alpha}q_{\beta}$.  Now we consider
the expression
\begin{eqnarray*}
\hspace{-0.3in}&&{\cal M}_{\rm Fig.\,\ref{fig:fig3}c}(n \to p e^-
  \bar{\nu}_e\gamma)_{\lambda'} - {\cal M}^{(1)}_{\rm
    Fig.\,\ref{fig:fig3}c}(n \to p e^-
  \bar{\nu}_e\gamma)_{\lambda'}=\nonumber\\ 
\hspace{-0.3in}&&= -
  \,e^2\int\frac{d^4q}{(2\pi)^4i}\,D_{\alpha\beta}(q)\,
  \frac{[\bar{u}_p(\vec{k}_p,\sigma_p)\,(2 k^{\alpha}_p -
      q^{\alpha})\,\gamma^{\mu}(1 + \lambda \gamma^5)
      u_n(\vec{k}_n,\sigma_n)]}{q^2 - 2 k_p\cdot q +
    i0}\nonumber\\ 
\end{eqnarray*}
\begin{eqnarray}\label{eq:C.22}
\hspace{-0.3in}&&\times
  \,\Big[\bar{u}_e(\vec{k}_e,\sigma_e)\,\hat{\varepsilon}^*_{\lambda'}\,
    \frac{1}{m_e - \hat{k}_e - \hat{k} - i
      0}\,\gamma^{\beta}\frac{1}{m_e - \hat{k}_e - \hat{k}- \hat{q} -
      i0}\,\gamma_{\mu}(1 - \gamma^5) v_{\nu}(\vec{k}_{\nu}, +
    \frac{1}{2}\,)\Big],
\end{eqnarray}
which we transcribe into the form
\begin{eqnarray}\label{eq:C.23}
\hspace{-0.3in}&&{\cal M}_{\rm Fig.\,\ref{fig:fig3}c}(n \to p e^-
  \bar{\nu}_e\gamma)_{\lambda'} - {\cal M}^{(1)}_{\rm
    Fig.\,\ref{fig:fig3}c}(n \to p e^-
  \bar{\nu}_e\gamma)_{\lambda'}=\nonumber\\ 
\hspace{-0.3in}&&=
  \,e^2\int\frac{d^4q}{(2\pi)^4i}\,D_{\alpha\beta}(q)\,
  \frac{[\bar{u}_p(\vec{k}_p,\sigma_p)\,(2 k^{\alpha}_p -
      q^{\alpha})\,\gamma^{\mu}(1 + \lambda \gamma^5)
      u_n(\vec{k}_n,\sigma_n)]}{q^2 - 2 k_p\cdot q +
    i0}\nonumber\\ 
\hspace{-0.3in}&&\times
\,\Big[\bar{u}_e(\vec{k}_e,\sigma_e)\,\hat{\varepsilon}^*_{\lambda'}\,
  \frac{1}{m_e - \hat{k}_e - \hat{k} - i 0}\,\frac{(m_e - \hat{k}_e -
    \hat{k})\,\gamma^{\beta} + 2\,(k_e + k)^{\beta} + q^{\beta} -
    i\,\sigma^{\beta\varphi}q_{\varphi}}{q^2 + 2\,(k_e + k)\cdot q + 2
    k_e \cdot k + i0}\,\gamma_{\mu}(1 - \gamma^5)
  v_{\nu}(\vec{k}_{\nu}, + \frac{1}{2}\,)\Big].
\end{eqnarray}
One may see that the amplitude
\begin{eqnarray}\label{eq:C.24}
\hspace{-0.3in}&&{\cal M}^{(2)}_{\rm Fig.\,\ref{fig:fig3}c}(n \to p
e^- \bar{\nu}_e\gamma)_{\lambda'} = -
\,e^2\int\frac{d^4q}{(2\pi)^4i}\,D_{\alpha\beta}(q)\,
\frac{[\bar{u}_p(\vec{k}_p,\sigma_p)\,(2 k^{\alpha}_p -
    q^{\alpha})\,\gamma^{\mu}(1 + \lambda \gamma^5)
    u_n(\vec{k}_n,\sigma_n)]}{q^2 - 2 k_p\cdot q + i0}\nonumber\\
\hspace{-0.3in}&&\times
\,\Big[\bar{u}_e(\vec{k}_e,\sigma_e)\,\hat{\varepsilon}^*_{\lambda'}\,
  \frac{1}{m_e - \hat{k}_e - \hat{k} - i 0}\,\frac{
    i\,\sigma^{\beta\varphi}q_{\varphi}}{q^2 + 2\,(k_e + k)\cdot q + 2
    k_e \cdot k+ i0}\gamma_{\mu}(1 - \gamma^5) v_{\nu}(\vec{k}_{\nu},
  + \frac{1}{2}\,)\Big]
\end{eqnarray}
is also invariant under a gauge transformation $D_{\alpha\beta}(q)
\to D_{\alpha\beta}(q) + c(q^2)\,q_{\alpha}q_{\beta}$. Now we discuss
the following expression
\begin{eqnarray}\label{eq:C.25}
\hspace{-0.3in}&&{\cal M}_{\rm Fig.\,\ref{fig:fig3}c}(n \to p e^-
\bar{\nu}_e\gamma)_{\lambda'} - {\cal M}^{(1)}_{\rm
  Fig.\,\ref{fig:fig3}c}(n \to p e^- \bar{\nu}_e\gamma)_{\lambda'} -
    {\cal M}^{(2)}_{\rm Fig.\,\ref{fig:fig3}c}(n \to p e^-
    \bar{\nu}_e\gamma)_{\lambda'} =\nonumber\\
\hspace{-0.3in}&&= -
  \,e^2\int\frac{d^4q}{(2\pi)^4i}\,D_{\alpha\beta}(q)\,
  \frac{[\bar{u}_p(\vec{k}_p,\sigma_p)\,(2 k^{\alpha}_p -
      q^{\alpha})\,\gamma^{\mu}(1 + \lambda \gamma^5)
      u_n(\vec{k}_n,\sigma_n)]}{q^2 - 2 k_p\cdot q +
    i0}\nonumber\\ 
\hspace{-0.3in}&&\times
\,\Big[\bar{u}_e(\vec{k}_e,\sigma_e)\,\hat{\varepsilon}^*_{\lambda'}\,
  \frac{1}{m_e - \hat{k}_e - \hat{k} - i 0}\,\frac{(m_e - \hat{k}_e -
    \hat{k})\,\gamma^{\beta} + 2\,(k_e + k)^{\beta} + q^{\beta}}{q^2 +
    2\, (k_e + k)\cdot q + 2 k_e \cdot k + i0}\,\gamma_{\mu}(1 -
  \gamma^5) v_{\nu}(\vec{k}_{\nu}, + \frac{1}{2}\,\Big)\Big],
\end{eqnarray}
which we transcribe into the form
\begin{eqnarray}\label{eq:C.26}
\hspace{-0.3in}&&{\cal M}_{\rm Fig.\,\ref{fig:fig3}c}(n \to p e^-
\bar{\nu}_e\gamma)_{\lambda'} - {\cal M}^{(1)}_{\rm
  Fig.\,\ref{fig:fig3}c}(n \to p e^- \bar{\nu}_e\gamma)_{\lambda'} -
    {\cal M}^{(2)}_{\rm Fig.\,\ref{fig:fig3}c}(n \to p e^-
    \bar{\nu}_e\gamma)_{\lambda'} = 
    \Big[\bar{u}_p(\vec{k}_p,\sigma_p)\, \gamma^{\mu}(1 + \lambda
      \gamma^5) u_n(\vec{k}_n,\sigma_n)\Big]\nonumber\\
\hspace{-0.3in}&&\times \,\Big[\bar{u}_e(\vec{k}_e,\sigma_e)\,
  \hat{\varepsilon}^*_{\lambda'}\, \gamma^{\beta}\,\gamma_{\mu}(1 -
  \gamma^5) v_{\nu}(\vec{k}_{\nu}, + \frac{1}{2}\,\Big)\Big]
\,( -\,e^2)\int\frac{d^4q}{(2\pi)^4i}\,D_{\alpha\beta}(q)\, \frac{(2
  k^{\alpha}_p - q^{\alpha})}{q^2 - 2 k_p\cdot q + i0}\,\frac{1}{q^2 +
  2\, (k_e + k)\cdot q + 2
    k_e \cdot k + i0}\nonumber\\
\hspace{-0.3in}&&+ \Big[\bar{u}_p(\vec{k}_p,\sigma_p)\,\gamma^{\mu}(1
  + \lambda \gamma^5)
  u_n(\vec{k}_n,\sigma_n)\Big]\,\Big[\bar{u}_e(\vec{k}_e,\sigma_e)\,
  \hat{\varepsilon}^*_{\lambda'}\, \frac{1}{m_e - \hat{k}_e - \hat{k}
    - i 0}\,\gamma_{\mu}(1 - \gamma^5) v_{\nu}(\vec{k}_{\nu}, +
  \frac{1}{2}\,\Big)\Big]\nonumber\\
\hspace{-0.3in}&&\times \,(- \,e^2)
\int\frac{d^4q}{(2\pi)^4i}\,D_{\alpha\beta}(q)\, \frac{(2 k^{\alpha}_p
  - q^{\alpha})}{q^2 - 2 k_p\cdot q + i0}\,\frac{2\,(k_e + k)^{\beta}
  + q^{\beta}}{q^2 + 2\, (k_e + k)\cdot q + 2 k_e \cdot k + i0}.
\end{eqnarray}
Then, we propose to rewrite Eq.(\ref{eq:C.26}) as follows
\begin{eqnarray}\label{eq:C.27}
\hspace{-0.3in}&&{\cal M}_{\rm Fig.\,\ref{fig:fig3}c}(n \to p e^-
\bar{\nu}_e\gamma)_{\lambda'} - \sum^3_{j =1}{\cal M}^{(j)}_{\rm
  Fig.\,\ref{fig:fig3}c}(n \to p e^- \bar{\nu}_e\gamma)_{\lambda'} = \Big[\bar{u}_p(\vec{k}_p,\sigma_p)\,\gamma^{\mu}(1
  + \lambda \gamma^5)
  u_n(\vec{k}_n,\sigma_n)\Big]\nonumber\\
\hspace{-0.3in}&&\times \,\Big[\bar{u}_e(\vec{k}_e,\sigma_e)\,
  \hat{\varepsilon}^*_{\lambda'}\, \frac{1}{m_e - \hat{k}_e - \hat{k}
    - i 0}\,\gamma_{\mu}(1 - \gamma^5) v_{\nu}(\vec{k}_{\nu}, +
  \frac{1}{2}\,\Big)\Big]\,e^2
\int\frac{d^4q}{(2\pi)^4i}\,D_{\alpha\beta}(q)\, \frac{(2 k^{\alpha}_p
  - q^{\alpha})(2 k^{\beta}_p - q^{\beta})}{(q^2 - 2 k_p\cdot q +
  i0)^2}\nonumber\\
\hspace{-0.3in}&& - \Big[\bar{u}_p(\vec{k}_p,\sigma_p)\,
  \gamma^{\mu}(1 + \lambda \gamma^5)
  u_n(\vec{k}_n,\sigma_n)\Big]\,\Big[\bar{u}_e(\vec{k}_e,\sigma_e)\,
  \hat{\varepsilon}^*_{\lambda'}\, \gamma^{\beta}\,\gamma_{\mu}(1 -
  \gamma^5) v_{\nu}(\vec{k}_{\nu}, + \frac{1}{2}\,\Big)\Big]\nonumber\\
\hspace{-0.3in}&&\times \,e^2
\int\frac{d^4q}{(2\pi)^4i}\,D_{\alpha\beta}(q)\, \frac{(2 k^{\alpha}_p
  - q^{\alpha})}{q^2 - 2 k_p\cdot q + i0}\,\frac{1}{q^2 + 2\, (k_e +
  k)\cdot q + 2 k_e \cdot k + i0}\nonumber\\
\hspace{-0.3in}&& + \Big[\bar{u}_p(\vec{k}_p,\sigma_p)\,
  \gamma^{\mu}(1 + \lambda \gamma^5)
  u_n(\vec{k}_n,\sigma_n)\Big]\,\Big[\bar{u}_e(\vec{k}_e,\sigma_e)\,
  \hat{\varepsilon}^*_{\lambda'}\, \frac{1}{m_e - \hat{k}_e - \hat{k}
    - i 0}\,\gamma_{\mu}(1 - \gamma^5) v_{\nu}(\vec{k}_{\nu}, +
  \frac{1}{2}\,\Big)\Big]\nonumber\\
\hspace{-0.3in}&&\times\,e^2\,
\int\frac{d^4q}{(2\pi)^4i}\,D_{\alpha\beta}(q)\,\frac{q^{\beta}}{q^2}\,
  \frac{(2 k^{\alpha}_p - q^{\alpha})}{q^2 - 2 k_p\cdot q +
    i0}\,\frac{2 k_e\cdot k}{q^2
    + 2\, (k_e + k)\cdot q + 2 k_e \cdot k + i0},
\end{eqnarray}
where ${\cal M}^{(3)}_{\rm Fig.\,\ref{fig:fig3}c}(n \to p e^-
\bar{\nu}_e\gamma)_{\lambda'}$ is the amplitude, invariant under a
gauge transformation $D_{\alpha\beta}(q) \to D_{\alpha\beta}(q) +
c(q^2)\,q_{\alpha}q_{\beta}$, defined by
\begin{eqnarray}\label{eq:C.28}
\hspace{-0.3in}{\cal M}^{(3)}_{\rm Fig.\,\ref{fig:fig3}c}(n \to p
e^- \bar{\nu}_e\gamma)_{\lambda'} &=&
- \Big[\bar{u}_p(\vec{k}_p,\sigma_p)\, \gamma^{\mu}(1 + \lambda
  \gamma^5) u_n(\vec{k}_n,\sigma_n)\Big]\nonumber\\
\hspace{-0.3in}&&\times\,\Big[\bar{u}_e(\vec{k}_e,\sigma_e)\,
  \hat{\varepsilon}^*_{\lambda'}\, \frac{1}{m_e - \hat{k}_e - \hat{k}
    - i 0}\,\gamma_{\mu}(1 - \gamma^5) v_{\nu}(\vec{k}_{\nu}, +
  \frac{1}{2}\,\Big)\Big]\nonumber\\
\hspace{-0.3in}&&\times\,e^2\,
\Big[\int\frac{d^4q}{(2\pi)^4i}\,D_{\alpha\beta}(q)\, \frac{(2
    k^{\alpha}_p - q^{\alpha})}{q^2 - 2 k_p\cdot q +
    i0}\,\frac{\displaystyle 2\,(k_e + k)^{\beta} + q^{\beta} +
    \frac{q^{\beta}}{q^2}\,2k_e\cdot k}{q^2 + 2\, (k_e + k)\cdot q + 2
    k_e \cdot k + i0}\nonumber\\
\hspace{-0.3in}&&+ \int\frac{d^4q}{(2\pi)^4i}\,D_{\alpha\beta}(q)\,
\frac{(2 k^{\alpha}_p - q^{\alpha})(2 k^{\beta}_p - q^{\beta})}{(q^2 -
  2 k_p\cdot q + i0)^2}\Big].
\end{eqnarray}
The r.h.s. of Eq.(\ref{eq:C.27}) is not invariant under a gauge
transformation $D_{\alpha\beta}(q) \to D_{\alpha\beta}(q) +
c(q^2)\,q_{\alpha}q_{\beta}$. It is important to emphasize that unlike
a gauge non--invariant part of the Feynman diagram
Fig.\,\ref{fig:fig6}a, which is independent of the electron energy, a
gauge non--invariant part of the diagram Fig.\,\ref{fig:fig3}c, given
by the r.h.s. of Eq.(\ref{eq:C.27}), has a constant part and a part
dependent on the electron and photon energies and momenta. The results
of the calculation are given by 
\begin{eqnarray}\label{eq:C.29}
\hspace{-0.3in}e^2 \int\frac{d^4q}{(2\pi)^4i}\,D_{\alpha\beta}(q)\,
\frac{(2 k^{\alpha}_p - q^{\alpha})(2 k^{\beta}_p - q^{\beta})}{(q^2 -
  2 k_p\cdot q + i0)^2} = \frac{e^2}{8\pi^2}\,\Big(\xi\,{\ell
  n}\Big(\frac{\Lambda}{m_p}\Big) + (3 - \xi)\,{\ell
  n}\Big(\frac{\mu}{m_p}\Big) + \frac{3}{2}\Big).
\end{eqnarray}
For the second integral we obtain the following expression
\begin{eqnarray}\label{eq:C.30}
\hspace{-0.3in}&&e^2 \int\frac{d^4q}{(2\pi)^4i}\,D_{\alpha\beta}(q)\,
\frac{(2 k^{\alpha}_p - q^{\alpha})}{q^2 - 2 k_p\cdot q +
  i0}\,\frac{1}{q^2 + 2\, (k_e + k)\cdot q + 2k_e\cdot k + i0}
= -
\frac{e^2}{8\pi^2}\,k_{p\beta}\int^1_0\frac{dx}{k^2_p(x)}\,{\ell
  n}\Big(- \frac{k^2_p(x)}{2 k_e\cdot k}\Big)\nonumber\\
\hspace{-0.3in}&& +
\frac{e^2}{16\pi^2}\,\int^1_0\frac{dx\,k_{p\beta}(x)}{k^2_p(x)} +
\frac{e^2}{16\pi^2}\,(1 - \xi)\,\frac{(k_e + k)_{\beta}}{(k_e +
  k)^2}\Big[1 + \frac{m^2_e}{(k_e + k)^2}\,{\ell n}\Big(- \frac{2 k_e
    \cdot k}{m^2_e}\Big)\Big],
\end{eqnarray}
where $k_p(x) = k_p x - (k_e + k)(1 - x)$. The third integral is equal
to
\begin{eqnarray}\label{eq:C.31}
\hspace{-0.3in}&&e^2\,
\int\frac{d^4q}{(2\pi)^4i}\,D_{\alpha\beta}(q)\,\frac{q^{\beta}}{q^2}\,
\frac{(2 k^{\alpha}_p - q^{\alpha})}{q^2 - 2 k_p\cdot q + i0}\,\frac{2
  k_e\cdot k}{q^2 + 2\, (k_e + k)\cdot q + 2 k_e \cdot k + i0}
=\nonumber\\ 
\hspace{-0.3in}&&= - \frac{e^2}{8\pi^2}\,\xi\,\Big[{\ell
    n}\Big(\frac{m_e}{\mu}\Big) + \Big(1 - \frac{k_e\cdot k}{(k_e +
    k)^2}\Big)\,{\ell n}\Big(- \frac{2k_e\cdot k}{m^2_e}\Big)\Big].
\end{eqnarray}
Thus, a part of the amplitude ${\cal M}_{\rm Fig.\,\ref{fig:fig3}c}(n
\to p\,e^-\,\bar{\nu}_e \gamma)_{\lambda'}$, invariant under a gauge
transformation $D_{\alpha\beta}(q) \to D_{\alpha\beta}(q) +
c(q^2)\,q_{\alpha}q_{\beta}$, is given by
\begin{eqnarray}\label{eq:C.32}
\hspace{-0.3in}&&{\cal M}_{\rm Fig.\,\ref{fig:fig3}c}(n \to p e^-
\bar{\nu}_e\gamma)_{\lambda'} = \nonumber\\
\hspace{-0.3in}&&= - \frac{e^2}{16\pi^2}
\,\Big[\bar{u}_p(\vec{k}_p,\sigma_p)\, \gamma^{\mu}(1 + \lambda
  \gamma^5) u_n(\vec{k}_n,\sigma_n)\Big]
\,\Big[\bar{u}_e(\vec{k}_e,\sigma_e)\,\hat{\varepsilon}^*_{\lambda'}\,
  \frac{1}{m_e - \hat{k}_e - \hat{k} - i 0}\,\gamma_{\mu}(1 -
  \gamma^5) v_{\nu}(\vec{k}_{\nu}, + \frac{1}{2}\,)\Big]\nonumber\\
\hspace{-0.3in}&&\times\, f_1(E_e,\vec{k}_e,\omega,
\vec{k}\,)\nonumber\\
\hspace{-0.3in}&&-\,\frac{e^2}{16\pi^2}
\,\Big[\bar{u}_p(\vec{k}_p,\sigma_p)\, \gamma^{\mu}(1 + \lambda
  \gamma^5) u_n(\vec{k}_n,\sigma_n)\Big]
\,\Big[\bar{u}_e(\vec{k}_e,\sigma_e)\,\hat{\varepsilon}^*_{\lambda'}\,
  \frac{1}{m_e - \hat{k}_e - \hat{k} - i 0}\,\gamma^0\,(\hat{k}_e +
  \hat{k})\,\gamma_{\mu}(1 - \gamma^5) v_{\nu}(\vec{k}_{\nu}, +
  \frac{1}{2}\,)\Big]\nonumber\\
\hspace{-0.3in}&&\times\, f_2(E_e,\vec{k}_e,\omega, \vec{k}\,),
\end{eqnarray}
where the functions $f_1(E_e,\vec{k}_e,\omega, \vec{k}\,)$ and
$f_2(E_e,\vec{k}_e,\omega, \vec{k}\,)$ are defined by
\begin{eqnarray}\label{eq:C.33}
\hspace{-0.3in}f_1(E_e,\vec{k}_e,\omega, \vec{k}\,) &=& {\ell
  n}\Big(\frac{m^2_p}{m^2_e + 2 k_e\cdot k}\Big) - 2\,\Big(1 -
\frac{k_e\cdot k}{(k_e + k)^2}\Big)\,{\ell n}\Big(- \frac{2k_e\cdot
  k}{m^2_e}\Big) - \frac{E_e + \omega}{|\vec{k}_e + \vec{k}\,|}\,{\ell
  n}\Big(\frac{E_e + \omega + |\vec{k}_e + \vec{k}\,|}{E_e + \omega -
  |\vec{k}_e + \vec{k}\,|}\Big),\nonumber\\
\hspace{-0.3in}f_2(E_e,\vec{k}_e,\omega, \vec{k}\,) &=&
\frac{1}{|\vec{k}_e + \vec{k}\,|}\,{\ell n}\Big(\frac{E_e + \omega +
  |\vec{k}_e + \vec{k}\,|}{E_e + \omega - |\vec{k}_e +
  \vec{k}\,|}\Big).
\end{eqnarray}
Independent of electron and photon energies and momenta contributions
of ${\cal M}_{\rm Fig.\,\ref{fig:fig3}c}(n \to p e^- \bar{\nu}_e
\gamma)_{\lambda'}$ are removed by renormalization of the Fermi weak
$G_F$ and axial $\lambda$ coupling constant \cite{Sirlin1967}.  In the
non--relativistic limit for the proton the amplitude
Eq.(\ref{eq:C.32}) is equal to
\begin{eqnarray}\label{eq:C.34}
\hspace{-0.3in}&&\frac{1}{2m_n}\,{\cal M}_{\rm
  Fig.\,\ref{fig:fig3}c}(n \to p e^- \bar{\nu}_e\gamma)_{\lambda'} =\nonumber\\
\hspace{-0.3in}&&= 
\,\frac{e^2}{16\pi^2} \,[\varphi^{\dagger}_p\varphi_n]
\,\Big\{\Big[\bar{u}_e(\vec{k}_e,\sigma_e)\,(2 k_e \cdot
  \varepsilon^*_{\lambda'} + \hat{\varepsilon}^*_{\lambda'}\hat{k})
  \,\gamma^0\,(1 - \gamma^5) v_{\nu}(\vec{k}_{\nu}, +
  \frac{1}{2}\,)\Big]\,\frac{1}{2k_e \cdot k}\,
f_1(E_e,\vec{k}_e,\omega, \vec{k}\,)\nonumber\\
\hspace{-0.3in}&&+ \Big[\bar{u}_e(\vec{k}_e,\sigma_e)\,(2 k_e \cdot
  \varepsilon^*_{\lambda'} +
  \hat{\varepsilon}^*_{\lambda'}\hat{k})\,\gamma^0\,(\hat{k}_e +
  \hat{k})\,\gamma^0\,(1 - \gamma^5) v_{\nu}(\vec{k}_{\nu}, +
  \frac{1}{2}\,)\Big]\,\frac{1}{2k_e\cdot k}\,
f_2(E_e,\vec{k}_e,\omega, \vec{k}\,)\Big\}\nonumber\\
\hspace{-0.3in}&&- \,\frac{e^2}{16\pi^2}\,\lambda
\,[\varphi^{\dagger}_p\varphi_n]\cdot \Big\{
\Big[\bar{u}_e(\vec{k}_e,\sigma_e)\,(2 k_e \cdot
  \varepsilon^*_{\lambda'} + \hat{\varepsilon}^*_{\lambda'}\hat{k})
  \,\vec{\gamma}\,(1 - \gamma^5) v_{\nu}(\vec{k}_{\nu}, +
  \frac{1}{2}\,)\Big]\,\frac{1}{2 k_e \cdot k}\,
f_1(E_e,\vec{k}_e,\omega, \vec{k}\,)\nonumber\\
\hspace{-0.3in}&&+ \Big[\bar{u}_e(\vec{k}_e,\sigma_e)\,(2 k_e \cdot
  \varepsilon^*_{\lambda'} + \hat{\varepsilon}^*_{\lambda'}\hat{k})\,
  \gamma^0\,(\hat{k}_e + \hat{k})\,\vec{\gamma}\, (1 - \gamma^5)
  v_{\nu}(\vec{k}_{\nu}, + \frac{1}{2}\,)\Big]\,\frac{1}{2k_e\cdot
  k}\, f_2(E_e,\vec{k}_e,\omega, \vec{k}\,)\Big\}.
\end{eqnarray}
The hermitian conjugate amplitude is 
\begin{eqnarray}\label{eq:C.35}
\hspace{-0.3in}&&\frac{1}{2m_n}\,{\cal M}^{\dagger}_{\rm
  Fig.\,\ref{fig:fig3}c}(n \to p e^- \bar{\nu}_e\gamma)_{\lambda'} =
\nonumber\\
\hspace{-0.3in}&&= \frac{e^2}{16\pi^2}
\,[\varphi^{\dagger}_n\varphi_p]
\,\Big\{\Big[\bar{v}_{\nu}(\vec{k}_{\nu},+ \frac{1}{2})\,\gamma^0\,(2
  k_e \cdot \varepsilon_{\lambda'} +
  \hat{k}\hat{\varepsilon}_{\lambda'}) \,(1 - \gamma^5)\,
  u_e(\vec{k}_e, \sigma_e)\Big]\,\frac{1}{2k_e \cdot k}\,
f^*_1(E_e,\vec{k}_e,\omega, \vec{k}\,)\nonumber\\
\hspace{-0.3in}&&+ \Big[\bar{v}_{\nu}(\vec{k}_{\nu},+ \frac{1}{2})\,
  \gamma^0\,(\hat{k}_e + \hat{k})\,\gamma^0\,(2 k_e \cdot
  \varepsilon_{\lambda'} + \hat{k}\,\hat{\varepsilon}_{\lambda'})\,(1
  - \gamma^5)\,u_e(\vec{k}_e, \sigma_e)\Big]\,\frac{1}{2k_e\cdot k}\,
f^*_2(E_e,\vec{k}_e,\omega, \vec{k}\,)\Big\}\nonumber\\
\hspace{-0.3in}&&- \,\frac{e^2}{16\pi^2}\,\lambda
\,[\varphi^{\dagger}_p\varphi_n]\cdot \Big\{\Big[\bar{v}_{\nu}(\vec{k}_{\nu},+
  \frac{1}{2})\,\vec{\gamma}\,(2 k_e \cdot \varepsilon_{\lambda'} +
  \hat{k}\,\hat{\varepsilon}_{\lambda'})\,(1 - \gamma^5)\,
  u_e(\vec{k}_e, \sigma_e)\Big]\,\frac{1}{2 k_e \cdot k}\,
f^*_1(E_e,\vec{k}_e,\omega, \vec{k}\,)\nonumber\\
\hspace{-0.3in}&& + \Big[\bar{v}_{\nu}(\vec{k}_{\nu},+
  \frac{1}{2})\,\vec{\gamma}\,(\hat{k}_e + \hat{k})\,
  \gamma^0\,(2\,k_e\cdot \varepsilon_{\lambda'} +
  \hat{k}\,\hat{\varepsilon}_{\lambda'})\, (1 - \gamma^5)
  u_e(\vec{k}_e, \sigma_e) \Big]\,\frac{1}{2k_e\cdot k}\,
f^*_2(E_e,\vec{k}_e,\omega, \vec{k}\,)\Big\}.
\end{eqnarray}
The contribution of the diagrams in Fig.\,\ref{fig:fig3} to the rate
of the neutron radiative $\beta^-$--decay is defined by
\begin{eqnarray}\label{eq:C.36}
\hspace{-0.3in}&&\lambda^{(\rm
  Fig.\,\ref{fig:fig3})}_{\beta\gamma}(\omega_{\rm max},\omega_{\rm
  min}) = (1 +
3\lambda^2)\,\frac{\alpha^2}{\pi^2}\,\frac{G^2_FV_{ud}|^2}{64\pi^3}
\int^{\omega_{\rm max}}_{\omega_{\rm min}} d\omega\int^{E_0 -
  \omega}_{m_e} dE_e\,F(E_e, Z = 1)\,(E_0 - E_e - \omega)\,\sqrt{E^2_e
  - m^2_e}\;\omega\nonumber\\
\hspace{-0.3in}&&\times \,
\int\frac{d\Omega_{e\gamma}}{4\pi}\int\frac{d\Omega_{\nu}}{4\pi}\,
\Big[\frac{1}{2}\sum_{\rm pol, \lambda'}\frac{1}{1 +
    3\lambda^2}\Big({\cal M}^{\dagger}_{\rm Fig.\,\ref{fig:fig1}}(n
  \to p e^- \bar{\nu}_e \gamma)_{\lambda'} \tilde{\cal M}_{\rm
    Fig.\,\ref{fig:fig3}}(n \to p e^- \bar{\nu}_e \gamma)_{\lambda'} +
      {\rm h.c.}\Big)\Big]\Big|_{E_{\nu} = E_0 - E_e - \omega},
\nonumber\\
\hspace{-0.3in}&&
\end{eqnarray}
where $\tilde{\cal M}_{\rm Fig.\,\ref{fig:fig3}} = (16\pi^2/2m_n
e^2)\,{\cal M}_{\rm Fig.\,\ref{fig:fig3}c}$.  The sum over
polarizations of the proton, electron and photon, averaged over
polarizations of the neutron, is defined by the following traces over
Dirac matrices
\begin{eqnarray}\label{eq:C.37}
\hspace{-0.3in}&&\frac{1}{2}\sum_{\rm pol, \lambda'}\frac{1}{1 +
  3\lambda^2}\Big({\cal M}^{\dagger}_{\rm Fig.\,\ref{fig:fig1}}(n \to
p e^- \bar{\nu}_e \gamma)_{\lambda'} \tilde{\cal M}_{\rm
  Fig.\,\ref{fig:fig3}}(n \to p e^- \bar{\nu}_e \gamma)_{\lambda'} +
{\rm h. c.}\Big) = \frac{1}{1 + 3\lambda^2}\, \frac{1}{2 (k_e\cdot
  k)^2}\nonumber\\
\hspace{-0.3in}&&\times\,\Big\{f_1\, {\rm tr}\{(m_e + \hat{k}_e)\,(2
k_e \cdot \varepsilon^*_{\lambda'} +
\hat{\varepsilon}^*_{\lambda'}\,\hat{k})\, \gamma^0 \,\hat{k}_{\nu}
\gamma^0\, (2 k_e \cdot \varepsilon_{\lambda'} + \hat{k}\,
\hat{\varepsilon}_{\lambda'})\, (1 - \gamma^5)\}\nonumber\\
\hspace{-0.3in}&& + \lambda^2\, \delta^{ij}\, f_1 \, {\rm tr}\{(m_e +
\hat{k}_e)\, (2 k_e \cdot \varepsilon^*_{\lambda'} +
\hat{\varepsilon}^*_{\lambda'}\,\hat{k})\, \vec{\gamma}^{\,i}
\,\hat{k}_{\nu} \, \vec{\gamma}^{\,j}\, (2 k_e \cdot
\varepsilon_{\lambda'} + \hat{k}\, \hat{\varepsilon}_{\lambda'})\, (1
- \gamma^5)\}\Big\} + \frac{1}{1 + 3\lambda^2}\, \frac{1}{2 (k_e\cdot
  k)^2}\nonumber\\
\hspace{-0.3in}&&\times\,\Big\{f_2\, {\rm tr}\{(m_e + \hat{k}_e)\,(2
k_e \cdot \varepsilon^*_{\lambda'} +
\hat{\varepsilon}^*_{\lambda'}\,\hat{k})\, \gamma^0 \, (\hat{k}_e +
\hat{k}\,)\, \gamma^0 \,\hat{k}_{\nu} \, \gamma^0 \, (2 k_e \cdot
\varepsilon_{\lambda'} + \hat{k}\, \hat{\varepsilon}_{\lambda'})\, (1
- \gamma^5)\} \nonumber\\
\hspace{-0.3in}&& + \lambda^2\, \delta^{ij}\, f_2 \, {\rm tr}\{(m_e +
\hat{k}_e)\, (2 k_e \cdot \varepsilon^*_{\lambda'} +
\hat{\varepsilon}^*_{\lambda'}\,\hat{k})\, \gamma^0 \, (\hat{k}_e +
\hat{k}\,)\, \vec{\gamma}^{\,i} \,\hat{k}_{\nu} \,
\vec{\gamma}^{\,j}\, (2 k_e \cdot \varepsilon_{\lambda'} + \hat{k}\,
\hat{\varepsilon}_{\lambda'})\, (1 - \gamma^5)\}\Big\} + {\rm h. c.}\,.
\end{eqnarray}
Having integrated over directions of the antineutrino momentum
$\vec{k}_{\nu}$ we arrive at the expression
\begin{eqnarray*}
\hspace{-0.3in}&&\int \frac{d\Omega_{\nu}}{4\pi} \,
\frac{1}{2}\sum_{\rm pol, \lambda'}\frac{1}{1 + 3\lambda^2}\Big({\cal
  M}^{\dagger}_{\rm Fig.\,\ref{fig:fig1}}(n \to p e^- \bar{\nu}_e
\gamma)_{\lambda'} \tilde{\cal M}_{\rm Fig.\,\ref{fig:fig3}}(n \to p
e^- \bar{\nu}_e \gamma)_{\lambda'} + {\rm h. c.}\Big) = \frac{E_{\nu}}{1 +
  3\lambda^2}\, \frac{1}{2 (k_e\cdot k)^2}\nonumber\\
\hspace{-0.3in}&&\times\,\Big\{f_1\, {\rm tr}\{(m_e + \hat{k}_e)\,(2
k_e \cdot \varepsilon^*_{\lambda'} +
\hat{\varepsilon}^*_{\lambda'}\,\hat{k})\, \gamma^0 \, (2 k_e \cdot
\varepsilon_{\lambda'} + \hat{k}\, \hat{\varepsilon}_{\lambda'})\, (1
- \gamma^5)\}\nonumber\\
\hspace{-0.3in}&& + \lambda^2\, \delta^{ij}\, f_1 \, {\rm tr}\{(m_e +
\hat{k}_e)\, (2 k_e \cdot \varepsilon^*_{\lambda'} +
\hat{\varepsilon}^*_{\lambda'}\,\hat{k})\, \vec{\gamma}^{\,i}
\,\gamma^0\, \vec{\gamma}^{\,j}\, (2 k_e \cdot \varepsilon_{\lambda'}
+ \hat{k}\, \hat{\varepsilon}_{\lambda'})\, (1 - \gamma^5)\}\Big\} +
\frac{E_{\nu}}{1 + 3\lambda^2}\, \frac{1}{2 (k_e\cdot k)^2}\nonumber\\
\end{eqnarray*}
\begin{eqnarray}\label{eq:C.38}
\hspace{-0.3in}&&\times\,\Big\{f_2\, {\rm tr}\{(m_e + \hat{k}_e)\,(2
k_e \cdot \varepsilon^*_{\lambda'} +
\hat{\varepsilon}^*_{\lambda'}\,\hat{k})\, \gamma^0 \, (\hat{k}_e +
\hat{k}\,)\, \gamma^0 \, (2 k_e \cdot
\varepsilon_{\lambda'} + \hat{k}\, \hat{\varepsilon}_{\lambda'})\, (1
- \gamma^5)\} \nonumber\\
\hspace{-0.3in}&& + \lambda^2\, \delta^{ij}\, f_2 \, {\rm tr}\{(m_e +
\hat{k}_e)\, (2 k_e \cdot \varepsilon^*_{\lambda'} +
\hat{\varepsilon}^*_{\lambda'}\,\hat{k})\, \gamma^0 \, (\hat{k}_e +
\hat{k}\,)\, \vec{\gamma}^{\,i} \,\gamma^0 \,
\vec{\gamma}^{\,j}\, (2 k_e \cdot \varepsilon_{\lambda'} + \hat{k}\,
\hat{\varepsilon}_{\lambda'})\, (1 - \gamma^5)\}\Big\} + {\rm h. c.}\,.
\end{eqnarray}
Since $\delta^{ij} \vec{\gamma}^{\,i} \,\gamma^0 \vec{\gamma}^{\,j} =
\vec{\gamma} \cdot \, \vec{\gamma} = 3\,\gamma^0$, we get
\begin{eqnarray}\label{eq:C.39}
\hspace{-0.3in}&&\int \frac{d\Omega_{\nu}}{4\pi} \,
\frac{1}{2}\sum_{\rm pol, \lambda'}\frac{1}{1 + 3\lambda^2}\Big({\cal
  M}^{\dagger}_{\rm Fig.\,\ref{fig:fig1}}(n \to p e^- \bar{\nu}_e
\gamma)_{\lambda'} \tilde{\cal M}_{\rm Fig.\,\ref{fig:fig3}}(n \to p
e^- \bar{\nu}_e \gamma)_{\lambda'} + {\rm h. c.}\Big) =\nonumber\\
\hspace{-0.3in}&&= \frac{E_{\nu}}{2 (k_e\cdot k)^2}\,\Big\{f_1\,
\sum_{ \lambda'}{\rm tr}\{(m_e + \hat{k}_e)\,(2 k_e \cdot
\varepsilon^*_{\lambda'} + \hat{\varepsilon}^*_{\lambda'}\,\hat{k})\,
\gamma^0 \, (2 k_e \cdot \varepsilon_{\lambda'} + \hat{k}\,
\hat{\varepsilon}_{\lambda'})\, (1 - \gamma^5)\}\nonumber\\
\hspace{-0.3in}&& + f_2\, \sum_{ \lambda'}{\rm tr}\{(m_e +
\hat{k}_e)\,(2 k_e \cdot \varepsilon^*_{\lambda'} +
\hat{\varepsilon}^*_{\lambda'}\,\hat{k})\, \gamma^0 \, (\hat{k}_e +
\hat{k}\,)\, \gamma^0 \, (2 k_e \cdot \varepsilon_{\lambda'} +
\hat{k}\, \hat{\varepsilon}_{\lambda'})\, (1 - \gamma^5)\} + {\rm
  h. c.}\Big\}.
\end{eqnarray}
The traces are equal to
\begin{eqnarray}\label{eq:C.40}
\hspace{-0.3in}&&\sum_{ \lambda'}{\rm tr}\{(m_e + \hat{k}_e)\,(2 k_e
\cdot \varepsilon^*_{\lambda'} +
\hat{\varepsilon}^*_{\lambda'}\,\hat{k})\, \gamma^0 \, (2 k_e \cdot
\varepsilon_{\lambda'} + \hat{k}\, \hat{\varepsilon}_{\lambda'})\, (1
- \gamma^5)\} = 16\, (E_e + \omega) \,\Big(k^2_e - (\vec{k}_e \cdot
\vec{n}_{\vec{k}})^2\Big)\nonumber\\
\hspace{-0.3in}&& + 16\, \omega^2 \, \Big(E_e - \vec{k}_e \cdot
\vec{n}_{\vec{k}}\Big),\nonumber\\
\hspace{-0.3in}&&\sum_{ \lambda'}{\rm tr}\{(m_e + \hat{k}_e)\,(2 k_e
\cdot \varepsilon^*_{\lambda'} +
\hat{\varepsilon}^*_{\lambda'}\,\hat{k})\, \gamma^0 \, (\hat{k}_e +
\hat{k}\,)\, \gamma^0 \, (2 k_e \cdot \varepsilon_{\lambda'} +
\hat{k}\, \hat{\varepsilon}_{\lambda'})\, (1 - \gamma^5)\} = \nonumber\\
\hspace{-0.3in}&& = 16\,
\Big(2 E_e (E_e + \omega) - \omega\, (E_e - \vec{k}_e \cdot
\vec{n}_{\vec{k}}) - m^2_e\Big)\, \Big(k^2_e - (\vec{k}_e \cdot
\vec{n}_{\vec{k}})^2\Big) + 32\, \omega \,(E_e + \omega)\, \Big(k^2_e
- (\vec{k}_e \cdot \vec{n}_{\vec{k}})^2\Big)\nonumber\\
\hspace{-0.3in}&& + 32\, \omega^2 \, (E_e + \omega) \, (E_e -
\vec{k}_e \cdot \vec{n}_{\vec{k}}) - 16\, \omega^2 \,(E_e - \vec{k}_e
\cdot \vec{n}_{\vec{k}})^2.
\end{eqnarray}
Plugging Eq.(\ref{eq:C.40}) into Eq.(\ref{eq:C.36}) for the
contribution of the diagram in Fig.\,\ref{fig:fig3} to the rate of the
neutron radiative $\beta^-$--decay with a photon from the energy
region $\omega_{\rm min} \le \omega \le \omega_{\rm max}$ we obtain
the following expression
\begin{eqnarray}\label{eq:C.41}
\hspace{-0.3in}&&\lambda^{(\rm
  Fig.\,\ref{fig:fig3})}_{\beta\gamma}(\omega_{\rm max},\omega_{\rm
  min}) = (1 +
3\lambda^2)\,\frac{\alpha^2}{\pi^2}\,\frac{G^2_F|V_{ud}|^2}{4\pi^3}
\int^{\omega_{\rm max}}_{\omega_{\rm min}} \frac{d\omega}{\omega}\int^{E_0 -
  \omega}_{m_e} dE_e\,F(E_e, Z = 1)\,(E_0 - E_e -
\omega)^2\,\sqrt{E^2_e - m^2_e}\nonumber\\
\hspace{-0.3in}&&\times \,
\int\frac{d\Omega_{e\gamma}}{4\pi}\,\Big\{{\rm Re}f_1(E_e, \vec{k}_e,
\omega, \vec{k}\,)\, \Big[(E_e + \omega)\, \frac{k^2_e - (\vec{k}_e
    \cdot \vec{n}_{\vec{k}})^2}{(E_e - \vec{k}_e \cdot
    \vec{n}_{\vec{k}})^2} + \frac{\omega^2}{E_e - \vec{k}_e \cdot
    \vec{n}_{\vec{k}}}\Big] + {\rm Re}f_2(E_e, \vec{k}_e, \omega,
\vec{k}\,)\, \Big[\Big(2\, (E_e + \omega)^2 - m^2_e\nonumber\\
\hspace{-0.3in}&& - \omega \,(E_e - \vec{k}_e \cdot
\vec{n}_{\vec{k}})\Big)\, \frac{k^2_e - (\vec{k}_e \cdot
  \vec{n}_{\vec{k}})^2}{(E_e - \vec{k}_e \cdot \vec{n}_{\vec{k}})^2} +
2\, (E_e + \omega)\, \frac{\omega^2}{E_e - \vec{k}_e \cdot
  \vec{n}_{\vec{k}}} - \omega^2\Big]\Big\},
\end{eqnarray}
where the functions $f_1(E_e, \vec{k}_e, \omega, \vec{k}\,)$ and
$f_2(E_e, \vec{k}_e, \omega, \vec{k}\,)$ are given in
Eq.(\ref{eq:C.33}).

\section*{Appendix D: The amplitude of the neutron 
radiative $\beta^-$--decay, described by Feynman diagrams in
Fig.\,\ref{fig:fig4}} \renewcommand{\theequation}{D-\arabic{equation}}
\setcounter{equation}{0}

The analytical expressions for the diagrams Fig.\,\ref{fig:fig4}f and
Fig.\,\ref{fig:fig4}g are given by
\begin{eqnarray}\label{eq:D.1}
\hspace{-0.3in}&&{\cal M}_{\rm Fig.\,\ref{fig:fig4}a}(n \to p e^-
\bar{\nu}_e\gamma)_{\lambda'} =\nonumber\\
\hspace{-0.3in}&&= - e^2\int \frac{d^4q}{(2\pi)^4
  i}\,\Big[\bar{u}_p(\vec{k}_p,
  \sigma_p)\,\gamma^{\alpha}\,\frac{1}{m_p - \hat{k}_p + \hat{q} - i
    0}\,\hat{\varepsilon}^*_{\lambda'}\,\frac{1}{m_p - \hat{k}_p -
    \hat{k} + \hat{q} - i 0}\,\gamma^{\mu}(1 + \lambda
  \gamma^5)\,u_n(\vec{k}_n, \sigma_n)\Big]\nonumber\\
\hspace{-0.3in}&&\times\,\Big[\bar{u}_e(\vec{k}_e,
  \sigma_e)\,\gamma^{\alpha}\,\frac{1}{m_e - \hat{k}_e - \hat{q} - i
    0}\,\gamma^{\mu}(1 + \lambda \gamma^5)\,v_{\nu}(\vec{k}_{\nu}, +
  \frac{1}{2})\Big]\,\frac{1}{q^2 + i 0}
\end{eqnarray}
and 
\begin{eqnarray}\label{eq:D.2}
\hspace{-0.3in}&&{\cal M}_{\rm Fig.\,\ref{fig:fig4}b}(n \to p e^-
\bar{\nu}_e\gamma)_{\lambda'} =\nonumber\\
\hspace{-0.3in}&&= e^2\int \frac{d^4q}{(2\pi)^4
  i}\,\Big[\bar{u}_p(\vec{k}_p,
  \sigma_p)\,\gamma^{\alpha}\,\frac{1}{m_p - \hat{k}_p + \hat{q} - i
    0}\,\gamma^{\mu}(1 + \lambda
  \gamma^5)\,u_n(\vec{k}_n, \sigma_n)\Big]\nonumber\\
\hspace{-0.3in}&&\times\,\Big[\bar{u}_e(\vec{k}_e,
  \sigma_e)\,\gamma_{\alpha}\,\frac{1}{m_e - \hat{k}_e - \hat{q} - i
    0}\,\hat{\varepsilon}^*_{\lambda'}\,\frac{1}{m_e - \hat{k}_e -
    \hat{k} - \hat{q} - i 0}\,\gamma_{\mu}(1 -
  \gamma^5)\,v_{\nu}(\vec{k}_{\nu}, + \frac{1}{2})\Big]\,\frac{1}{q^2 +
  i 0}
\end{eqnarray}
For the calculation of the integrals over $q$ we rewrite Eq.(\ref{eq:D.1}) and Eq.(\ref{eq:D.2}) as follows 
\begin{eqnarray}\label{eq:D.3}
\hspace{-0.3in}&&{\cal M}_{\rm Fig.\,\ref{fig:fig4}a}(n \to p e^-
\bar{\nu}_e\gamma)_{\lambda'} =\nonumber\\
\hspace{-0.3in}&&= e^2\int \frac{d^4q}{(2\pi)^4
  i}\,\Big[\bar{u}_p(\vec{k}_p, \sigma_p)\,\frac{(2k^{\alpha}_p -
    \gamma^{\alpha}\hat{q})}{q^2 - 2k_p\cdot q +
    i0}\,\hat{\varepsilon}^*_{\lambda'}\,\frac{(m_p + \hat{k}_p +
    \hat{k} - \hat{q})}{q^2 - 2(k_p + k)\cdot q + 2 k_p\cdot k + i
    0}\,\gamma^{\mu}(1 + \lambda \gamma^5)\,u_n(\vec{k}_n,
  \sigma_n)\Big]\nonumber\\
\hspace{-0.3in}&&\times\,\Big[\bar{u}_e(\vec{k}_e,
  \sigma_e)\,\frac{(2k_{e\alpha} +
    \gamma_{\alpha}\hat{q})}{q^2 + 2k_e\cdot q + i 0}\,\gamma^{\mu}(1
  + \lambda \gamma^5)\,v_{\nu}(\vec{k}_{\nu}, +
  \frac{1}{2})\Big]\,\frac{1}{q^2 + i 0}
\end{eqnarray}
and 
\begin{eqnarray}\label{eq:D.4}
\hspace{-0.3in}&&{\cal M}_{\rm Fig.\,\ref{fig:fig4}b}(n \to p e^-
\bar{\nu}_e\gamma)_{\lambda'} =\nonumber\\
\hspace{-0.3in}&&= -\, e^2\int \frac{d^4q}{(2\pi)^4
  i}\,\Big[\bar{u}_p(\vec{k}_p, \sigma_p)\,\frac{(2 k^{\alpha}_p -
    \gamma^{\alpha}\hat{q})}{q^2 - 2 k_p\cdot q + i 0}\,\gamma^{\mu}(1
  + \lambda \gamma^5)\,u_n(\vec{k}_n, \sigma_n)\Big]\nonumber\\
\hspace{-0.3in}&&\times\,\Big[\bar{u}_e(\vec{k}_e, \sigma_e)\,\frac{(2
    k_{e\alpha} + \gamma_{\alpha}\hat{q})}{q^2 + 2 k_e\cdot q + i
    0}\,\hat{\varepsilon}^*_{\lambda'}\,\frac{(m_e+ \hat{k}_e +
    \hat{k} + \hat{q})}{q^2 + 2 (k_e + k)\cdot q + 2 k_e \cdot k + i
    0}\,\gamma_{\mu}(1 - \gamma^5)\,v_{\nu}(\vec{k}_{\nu}, +
  \frac{1}{2})\Big]\,\frac{1}{q^2 + i 0},
\end{eqnarray}
where we have used the properties of the Dirac $\gamma$--matrices and
Dirac equations for the free proton and electron \cite{Itzykson1980},
keeping the proton and electron on--mass shell $k^2_p = m^2_p$ and
$k^2_e = m^2_e$. Then, we merge the denominators \cite{Ivanov2013}
\begin{eqnarray}\label{eq:D.5}
\hspace{-0.3in}&&\frac{1}{q^2 - 2k_p\cdot q + i0}\,\frac{1}{q^2 -
  2(k_p + k)\cdot q + 2 k_p \cdot k + i 0}\,\frac{1}{q^2 + 2k_e\cdot q
  + i 0}\,\frac{1}{q^2 - \mu^2 + i 0} =\nonumber\\
\hspace{-0.3in}&&=
\int^1_0dx\int^1_0dy\,2\,y\int^1_0dz\,3\,z^2\,\frac{1}{[(q -
    k_p(x,y)z)^2 - k^2_p(x,y)z^2 + 2 k_p\cdot k \, x y z - \mu^2 (1 - z)
    + i 0]^4},
\end{eqnarray}
 and
\begin{eqnarray}\label{eq:D.6}
\hspace{-0.3in}&&\frac{1}{q^2 - 2 k_p\cdot q + i 0}\,\frac{1}{q^2 + 2
  k_e\cdot q + i 0}\,\frac{1}{q^2 + 2 (k_e + k)\cdot q + 2 k_e \cdot k
  + i 0}\,\frac{1}{q^2 - \mu^2 + i 0} = \nonumber\\
\hspace{-0.3in}&&=
\int^1_0dx\int^1_0dy\,2\,y\int^1_0dz\,3\,z^2\,\frac{1}{[(q +
    k_e(x,y)z)^2 - k^2_e(x,y)z^2 + 2 k_e \cdot k\, x y z - \mu^2 (1 - z)
    + i 0]^4},
\end{eqnarray}
where $k_p(x,y) = k_p y - k_e(1 - y) + k\,x y$, $k_e(x,y) = k_e y -
k_p (1 - y) + k\,x y$ and $\mu$ is a photon mass, regularizing
infrared divergences \cite{Sirlin1967}. Making the shifts of variables
$q - k_p(x,y)z \to q$ and $q + k_e(x,y)z \to q$ in Eq.(\ref{eq:D.3})
and Eq.(\ref{eq:D.4}), respectively, and integrating over the
directions of the virtual 4--momentum $q$ we arrive at the expressions
\begin{eqnarray}\label{eq:D.7}
\hspace{-0.3in}&&{\cal M}_{\rm Fig.\,\ref{fig:fig4}a}(n \to p e^-
\bar{\nu}_e\gamma)_{\lambda'} =\nonumber\\
\hspace{-0.3in}&&= e^2
\int^1_0dx\int^1_0dy\,2\,y\int^1_0dz\,3\,z^2\,\int
\frac{d^4q}{(2\pi)^4 i}\,\frac{1}{[q^2 - k^2_p(x,y)z^2 + 2 k_p\cdot k
    \, x y z - \mu^2 (1 - z) + i 0]^4}\nonumber\\
\hspace{-0.3in}&&\times \,\Big\{\Big[\bar{u}_p(\vec{k}_p,
  \sigma_p)\,\Big(2 k^{\alpha}_p -
  \gamma^{\alpha}\hat{k}_p(x)z\Big)\,\hat{\varepsilon}^*_{\lambda'}\,\Big(m_p
  + \hat{k}_p + \hat{k} - \hat{k}_p(x,y)z\Big)\,\gamma^{\mu}(1 + \lambda
  \gamma^5)\,u_n(\vec{k}_n, \sigma_n)\Big]\nonumber\\
\hspace{-0.3in}&&\times\,\Big[\bar{u}_e(\vec{k}_e, \sigma_e)\,\Big(2
  k_{e\alpha} + \gamma_{\alpha}\hat{k}_p(x,y)z\Big)\,\gamma^{\mu}(1 +
  \lambda \gamma^5)\,v_{\nu}(\vec{k}_{\nu}, + \frac{1}{2})\Big]\nonumber\\
\hspace{-0.3in}&& - \frac{1}{2}\,q^2\,\Big[\bar{u}_p(\vec{k}_p,
  \sigma_p)\,
  \gamma^{\alpha}\,\hat{\varepsilon}^*_{\lambda'}\,\gamma^{\mu}(1 +
  \lambda \gamma^5)\,u_n(\vec{k}_n,
  \sigma_n)\Big]\,\Big[\bar{u}_e(\vec{k}_e, \sigma_e)\,\Big(2
  k_{e\alpha} + \gamma_{\alpha}\hat{k}_p(x,y)z\Big)\,\gamma^{\mu}(1 +
  \lambda \gamma^5)\,v_{\nu}(\vec{k}_{\nu}, +
  \frac{1}{2})\Big]\nonumber\\
\hspace{-0.3in}&& - \frac{1}{4}\,q^2\,\Big[\bar{u}_p(\vec{k}_p,
  \sigma_p)\,
  \gamma^{\alpha}\,\gamma^{\beta}\,\hat{\varepsilon}^*_{\lambda'}\,\Big(m_p
  + \hat{k}_p + \hat{k} - \hat{k}_p(x,y)z\Big)\,\gamma^{\mu}(1 +
  \lambda \gamma^5)\,u_n(\vec{k}_n,
  \sigma_n)\Big]\nonumber\\
\hspace{-0.3in}&&\times\,\Big[\bar{u}_e(\vec{k}_e,
  \sigma_e)\,\gamma_{\alpha}\,\gamma_{\beta}\,\gamma^{\mu}(1 + \lambda
  \gamma^5)\,v_{\nu}(\vec{k}_{\nu}, + \frac{1}{2})\Big]\nonumber\\
\hspace{-0.3in}&& - \frac{1}{4}\,q^2\,\Big[\bar{u}_p(\vec{k}_p,
  \sigma_p)\,\Big(2 k^{\alpha}_p -
  \gamma^{\alpha}\hat{k}_p(x,y)z\Big)\,\hat{\varepsilon}^*_{\lambda'}\,
  \gamma^{\beta} \,\gamma^{\mu}(1 + \lambda \gamma^5)\,u_n(\vec{k}_n,
  \sigma_n)\Big]\,\Big[\bar{u}_e(\vec{k}_e,
  \sigma_e)\,\gamma_{\alpha}\,\gamma_{\beta}\,\gamma_{\mu}(1 -
  \gamma^5)\,v_{\nu}(\vec{k}_{\nu}, +
  \frac{1}{2})\Big]\Big\}\nonumber\\
\hspace{-0.3in}&&
\end{eqnarray}
and 
\begin{eqnarray*}
\hspace{-0.3in}&&{\cal M}_{\rm Fig.\,\ref{fig:fig4}b}(n \to p e^-
\bar{\nu}_e\gamma)_{\lambda'} =\nonumber\\
\hspace{-0.3in}&&= -\, e^2
\int^1_0dx\int^1_0dy\,2\,y\int^1_0dz\,3\,z^2\,\int
\frac{d^4q}{(2\pi)^4 i}\,\frac{1}{[q^2 - k^2_e(x,y)z^2 + 2 k_e\cdot k
    \, x y z - \mu^2 (1 - z) + i 0]^4}\nonumber\\
\hspace{-0.3in}&&\times \,\Big\{\Big[\bar{u}_p(\vec{k}_p,
  \sigma_p)\,\Big(2 k^{\alpha}_p +
  \gamma^{\alpha}\hat{k}_e(x,y)z\Big)\,\gamma^{\mu}(1 + \lambda
  \gamma^5)\,u_n(\vec{k}_n, \sigma_n)\Big]\nonumber\\
\end{eqnarray*}
\begin{eqnarray}\label{eq:D.8}
\hspace{-0.3in}&&\times\,\Big[\bar{u}_e(\vec{k}_e, \sigma_e)\,\Big(2
  k_{e\alpha} - \gamma_{\alpha}\hat{k}_e(x,y)z\Big)\,
  \hat{\varepsilon}^*_{\lambda'}\,\Big(m_e + \hat{k}_e + \hat{k} -
  \hat{k}_e(x,y)z\Big)\,\gamma_{\mu}(1 -
  \gamma^5)\,v_{\nu}(\vec{k}_{\nu}, + \frac{1}{2})\Big]\nonumber\\
\hspace{-0.3in}&& - \frac{1}{2}\,q^2\,\Big[\bar{u}_p(\vec{k}_p,
  \sigma_p)\,\Big(2 k^{\alpha}_p +
  \gamma^{\alpha}\hat{k}_e(x,y)z\Big)\,\gamma^{\mu}(1 + \lambda
  \gamma^5)\,u_n(\vec{k}_n, \sigma_n)\Big]\,\Big[\bar{u}_e(\vec{k}_e,
  \sigma_e)\,\gamma_{\alpha}\, \hat{\varepsilon}^*_{\lambda'}\,
  \gamma^{\mu}(1 + \lambda \gamma^5)\,v_{\nu}(\vec{k}_{\nu}, +
  \frac{1}{2})\Big]\nonumber\\
\hspace{-0.3in}&& - \frac{1}{4}\,q^2\,\Big[\bar{u}_p(\vec{k}_p,
  \sigma_p)\, \gamma^{\alpha}\,\gamma^{\beta}\,\gamma^{\mu}(1 +
  \lambda \gamma^5)\,u_n(\vec{k}_n,
  \sigma_n)\Big]\nonumber\\
\hspace{-0.3in}&&\times\,\Big[\bar{u}_e(\vec{k}_e,
  \sigma_e)\,\gamma_{\alpha}\gamma_{\beta}\,
  \hat{\varepsilon}^*_{\lambda'}\,\Big(m_e + \hat{k}_e + \hat{k} -
  \hat{k}_e(x,y)z\Big)\, \gamma^{\mu}(1 + \lambda
  \gamma^5)\,v_{\nu}(\vec{k}_{\nu}, + \frac{1}{2})\Big]\nonumber\\
\hspace{-0.3in}&& - \frac{1}{4}\,q^2\,\Big[\bar{u}_p(\vec{k}_p,
  \sigma_p)\, \gamma^{\alpha}\,\gamma^{\beta}\,\gamma^{\mu}(1 +
  \lambda \gamma^5)\,u_n(\vec{k}_n,
  \sigma_n)\Big]\,\Big[\bar{u}_e(\vec{k}_e, \sigma_e)\,\Big(2
  k_{e\alpha} - \gamma_{\alpha} \hat{k}_e(x,y)z\Big) \,
  \hat{\varepsilon}^*_{\lambda'}\,\gamma_{\beta}\, \gamma^{\mu}(1 +
  \lambda \gamma^5)\,v_{\nu}(\vec{k}_{\nu}, + \frac{1}{2})\Big].\nonumber\\
\hspace{-0.3in}&&
\end{eqnarray}
Making a Wick rotation and integrating over $q^2$ we arrive at the
following expressions for the diagrams Fig.\,\ref{fig:fig4}f and
Fig.\,\ref{fig:fig4}g
\begin{eqnarray}\label{eq:D.9}
\hspace{-0.3in}&&{\cal M}_{\rm Fig.\,\ref{fig:fig4}a}(n \to p e^-
\bar{\nu}_e\gamma)_{\lambda'} = \frac{e^2}{96\pi^2}
\int^1_0dx\int^1_0dy\,2\,y\int^1_0dz\,3\,z^2\,\frac{1}{[k^2_p(x,y)z^2
    - 2 k_p\cdot k \, x y z+ \mu^2 (1 - z)]^2}\nonumber\\
\hspace{-0.3in}&&\times \,\Big[\bar{u}_p(\vec{k}_p, \sigma_p)\,\Big(2
  k^{\alpha}_p - \gamma^{\alpha}\hat{k}_p(x, y)
  z\Big)\,\hat{\varepsilon}^*_{\lambda'}\,\Big(m_p + \hat{k}_p +
  \hat{k} - \hat{k}_p(x,y)z\Big)\,\gamma^{\mu}(1 + \lambda
  \gamma^5)\,u_n(\vec{k}_n, \sigma_n)\Big]\nonumber\\
\hspace{-0.3in}&&\times\,\Big[\bar{u}_e(\vec{k}_e, \sigma_e)\,\Big(2
  k_{e\alpha} + \gamma_{\alpha}\hat{k}_p(x,y)z\Big)\,\gamma^{\mu}(1 +
  \lambda \gamma^5)\,v_{\nu}(\vec{k}_{\nu}, + \frac{1}{2})\Big]\nonumber\\
\hspace{-0.3in}&& +
\frac{e^2}{48\pi^2}\int^1_0dx\int^1_0dy\,2\,y\int^1_0dz\,3\,z^2\,\frac{1}{k^2_p(x,y)z^2 - 2 k_p \cdot k\, xyz 
  + \mu^2 (1 - z)}\nonumber\\
\hspace{-0.3in}&&\times \,\Big\{\frac{1}{2}\,\Big[\bar{u}_p(\vec{k}_p,
  \sigma_p)\,
  \gamma^{\alpha}\,\hat{\varepsilon}^*_{\lambda'}\,\gamma^{\mu}(1 +
  \lambda \gamma^5)\,u_n(\vec{k}_n,
  \sigma_n)\Big]\,\Big[\bar{u}_e(\vec{k}_e, \sigma_e)\,\Big(2
  k_{e\alpha} + \gamma_{\alpha}\hat{k}_p(x,y)z\Big)\,\gamma^{\mu}(1 +
  \lambda \gamma^5)\,v_{\nu}(\vec{k}_{\nu}, +
  \frac{1}{2})\Big]\nonumber\\
\hspace{-0.3in}&& + \frac{1}{4}\,\Big[\bar{u}_p(\vec{k}_p,
  \sigma_p)\,
  \gamma^{\alpha}\,\gamma^{\beta}\,\hat{\varepsilon}^*_{\lambda'}\,\Big(m_p
  + \hat{k}_p + \hat{k} - \hat{k}_p(x,y)z\Big)\,\gamma^{\mu}(1 +
  \lambda \gamma^5)\,u_n(\vec{k}_n,
  \sigma_n)\Big]\nonumber\\
\hspace{-0.3in}&&\times\,\Big[\bar{u}_e(\vec{k}_e,
  \sigma_e)\,\gamma_{\alpha}\,\gamma_{\beta}\,\gamma^{\mu}(1 + \lambda
  \gamma^5)\,v_{\nu}(\vec{k}_{\nu}, + \frac{1}{2})\Big]\nonumber\\
\hspace{-0.3in}&&+ \frac{1}{4}\,\Big[\bar{u}_p(\vec{k}_p,
  \sigma_p)\,\Big(2 k^{\alpha}_p -
  \gamma^{\alpha}\hat{k}_p(x,y)z\Big)\,\hat{\varepsilon}^*_{\lambda'}\,
  \gamma^{\beta} \,\gamma^{\mu}(1 + \lambda \gamma^5)\,u_n(\vec{k}_n,
  \sigma_n)\Big]\,\Big[\bar{u}_e(\vec{k}_e,
  \sigma_e)\,\gamma_{\alpha}\,\gamma_{\beta}\,\gamma_{\mu}(1 -
  \gamma^5)\,v_{\nu}(\vec{k}_{\nu}, +
  \frac{1}{2})\Big]\Big\}\nonumber\\
\hspace{-0.3in}&&
\end{eqnarray}
and 
\begin{eqnarray}\label{eq:D.10}
\hspace{-0.3in}&&{\cal M}_{\rm Fig.\,\ref{fig:fig4}b}(n \to p e^-
\bar{\nu}_e\gamma)_{\lambda'} = - \frac{e^2}{96\pi^2}
\int^1_0dx\int^1_ody\,2\,y\int^1_0dz\,3\,z^2\,\frac{1}{[k^2_e(x,y)z^2
  - 2 k_e\cdot k\, xyz  + \mu^2 (1 - z)]^2}\nonumber\\
\hspace{-0.3in}&&\times \,\Big\{\Big[\bar{u}_p(\vec{k}_p,
  \sigma_p)\,\Big(2 k^{\alpha}_p +
  \gamma^{\alpha}\hat{k}_e(x,y)z\Big)\,\gamma^{\mu}(1 + \lambda
  \gamma^5)\,u_n(\vec{k}_n, \sigma_n)\Big]\nonumber\\
\hspace{-0.3in}&&\times\,\Big[\bar{u}_e(\vec{k}_e, \sigma_e)\,\Big(2
  k_{e\alpha} - \gamma_{\alpha}\hat{k}_e(x,y)z\Big)\,
  \hat{\varepsilon}^*_{\lambda'}\,\Big(m_e + \hat{k}_e + \hat{k} -
  \hat{k}_e(x,y)z\Big)\,\gamma_{\mu}(1 -
  \gamma^5)\,v_{\nu}(\vec{k}_{\nu}, + \frac{1}{2})\Big]\nonumber\\
\hspace{-0.3in}&& - \frac{e^2}{48\pi^2}\int^1_0dx\int^1_0
dy\,2\,y\int^1_0dz\,3\,z^2\,\frac{1}{k^2_e(x,y)z^2 - 2 k_e\cdot k\,
  xyz + \mu^2 (1 - z)}\nonumber\\
\hspace{-0.3in}&&\times \,\Big\{\frac{1}{2}\,\Big[\bar{u}_p(\vec{k}_p,
  \sigma_p)\,\Big(2 k^{\alpha}_p +
  \gamma^{\alpha}\hat{k}_e(x,y)z\Big)\,\gamma^{\mu}(1 + \lambda
  \gamma^5)\,u_n(\vec{k}_n, \sigma_n)\Big]\,\Big[\bar{u}_e(\vec{k}_e,
  \sigma_e)\,\gamma_{\alpha}\, \hat{\varepsilon}^*_{\lambda'}\,
  \gamma^{\mu}(1 + \lambda \gamma^5)\,v_{\nu}(\vec{k}_{\nu}, +
  \frac{1}{2})\Big]\nonumber\\
\hspace{-0.3in}&& + \frac{1}{4}\,\Big[\bar{u}_p(\vec{k}_p,
  \sigma_p)\, \gamma^{\alpha}\,\gamma^{\beta}\,\gamma^{\mu}(1 +
  \lambda \gamma^5)\,u_n(\vec{k}_n,
  \sigma_n)\Big]\nonumber\\
\hspace{-0.3in}&&\times\,\Big[\bar{u}_e(\vec{k}_e,
  \sigma_e)\,\gamma_{\alpha}\gamma_{\beta}\,
  \hat{\varepsilon}^*_{\lambda'}\,\Big(m_e + \hat{k}_e + \hat{k} -
  \hat{k}_e(x,y)z\Big)\, \gamma^{\mu}(1 + \lambda
  \gamma^5)\,v_{\nu}(\vec{k}_{\nu}, + \frac{1}{2})\Big]\nonumber\\
\hspace{-0.3in}&& + \frac{1}{4}\,\Big[\bar{u}_p(\vec{k}_p, \sigma_p)\,
  \gamma^{\alpha}\,\gamma^{\beta}\,\gamma^{\mu}(1 + \lambda
  \gamma^5)\,u_n(\vec{k}_n, \sigma_n)\Big]\,\Big[\bar{u}_e(\vec{k}_e,
  \sigma_e)\,\Big(2 k_{e\alpha} - \gamma_{\alpha} \hat{k}_e(x,y)z\Big)
  \, \hat{\varepsilon}^*_{\lambda'}\,\gamma_{\beta}\, \gamma^{\mu}(1 +
  \lambda \gamma^5)\,v_{\nu}(\vec{k}_{\nu}, +
  \frac{1}{2})\Big]\Big\}.\nonumber\\
\hspace{-0.3in}&&
\end{eqnarray}
For the analysis of the integrals over the Feynman parameters $x,y$
and $z$ it is convenient to rewrite Eq.(\ref{eq:D.9}) and
Eq.(\ref{eq:D.10}) as follows 
\begin{eqnarray*}
\hspace{-0.3in}&&{\cal M}_{\rm Fig.\,\ref{fig:fig4}a}(n \to p e^-
\bar{\nu}_e\gamma)_{\lambda'} = \frac{e^2}{96\pi^2}
\int^1_0dx\int^1_0dy\,2\,y\int^1_0dz\,3\,z^2\,\frac{1}{[k^2_p(x,y)z^2
    - 2 k_p\cdot k \, x y z+ \mu^2 (1 - z)]^2}\nonumber\\
\end{eqnarray*}
\begin{eqnarray}\label{eq:D.11}
\hspace{-0.3in}&&\times \,\Big\{ 4 (k_p\cdot
k_e)\Big[\bar{u}_p(\vec{k}_p,
  \sigma_p)\,\hat{\varepsilon}^*_{\lambda'}\,\Big(m_p + \hat{k}_p +
  \hat{k}\Big)\,\gamma^{\mu}(1 + \lambda \gamma^5)\,u_n(\vec{k}_n,
  \sigma_n)\Big]\,\Big[\bar{u}_e(\vec{k}_e, \sigma_e)\,\gamma^{\mu}(1
  + \lambda \gamma^5)\,v_{\nu}(\vec{k}_{\nu}, +
  \frac{1}{2})\Big]\nonumber\\
\hspace{-0.3in}&& + 2 z\,\Big[\bar{u}_p(\vec{k}_p,
  \sigma_p)\,\hat{\varepsilon}^*_{\lambda'}\,\Big(m_p + \hat{k}_p +
  \hat{k}\Big)\,\gamma^{\mu}(1 + \lambda \gamma^5)\,u_n(\vec{k}_n,
  \sigma_n)\Big]\,\Big[\bar{u}_e(\vec{k}_e,
  \sigma_e)\,\hat{k}_p\hat{k}_p(x,y)\,\gamma^{\mu}(1 + \lambda
  \gamma^5)\,v_{\nu}(\vec{k}_{\nu}, + \frac{1}{2})\Big]\nonumber\\
\hspace{-0.3in}&&- 4\,z\,(k_p\cdot k_e)\Big[\bar{u}_p(\vec{k}_p,
  \sigma_p)\,\hat{\varepsilon}^*_{\lambda'}\,\hat{k}_p(x,y)z\,\gamma^{\mu}(1
  + \lambda \gamma^5)\,u_n(\vec{k}_n,
  \sigma_n)\Big]\,\Big[\bar{u}_e(\vec{k}_e, \sigma_e)\,\gamma^{\mu}(1
  + \lambda \gamma^5)\,v_{\nu}(\vec{k}_{\nu}, +
  \frac{1}{2})\Big]\nonumber\\
\hspace{-0.3in}&& - 2\,z\, \Big[\bar{u}_p(\vec{k}_p,
  \sigma_p)\,\hat{k}_e\hat{k}_p(x,
  y)\,\hat{\varepsilon}^*_{\lambda'}\,\Big(m_p + \hat{k}_p +
  \hat{k}\Big)\,\gamma^{\mu}(1 + \lambda \gamma^5)\,u_n(\vec{k}_n,
  \sigma_n)\Big]\,\Big[\bar{u}_e(\vec{k}_e, \sigma_e)\,\gamma^{\mu}(1
  + \lambda \gamma^5)\,v_{\nu}(\vec{k}_{\nu}, +
  \frac{1}{2})\Big]\nonumber\\
\hspace{-0.3in}&& + 2\,z^2 \Big[\bar{u}_p(\vec{k}_p,
  \sigma_p)\,\hat{k}_e\hat{k}_p(x, y)\,
  \hat{\varepsilon}^*_{\lambda'}\, \hat{k}_p(x,y)\,\gamma^{\mu}(1 +
  \lambda \gamma^5)\,u_n(\vec{k}_n,
  \sigma_n)\Big]\,\Big[\bar{u}_e(\vec{k}_e, \sigma_e)\,\gamma^{\mu}(1
  + \lambda \gamma^5)\,v_{\nu}(\vec{k}_{\nu}, +
  \frac{1}{2})\Big]\nonumber\\
\hspace{-0.3in}&& - z^2\Big[\bar{u}_p(\vec{k}_p,
  \sigma_p)\,\gamma^{\alpha}\hat{k}_p(x,
  y)\,\hat{\varepsilon}^*_{\lambda'}\,\Big(m_p + \hat{k}_p +
  \hat{k}\Big)\,\gamma^{\mu}(1 + \lambda \gamma^5)\,u_n(\vec{k}_n,
  \sigma_n)\Big]\,\Big[\bar{u}_e(\vec{k}_e,
  \sigma_e)\,\gamma_{\alpha}\hat{k}_p(x,y)\,\gamma^{\mu}(1 + \lambda
  \gamma^5)\,v_{\nu}(\vec{k}_{\nu}, + \frac{1}{2})\Big]\nonumber\\
\hspace{-0.3in}&& - 2\,z^2 \Big[\bar{u}_p(\vec{k}_p,
  \sigma_p)\,\hat{\varepsilon}^*_{\lambda'}\,\hat{k}_p(x,y)\,\gamma^{\mu}(1
  + \lambda \gamma^5)\,u_n(\vec{k}_n,
  \sigma_n)\Big]\,\Big[\bar{u}_e(\vec{k}_e, \sigma_e)\,\hat{k}_p
  \hat{k}_p(x,y)\,\gamma^{\mu}(1 + \lambda
  \gamma^5)\,v_{\nu}(\vec{k}_{\nu}, + \frac{1}{2})\Big]\nonumber\\
\hspace{-0.3in}&& + z^3 \Big[\bar{u}_p(\vec{k}_p, \sigma_p)\,
  \gamma^{\alpha}\hat{k}_p(x,
  y)\,\hat{\varepsilon}^*_{\lambda'}\,\hat{k}_p(x,y)\,\gamma^{\mu}(1 +
  \lambda \gamma^5)\,u_n(\vec{k}_n,
  \sigma_n)\Big]\,\Big[\bar{u}_e(\vec{k}_e, \sigma_e)\,
  \gamma_{\alpha}\hat{k}_p(x,y)\,\gamma^{\mu}(1 + \lambda
  \gamma^5)\,v_{\nu}(\vec{k}_{\nu}, +
  \frac{1}{2})\Big]\Big\}\nonumber\\
\hspace{-0.3in}&& +
\frac{e^2}{48\pi^2}\int^1_0dx\int^1_0dy\,2\,y\int^1_0dz\,3\,z^2\,\frac{1}{k^2_p(x,y)z^2 - 2 k_p \cdot k\, xyz 
  + \mu^2 (1 - z)}\nonumber\\
\hspace{-0.3in}&&\times \,\Big\{\Big[\bar{u}_p(\vec{k}_p, \sigma_p)\,
  \hat{k}_e\,\hat{\varepsilon}^*_{\lambda'}\,\gamma^{\mu}(1 + \lambda
  \gamma^5)\,u_n(\vec{k}_n, \sigma_n)\Big]\,\Big[\bar{u}_e(\vec{k}_e,
  \sigma_e)\,\gamma^{\mu}(1 + \lambda
  \gamma^5)\,v_{\nu}(\vec{k}_{\nu}, + \frac{1}{2})\Big]\nonumber\\
\hspace{-0.3in}&& + \frac{1}{4}\,\Big[\bar{u}_p(\vec{k}_p, \sigma_p)\,
  \gamma^{\alpha}\,\gamma^{\beta}\,\hat{\varepsilon}^*_{\lambda'}\,\Big(m_p
  + \hat{k}_p + \hat{k}\Big)\,\gamma^{\mu}(1 + \lambda
  \gamma^5)\,u_n(\vec{k}_n, \sigma_n)\Big]\,\Big[\bar{u}_e(\vec{k}_e,
  \sigma_e)\,\gamma_{\alpha}\,\gamma_{\beta}\,\gamma^{\mu}(1 + \lambda
  \gamma^5)\,v_{\nu}(\vec{k}_{\nu}, + \frac{1}{2})\Big]\nonumber\\
\hspace{-0.3in}&&+ \frac{1}{2}\,\Big[\bar{u}_p(\vec{k}_p,
  \sigma_p)\,\hat{\varepsilon}^*_{\lambda'}\, \gamma^{\beta}
  \,\gamma^{\mu}(1 + \lambda \gamma^5)\,u_n(\vec{k}_n,
  \sigma_n)\Big]\,\Big[\bar{u}_e(\vec{k}_e,
  \sigma_e)\,\hat{k}_p\,\gamma_{\beta}\,\gamma_{\mu}(1 -
  \gamma^5)\,v_{\nu}(\vec{k}_{\nu}, + \frac{1}{2})\Big]\nonumber\\
\hspace{-0.3in}&&+ \frac{1}{2}\,z \Big[\bar{u}_p(\vec{k}_p,
  \sigma_p)\,
  \gamma^{\alpha}\,\hat{\varepsilon}^*_{\lambda'}\,\gamma^{\mu}(1 +
  \lambda \gamma^5)\,u_n(\vec{k}_n,
  \sigma_n)\Big]\,\Big[\bar{u}_e(\vec{k}_e,
  \sigma_e)\,\gamma_{\alpha}\hat{k}_p(x,y)\,\gamma^{\mu}(1 + \lambda
  \gamma^5)\,v_{\nu}(\vec{k}_{\nu}, + \frac{1}{2})\Big]\nonumber\\
\hspace{-0.3in}&& - \frac{1}{4}\,z \Big[\bar{u}_p(\vec{k}_p,
  \sigma_p)\,
  \gamma^{\alpha}\,\gamma^{\beta}\,\hat{\varepsilon}^*_{\lambda'}\,
  \hat{k}_p(x,y)\,\gamma^{\mu}(1 + \lambda \gamma^5)\,u_n(\vec{k}_n,
  \sigma_n)\Big]\,\Big[\bar{u}_e(\vec{k}_e,
  \sigma_e)\,\gamma_{\alpha}\,\gamma_{\beta}\,\gamma^{\mu}(1 + \lambda
  \gamma^5)\,v_{\nu}(\vec{k}_{\nu}, + \frac{1}{2})\Big]\nonumber\\
\hspace{-0.3in}&&- \frac{1}{4}\,z \Big[\bar{u}_p(\vec{k}_p,
  \sigma_p)\, \gamma^{\alpha}\hat{k}_p(x,y)
  \,\hat{\varepsilon}^*_{\lambda'}\, \gamma^{\beta} \,\gamma^{\mu}(1 +
  \lambda \gamma^5)\,u_n(\vec{k}_n,
  \sigma_n)\Big]\,\Big[\bar{u}_e(\vec{k}_e,
  \sigma_e)\,\gamma_{\alpha}\,\gamma_{\beta}\,\gamma_{\mu}(1 -
  \gamma^5)\,v_{\nu}(\vec{k}_{\nu}, + \frac{1}{2})\Big]\Big\}
\end{eqnarray}
and
\begin{eqnarray*}
\hspace{-0.3in}&&{\cal M}_{\rm Fig.\,\ref{fig:fig4}b}(n \to p e^-
\bar{\nu}_e\gamma)_{\lambda'} = - \frac{e^2}{96\pi^2}
\int^1_0dx\int^1_ody\,2\,y\int^1_0dz\,3\,z^2\,\frac{1}{[k^2_e(x,y)z^2
  - 2 k_e\cdot k\, xyz  + \mu^2 (1 - z)]^2}\nonumber\\
\hspace{-0.3in}&&\times \,\Big\{4 (k_p\cdot k_e)
\Big[\bar{u}_p(\vec{k}_p, \sigma_p)\,\gamma^{\mu}(1 + \lambda
  \gamma^5)\,u_n(\vec{k}_n, \sigma_n)\Big]\,\Big[\bar{u}_e(\vec{k}_e,
  \sigma_e)\, \hat{\varepsilon}^*_{\lambda'}\,\Big(m_e + \hat{k}_e +
  \hat{k}\Big)\,\gamma_{\mu}(1 - \gamma^5)\,v_{\nu}(\vec{k}_{\nu}, +
  \frac{1}{2})\Big]\nonumber\\
\hspace{-0.3in}&&- 2\,z \Big[\bar{u}_p(\vec{k}_p, \sigma_p)\,
  \gamma^{\mu}(1 + \lambda \gamma^5)\,u_n(\vec{k}_n,
  \sigma_n)\Big]\,\Big[\bar{u}_e(\vec{k}_e, \sigma_e)\,\hat{k}_p
  \hat{k}_e(x,y) \, \hat{\varepsilon}^*_{\lambda'}\,\Big(m_e +
  \hat{k}_e + \hat{k}\Big)\,\gamma_{\mu}(1 -
  \gamma^5)\,v_{\nu}(\vec{k}_{\nu}, + \frac{1}{2})\Big]\nonumber\\
\hspace{-0.3in}&&+ 2\,z \Big[\bar{u}_p(\vec{k}_p, \sigma_p)\,
  \hat{k}_e \hat{k}_e(x,y)z\Big)\,\gamma^{\mu}(1 + \lambda
  \gamma^5)\,u_n(\vec{k}_n, \sigma_n)\Big]\,\Big[\bar{u}_e(\vec{k}_e,
  \sigma_e)\, \hat{\varepsilon}^*_{\lambda'}\,\Big(m_e + \hat{k}_e +
  \hat{k}\Big)\,\gamma_{\mu}(1 - \gamma^5)\,v_{\nu}(\vec{k}_{\nu}, +
  \frac{1}{2})\Big]\nonumber\\
\hspace{-0.3in}&&- 4 z\,(k_p\cdot k_e) \Big[\bar{u}_p(\vec{k}_p,
  \sigma_p)\,\gamma^{\mu}(1 + \lambda \gamma^5)\,u_n(\vec{k}_n,
  \sigma_n)\Big]\,\Big[\bar{u}_e(\vec{k}_e, \sigma_e)\,
  \hat{\varepsilon}^*_{\lambda'}\, \hat{k}_e(x,y)\,\gamma_{\mu}(1 -
  \gamma^5)\,v_{\nu}(\vec{k}_{\nu}, + \frac{1}{2})\Big]\nonumber\\
\hspace{-0.3in}&&- 2\,z^2 \Big[\bar{u}_p(\vec{k}_p,
  \sigma_p)\,\hat{k}_e \hat{k}_e(x,y)z\Big)\,\gamma^{\mu}(1 + \lambda
  \gamma^5)\,u_n(\vec{k}_n, \sigma_n)\Big]\,\Big[\bar{u}_e(\vec{k}_e,
  \sigma_e)\, \hat{\varepsilon}^*_{\lambda'}\,
  \hat{k}_e(x,y)\,\gamma_{\mu}(1 - \gamma^5)\,v_{\nu}(\vec{k}_{\nu}, +
  \frac{1}{2})\Big]\nonumber\\
\hspace{-0.3in}&&+ 2\,z^2 \Big[\bar{u}_p(\vec{k}_p,
  \sigma_p)\,\gamma^{\mu}(1 + \lambda \gamma^5)\,u_n(\vec{k}_n,
  \sigma_n)\Big] \,\Big[\bar{u}_e(\vec{k}_e, \sigma_e)\,\hat{k}_p
  \hat{k}_e(x,y)\, \hat{\varepsilon}^*_{\lambda'}\,\hat{k}_e(x,y)
  \,\gamma_{\mu}(1 - \gamma^5)\,v_{\nu}(\vec{k}_{\nu}, +
  \frac{1}{2})\Big]\nonumber\\
\hspace{-0.3in}&&- z^2 \Big[\bar{u}_p(\vec{k}_p,
  \sigma_p)\,\gamma^{\alpha}\hat{k}_e(x,y)\,\gamma^{\mu}(1 + \lambda
  \gamma^5)\,u_n(\vec{k}_n, \sigma_n)\Big] \,\Big[\bar{u}_e(\vec{k}_e,
  \sigma_e)\, \gamma_{\alpha}\hat{k}_e(x,y) \,
  \hat{\varepsilon}^*_{\lambda'}\, \Big(m_e + \hat{k}_e +
  \hat{k}\Big)\, \gamma_{\mu}(1 - \gamma^5)\,v_{\nu}(\vec{k}_{\nu}, +
  \frac{1}{2})\Big]\nonumber\\
\hspace{-0.3in}&&+ z^3 \Big[\bar{u}_p(\vec{k}_p, \sigma_p)\,
  \gamma^{\alpha}\hat{k}_e(x,y)\,\gamma^{\mu}(1 + \lambda
  \gamma^5)\,u_n(\vec{k}_n, \sigma_n)\Big] \,\Big[\bar{u}_e(\vec{k}_e,
  \sigma_e)\, \gamma_{\alpha}\hat{k}_e(x,y) \,
  \hat{\varepsilon}^*_{\lambda'}\, \hat{k}_e(x,y) \,\gamma_{\mu}(1 -
  \gamma^5)\,v_{\nu}(\vec{k}_{\nu}, + \frac{1}{2})\Big]\nonumber\\
\hspace{-0.3in}&& - \frac{e^2}{48\pi^2}\int^1_0dx\int^1_0
dy\,2\,y\int^1_0dz\,3\,z^2\,\frac{1}{k^2_e(x,y)z^2 - 2 k_e\cdot k\,
  xyz + \mu^2 (1 - z)}\nonumber\\
\hspace{-0.3in}&&\times \,\Big\{\Big[\bar{u}_p(\vec{k}_p,
  \sigma_p)\,\gamma^{\mu}(1 + \lambda \gamma^5)\,u_n(\vec{k}_n,
  \sigma_n)\Big]\,\Big[\bar{u}_e(\vec{k}_e, \sigma_e)\, \hat{k}_p \,
  \hat{\varepsilon}^*_{\lambda'}\, \gamma^{\mu}(1 + \lambda
  \gamma^5)\,v_{\nu}(\vec{k}_{\nu}, + \frac{1}{2})\Big]\nonumber\\
\end{eqnarray*}
\begin{eqnarray}\label{eq:D.12}
\hspace{-0.3in}&& + \frac{1}{4}\,\Big[\bar{u}_p(\vec{k}_p, \sigma_p)\,
  \gamma^{\alpha}\,\gamma^{\beta}\,\gamma^{\mu}(1 + \lambda
  \gamma^5)\,u_n(\vec{k}_n, \sigma_n)\Big] \,\Big[\bar{u}_e(\vec{k}_e,
  \sigma_e)\,\gamma_{\alpha}\gamma_{\beta}\,
  \hat{\varepsilon}^*_{\lambda'}\,\Big(m_e + \hat{k}_e +
  \hat{k}\Big)\, \gamma^{\mu}(1 + \lambda
  \gamma^5)\,v_{\nu}(\vec{k}_{\nu}, + \frac{1}{2})\Big]\nonumber\\
\hspace{-0.3in}&& + \frac{1}{2}\,\Big[\bar{u}_p(\vec{k}_p, \sigma_p)\,
  \hat{k}_e \,\gamma^{\beta}\,\gamma^{\mu}(1 + \lambda
  \gamma^5)\,u_n(\vec{k}_n, \sigma_n)\Big]\,\Big[\bar{u}_e(\vec{k}_e,
  \sigma_e)\, \hat{\varepsilon}^*_{\lambda'}\,\gamma_{\beta}\,
  \gamma^{\mu}(1 + \lambda \gamma^5)\,v_{\nu}(\vec{k}_{\nu}, +
  \frac{1}{2})\Big]\nonumber\\
\hspace{-0.3in}&& + \frac{1}{2}\,z \Big[\bar{u}_p(\vec{k}_p,
  \sigma_p)\, \gamma^{\alpha}\hat{k}_e(x,y) \,\gamma^{\mu}(1 + \lambda
  \gamma^5)\,u_n(\vec{k}_n, \sigma_n)\Big]\,\Big[\bar{u}_e(\vec{k}_e,
  \sigma_e)\,\gamma_{\alpha}\, \hat{\varepsilon}^*_{\lambda'}\,
  \gamma^{\mu}(1 + \lambda \gamma^5)\,v_{\nu}(\vec{k}_{\nu}, +
  \frac{1}{2})\Big]\nonumber\\
\hspace{-0.3in}&& - \frac{1}{4}\,z \Big[\bar{u}_p(\vec{k}_p,
  \sigma_p)\, \gamma^{\alpha}\,\gamma^{\beta}\,\gamma^{\mu}(1 +
  \lambda \gamma^5)\,u_n(\vec{k}_n, \sigma_n)\Big]
\,\Big[\bar{u}_e(\vec{k}_e, \sigma_e)\,\gamma_{\alpha}\gamma_{\beta}\,
  \hat{\varepsilon}^*_{\lambda'}\,
  \hat{k}_e(x,y) \, \gamma^{\mu}(1 + \lambda
  \gamma^5)\,v_{\nu}(\vec{k}_{\nu}, + \frac{1}{2})\Big]\nonumber\\
\hspace{-0.3in}&& - \frac{1}{4}\,z \Big[\bar{u}_p(\vec{k}_p,
  \sigma_p)\, \gamma^{\alpha}\,\gamma^{\beta}\,\gamma^{\mu}(1 +
  \lambda \gamma^5)\,u_n(\vec{k}_n,
  \sigma_n)\Big]\,\Big[\bar{u}_e(\vec{k}_e, \sigma_e)\,\gamma_{\alpha}
  \hat{k}_e(x,y) \, \hat{\varepsilon}^*_{\lambda'}\,\gamma_{\beta}\,
  \gamma^{\mu}(1 + \lambda \gamma^5)\,v_{\nu}(\vec{k}_{\nu}, +
  \frac{1}{2})\Big]\Big\}.
\end{eqnarray}
One may show that in the large proton mass expansion the amplitudes
Eq.(\ref{eq:D.11}) and Eq.(\ref{eq:D.12}) behave as $O(1/m_p)$ or even
faster and vanish at $m_p \to \infty$. We assume that the
contributions of other diagrams in Fig.\,\ref{fig:fig4} either cancel
a gauge--non-invariant part of the diagram Fig.\,\ref{fig:fig3}c or is
a constant, which can be removed by renormalization of the Fermi weak
coupling constant $G_F$ and the axial coupling constant $\lambda$
\cite{Sirlin1967}.

\section*{Appendix E: The amplitude of the neutron 
radiative $\beta^-$--decay, described by Feynman diagrams in
Fig.\,\ref{fig:fig5}} \renewcommand{\theequation}{E-\arabic{equation}}
\setcounter{equation}{0}

In this Appendix we calculate the diagrams in
Fig.\,\ref{fig:fig5}. These diagrams describe the process of the
neutron radiative $\beta^-$--decay with emission of two real photons
$n \to p + e^- + \bar{\nu}_e + \gamma + \gamma$. The contribution of
these diagrams to the rate of the neutron $\beta^-$--decay is of order
$O(\alpha^2/\pi^2)$ and after the integration over degrees of freedom
of one of the photons one may hardly distinguish such a contribution
from that of the neutron radiative $\beta^-$--decay with an emission
of one real photon. Since the contribution of the process, when
photons are emitted by the proton, is suppressed to leading order in
the large proton mass expansion, we take into account only the
emission of photons by the electron. The analytical expression of the
diagrams in Fig.\,\ref{fig:fig5} is given by
\begin{eqnarray}\label{eq:E.1}
\hspace{-0.3in}&&{\cal M}_{\rm Fig.\,\ref{fig:fig5}}(n \to p e^-
\bar{\nu}_e\gamma\gamma)_{\lambda'\lambda''} =
-e\,\Big[\bar{u}_p(\vec{k}_p, \sigma_p)\,\gamma^{\mu}(1 + \lambda
  \gamma^5)\,u_n(\vec{k}_n, \sigma_n)\Big]\nonumber\\
\hspace{-0.3in}&&\times\,\Big\{\Big[\bar{u}_e(\vec{k}_e,\sigma_e)\,
  \hat{\varepsilon}^*_{\lambda'}(k)\,\frac{1}{m_e - \hat{k}_e - \hat{k} -
    i 0}\,\hat{\varepsilon}^*_{\lambda''}(q)\,\frac{1}{m_e - \hat{k}_e -
    \hat{k} - \hat{q} - i 0}\,\gamma_{\mu}(1 - \gamma^5)
  v_{\nu}(\vec{k}_{\nu}, + \frac{1}{2})\Big]\nonumber\\
\hspace{-0.3in}&& + \Big[\bar{u}_e(\vec{k}_e,\sigma_e)\,
  \hat{\varepsilon}^*_{\lambda''}(q)\,\frac{1}{m_e - \hat{k}_e - \hat{q}
    - i 0}\,\hat{\varepsilon}^*_{\lambda'}(k)\,\frac{1}{m_e - \hat{k}_e -
    \hat{k} - \hat{q} - i 0}\,\gamma_{\mu}(1 - \gamma^5)
  v_{\nu}(\vec{k}_{\nu}, + \frac{1}{2})\Big]\Big\},
\end{eqnarray}
where the polarization vectors $\varepsilon^*_{\lambda'}(k)$ and
$\varepsilon^*_{\lambda''}(q)$ are taken in the physical gauge and
obey the constraints $\vec{k}\cdot
\vec{\varepsilon}^{\,*}_{\lambda'}(k) = 0$ and $\vec{q}\cdot
\vec{\varepsilon}^{\,*}_{\lambda''}(q) = 0$, respectively, with $k^2 =
q^2 = 0$.  Then, we rewrite Eq.(\ref{eq:E.1}) as follows
\begin{eqnarray}\label{eq:E.2}
\hspace{-0.3in}&&{\cal M}_{\rm Fig.\,\ref{fig:fig5}}(n \to p e^-
\bar{\nu}_e\gamma\gamma)_{\lambda'\lambda''} = - e\,
\Big[\bar{u}_p(\vec{k}_p, \sigma_p)\,\gamma^{\mu}(1 + \lambda
  \gamma^5)\,u_n(\vec{k}_n, \sigma_n)\Big]\nonumber\\
\hspace{-0.3in}&&\times\,\Big\{\frac{1}{2k_e\cdot k +
  i0}\,\frac{1}{2k_e\cdot (k + q) + 2 k\cdot q + i0}\nonumber\\
\hspace{-0.3in}&&\times\,\Big[\bar{u}_e(\vec{k}_e,\sigma_e)\,
  \Big(2k_e\cdot \varepsilon^*_{\lambda'} +
  \hat{\varepsilon}^*_{\lambda'}\,\hat{k}\Big)\,\hat{\varepsilon}^*_{\lambda''}
  \, \Big(m_e + \hat{k}_e + \hat{k} + \hat{q}\Big) \,\gamma_{\mu}(1 -
  \gamma^5) v_{\nu}(\vec{k}_{\nu}, + \frac{1}{2})\Big]\nonumber\\
\hspace{-0.3in}&& + \frac{1}{2k_e\cdot q + i0}\,\frac{1}{2k_e\cdot (k
  + q) + 2 k\cdot q + i0}\nonumber\\
\hspace{-0.3in}&&\times\,\Big[\bar{u}_e(\vec{k}_e,\sigma_e)\,
  \Big(2k_e\cdot \varepsilon^*_{\lambda''} +
  \hat{\varepsilon}^*_{\lambda''}\,\hat{q}\Big)
  \,\hat{\varepsilon}^*_{\lambda'}\, \Big(m_e + \hat{k}_e + \hat{k} +
  \hat{q}\Big) \,\gamma_{\mu}(1 - \gamma^5) v_{\nu}(\vec{k}_{\nu}, +
  \frac{1}{2})\Big]\Big\}.
\end{eqnarray}
In the non--relativistic proton approximation Eq.(\ref{eq:E.2}) takes
the form
\begin{eqnarray*}
\hspace{-0.3in}&&\frac{1}{2m_n e}\,{\cal M}_{\rm Fig.\,\ref{fig:fig5}}(n
\to p e^- \bar{\nu}_e\gamma\gamma)_{\lambda'\lambda''} = -
\frac{1}{2k_e\cdot (k + q) + 2 k\cdot q + i0}\nonumber\\
\hspace{-0.3in}&&\times\,\left\{\Big\{[\varphi^{\dagger}_p\varphi_n]\,\Big(
\frac{1}{2k_e\cdot k + i0}\,\Big[\bar{u}_e(\vec{k}_e,\sigma_e)\,
  \Big(2k_e\cdot \varepsilon^*_{\lambda'} +
  \hat{\varepsilon}^*_{\lambda'}\,\hat{k}\Big)\,
  \hat{\varepsilon}^*_{\lambda''}\, \Big(m_e + \hat{k}_e + \hat{k} +
  \hat{q}\Big) \,\gamma^0\,(1 - \gamma^5) v_{\nu}(\vec{k}_{\nu}, +
  \frac{1}{2})\Big]\right.\nonumber\\
\hspace{-0.3in}&& + \frac{1}{2k_e\cdot q + q^2 +
  i0}\,\Big[\bar{u}_e(\vec{k}_e,\sigma_e)\, \Big(2k_e\cdot
  \varepsilon^*_{\lambda''} +
  \hat{\varepsilon}^*_{\lambda''}\,\hat{q}\Big)
  \,\hat{\varepsilon}^*_{\lambda'}\, \Big(m_e + \hat{k}_e + \hat{k} +
  \hat{q}\Big) \,\gamma^0\,(1 - \gamma^5) v_{\nu}(\vec{k}_{\nu}, +
  \frac{1}{2})\Big]\Big)\Big\}\nonumber\\
\end{eqnarray*}
\begin{eqnarray}\label{eq:E.3}
\hspace{-0.3in}&&-
\lambda\,\Big\{[\varphi^{\dagger}_p\,\vec{\sigma}\,\varphi_n]\cdot
\Big( \frac{1}{2k_e\cdot k + i0}\,\Big[\bar{u}_e(\vec{k}_e,\sigma_e)\,
  \Big(2k_e\cdot \varepsilon^*_{\lambda'} +
  \hat{\varepsilon}^*_{\lambda'}\,\hat{k}\Big)\,
  \hat{\varepsilon}^*_{\lambda''}\, \Big(m_e + \hat{k}_e + \hat{k} +
  \hat{q}\Big) \,\vec{\gamma}\,(1 - \gamma^5) v_{\nu}(\vec{k}_{\nu}, +
  \frac{1}{2})\Big]\nonumber\\
\hspace{-0.3in}&&\left. + \frac{1}{2k_e\cdot q +
  i0}\,\Big[\bar{u}_e(\vec{k}_e,\sigma_e)\, \Big(2k_e\cdot
  \varepsilon^*_{\lambda''} +
  \hat{\varepsilon}^*_{\lambda''}\,\hat{q}\Big)
  \,\hat{\varepsilon}^*_{\lambda'}\, \Big(m_e + \hat{k}_e + \hat{k} +
  \hat{q}\Big) \,\vec{\gamma}\,(1 - \gamma^5) v_{\nu}(\vec{k}_{\nu}, +
  \frac{1}{2})\Big]\Big)\Big\}\right\}.
\end{eqnarray}
The hermitian conjugate amplitude is equal to
\begin{eqnarray}\label{eq:E.4}
\hspace{-0.3in}&&\frac{1}{2 m_n e}\,{\cal M}^{\dagger}_{\rm
  Fig.\,\ref{fig:fig5}}(n \to p e^-
\bar{\nu}_e\gamma\gamma)_{\lambda'\lambda''} = - \frac{1}{2 k_e\cdot
  (k + q) + 2 k\cdot q - i 0}\nonumber\\
\hspace{-0.3in}&&\times\,\left\{\Big\{[\varphi^{\dagger}_n\varphi_p]\,
\Big(\frac{1}{2k_e\cdot k - i0}\, \Big[ \bar{v}_{\nu}(\vec{k}_{\nu}, +
  \frac{1}{2})\gamma^0 (1 - \gamma^5)\, \Big(m_e + \hat{k}_e + \hat{k}
  + \hat{q}\Big)\, \hat{\varepsilon}_{\lambda''}\,\Big(2 k_e \cdot
  \varepsilon_{\lambda'} +
  \hat{k}\,\hat{\varepsilon}_{\lambda'}\Big)\,
  u_e(\vec{k}_e,\sigma_e)\Big]\right.\nonumber\\
\hspace{-0.3in}&& + \frac{1}{2k_e\cdot q - i0}\, \Big[
  \bar{v}_{\nu}(\vec{k}_{\nu}, + \frac{1}{2})\gamma^0 (1 - \gamma^5)\,
  \Big(m_e + \hat{k}_e + \hat{k} + \hat{q}\Big)\,
  \hat{\varepsilon}_{\lambda'}\,\Big(2 k_e \cdot
  \varepsilon_{\lambda''} +
  \hat{q}\,\hat{\varepsilon}_{\lambda''}\Big)\,
  u_e(\vec{k}_e,\sigma_e)\Big]\Big)\Big\}\nonumber\\
\hspace{-0.3in}&&-
\lambda\,\Big\{[\varphi^{\dagger}_p\,\vec{\sigma}\,\varphi_n]\cdot
\Big( \frac{1}{2k_e\cdot k - i0}\,\Big[\bar{v}_{\nu}(\vec{k}_{\nu}, +
  \frac{1}{2})\,\vec{\gamma}\, (1 - \gamma^5)\, \Big(m_e + \hat{k}_e +
  \hat{k} + \hat{q}\Big)\, \hat{\varepsilon}_{\lambda''}\,\Big(2 k_e
  \cdot \varepsilon_{\lambda'} +
  \hat{k}\,\hat{\varepsilon}_{\lambda'}\Big)\, u_e(\vec{k}_e,\sigma_e)
  \Big]\nonumber\\
\hspace{-0.3in}&&\left. + \frac{1}{2k_e\cdot q -
  i0}\,\Big[\bar{v}_{\nu}(\vec{k}_{\nu}, +
  \frac{1}{2})\,\vec{\gamma}\, (1 - \gamma^5)\, \Big(m_e + \hat{k}_e +
  \hat{k} + \hat{q}\Big)\, \hat{\varepsilon}_{\lambda'}\,\Big(2 k_e
  \cdot \varepsilon_{\lambda''} +
  \hat{q}\,\hat{\varepsilon}_{\lambda''}\Big)\,
  u_e(\vec{k}_e,\sigma_e) \Big]\Big)\Big\}\right\}.
\end{eqnarray}
The squared absolute value of the amplitude Eq.(\ref{eq:E.3}) averaged
over the neutron spin and summed over the polarizations of the proton
and electron 
\begin{eqnarray}\label{eq:E.5}
\hspace{-0.3in}&&\frac{1}{2}\sum_{\rm pol}\frac{|{\cal M}_{\rm
    Fig.\,\ref{fig:fig5}}(n \to p e^-
  \bar{\nu}_e\gamma\gamma)_{\lambda'\lambda''}|^2}{4 m^2_n e^2} =
\frac{1}{(2 k_e\cdot (k + q) + 2 k\cdot q)^2} \Big\{\frac{1}{(2k_e\cdot
  k)^2} {\rm tr}\{(m_e + \hat{k}_e)(2k_e\cdot
\varepsilon^*_{\lambda'} + \hat{\varepsilon}^*_{\lambda'}
\hat{k})\hat{\varepsilon}^*_{\lambda''}\nonumber\\
\hspace{-0.3in}&& \times(m_e + \hat{k}_e + \hat{k} +
\hat{q})\gamma^0(1 - \gamma^5)\hat{k}_{\nu}\gamma^0(1 - \gamma^5)(m_e
+ \hat{k}_e + \hat{k} +
\hat{q})\hat{\varepsilon}_{\lambda''}(2k_e\cdot \varepsilon_{\lambda'}
+ \hat{k}\hat{\varepsilon}_{\lambda'})\} + \frac{1}{2k_e\cdot k}\,
\frac{1}{2k_e\cdot q}\nonumber\\
\hspace{-0.3in}&&\times {\rm tr}\{(m_e + \hat{k}_e)(2k_e\cdot
\varepsilon^*_{\lambda''} + \hat{\varepsilon}^*_{\lambda''}
\hat{q})\hat{\varepsilon}^*_{\lambda'} (m_e + \hat{k}_e + \hat{k} +
\hat{q}) \gamma^0 (1 - \gamma^5) \hat{k}_{\nu} \gamma^0(1 - \gamma^5)
(m_e + \hat{k}_e + \hat{k} +
\hat{q})\hat{\varepsilon}_{\lambda''}(2k_e\cdot \varepsilon_{\lambda'}
+ \hat{k}\hat{\varepsilon}_{\lambda'})\}\nonumber\\
\hspace{-0.3in}&& + \frac{1}{2k_e\cdot k}\, \frac{1}{2k_e\cdot q} {\rm
  tr}\{(m_e + \hat{k}_e)(2k_e\cdot \varepsilon^*_{\lambda'} +
\hat{\varepsilon}^*_{\lambda'} \hat{k})\hat{\varepsilon}^*_{\lambda''}
(m_e + \hat{k}_e + \hat{k} + \hat{q}) \gamma^0 (1 - \gamma^5)
\hat{k}_{\nu} \gamma^0(1 - \gamma^5) (m_e + \hat{k}_e + \hat{k} +
\hat{q})\nonumber\\
\hspace{-0.3in}&&\times \hat{\varepsilon}_{\lambda'}(2k_e\cdot
\varepsilon_{\lambda''} + \hat{q}\hat{\varepsilon}_{\lambda''})\}+
\frac{1}{(2k_e\cdot q)^2} {\rm tr}\{(m_e + \hat{k}_e)(2k_e\cdot
\varepsilon^*_{\lambda''} + \hat{\varepsilon}^*_{\lambda''}
\hat{q})\hat{\varepsilon}^*_{\lambda'} (m_e + \hat{k}_e + \hat{k} +
\hat{q})\gamma^0(1 - \gamma^5)\hat{k}_{\nu}\gamma^0(1 -
\gamma^5)\nonumber\\
\hspace{-0.3in}&&\times (m_e + \hat{k}_e + \hat{k} +
\hat{q})\hat{\varepsilon}_{\lambda'}(2k_e\cdot \varepsilon_{\lambda''}
+ \hat{q}\hat{\varepsilon}_{\lambda''})\}\Big\} +
\frac{\lambda^2\,\delta^{ij}}{(2 k_e \cdot (k + q) + 2 k\cdot q)^2}
\Big\{\frac{1}{(2k_e\cdot k)^2} {\rm tr}\{(m_e + \hat{k}_e)(2k_e\cdot
\varepsilon^*_{\lambda'} + \hat{\varepsilon}^*_{\lambda'}
\hat{k})\hat{\varepsilon}^*_{\lambda''}\nonumber\\
\hspace{-0.3in}&& \times(m_e + \hat{k}_e + \hat{k} +
\hat{q})\gamma^i(1 - \gamma^5)\hat{k}_{\nu}\gamma^j(1 - \gamma^5)(m_e
+ \hat{k}_e + \hat{k} +
\hat{q})\hat{\varepsilon}_{\lambda''}(2k_e\cdot \varepsilon_{\lambda'}
+ \hat{k}\hat{\varepsilon}_{\lambda'})\} + \frac{1}{2k_e\cdot k}
\,\frac{1}{2k_e\cdot q}\nonumber\\
\hspace{-0.3in}&&\times {\rm tr}\{(m_e + \hat{k}_e)(2k_e\cdot
\varepsilon^*_{\lambda''} + \hat{\varepsilon}^*_{\lambda''}
\hat{q})\hat{\varepsilon}^*_{\lambda'} (m_e + \hat{k}_e + \hat{k} +
\hat{q}) \gamma^i (1 - \gamma^5) \hat{k}_{\nu} \gamma^j(1 - \gamma^5)
(m_e + \hat{k}_e + \hat{k} +
\hat{q})\hat{\varepsilon}_{\lambda''}(2k_e\cdot \varepsilon_{\lambda'}
+ \hat{k}\hat{\varepsilon}_{\lambda'})\} \nonumber\\
\hspace{-0.3in}&& + \frac{1}{2k_e\cdot k}\, \frac{1}{2k_e\cdot q} {\rm
  tr}\{(m_e + \hat{k}_e)(2k_e\cdot \varepsilon^*_{\lambda'} +
\hat{\varepsilon}^*_{\lambda'} \hat{k})\hat{\varepsilon}^*_{\lambda''}
(m_e + \hat{k}_e + \hat{k} + \hat{q}) \gamma^j (1 - \gamma^5)
\hat{k}_{\nu} \gamma^i (1 - \gamma^5) (m_e + \hat{k}_e + \hat{k} +
\hat{q})\nonumber\\
\hspace{-0.3in}&&\times \hat{\varepsilon}_{\lambda'}(2k_e\cdot
\varepsilon_{\lambda''} + \hat{q}\hat{\varepsilon}_{\lambda''})\} +
\frac{1}{(2k_e\cdot q)^2} {\rm tr}\{(m_e + \hat{k}_e)(2k_e\cdot
\varepsilon^*_{\lambda''} + \hat{\varepsilon}^*_{\lambda''}
\hat{q})\hat{\varepsilon}^*_{\lambda'} (m_e + \hat{k}_e + \hat{k} +
\hat{q})\gamma^i (1 - \gamma^5)\hat{k}_{\nu}\gamma^j (1 -
\gamma^5)\nonumber\\
\hspace{-0.3in}&&\times (m_e + \hat{k}_e + \hat{k} +
\hat{q})\hat{\varepsilon}_{\lambda'}(2k_e\cdot \varepsilon_{\lambda''}
+ \hat{q}\hat{\varepsilon}_{\lambda''})\}\Big\}.
\end{eqnarray}
For the subsequent calculation we omit the traces with the
$\gamma^5$--matrix, which should not contribute to the rate of the
neutron radiative $\beta^-$-decay with two photons in the final
state. This gives
\begin{eqnarray*}
\hspace{-0.3in}&&\frac{1}{2}\sum_{\rm pol}\frac{|{\cal M}_{\rm
    Fig.\,\ref{fig:fig5}}(n \to p e^-
  \bar{\nu}_e\gamma\gamma)_{\lambda'\lambda''}|^2}{8 m^2_n e^2} =
\frac{1}{(2 k_e\cdot (k + q) + 2 k\cdot q)^2}
\Big\{\frac{1}{(2k_e\cdot k)^2} {\rm tr}\{(m_e + \hat{k}_e)(2k_e\cdot
\varepsilon^*_{\lambda'} + \hat{\varepsilon}^*_{\lambda'}
\hat{k})\hat{\varepsilon}^*_{\lambda''}\nonumber\\
\hspace{-0.3in}&& \times(m_e + \hat{k}_e + \hat{k} + \hat{q})\gamma^0
\hat{k}_{\nu}\gamma^0 (m_e + \hat{k}_e + \hat{k} +
\hat{q})\hat{\varepsilon}_{\lambda''}(2k_e\cdot \varepsilon_{\lambda'}
+ \hat{k}\hat{\varepsilon}_{\lambda'})\} + \frac{1}{2k_e\cdot k}\,
\frac{1}{2k_e\cdot q }\nonumber\\
\hspace{-0.3in}&&\times {\rm tr}\{(m_e + \hat{k}_e)(2k_e\cdot
\varepsilon^*_{\lambda''} + \hat{\varepsilon}^*_{\lambda''}
\hat{q})\hat{\varepsilon}^*_{\lambda'} (m_e + \hat{k}_e + \hat{k} +
\hat{q}) \gamma^0 \hat{k}_{\nu} \gamma^0 (m_e + \hat{k}_e + \hat{k} +
\hat{q})\hat{\varepsilon}_{\lambda''}(2k_e\cdot \varepsilon_{\lambda'}
+ \hat{k}\hat{\varepsilon}_{\lambda'})\}\nonumber\\
\end{eqnarray*}
\begin{eqnarray}\label{eq:E.6}
\hspace{-0.3in}&& + \frac{1}{2k_e\cdot k}\, \frac{1}{2k_e\cdot q} {\rm
  tr}\{(m_e + \hat{k}_e)(2k_e\cdot \varepsilon^*_{\lambda'} +
\hat{\varepsilon}^*_{\lambda'} \hat{k})\hat{\varepsilon}^*_{\lambda''}
(m_e + \hat{k}_e + \hat{k} + \hat{q}) \gamma^0 \hat{k}_{\nu} \gamma^0
(m_e + \hat{k}_e + \hat{k} + \hat{q})\nonumber\\
\hspace{-0.3in}&&\times \hat{\varepsilon}_{\lambda'}(2k_e\cdot
\varepsilon_{\lambda''} + \hat{q}\hat{\varepsilon}_{\lambda''})\}+
\frac{1}{(2k_e\cdot q)^2} {\rm tr}\{(m_e + \hat{k}_e)(2k_e\cdot
\varepsilon^*_{\lambda''} + \hat{\varepsilon}^*_{\lambda''}
\hat{q})\hat{\varepsilon}^*_{\lambda'} (m_e + \hat{k}_e + \hat{k} + \hat{q})\gamma^0 \hat{k}_{\nu}\gamma^0 \nonumber\\
\hspace{-0.3in}&&\times (m_e + \hat{k}_e + \hat{k} + \hat{q})
\hat{\varepsilon}_{\lambda'} (2k_e\cdot \varepsilon_{\lambda''} +
\hat{q}\hat{\varepsilon}_{\lambda''})\}\Big\}\nonumber\\
\hspace{-0.3in}&& + \frac{\lambda^2\,\delta^{ij}}{(2 k_e \cdot (k + q)
  + 2 k\cdot q)^2}\, \Big\{\frac{1}{(2k_e\cdot k )^2} {\rm tr}\{(m_e +
\hat{k}_e)(2k_e\cdot \varepsilon^*_{\lambda'} +
\hat{\varepsilon}^*_{\lambda'}
\hat{k})\hat{\varepsilon}^*_{\lambda''}\nonumber\\
\hspace{-0.3in}&& \times(m_e + \hat{k}_e + \hat{k} + \hat{q})\gamma^i
\hat{k}_{\nu}\gamma^j (m_e + \hat{k}_e + \hat{k} +
\hat{q})\hat{\varepsilon}_{\lambda''}(2k_e\cdot \varepsilon_{\lambda'}
+ \hat{k}\hat{\varepsilon}_{\lambda'})\} + \frac{1}{2k_e\cdot k}\,
\frac{1}{2k_e\cdot q}\nonumber\\
\hspace{-0.3in}&&\times {\rm tr}\{(m_e + \hat{k}_e)(2k_e\cdot
\varepsilon^*_{\lambda''} + \hat{\varepsilon}^*_{\lambda''}
\hat{q})\hat{\varepsilon}^*_{\lambda'} (m_e + \hat{k}_e + \hat{k} +
\hat{q}) \vec{\gamma}^{\,i} \hat{k}_{\nu}\vec{\gamma}^{\,j} (m_e +
\hat{k}_e + \hat{k} + \hat{q}) \hat{\varepsilon}_{\lambda''}
(2k_e\cdot \varepsilon_{\lambda'} +
\hat{k}\hat{\varepsilon}_{\lambda'})\}\nonumber\\
\hspace{-0.3in}&& + \frac{1}{2k_e\cdot k}\, \frac{1}{2k_e\cdot q} {\rm
  tr}\{(m_e + \hat{k}_e)(2k_e\cdot \varepsilon^*_{\lambda'} +
\hat{\varepsilon}^*_{\lambda'} \hat{k})\hat{\varepsilon}^*_{\lambda''}
(m_e + \hat{k}_e + \hat{k} + \hat{q}) \vec{\gamma}^{\,j} \hat{k}_{\nu}
\vec{\gamma}^{\,i} (m_e + \hat{k}_e + \hat{k} + \hat{q})\nonumber\\
\hspace{-0.3in}&&\times \hat{\varepsilon}_{\lambda'}(2k_e\cdot
\varepsilon_{\lambda''} + \hat{q}\hat{\varepsilon}_{\lambda''})\} +
\frac{1}{(2k_e\cdot q)^2} {\rm tr}\{(m_e + \hat{k}_e)(2k_e\cdot
\varepsilon^*_{\lambda''} + \hat{\varepsilon}^*_{\lambda''}
\hat{q})\hat{\varepsilon}^*_{\lambda'} (m_e + \hat{k}_e + \hat{k} +
\hat{q}) \vec{\gamma}^{\,i} \hat{k}_{\nu}
\vec{\gamma}^{\,j}\nonumber\\
\hspace{-0.3in}&&\times (m_e + \hat{k}_e + \hat{k} +
\hat{q})\hat{\varepsilon}_{\lambda'}(2k_e\cdot \varepsilon_{\lambda''}
+ \hat{q}\hat{\varepsilon}_{\lambda''})\}\Big\}.
\end{eqnarray}
Then, we average over directions of the antineutrino momentum
$\vec{k}_{\nu}$ and sum over polarizations of photons. This gives
\begin{eqnarray}\label{eq:E.7}
\hspace{-0.3in}&&\int \frac{d\Omega_{\nu}}{4\pi}\,\frac{1}{2}\sum_{\rm
  pol, \lambda', \lambda''}\frac{|{\cal M}_{\rm
    Fig.\,\ref{fig:fig5}}(n \to p e^-
  \bar{\nu}_e\gamma\gamma)_{\lambda'\lambda''}|^2}{8 m^2_n e^2 (1 + 3
  \lambda^2) E_{\nu}} = \frac{1}{(2 k_e\cdot (k + q) + 2 k \cdot
  q)^2}\nonumber\\
\hspace{-0.3in}&&\times \Big\{\frac{1}{(2k_e\cdot k)^2}
\sum_{\lambda', \lambda''} {\rm tr}\{(m_e + \hat{k}_e)\, (2k_e\cdot
\varepsilon^*_{\lambda'} + \hat{\varepsilon}^*_{\lambda'} \hat{k})\,
\hat{\varepsilon}^*_{\lambda''} \, (m_e + \hat{k}_e + \hat{k} +
\hat{q})\, \gamma^0 \, (m_e + \hat{k}_e + \hat{k} + \hat{q})\,
\hat{\varepsilon}_{\lambda''}\, (2k_e\cdot \varepsilon_{\lambda'} +
\hat{k}\, \hat{\varepsilon}_{\lambda'})\}\nonumber\\
\hspace{-0.3in}&& + \frac{1}{2k_e\cdot k}\, \frac{1}{2k_e\cdot q }
\sum_{\lambda', \lambda''} {\rm tr}\{(m_e + \hat{k}_e)\, (2k_e\cdot
\varepsilon^*_{\lambda''} + \hat{\varepsilon}^*_{\lambda''}
\hat{q})\, \hat{\varepsilon}^*_{\lambda'}\, (m_e + \hat{k}_e + \hat{k} +
\hat{q})\, \gamma^0 \,(m_e + \hat{k}_e + \hat{k} +
\hat{q})\, \hat{\varepsilon}_{\lambda''} \,  (2k_e\cdot \varepsilon_{\lambda'}
+ \hat{k}\, \hat{\varepsilon}_{\lambda'})\}\nonumber\\
\hspace{-0.3in}&& + \frac{1}{2k_e\cdot k}\, \frac{1}{2k_e\cdot q}
\sum_{\lambda', \lambda''} {\rm tr}\{(m_e + \hat{k}_e)\, (2k_e\cdot
\varepsilon^*_{\lambda'} + \hat{\varepsilon}^*_{\lambda'}
\hat{k})\, \hat{\varepsilon}^*_{\lambda''}\, (m_e + \hat{k}_e + \hat{k} +
\hat{q})\, \gamma^0 \, (m_e + \hat{k}_e + \hat{k} + \hat{q})\,
\hat{\varepsilon}_{\lambda'}(2k_e\cdot \varepsilon_{\lambda''} +
\hat{q}\, \hat{\varepsilon}_{\lambda''})\}\nonumber\\
\hspace{-0.3in}&&+ \frac{1}{(2k_e\cdot q)^2} \sum_{\lambda',
  \lambda''} {\rm tr}\{(m_e + \hat{k}_e)(2k_e\cdot
\varepsilon^*_{\lambda''} + \hat{\varepsilon}^*_{\lambda''} \hat{q})\,
\hat{\varepsilon}^*_{\lambda'} \, (m_e + \hat{k}_e + \hat{k} +
\hat{q})\, \gamma^0 \, (m_e + \hat{k}_e + \hat{k} + \hat{q})
\hat{\varepsilon}_{\lambda'} \, (2k_e\cdot \varepsilon_{\lambda''} +
\hat{q}\, \hat{\varepsilon}_{\lambda''})\}\Big\}.\nonumber\\
\hspace{-0.3in}&&
\end{eqnarray}
For the calculation of the contribution of the diagram
Fig.\,\ref{fig:fig5} to the rate of the neutron radiative
$\beta^-$--decay with two photons in the final state we have to sum
over the photon physical degrees of freedom only.  For this aim we use
the following relations \cite{Ivanov2013,Ivanov2017}
\begin{eqnarray}\label{eq:E.8}
\hspace{-0.3in}\vec{k}\cdot \vec{\varepsilon}^{\,*}_{\lambda'} &=&
\vec{k}\cdot \vec{\varepsilon}_{\lambda'} =
0\;,\;\vec{\varepsilon}^{\,*}_{\lambda'}\cdot
\vec{\varepsilon}_{\bar{\lambda}'} =
\delta_{\lambda'\bar{\lambda}'}\;,\; \vec{q}\cdot
\vec{\varepsilon}^{\,*}_{\lambda''} = \vec{q}\cdot
\vec{\varepsilon}_{\lambda''} =
0\;,\;\vec{\varepsilon}^{\,*}_{\lambda''}\cdot
\vec{\varepsilon}_{\bar{\lambda}''} = \delta_{\lambda''
  \bar{\lambda}''},\nonumber\\
\hspace{-0.3in}\sum_{\lambda' = 1,2}\vec{\varepsilon}^{\,i
  *}_{\lambda'}\vec{\varepsilon}^{\,j}_{\lambda'} &=& \delta^{ij} -
\frac{\vec{k}^{\,i} \vec{k}^{\,j}}{\omega^2} = \delta^{ij} -
\vec{n}^{\,i}_{\vec{k}}\, \vec{n}^{\,j}_{\vec{k}}\;,\;\sum_{\lambda'' =
  1,2}\vec{\varepsilon}^{\,i
  *}_{\lambda''}\vec{\varepsilon}^{\,j}_{\lambda''} = \delta^{ij} -
\frac{\vec{q}^{\,i} \vec{q}^{\,j}}{q^2_0} = \delta^{ij} -
\vec{n}^{\,i}_{\vec{q}}\, \vec{n}^{\,j}_{\vec{q}},
\end{eqnarray}
where $\omega$ and $q_0$ are photon energies and $\vec{n}_{\vec{k}} =
\vec{k}/\omega$ and $\vec{n}_{\vec{q}} = \vec{q}/q_0$ are unit vectors
directed along photon momenta. For the traces in Eq.(\ref{eq:E.7}) we
obtain the following expressions
\begin{eqnarray*}
\hspace{-0.3in}&&\sum_{\lambda', \lambda''} {\rm tr}\{(m_e +
\hat{k}_e)\, (2k_e\cdot \varepsilon^*_{\lambda'} +
\hat{\varepsilon}^*_{\lambda'} \hat{k})\,
\hat{\varepsilon}^*_{\lambda''} \, (m_e + \hat{k}_e + \hat{k} +
\hat{q})\, \gamma^0 \, (m_e + \hat{k}_e + \hat{k} + \hat{q})\,
\hat{\varepsilon}_{\lambda''}\, (2k_e\cdot \varepsilon_{\lambda'} +
\hat{k}\, \hat{\varepsilon}_{\lambda'})\} = \nonumber\\
\hspace{-0.3in}&& = 64\,q_0\,\Big(\omega\, (E_e - \vec{k}_e \cdot
\vec{n}_{\vec{k}}) + q_0\, (E_e - \vec{k}_e \cdot \vec{n}_{\vec{q}}) -
(E_e + \omega)\, \omega \,(1 - \vec{n}_{\vec{k}}\cdot
\vec{n}_{\vec{q}})\Big)\, \Big(k^2_e - (\vec{k}_e \cdot
\vec{n}_{\vec{k}})^2\Big) + 64\, (E_e + \omega + q_0)\nonumber\\
\hspace{-0.3in}&&\times \,\Big(k^2_e - (\vec{k}_e \cdot
\vec{n}_{\vec{k}})^2\Big) \, \Big(k^2_e - (\vec{k}_e \cdot
\vec{n}_{\vec{q}})^2\Big) + 128\, (E_e + \omega + q_0)\, \omega
\,\Big(k^2_e - (\vec{k}_e \cdot \vec{n}_{\vec{k}})^2\Big) \,
\Big((\vec{k}_e \cdot \vec{n}_{\vec{k}}) - (\vec{k}_e \cdot
\vec{n}_{\vec{q}})(\vec{n}_{\vec{k}}\cdot \vec{n}_{\vec{q}})\Big)\nonumber\\
\hspace{-0.3in}&&+ 64\, (E_e + \omega + q_0)\, \omega^2 \,\Big(k^2_e -
(\vec{k}_e \cdot \vec{n}_{\vec{k}})^2\Big) \, \Big(1 -
(\vec{n}_{\vec{k}}\cdot \vec{n}_{\vec{q}})^2\Big) - 64\, (E_e + \omega
+ q_0)\, \omega \,\Big(E_e - \vec{k}_e \cdot \vec{n}_{\vec{k}}\Big)\,
\Big(k^2_e - (\vec{k}_e \cdot \vec{n}_{\vec{k}})^2\nonumber\\
\hspace{-0.3in}&& - (\vec{k}_e \cdot \vec{n}_{\vec{q}})^2 + (\vec{k}_e
\cdot \vec{n}_{\vec{k}}) (\vec{k}_e \cdot \vec{n}_{\vec{q}})
(\vec{n}_{\vec{k}}\cdot \vec{n}_{\vec{q}})\Big) + 64\, (E_e + \omega +
q_0)\, \omega^2 \,\Big(E_e - \vec{k}_e \cdot \vec{n}_{\vec{k}}\Big)\,
(\vec{n}_{\vec{k}}\cdot \vec{n}_{\vec{q}})\, \Big((\vec{k}_e \cdot
\vec{n}_{\vec{q}}) - (\vec{k}_e \cdot \vec{n}_{\vec{k}})\nonumber\\
\end{eqnarray*}
\begin{eqnarray}\label{eq:E.9}
\hspace{-0.3in}&&\times\, (\vec{n}_{\vec{k}}\cdot
\vec{n}_{\vec{q}})\Big) - 64\, (E_e + \omega + q_0)\, \omega \, q_0 \,
(E_e - \vec{k}_e \cdot \vec{n}_{\vec{k}}) \,\Big((\vec{k}_e \cdot
\vec{n}_{\vec{q}}) - (\vec{k}_e \cdot \vec{n}_{\vec{k}})
(\vec{n}_{\vec{q}}\cdot \vec{n}_{\vec{k}})\Big) + 64\, (E_e + \omega +
q_0)\, \omega^2 \nonumber\\
\hspace{-0.3in}&&\times\,(E_e - \vec{k}_e \cdot \vec{n}_{\vec{k}}) \,
\Big(E_e - (\vec{k}_e \cdot \vec{n}_{\vec{q}}) (\vec{n}_{\vec{k}}\cdot
\vec{n}_{\vec{q}}) + \omega \, (1 - (\vec{n}_{\vec{k}}\cdot
\vec{n}_{\vec{q}})^2) + q_0 \,(1 - \vec{n}_{\vec{k}}\cdot
\vec{n}_{\vec{q}})\Big) - 64\, \omega^2 \, \Big(\omega \,(E_e -
\vec{k}_e \cdot \vec{n}_{\vec{k}})\nonumber\\
\hspace{-0.3in}&& + q_0\,(E_e - \vec{k}_e \cdot \vec{n}_{\vec{q}}) +
\omega \, q_0 \, (1 - \vec{n}_{\vec{k}}\cdot \vec{n}_{\vec{q}})\Big)
\end{eqnarray}
and 
\begin{eqnarray*}
\hspace{-0.3in}&&\sum_{\lambda', \lambda''} {\rm tr}\{(m_e +
\hat{k}_e)\, (2k_e\cdot \varepsilon^*_{\lambda''} +
\hat{\varepsilon}^*_{\lambda''} \hat{q})\,
\hat{\varepsilon}^*_{\lambda'}\, (m_e + \hat{k}_e + \hat{k} +
\hat{q})\, \gamma^0 \,(m_e + \hat{k}_e + \hat{k} + \hat{q})\,
\hat{\varepsilon}_{\lambda''} \, (2k_e\cdot \varepsilon_{\lambda'} +
\hat{k}\, \hat{\varepsilon}_{\lambda'})\} = \nonumber\\
\hspace{-0.3in}&&= - 32\, m^2_e \, (E_e + \omega + q_0)\, \Big(k^2_e -
(\vec{k}_e \cdot \vec{n}_{\vec{k}})^2 - (\vec{k}_e \cdot
\vec{n}_{\vec{q}})^2 + (\vec{k}_e \cdot \vec{n}_{\vec{k}})(\vec{k}_e
\cdot \vec{n}_{\vec{q}})(\vec{n}_{\vec{k}}\cdot
\vec{n}_{\vec{q}})\Big) - 32\, m^2_e \, \omega\, (E_e + \omega +
q_0)\nonumber\\
\hspace{-0.3in}&&\times\, \Big((\vec{k}_e \cdot \vec{n}_{\vec{k}}) -
(\vec{k}_e \cdot \vec{n}_{\vec{q}})(\vec{n}_{\vec{k}}\cdot
\vec{n}_{\vec{q}})\Big) - 32\, m^2_e \, q_0\, (E_e + \omega + q_0)\,
\Big((\vec{k}_e \cdot \vec{n}_{\vec{q}}) - (\vec{k}_e \cdot
\vec{n}_{\vec{k}})(\vec{n}_{\vec{q}}\cdot \vec{n}_{\vec{k}})\Big) -
32\, m^2_e \, \omega\, q_0\nonumber\\
\hspace{-0.3in}&&\times\, (E_e + \omega + q_0)\, \Big(1 -
\vec{n}_{\vec{k}}\cdot \vec{n}_{\vec{q}}\Big) - 32\, E_e \,
\Big(\omega \,(E_e - \vec{k}_e \cdot \vec{n}_{\vec{k}}) + q_0 \,(E_e -
\vec{k}_e \cdot \vec{n}_{\vec{q}}) + \omega \, q_0 \,(1 -
\vec{n}_{\vec{k}}\cdot \vec{n}_{\vec{q}})\Big)\, \Big(k^2_e - (\vec{k}_e
\cdot \vec{n}_{\vec{k}})^2\nonumber\\
\hspace{-0.3in}&& - (\vec{k}_e \cdot \vec{n}_{\vec{q}})^2 + (\vec{k}_e
\cdot \vec{n}_{\vec{k}})(\vec{k}_e \cdot
\vec{n}_{\vec{q}})(\vec{n}_{\vec{k}}\cdot \vec{n}_{\vec{q}})\Big) +
32\, \omega\, (E_e + \omega + q_0)\, \Big((E_e - \vec{k}_e \cdot
\vec{n}_{\vec{k}}) + q_0 \,(1 - \vec{n}_{\vec{k}}\cdot
\vec{n}_{\vec{q}})\Big)\, \Big(k^2_e - (\vec{k}_e \cdot
\vec{n}_{\vec{k}})^2\nonumber\\
\hspace{-0.3in}&& - (\vec{k}_e \cdot \vec{n}_{\vec{q}})^2 + (\vec{k}_e
\cdot \vec{n}_{\vec{k}})(\vec{k}_e \cdot
\vec{n}_{\vec{q}})(\vec{n}_{\vec{k}}\cdot \vec{n}_{\vec{q}})\Big) +
64\, (E_e + \omega + q_0)\, \Big(k^2_e - (\vec{k}_e \cdot
\vec{n}_{\vec{k}})^2\Big)\, \Big[\Big(k^2_e - (\vec{k}_e \cdot
  \vec{n}_{\vec{q}})^2\Big) + \omega \, \Big((\vec{k}_e \cdot
  \vec{n}_{\vec{k}}) \nonumber\\
\hspace{-0.3in}&& - (\vec{k}_e \cdot \vec{n}_{\vec{q}})\,
(\vec{n}_{\vec{q}}\cdot \vec{n}_{\vec{k}})\Big)\Big] + 32\, \omega
\, \Big(\omega \, (E_e - \vec{k}_e \cdot \vec{n}_{\vec{k}}) + q_0
\,(E_e - \vec{k}_e \cdot \vec{n}_{\vec{q}}) + \omega \, q_0 \, (1 -
\vec{n}_{\vec{k}}\cdot \vec{n}_{\vec{q}})\Big) \,\Big(k^2_e -
(\vec{k}_e \cdot \vec{n}_{\vec{q}})^2\Big)\nonumber\\
\hspace{-0.3in}&& - 32\, \omega \, E_e \, \Big(\omega \,(E_e -
\vec{k}_e \cdot \vec{n}_{\vec{k}}) + q_0 \,(E_e - \vec{k}_e \cdot
\vec{n}_{\vec{q}}) + \omega \, q_0 \, (1 - \vec{n}_{\vec{k}}\cdot
\vec{n}_{\vec{q}})\Big) \, \Big((\vec{k}_e \cdot \vec{n}_{\vec{k}}) -
(\vec{k}_e \cdot \vec{n}_{\vec{q}}) (\vec{n}_{\vec{k}}\cdot
\vec{n}_{\vec{q}})\Big)\nonumber\\
\hspace{-0.3in}&& - 32\, \omega \, \Big(\omega \,(E_e - \vec{k}_e
\cdot \vec{n}_{\vec{k}}) + q_0 \,(E_e - \vec{k}_e \cdot
\vec{n}_{\vec{q}}) + \omega \, q_0 \, (1 - \vec{n}_{\vec{k}}\cdot
\vec{n}_{\vec{q}})\Big) \, \Big(k^2_e - (\vec{k}_e \cdot
\vec{n}_{\vec{k}})^2 - (\vec{k}_e \cdot \vec{n}_{\vec{q}})^2 +
(\vec{k}_e \cdot \vec{n}_{\vec{k}})(\vec{k}_e \cdot
\vec{n}_{\vec{q}})\nonumber\\
\hspace{-0.3in}&&\times \,(\vec{n}_{\vec{k}}\cdot
\vec{n}_{\vec{q}})\Big) + 32\, \omega \, (E_e + \omega + q_0)\,
\Big((E_e - \vec{k}_e \cdot \vec{n}_{\vec{k}}) + q_0 \, (1 -
\vec{n}_{\vec{k}}\cdot \vec{n}_{\vec{q}})\Big)\,\Big(k^2_e -
(\vec{k}_e \cdot \vec{n}_{\vec{k}})^2 - (\vec{k}_e \cdot
\vec{n}_{\vec{q}})^2 + (\vec{k}_e \cdot \vec{n}_{\vec{k}})\nonumber\\
\hspace{-0.3in}&&\times \,(\vec{k}_e \cdot
\vec{n}_{\vec{q}})(\vec{n}_{\vec{k}}\cdot \vec{n}_{\vec{q}})\Big) +
32\, \omega\, (E_e + \omega + q_0)\,\Big((\vec{k}_e \cdot
\vec{n}_{\vec{k}}) - (\vec{k}_e \cdot \vec{n}_{\vec{q}})
(\vec{n}_{\vec{k}}\cdot \vec{n}_{\vec{q}})\Big)\, \Big[\Big(k^2_e -
  (\vec{k}_e \cdot \vec{n}_{\vec{k}})^2\Big) + q_0\, \Big((\vec{k}_e
  \cdot \vec{n}_{\vec{q}}) - (\vec{k}_e \cdot
  \vec{n}_{\vec{k}})\nonumber\\
\hspace{-0.3in}&& \times\, (\vec{n}_{\vec{q}}\cdot
\vec{n}_{\vec{k}})\Big)\Big] + 32\, \omega \, (E_e + \omega +
q_0)\, \Big(m^2_e + \omega \,(E_e - \vec{k}_e \cdot \vec{n}_{\vec{k}})
+ q_0 \,(E_e - \vec{k}_e \cdot \vec{n}_{\vec{q}})\Big)
\,\Big((\vec{k}_e \cdot \vec{n}_{\vec{k}}) - (\vec{k}_e \cdot
\vec{n}_{\vec{q}})(\vec{n}_{\vec{k}}\cdot \vec{n}_{\vec{q}})\Big)\nonumber\\
\hspace{-0.3in}&& - 32\, \omega \, (E_e + \omega + q_0)\, \Big((E_e -
\vec{k}_e \cdot \vec{n}_{\vec{k}}) + q_0\, (1 - \vec{n}_{\vec{k}}\cdot
\vec{n}_{\vec{q}})\Big)\, \Big(k^2_e - (\vec{k}_e \cdot
\vec{n}_{\vec{q}})^2\Big) + 32\, \omega \, (E_e + \omega + q_0)\,(E_e
- \vec{k}_e \cdot \vec{n}_{\vec{k}})\nonumber\\
\hspace{-0.3in}&& \times\,\Big[ \Big(k^2_e - (\vec{k}_e \cdot
  \vec{n}_{\vec{q}})^2\Big) + \omega\, \Big((\vec{k}_e \cdot
  \vec{n}_{\vec{k}}) - (\vec{k}_e \cdot
  \vec{n}_{\vec{q}})(\vec{n}_{\vec{k}}\cdot
  \vec{n}_{\vec{q}})\Big)\Big]- 32\, \, q_0 \,E_e \, \Big(\omega\,
(E_e - \vec{k}_e \cdot \vec{n}_{\vec{k}}) + q_0\, (E_e - \vec{k}_e
\cdot \vec{n}_{\vec{q}}) \nonumber\\
\hspace{-0.3in}&&+ \omega \, q_0 \, (1 - \vec{n}_{\vec{k}}\cdot
\vec{n}_{\vec{q}})\Big)\, \Big((\vec{k}_e \cdot \vec{n}_{\vec{q}}) -
(\vec{k}_e \cdot \vec{n}_{\vec{k}})(\vec{n}_{\vec{q}}\cdot
\vec{n}_{\vec{k}})\Big) + 32\,q_0 \, \Big(\omega\, (E_e - \vec{k}_e
\cdot \vec{n}_{\vec{k}}) + q_0\, (E_e - \vec{k}_e \cdot
\vec{n}_{\vec{q}})\nonumber\\
\hspace{-0.3in}&& + \omega \, q_0 \, (1 - \vec{n}_{\vec{k}}\cdot
\vec{n}_{\vec{q}})\Big)\, \Big(k^2_e - (\vec{k}_e \cdot
\vec{n}_{\vec{k}})^2\Big) - 32\,q_0 \, \Big(\omega\, (E_e -
\vec{k}_e \cdot \vec{n}_{\vec{k}}) + q_0\, (E_e - \vec{k}_e \cdot
\vec{n}_{\vec{q}}) + \omega \, q_0 \, (1 - \vec{n}_{\vec{k}}\cdot
\vec{n}_{\vec{q}})\Big) \nonumber\\
\hspace{-0.3in}&&\times \, \Big(k^2_e - (\vec{k}_e \cdot
\vec{n}_{\vec{k}})^2- (\vec{k}_e \cdot \vec{n}_{\vec{q}})^2 +
(\vec{k}_e \cdot \vec{n}_{\vec{k}})(\vec{k}_e \cdot
\vec{n}_{\vec{q}})(\vec{n}_{\vec{k}}\cdot \vec{n}_{\vec{q}})\Big) +
32\, q_0 \, (E_e + \omega + q_0)\, \Big(m^2_e + \omega \,(E_e -
\vec{k}_e \cdot \vec{n}_{\vec{k}})\nonumber\\ 
\hspace{-0.3in}&& + q_0 \,(E_e -
\vec{k}_e \cdot \vec{n}_{\vec{q}})\Big)\, \Big((\vec{k}_e \cdot
\vec{n}_{\vec{q}}) - (\vec{k}_e \cdot
\vec{n}_{\vec{k}})(\vec{n}_{\vec{q}}\cdot \vec{n}_{\vec{k}})\Big) +
32\, q_0 \, (E_e + \omega + q_0)\, (E_e - \vec{k}_e \cdot
\vec{n}_{\vec{k}})\nonumber\\
\hspace{-0.3in}&&\times \,\Big[\Big(k^2_e - (\vec{k}_e \cdot
  \vec{n}_{\vec{k}})^2\Big) + q_0\, \Big((\vec{k}_e \cdot
  \vec{n}_{\vec{q}}) - (\vec{k}_e \cdot
  \vec{n}_{\vec{k}})(\vec{n}_{\vec{q}}\cdot
  \vec{n}_{\vec{k}})\Big)\Big] - 32\, q_0 \, (E_e + \omega + q_0)\,
\Big((E_e - \vec{k}_e \cdot \vec{n}_{\vec{q}})\nonumber\\
\hspace{-0.3in}&& + \omega \, (1 - \vec{n}_{\vec{k}}\cdot
\vec{n}_{\vec{q}})\Big)\, \Big(k^2_e - (\vec{k}_e \cdot
\vec{n}_{\vec{k}})^2 \Big) + 32\, q_0 \, (E_e + \omega + q_0)\,
\Big((\vec{k}_e \cdot \vec{n}_{\vec{q}}) - (\vec{k}_e \cdot
\vec{n}_{\vec{k}})(\vec{n}_{\vec{q}}\cdot \vec{n}_{\vec{k}})\Big) \,
\Big[\Big(k^2_e - (\vec{k}_e \cdot \vec{n}_{\vec{q}})^2\Big)\nonumber\\
\hspace{-0.3in}&& + \omega\, \Big((\vec{k}_e \cdot \vec{n}_{\vec{k}})
- (\vec{k}_e \cdot \vec{n}_{\vec{q}})\,(\vec{n}_{\vec{k}}\cdot
\vec{n}_{\vec{q}})\Big)\Big] + 32\, q_0 \, (E_e + \omega + q_0)\,
\Big((E_e - \vec{k}_e \cdot \vec{n}_{\vec{q}}) + \omega \, (1 -
\vec{n}_{\vec{k}}\cdot \vec{n}_{\vec{q}})\Big)\, \Big(k^2_e -
(\vec{k}_e \cdot \vec{n}_{\vec{k}})^2\nonumber\\
\hspace{-0.3in}&& - (\vec{k}_e \cdot \vec{n}_{\vec{q}})^2 + (\vec{k}_e
\cdot \vec{n}_{\vec{k}})(\vec{k}_e \cdot \vec{n}_{\vec{q}})
\,(\vec{n}_{\vec{k}}\cdot \vec{n}_{\vec{q}})\Big) - 8\, m^2_e \,
\omega \, q_0 \, \Big(1 - (\vec{n}_{\vec{k}}\cdot
\vec{n}_{\vec{q}})^2\Big)\, \Big(E_e (1 + \vec{n}_{\vec{k}}\cdot
\vec{n}_{\vec{q}}) - (\vec{k}_e \cdot \vec{n}_{\vec{k}}) - (\vec{k}_e
\cdot \vec{n}_{\vec{q}})\Big)\nonumber\\
\hspace{-0.3in}&& - 8\, m^2_e \, \omega \, q_0 \,
(\vec{n}_{\vec{k}}\cdot \vec{n}_{\vec{q}})\, \Big((\vec{k}_e \cdot
\vec{n}_{\vec{q}}) - (\vec{k}_e \cdot
\vec{n}_{\vec{k}})\,(\vec{n}_{\vec{q}}\cdot \vec{n}_{\vec{k}})\Big) +
8\, m^2_e \, \omega \, q_0 \, (E_e - \omega)\, (\vec{n}_{\vec{q}}\cdot
\vec{n}_{\vec{k}}) \, \Big(1 - (\vec{n}_{\vec{q}}\cdot
\vec{n}_{\vec{k}})^2\Big)\nonumber\\
\hspace{-0.3in}&& + 8\,\omega \, q_0\, \Big(1 -
(\vec{n}_{\vec{k}}\cdot \vec{n}_{\vec{q}})^2\Big)\, \Big(m^2_e +
2\,\omega \, (E_e - \vec{k}_e \cdot \vec{n}_{\vec{k}}) + 2\, q_0\,
(E_e - \vec{k}_e \cdot \vec{n}_{\vec{q}}) + 2\, \omega\, q_0 \,(1 -
\vec{n}_{\vec{k}}\cdot \vec{n}_{\vec{q}})\Big)\nonumber\\
\hspace{-0.3in}&&+ \Big[8\, q_0 \,\Big(m^2_e + 2\,\omega \, (E_e -
  \vec{k}_e \cdot \vec{n}_{\vec{k}}) + 2\, q_0\, (E_e - \vec{k}_e
  \cdot \vec{n}_{\vec{q}}) + 2\, \omega\, q_0 \,(1 -
  \vec{n}_{\vec{k}}\cdot \vec{n}_{\vec{q}})\Big) - 16\,\omega\, q_0\,
  (E_e + \omega + q_0)\nonumber\\
\hspace{-0.3in}&&\times \, \Big( (E_e - \vec{k}_e \cdot
\vec{n}_{\vec{q}}) + \omega\,(1 - \vec{n}_{\vec{k}}\cdot
\vec{n}_{\vec{q}})\Big)\Big] \,(\vec{n}_{\vec{k}}\cdot
\vec{n}_{\vec{q}})\, \Big((\vec{k}_e \cdot \vec{n}_{\vec{q}}) -
(\vec{k}_e \cdot \vec{n}_{\vec{k}}) ( \vec{n}_{\vec{q}}\cdot
\vec{n}_{\vec{k}})\Big) - 8\, \omega\, q_0 \, \Big[ (E_e -
  \omega)\nonumber\\
\end{eqnarray*}
\begin{eqnarray}\label{eq:E.10}
\hspace{-0.3in}&&\times \, \Big(m^2_e + 2\,\omega \, (E_e - \vec{k}_e
\cdot \vec{n}_{\vec{k}}) + 2\, q_0\, (E_e - \vec{k}_e \cdot
\vec{n}_{\vec{q}}) + 2\, \omega\, q_0 \,(1 - \vec{n}_{\vec{k}}\cdot
\vec{n}_{\vec{q}})\Big) - 2\, (E_e + \omega + q_0)\,\Big(m^2_e +
\omega \, (E_e - \vec{k}_e \cdot \vec{n}_{\vec{k}})\nonumber\\
\hspace{-0.3in}&& + q_0\, (E_e - \vec{k}_e \cdot
\vec{n}_{\vec{q}})\Big)\Big]\, ( \vec{n}_{\vec{k}}\cdot
\vec{n}_{\vec{q}}) \, \Big(1 - ( \vec{n}_{\vec{k}}\cdot
\vec{n}_{\vec{q}})^2\Big) - 16\, \omega\, q_0 \, (E_e + \omega +
q_0)\, \Big(1 - ( \vec{n}_{\vec{k}}\cdot \vec{n}_{\vec{q}})^2\Big)\,
\Big[(E_e - \vec{k}_e \cdot \vec{n}_{\vec{q}}) \nonumber\\
\hspace{-0.3in}&&\times \,\Big((E_e - \vec{k}_e \cdot
\vec{n}_{\vec{k}}) + q_0 \,(1 - \vec{n}_{\vec{k}}\cdot
\vec{n}_{\vec{q}})\Big) + (E_e - \vec{k}_e \cdot \vec{n}_{\vec{k}})\,
\Big((E_e - \vec{k}_e \cdot \vec{n}_{\vec{q}}) + \omega \,(1 -
\vec{n}_{\vec{k}}\cdot \vec{n}_{\vec{q}})\Big) - \Big(m^2_e + \omega\,
(E_e - \vec{k}_e \cdot \vec{n}_{\vec{k}})\nonumber\\
\hspace{-0.3in}&& + q_0 \,(E_e - \vec{k}_e \cdot
\vec{n}_{\vec{q}})\Big)\Big] + 32\, \omega \, q_0 \, (E_e + \omega
+ q_0)\, \Big(1 - \vec{n}_{\vec{k}}\cdot \vec{n}_{\vec{q}}\Big)\,
\Big[\Big(k^2_e - (\vec{k}_e \cdot \vec{n}_{\vec{k}})^2\Big) + q_0
  \,\Big((\vec{k}_e \cdot \vec{n}_{\vec{q}}) - (\vec{k}_e \cdot
  \vec{n}_{\vec{k}})\nonumber\\
\hspace{-0.3in}&& \times\, (\vec{n}_{\vec{q}}\cdot
\vec{n}_{\vec{k}})\Big)\Big] - 32\, \omega \, q_0 \, (E_e + \omega +
q_0)\, (E_e - \vec{k}_e \cdot \vec{n}_{\vec{q}})\, \Big[\omega
  \,\Big(1 - ( \vec{n}_{\vec{k}}\cdot \vec{n}_{\vec{q}})^2\Big) +
  \Big((\vec{k}_e \cdot \vec{n}_{\vec{k}}) - (\vec{k}_e \cdot
  \vec{n}_{\vec{q}}) (\vec{n}_{\vec{k}}\cdot
  \vec{n}_{\vec{q}})\Big)\Big]\nonumber\\
\hspace{-0.3in}&& + 32\, \omega \, q_0 \, (E_e + \omega + q_0)\,
\Big(1 - \vec{n}_{\vec{k}}\cdot
\vec{n}_{\vec{q}}\Big)\,\Big[\Big(k^2_e - (\vec{k}_e \cdot
  \vec{n}_{\vec{q}})^2\Big) + \omega \,\Big((\vec{k}_e \cdot
  \vec{n}_{\vec{k}}) - (\vec{k}_e \cdot \vec{n}_{\vec{q}}) \,
  (\vec{n}_{\vec{k}}\cdot \vec{n}_{\vec{q}})\Big)\Big] \nonumber\\
\hspace{-0.3in}&& - 16\, \omega \, q_0 \, (E_e + \omega + q_0)\,\Big(1
- \vec{n}_{\vec{k}}\cdot \vec{n}_{\vec{q}}\Big)\, \Big[\Big(k^2_e -
  (\vec{k}_e \cdot \vec{n}_{\vec{k}})^2 - (\vec{k}_e \cdot
  \vec{n}_{\vec{q}})^2 + (\vec{k}_e \cdot \vec{n}_{\vec{k}})(\vec{k}_e
  \cdot \vec{n}_{\vec{q}})\,(\vec{n}_{\vec{k}}\cdot
  \vec{n}_{\vec{q}})\Big) - \omega\, (\vec{n}_{\vec{k}}\cdot
  \vec{n}_{\vec{q}})\nonumber\\
\hspace{-0.3in}&&\times\,\Big((\vec{k}_e \cdot \vec{n}_{\vec{q}}) -
(\vec{k}_e \cdot \vec{n}_{\vec{k}})\Big)\Big] \nonumber\\
\hspace{-0.3in}&&- 16\, \omega \, q_0 \, (E_e + \omega + q_0)\, (E_e -
\vec{k}_e \cdot \vec{n}_{\vec{k}})\,(\vec{n}_{\vec{k}}\cdot
\vec{n}_{\vec{q}})\, \Big[\omega \,\Big(1 - (\vec{n}_{\vec{k}}\cdot
  \vec{n}_{\vec{q}})^2\Big) + \Big((\vec{k}_e \cdot \vec{n}_{\vec{q}})
  - (\vec{k}_e \cdot \vec{n}_{\vec{k}}) (\vec{n}_{\vec{q}}\cdot
  \vec{n}_{\vec{k}})\Big)\Big]\nonumber\\
\hspace{-0.3in}&& - 16\, \omega \, q_0 \, (E_e + \omega +
q_0)\,(\vec{n}_{\vec{k}}\cdot \vec{n}_{\vec{q}}) \, \Big((E_e -
\vec{k}_e \cdot \vec{n}_{\vec{k}}) + q_0\, (1 - \vec{n}_{\vec{k}}\cdot
\vec{n}_{\vec{q}})\Big)\, \Big((\vec{k}_e \cdot \vec{n}_{\vec{k}}) -
(\vec{k}_e \cdot \vec{n}_{\vec{q}}) (\vec{n}_{\vec{k}}\cdot
\vec{n}_{\vec{q}})\Big)\nonumber\\
\hspace{-0.3in}&& - 16\, \omega \, q_0 \, (E_e + \omega +
q_0)\,(\vec{n}_{\vec{k}}\cdot \vec{n}_{\vec{q}}) \,(E_e - \vec{k}_e
\cdot \vec{n}_{\vec{k}})\, \Big[q_0 \, \Big(1 -
  (\vec{n}_{\vec{k}}\cdot \vec{n}_{\vec{q}})^2\Big) + \Big((\vec{k}_e
  \cdot \vec{n}_{\vec{q}}) - (\vec{k}_e \cdot \vec{n}_{\vec{k}})
  (\vec{n}_{\vec{q}}\cdot \vec{n}_{\vec{k}})\Big)\Big] \nonumber\\
\hspace{-0.3in}&&- 16\, \omega \, q_0 \, (E_e + \omega + q_0)\,\Big(1
- \vec{n}_{\vec{k}}\cdot \vec{n}_{\vec{q}}\Big)\, \Big[\Big(k^2_e -
  (\vec{k}_e \cdot \vec{n}_{\vec{k}})^2 - (\vec{k}_e \cdot
  \vec{n}_{\vec{q}})^2 + (\vec{k}_e \cdot \vec{n}_{\vec{k}})(\vec{k}_e
  \cdot \vec{n}_{\vec{q}})(\vec{n}_{\vec{k}}\cdot
  \vec{n}_{\vec{q}})\Big)\nonumber\\
\hspace{-0.3in}&& - q_0\, (\vec{n}_{\vec{k}}\cdot \vec{n}_{\vec{q}})\,
\Big((\vec{k}_e \cdot \vec{n}_{\vec{k}}) - (\vec{k}_e \cdot
\vec{n}_{\vec{q}}) (\vec{n}_{\vec{k}}\cdot
\vec{n}_{\vec{q}})\Big)\Big] - 32\, \omega \, q_0 \, (E_e + \omega +
q_0)\, (E_e - \vec{k}_e \cdot \vec{n}_{\vec{k}})\, \Big[\Big((\vec{k}_e \cdot
  \vec{n}_{\vec{q}}) - (\vec{k}_e \cdot \vec{n}_{\vec{k}})
  (\vec{n}_{\vec{q}}\cdot \vec{n}_{\vec{k}})\Big) \nonumber\\
\hspace{-0.3in}&&   +    q_0\,   \Big(1    -   (\vec{n}_{\vec{k}}\cdot
\vec{n}_{\vec{q}})^2\Big)\Big].
\end{eqnarray}
The last two traces in Eq.(\ref{eq:E.7}) can be obtained from
Eq.(\ref{eq:E.9}) and Eq.(\ref{eq:E.10}) by a replacement $\omega
\longleftrightarrow q_0$ and $\vec{n}_{\vec{k}} \longleftrightarrow
\vec{n}_{\vec{q}}$, respectively.

The rate of the neutron radiative $\beta^-$--decay with two photons in
the final state is defined by
\begin{eqnarray}\label{eq:E.11} 
\lambda^{(\rm Fig.\,\ref{fig:fig5})}_{\beta \gamma\gamma} &=&
\frac{1}{2m_n}\,\frac{1}{2}\int \frac{1}{2}\,\sum_{\rm pol.,\lambda',
  \lambda''}|M(n \to
p\,e^-\,\bar{\nu}_e\,\gamma\gamma)_{\lambda'\lambda''}|^2\,(2\pi)^4\,
\delta^{(4)}(k_n - k_p - k_e - k_{\nu} - k - q)\nonumber\\
\hspace{-0.3in}&&\times\, \frac{d^3k_p}{(2\pi)^3 2 E_p}\,
\frac{d^3k_e}{(2\pi)^3 2 E_e}\, \frac{d^3k_{\nu}}{(2\pi)^3 2
  E_{\nu}}\, \frac{d^3k}{(2\pi)^3 2 \omega}\, \frac{d^3q}{(2\pi)^3 2
  q_0},
\end{eqnarray}
where the factor $1/2$ in front of the integral takes into account the
identity of photons in the final state. A relation of the amplitude
$M(n \to p\,e^-\,\bar{\nu}_e\,\gamma\gamma)_{\lambda'\lambda''}$ to
the amplitude ${\cal M}(n \to
p\,e^-\,\bar{\nu}_e\,\gamma\gamma)_{\lambda'\lambda''}$ is given by
Eq.(\ref{eq:A.1}).  Having integrated over the degrees of freedom of the
photon with 4--momentum $q$ and keeping the energy of the photon with
4--momentum $k$ within the interval $\omega_{\rm min} \le \omega \le
\omega_{\rm max}$ the rate of the two photon radiative
$\beta^-$--decay of the neutron is given by
\begin{eqnarray}\label{eq:E.12} 
\lambda^{(\rm Fig.\,\ref{fig:fig5})}_{\beta \gamma\gamma}(\omega_{\rm
  max}, \omega_{\rm min}) &=& (1 +
3\lambda^2)\,\frac{\alpha^2}{\pi^2}\,\frac{G^2_F|V_{ud}|^2}{16\pi^3}
\int^{\omega_{\rm max}}_{\omega_{\rm min}}\!\!\! d\omega\int^{E_0 -
  \omega}_{m_e}\!\!\! dE_e \,k_e\, F(E_e, Z = 1)\int^{E_0 - E_e -
  \omega}_0\!\!\! dq_0\,(E_0 - E_e - \omega -
q_0)^2\nonumber\\ &&\times
\int\frac{d\Omega_{e\gamma}}{4\pi}\int\frac{d\Omega_{e\gamma'}}{4\pi}\int
\frac{d\Omega_{\nu}}{4\pi}\,\frac{1}{2}\sum_{\rm pol, \lambda',
  \lambda''}\frac{|{\cal M}_{\rm Fig.\,\ref{fig:fig5}}(n \to p e^-
  \bar{\nu}_e\gamma\gamma)_{\lambda'\lambda''}|^2}{8 m^2_n e^2(1 +
  3\lambda^2) E_{\nu}},
\end{eqnarray}
where $d\Omega_{e\gamma}$ and $d\Omega_{e\gamma'}$ are the elements of
the solid angles of the electron--photon correlations of photons with
3--momenta $\vec{k}$ and $\vec{q}$, respectively. Then, we introduce
the notation
\begin{eqnarray}\label{eq:E.13} 
\int\frac{d\Omega_{\nu}}{4\pi}\,\frac{1}{2}\sum_{\rm pol, \lambda',
  \lambda''}\frac{|{\cal M}_{\rm Fig.\,\ref{fig:fig5}}(n \to p e^-
  \bar{\nu}_e\gamma\gamma)_{\lambda'\lambda''}|^2}{8 m^2_n e^2(1 +
  3\lambda^2) E_{\nu}} &=& \rho^{(1)}_{e\gamma\gamma'}(E_e,\vec{k}_e,
\omega, \vec{n}_{\vec{k}}, q_0, \vec{n}_{\vec{q}}) +
\rho^{(2)}_{e\gamma\gamma'}(E_e,\vec{k}_e, \omega, \vec{n}_{\vec{k}},
q_0, \vec{n}_{\vec{q}})\nonumber\\ &+&
\rho^{(2)}_{e\gamma\gamma'}(E_e,\vec{k}_e, q_0, \vec{n}_{\vec{q}},
\omega, \vec{n}_{\vec{k}}),
\end{eqnarray}
where we have denoted
\begin{eqnarray}\label{eq:E.14} 
\hspace{-0.3in}&& \rho^{(1)}_{e\gamma\gamma'}(E_e,\vec{k}_e, \omega,
\vec{n}_{\vec{k}}, q_0, \vec{n}_{\vec{q}}) =
\frac{1}{\omega}\,\frac{q_0}{\displaystyle \Big((\omega \, (E_e -
  \vec{k}_e\cdot \vec{n}_{\vec{k}}) + q_0 \,(E_e - \vec{k}_e\cdot
  \vec{n}_{\vec{q}}) + \omega\, q_0 \, (1 -
  \vec{n}_{\vec{k}}\cdot\vec{n}_{\vec{q}})\Big) ^2}\, \frac{1}{(E_e -
  \vec{k}_e\cdot \vec{n}_{\vec{k}})^2}\nonumber\\
\hspace{-0.3in}&&\times \,\frac{1}{8}\, \sum_{\lambda', \lambda''}
       {\rm tr}\{(m_e + \hat{k}_e)\, (2k_e\cdot
       \varepsilon^*_{\lambda'} + \hat{\varepsilon}^*_{\lambda'}
       \hat{k})\, \hat{\varepsilon}^*_{\lambda''} \, (m_e + \hat{k}_e
       + \hat{k} + \hat{q})\, \gamma^0 \, (m_e + \hat{k}_e + \hat{k} +
       \hat{q})\, \hat{\varepsilon}_{\lambda''}\, (2k_e\cdot
       \varepsilon_{\lambda'} + \hat{k}\,
       \hat{\varepsilon}_{\lambda'})\}
\end{eqnarray}
and
\begin{eqnarray}\label{eq:E.15} 
\hspace{-0.3in}&& \rho^{(2)}_{e\gamma\gamma'}(E_e,\vec{k}_e, \omega,
\vec{n}_{\vec{k}}, q_0, \vec{n}_{\vec{q}}) =
\frac{1}{\displaystyle \Big((\omega \, (E_e -
  \vec{k}_e\cdot \vec{n}_{\vec{k}}) + q_0 \,(E_e - \vec{k}_e\cdot
  \vec{n}_{\vec{q}}) + \omega\, q_0 \, (1 -
  \vec{n}_{\vec{k}}\cdot\vec{n}_{\vec{q}})\Big) ^2}\, \frac{1}{E_e -
  \vec{k}_e\cdot \vec{n}_{\vec{k}}} \frac{1}{E_e -
  \vec{k}_e\cdot \vec{n}_{\vec{q}}}\nonumber\\
\hspace{-0.3in}&&\times \,\frac{1}{16}\, \sum_{\lambda', \lambda''}
       {\rm tr}\{(m_e + \hat{k}_e)\, (2k_e\cdot
       \varepsilon^*_{\lambda''} + \hat{\varepsilon}^*_{\lambda''}
       \hat{q})\, \hat{\varepsilon}^*_{\lambda'}\, (m_e + \hat{k}_e +
       \hat{k} + \hat{q})\, \gamma^0 \,(m_e + \hat{k}_e + \hat{k} +
       \hat{q})\, \hat{\varepsilon}_{\lambda''} \, (2k_e\cdot
       \varepsilon_{\lambda'} + \hat{k}\,
       \hat{\varepsilon}_{\lambda'})\}
\end{eqnarray}
and
\begin{eqnarray}\label{eq:E.16} 
\hspace{-0.3in}&&\rho^{(2)}_{e\gamma\gamma'}(E_e,\vec{k}_e, q_0, \vec{n}_{\vec{q}},
\omega, \vec{n}_{\vec{k}}) =
\frac{1}{\displaystyle \Big((\omega \, (E_e -
  \vec{k}_e\cdot \vec{n}_{\vec{k}}) + q_0 \,(E_e - \vec{k}_e\cdot
  \vec{n}_{\vec{q}}) + \omega\, q_0 \, (1 -
  \vec{n}_{\vec{k}}\cdot\vec{n}_{\vec{q}})\Big) ^2}\, \frac{1}{E_e -
  \vec{k}_e\cdot \vec{n}_{\vec{k}}} \frac{1}{E_e -
  \vec{k}_e\cdot \vec{n}_{\vec{q}}}\nonumber\\
\hspace{-0.3in}&&\times \,\frac{1}{16}\, \sum_{\lambda', \lambda''}
       {\rm tr}\{(m_e + \hat{k}_e)\, (2k_e\cdot
       \varepsilon^*_{\lambda'} + \hat{\varepsilon}^*_{\lambda'}
       \hat{k})\, \hat{\varepsilon}^*_{\lambda''}\, (m_e + \hat{k}_e +
       \hat{k} + \hat{q})\, \gamma^0 \, (m_e + \hat{k}_e + \hat{k} +
       \hat{q})\, \hat{\varepsilon}_{\lambda'}(2k_e\cdot
       \varepsilon_{\lambda''} + \hat{q}\,
       \hat{\varepsilon}_{\lambda''})\}.
\end{eqnarray}
For the trace in Eq.(\ref{eq:E.16}) we obtain the following expression 
\begin{eqnarray*}
\hspace{-0.3in}&&\sum_{\lambda', \lambda''} {\rm tr}\{(m_e +
\hat{k}_e)\, (2k_e\cdot \varepsilon^*_{\lambda'} +
\hat{\varepsilon}^*_{\lambda'} \hat{k})\,
\hat{\varepsilon}^*_{\lambda''}\, (m_e + \hat{k}_e + \hat{k} +
\hat{q})\, \gamma^0 \, (m_e + \hat{k}_e + \hat{k} + \hat{q})\,
\hat{\varepsilon}_{\lambda'}(2k_e\cdot \varepsilon_{\lambda''} +
\hat{q}\, \hat{\varepsilon}_{\lambda''})\} = \nonumber\\
\hspace{-0.3in}&&= - 32\, m^2_e \, (E_e + \omega + q_0)\, \Big(k^2_e -
(\vec{k}_e \cdot \vec{n}_{\vec{k}})^2 - (\vec{k}_e \cdot
\vec{n}_{\vec{q}})^2 + (\vec{k}_e \cdot \vec{n}_{\vec{k}})(\vec{k}_e
\cdot \vec{n}_{\vec{q}})(\vec{n}_{\vec{k}}\cdot
\vec{n}_{\vec{q}})\Big) - 32\, m^2_e \, \omega\, (E_e + \omega +
q_0)\nonumber\\
\hspace{-0.3in}&&\times\, \Big((\vec{k}_e \cdot \vec{n}_{\vec{k}}) -
(\vec{k}_e \cdot \vec{n}_{\vec{q}})(\vec{n}_{\vec{k}}\cdot
\vec{n}_{\vec{q}})\Big) - 32\, m^2_e \, q_0\, (E_e + \omega + q_0)\,
\Big((\vec{k}_e \cdot \vec{n}_{\vec{q}}) - (\vec{k}_e \cdot
\vec{n}_{\vec{k}})(\vec{n}_{\vec{q}}\cdot \vec{n}_{\vec{k}})\Big) -
32\, m^2_e \, \omega\, q_0\nonumber\\
\hspace{-0.3in}&&\times\, (E_e + \omega + q_0)\, \Big(1 -
\vec{n}_{\vec{k}}\cdot \vec{n}_{\vec{q}}\Big) - 32\, E_e \,
\Big(\omega \,(E_e - \vec{k}_e \cdot \vec{n}_{\vec{k}}) + q_0 \,(E_e -
\vec{k}_e \cdot \vec{n}_{\vec{q}}) + \omega \, q_0 \,(1 -
\vec{n}_{\vec{k}}\cdot \vec{n}_{\vec{q}})\Big)\, \Big(k^2_e - (\vec{k}_e
\cdot \vec{n}_{\vec{k}})^2\nonumber\\
\hspace{-0.3in}&& - (\vec{k}_e \cdot \vec{n}_{\vec{q}})^2 + (\vec{k}_e
\cdot \vec{n}_{\vec{k}})(\vec{k}_e \cdot
\vec{n}_{\vec{q}})(\vec{n}_{\vec{k}}\cdot \vec{n}_{\vec{q}})\Big) +
32\, q_0\, (E_e + \omega + q_0)\, \Big((E_e - \vec{k}_e \cdot
\vec{n}_{\vec{q}}) + \omega \,(1 - \vec{n}_{\vec{k}}\cdot
\vec{n}_{\vec{q}})\Big)\, \Big(k^2_e - (\vec{k}_e \cdot
\vec{n}_{\vec{k}})^2\nonumber\\
\hspace{-0.3in}&& - (\vec{k}_e \cdot \vec{n}_{\vec{q}})^2 + (\vec{k}_e
\cdot \vec{n}_{\vec{k}})(\vec{k}_e \cdot
\vec{n}_{\vec{q}})(\vec{n}_{\vec{k}}\cdot \vec{n}_{\vec{q}})\Big) +
64\, (E_e + \omega + q_0)\, \Big(k^2_e - (\vec{k}_e \cdot
\vec{n}_{\vec{q}})^2\Big)\, \Big[\Big(k^2_e - (\vec{k}_e \cdot
  \vec{n}_{\vec{k}})^2\Big) + q_0 \, \Big((\vec{k}_e \cdot
  \vec{n}_{\vec{q}}) \nonumber\\
\hspace{-0.3in}&& - (\vec{k}_e \cdot \vec{n}_{\vec{k}})\,
(\vec{n}_{\vec{q}}\cdot \vec{n}_{\vec{k}})\Big)\Big] + 32\, q_0
\, \Big(\omega \, (E_e - \vec{k}_e \cdot \vec{n}_{\vec{k}}) + q_0
\,(E_e - \vec{k}_e \cdot \vec{n}_{\vec{q}}) + \omega \, q_0 \, (1 -
\vec{n}_{\vec{k}}\cdot \vec{n}_{\vec{q}})\Big) \,\Big(k^2_e -
(\vec{k}_e \cdot \vec{n}_{\vec{k}})^2\Big) \nonumber\\
\hspace{-0.3in}&& - 32\, q_0 \, E_e \, \Big(\omega \,(E_e -
\vec{k}_e \cdot \vec{n}_{\vec{k}}) + q_0 \,(E_e - \vec{k}_e \cdot
\vec{n}_{\vec{q}}) + \omega \, q_0 \, (1 - \vec{n}_{\vec{k}}\cdot
\vec{n}_{\vec{q}})\Big) \, \Big((\vec{k}_e \cdot \vec{n}_{\vec{q}}) -
(\vec{k}_e \cdot \vec{n}_{\vec{k}}) (\vec{n}_{\vec{q}}\cdot
\vec{n}_{\vec{k}})\Big) \nonumber\\
\hspace{-0.3in}&& - 32\, q_0  \, \Big(\omega \,(E_e - \vec{k}_e
\cdot \vec{n}_{\vec{k}}) + q_0 \,(E_e - \vec{k}_e \cdot
\vec{n}_{\vec{q}}) + \omega \, q_0 \, (1 - \vec{n}_{\vec{k}}\cdot
\vec{n}_{\vec{q}})\Big) \, \Big(k^2_e - (\vec{k}_e \cdot
\vec{n}_{\vec{k}})^2 - (\vec{k}_e \cdot \vec{n}_{\vec{q}})^2 +
(\vec{k}_e \cdot \vec{n}_{\vec{k}})(\vec{k}_e \cdot
\vec{n}_{\vec{q}})\nonumber\\
\hspace{-0.3in}&&\times \,(\vec{n}_{\vec{k}}\cdot
\vec{n}_{\vec{q}})\Big)  + 32\, q_0 \, (E_e + \omega + q_0)\,
\Big((E_e - \vec{k}_e \cdot \vec{n}_{\vec{q}}) + \omega \, (1 -
\vec{n}_{\vec{k}}\cdot \vec{n}_{\vec{q}})\Big)\,\Big(k^2_e -
(\vec{k}_e \cdot \vec{n}_{\vec{k}})^2 - (\vec{k}_e \cdot
\vec{n}_{\vec{q}})^2 + (\vec{k}_e \cdot \vec{n}_{\vec{k}})\nonumber\\
\hspace{-0.3in}&&\times \,(\vec{k}_e \cdot
\vec{n}_{\vec{q}})(\vec{n}_{\vec{k}}\cdot \vec{n}_{\vec{q}})\Big) +
32\, q_0 \, (E_e + \omega + q_0)\,\Big((\vec{k}_e \cdot
\vec{n}_{\vec{q}}) - (\vec{k}_e \cdot \vec{n}_{\vec{k}})
(\vec{n}_{\vec{q}}\cdot \vec{n}_{\vec{k}})\Big)\, \Big[\Big(k^2_e -
  (\vec{k}_e \cdot \vec{n}_{\vec{q}})^2\Big) + \omega \,
  \Big((\vec{k}_e \cdot \vec{n}_{\vec{k}}) - (\vec{k}_e \cdot
  \vec{n}_{\vec{q}})\nonumber\\
\hspace{-0.3in}&& \times\, (\vec{n}_{\vec{k}}\cdot
\vec{n}_{\vec{q}})\Big)\Big] + 32\, q_0 \, (E_e + \omega + q_0)\,
\Big(m^2_e + \omega \,(E_e - \vec{k}_e \cdot \vec{n}_{\vec{k}}) + q_0
\,(E_e - \vec{k}_e \cdot \vec{n}_{\vec{q}})\Big) \,\Big((\vec{k}_e
\cdot \vec{n}_{\vec{q}}) - (\vec{k}_e \cdot
\vec{n}_{\vec{k}})(\vec{n}_{\vec{q}}\cdot
\vec{n}_{\vec{k}})\Big) \nonumber\\
\hspace{-0.3in}&& - 32\, q_0 \, (E_e + \omega + q_0)\, \Big((E_e -
\vec{k}_e \cdot \vec{n}_{\vec{q}}) + \omega \, (1 -
\vec{n}_{\vec{k}}\cdot \vec{n}_{\vec{q}})\Big)\, \Big(k^2_e -
(\vec{k}_e \cdot \vec{n}_{\vec{k}})^2\Big) + 32\, q_0 \, (E_e +
\omega + q_0)\,(E_e - \vec{k}_e \cdot \vec{n}_{\vec{q}})\nonumber\\ 
\hspace{-0.3in}&& \times\,\Big[ \Big(k^2_e - (\vec{k}_e \cdot
  \vec{n}_{\vec{k}})^2\Big) + q_0\, \Big((\vec{k}_e \cdot
  \vec{n}_{\vec{q}}) - (\vec{k}_e \cdot
  \vec{n}_{\vec{k}})(\vec{n}_{\vec{q}}\cdot
  \vec{n}_{\vec{k}})\Big)\Big] - 32\, \, \omega \,E_e \, \Big(\omega\,
(E_e - \vec{k}_e \cdot \vec{n}_{\vec{k}}) + q_0\, (E_e - \vec{k}_e
\cdot \vec{n}_{\vec{q}}) \nonumber\\
\hspace{-0.3in}&&+ \omega \, q_0 \, (1 - \vec{n}_{\vec{k}}\cdot
\vec{n}_{\vec{q}})\Big)\, \Big((\vec{k}_e \cdot \vec{n}_{\vec{k}}) -
(\vec{k}_e \cdot \vec{n}_{\vec{q}})(\vec{n}_{\vec{k}}\cdot
\vec{n}_{\vec{q}})\Big) + 32\,\omega \, \Big(\omega\, (E_e - \vec{k}_e
\cdot \vec{n}_{\vec{k}}) + q_0\, (E_e - \vec{k}_e \cdot
\vec{n}_{\vec{q}})\nonumber\\
\hspace{-0.3in}&& + \omega \, q_0 \, (1 - \vec{n}_{\vec{k}}\cdot
\vec{n}_{\vec{q}})\Big)\, \Big(k^2_e - (\vec{k}_e \cdot
\vec{n}_{\vec{q}})^2\Big) - 32\,\omega \, \Big(\omega\, (E_e -
\vec{k}_e \cdot \vec{n}_{\vec{k}}) + q_0\, (E_e - \vec{k}_e \cdot
\vec{n}_{\vec{q}}) + \omega \, q_0 \, (1 - \vec{n}_{\vec{k}}\cdot
\vec{n}_{\vec{q}})\Big) \nonumber\\
\hspace{-0.3in}&&\times \, \Big(k^2_e - (\vec{k}_e \cdot
\vec{n}_{\vec{k}})^2- (\vec{k}_e \cdot \vec{n}_{\vec{q}})^2 +
(\vec{k}_e \cdot \vec{n}_{\vec{k}})(\vec{k}_e \cdot
\vec{n}_{\vec{q}})(\vec{n}_{\vec{k}}\cdot \vec{n}_{\vec{q}})\Big) +
32\, \omega \, (E_e + \omega + q_0)\, \Big(m^2_e + \omega \,(E_e -
\vec{k}_e \cdot \vec{n}_{\vec{k}})\nonumber\\
\end{eqnarray*}
\begin{eqnarray}\label{eq:E.17}
\hspace{-0.3in}&& + q_0 \,(E_e - \vec{k}_e \cdot
\vec{n}_{\vec{q}})\Big)\, \Big((\vec{k}_e \cdot \vec{n}_{\vec{k}}) -
(\vec{k}_e \cdot \vec{n}_{\vec{q}})(\vec{n}_{\vec{k}}\cdot
\vec{n}_{\vec{q}})\Big) + 32\, \omega \, (E_e + \omega + q_0)\, (E_e -
\vec{k}_e \cdot \vec{n}_{\vec{q}})\nonumber\\
\hspace{-0.3in}&&\times \,\Big[\Big(k^2_e - (\vec{k}_e \cdot
  \vec{n}_{\vec{q}})^2\Big) + \omega \, \Big((\vec{k}_e \cdot
  \vec{n}_{\vec{k}}) - (\vec{k}_e \cdot
  \vec{n}_{\vec{q}})(\vec{n}_{\vec{k}}\cdot
  \vec{n}_{\vec{q}})\Big)\Big] - 32\, \omega \, (E_e + \omega + q_0)\,
\Big((E_e - \vec{k}_e \cdot \vec{n}_{\vec{k}})\nonumber\\
\hspace{-0.3in}&& + q_0 \, (1 - \vec{n}_{\vec{k}}\cdot
\vec{n}_{\vec{q}})\Big)\, \Big(k^2_e - (\vec{k}_e \cdot
\vec{n}_{\vec{q}})^2 \Big) + 32\, \omega \, (E_e + \omega + q_0)\,
\Big((\vec{k}_e \cdot \vec{n}_{\vec{k}}) - (\vec{k}_e \cdot
\vec{n}_{\vec{q}})(\vec{n}_{\vec{k}}\cdot \vec{n}_{\vec{q}})\Big) \,
\Big[\Big(k^2_e - (\vec{k}_e \cdot
  \vec{n}_{\vec{k}})^2\Big)\nonumber\\
\hspace{-0.3in}&& + q_0\, \Big((\vec{k}_e \cdot \vec{n}_{\vec{q}}) -
(\vec{k}_e \cdot \vec{n}_{\vec{k}})\,(\vec{n}_{\vec{q}}\cdot
\vec{n}_{\vec{k}})\Big)\Big] + 32\, \omega \, (E_e + \omega +
q_0)\, \Big((E_e - \vec{k}_e \cdot \vec{n}_{\vec{k}}) + q_0 \, (1 -
\vec{n}_{\vec{k}}\cdot \vec{n}_{\vec{q}})\Big)\, \Big(k^2_e -
(\vec{k}_e \cdot \vec{n}_{\vec{k}})^2\nonumber\\
\hspace{-0.3in}&& - (\vec{k}_e \cdot \vec{n}_{\vec{q}})^2 + (\vec{k}_e
\cdot \vec{n}_{\vec{k}})(\vec{k}_e \cdot \vec{n}_{\vec{q}})
\,(\vec{n}_{\vec{k}}\cdot \vec{n}_{\vec{q}})\Big) - 8\, m^2_e \,
\omega \, q_0 \, \Big(1 - (\vec{n}_{\vec{k}}\cdot
\vec{n}_{\vec{q}})^2\Big)\, \Big(E_e (1 + \vec{n}_{\vec{k}}\cdot
\vec{n}_{\vec{q}}) - (\vec{k}_e \cdot \vec{n}_{\vec{k}}) - (\vec{k}_e
\cdot \vec{n}_{\vec{q}})\Big)\nonumber\\
\hspace{-0.3in}&& - 8\, m^2_e \, \omega \, q_0 \,
(\vec{n}_{\vec{k}}\cdot \vec{n}_{\vec{q}})\, \Big((\vec{k}_e \cdot
\vec{n}_{\vec{k}}) - (\vec{k}_e \cdot
\vec{n}_{\vec{q}}))\,(\vec{n}_{\vec{k}}\cdot \vec{n}_{\vec{q}})\Big) +
8\, m^2_e \, \omega \, q_0 \, (E_e - q_0)\, (\vec{n}_{\vec{q}}\cdot
\vec{n}_{\vec{k}}) \, \Big(1 - (\vec{n}_{\vec{q}}\cdot
\vec{n}_{\vec{k}})^2\Big)\nonumber\\
\hspace{-0.3in}&& + 8\,\omega \, q_0\, \Big(1 -
(\vec{n}_{\vec{k}}\cdot \vec{n}_{\vec{q}})^2\Big)\, \Big(m^2_e +
2\,\omega \, (E_e - \vec{k}_e \cdot \vec{n}_{\vec{k}}) + 2\, q_0\,
(E_e - \vec{k}_e \cdot \vec{n}_{\vec{q}}) + 2\, \omega\, q_0 \,(1 -
\vec{n}_{\vec{k}}\cdot \vec{n}_{\vec{q}})\Big) \nonumber\\
\hspace{-0.3in}&& + \Big[8\, \omega \,\Big(m^2_e + 2\,\omega \, (E_e -
  \vec{k}_e \cdot \vec{n}_{\vec{k}}) + 2\, q_0\, (E_e - \vec{k}_e
  \cdot \vec{n}_{\vec{q}}) + 2\, \omega\, q_0 \,(1 -
  \vec{n}_{\vec{k}}\cdot \vec{n}_{\vec{q}})\Big) - 16\,\omega\, q_0\,
  (E_e + \omega + q_0)\nonumber\\
\hspace{-0.3in}&&\times \, \Big( (E_e - \vec{k}_e \cdot
\vec{n}_{\vec{k}}) + q_0\, (1 - \vec{n}_{\vec{k}}\cdot
\vec{n}_{\vec{q}})\Big)\Big] \,(\vec{n}_{\vec{k}}\cdot
\vec{n}_{\vec{q}})\, \Big((\vec{k}_e \cdot \vec{n}_{\vec{k}}) -
(\vec{k}_e \cdot \vec{n}_{\vec{q}}) ( \vec{n}_{\vec{k}}\cdot
\vec{n}_{\vec{q}})\Big) - 8\, \omega\, q_0 \, \Big[ (E_e -
  q_0)\nonumber\\
\hspace{-0.3in}&&\times \, \Big(m^2_e + 2\,\omega \, (E_e - \vec{k}_e
\cdot \vec{n}_{\vec{k}}) + 2\, q_0\, (E_e - \vec{k}_e \cdot
\vec{n}_{\vec{q}}) + 2\, \omega\, q_0 \,(1 - \vec{n}_{\vec{k}}\cdot
\vec{n}_{\vec{q}})\Big) - 2\, (E_e + \omega + q_0)\,\Big(m^2_e +
\omega \, (E_e - \vec{k}_e \cdot \vec{n}_{\vec{k}})\nonumber\\
\hspace{-0.3in}&& + q_0\, (E_e - \vec{k}_e \cdot
\vec{n}_{\vec{q}})\Big)\Big]\, ( \vec{n}_{\vec{k}}\cdot
\vec{n}_{\vec{q}}) \, \Big(1 - ( \vec{n}_{\vec{k}}\cdot
\vec{n}_{\vec{q}})^2\Big) - 16\, \omega\, q_0 \, (E_e + \omega +
q_0)\, \Big(1 - ( \vec{n}_{\vec{k}}\cdot \vec{n}_{\vec{q}})^2\Big)\,
\Big[(E_e - \vec{k}_e \cdot \vec{n}_{\vec{q}}) \nonumber\\
\hspace{-0.3in}&&\times \,\Big((E_e - \vec{k}_e \cdot
\vec{n}_{\vec{k}}) + q_0 \,(1 - \vec{n}_{\vec{k}}\cdot
\vec{n}_{\vec{q}})\Big) + (E_e - \vec{k}_e \cdot \vec{n}_{\vec{k}})\,
\Big((E_e - \vec{k}_e \cdot \vec{n}_{\vec{q}}) + \omega \,(1 -
\vec{n}_{\vec{k}}\cdot \vec{n}_{\vec{q}})\Big) - \Big(m^2_e + \omega\,
(E_e - \vec{k}_e \cdot \vec{n}_{\vec{k}})\nonumber\\
\hspace{-0.3in}&& + q_0 \,(E_e - \vec{k}_e \cdot
\vec{n}_{\vec{q}})\Big)\Big] + 32\, \omega \, q_0 \, (E_e + \omega
+ q_0)\, \Big(1 - \vec{n}_{\vec{k}}\cdot \vec{n}_{\vec{q}}\Big)\,
\Big[\Big(k^2_e - (\vec{k}_e \cdot \vec{n}_{\vec{q}})^2\Big) + \omega
  \,\Big((\vec{k}_e \cdot \vec{n}_{\vec{k}}) - (\vec{k}_e \cdot
  \vec{n}_{\vec{q}})\nonumber\\
\hspace{-0.3in}&& \times\, (\vec{n}_{\vec{k}}\cdot
\vec{n}_{\vec{q}})\Big)\Big] - 32\, \omega \, q_0 \, (E_e + \omega
+ q_0)\, (E_e - \vec{k}_e \cdot \vec{n}_{\vec{k}})\, \Big[q_0 \,\Big(1
  - ( \vec{n}_{\vec{k}}\cdot \vec{n}_{\vec{q}})^2\Big) +
  \Big((\vec{k}_e \cdot \vec{n}_{\vec{q}}) - (\vec{k}_e \cdot
  \vec{n}_{\vec{k}}) (\vec{n}_{\vec{q}}\cdot
  \vec{n}_{\vec{k}})\Big)\Big] \nonumber\\
\hspace{-0.3in}&& + 32\, \omega \, q_0 \, (E_e + \omega + q_0)\,
\Big(1 - \vec{n}_{\vec{k}}\cdot
\vec{n}_{\vec{k}}\Big)\,\Big[\Big(k^2_e - (\vec{k}_e \cdot
  \vec{n}_{\vec{k}})^2\Big) + q_0 \,\Big((\vec{k}_e \cdot
  \vec{n}_{\vec{q}}) - (\vec{k}_e \cdot \vec{n}_{\vec{k}}) \,
  (\vec{n}_{\vec{q}}\cdot \vec{n}_{\vec{k}})\Big)\Big] \nonumber\\
\hspace{-0.3in}&& - 16\, \omega \, q_0 \, (E_e + \omega + q_0)\,\Big(1
- \vec{n}_{\vec{k}}\cdot \vec{n}_{\vec{q}}\Big)\, \Big[\Big(k^2_e -
  (\vec{k}_e \cdot \vec{n}_{\vec{k}})^2 - (\vec{k}_e \cdot
  \vec{n}_{\vec{q}})^2 + (\vec{k}_e \cdot \vec{n}_{\vec{k}})(\vec{k}_e
  \cdot \vec{n}_{\vec{q}})\,(\vec{n}_{\vec{k}}\cdot
  \vec{n}_{\vec{q}})\Big) - q_0 \, (\vec{n}_{\vec{k}}\cdot
  \vec{n}_{\vec{q}})\nonumber\\
\hspace{-0.3in}&&\times\,\Big((\vec{k}_e \cdot \vec{n}_{\vec{k}}) -
(\vec{k}_e \cdot \vec{n}_{\vec{q}})\Big)\Big] \nonumber\\
\hspace{-0.3in}&&- 16\, \omega \, q_0 \, (E_e + \omega + q_0)\, (E_e -
\vec{k}_e \cdot \vec{n}_{\vec{q}})\,(\vec{n}_{\vec{k}}\cdot
\vec{n}_{\vec{q}})\, \Big[q_0 \,\Big(1 - (\vec{n}_{\vec{k}}\cdot
  \vec{n}_{\vec{q}})^2\Big) + \Big((\vec{k}_e \cdot \vec{n}_{\vec{k}})
  - (\vec{k}_e \cdot \vec{n}_{\vec{q}}) (\vec{n}_{\vec{k}}\cdot
  \vec{n}_{\vec{q}})\Big)\Big]\nonumber\\
\hspace{-0.3in}&& - 16\, \omega \, q_0 \, (E_e + \omega +
q_0)\,(\vec{n}_{\vec{k}}\cdot \vec{n}_{\vec{q}}) \, \Big((E_e -
\vec{k}_e \cdot \vec{n}_{\vec{q}}) + \omega \, (1 - \vec{n}_{\vec{k}}\cdot
\vec{n}_{\vec{q}})\Big)\, \Big((\vec{k}_e \cdot \vec{n}_{\vec{q}}) -
(\vec{k}_e \cdot \vec{n}_{\vec{k}}) (\vec{n}_{\vec{q}}\cdot
\vec{n}_{\vec{k}})\Big)\Big] \nonumber\\
\hspace{-0.3in}&& - 16\, \omega \, q_0 \, (E_e + \omega + q_0) \,(E_e
- \vec{k}_e \cdot \vec{n}_{\vec{q}})\,(\vec{n}_{\vec{k}}\cdot
\vec{n}_{\vec{q}})\, \Big[\omega \, \Big(1 - (\vec{n}_{\vec{k}}\cdot
  \vec{n}_{\vec{q}})^2\Big) + \Big((\vec{k}_e \cdot \vec{n}_{\vec{k}})
  - (\vec{k}_e \cdot \vec{n}_{\vec{q}}) (\vec{n}_{\vec{k}}\cdot
  \vec{n}_{\vec{q}})\Big)\Big] \nonumber\\
\hspace{-0.3in}&&- 16\, \omega \, q_0 \, (E_e + \omega + q_0)\,\Big(1
- \vec{n}_{\vec{k}}\cdot \vec{n}_{\vec{q}}\Big)\, \Big[\Big(k^2_e -
  (\vec{k}_e \cdot \vec{n}_{\vec{k}})^2 - (\vec{k}_e \cdot
  \vec{n}_{\vec{q}})^2 + (\vec{k}_e \cdot \vec{n}_{\vec{k}})(\vec{k}_e
  \cdot \vec{n}_{\vec{q}})(\vec{n}_{\vec{k}}\cdot
  \vec{n}_{\vec{q}})\Big)\nonumber\\
\hspace{-0.3in}&& - \omega \, (\vec{n}_{\vec{k}}\cdot
\vec{n}_{\vec{q}})\, \Big((\vec{k}_e \cdot \vec{n}_{\vec{q}}) -
(\vec{k}_e \cdot \vec{n}_{\vec{k}}) (\vec{n}_{\vec{q}}\cdot
\vec{n}_{\vec{k}})\Big)\Big] - 32\, \omega \, q_0 \, (E_e + \omega +
q_0)\, (E_e - \vec{k}_e \cdot \vec{n}_{\vec{q}})\,
\Big[\Big((\vec{k}_e \cdot \vec{n}_{\vec{k}}) - (\vec{k}_e \cdot
  \vec{n}_{\vec{q}}) (\vec{n}_{\vec{k}}\cdot \vec{n}_{\vec{q}})\Big)
  \nonumber\\
\hspace{-0.3in}&& + \omega \, \Big(1 - (\vec{n}_{\vec{k}}\cdot
\vec{n}_{\vec{q}})^2\Big)\Big].
\end{eqnarray}
Thus, the rate of the neutron radiative $\beta^-$--decay with two
photons in the final state for one of the photons from the energy
region $\omega_{\rm min} \le \omega \le \omega_{\rm max}$ is given by
\begin{eqnarray}\label{eq:E.18} 
\hspace{-0.3in}&&\lambda^{(\rm Fig.\,\ref{fig:fig5})}_{\beta
  \gamma\gamma}(\omega_{\rm max}, \omega_{\rm min}) = (1 +
3\lambda^2)\,\frac{\alpha^2}{\pi^2}\,\frac{G^2_F|V_{ud}|^2}{16\pi^3}
\int^{\omega_{\rm max}}_{\omega_{\rm min}}\!\!\! d\omega\int^{E_0 -
  \omega}_{m_e}\!\!\! dE_e \,k_e\, F(E_e, Z = 1)\int^{E_0 - E_e -
  \omega}_0\!\!\! dq_0\,(E_0 - E_e - \omega - q_0)^2\nonumber\\
\hspace{-0.3in}&&\times
\int\frac{d\Omega_{e\gamma}}{4\pi}\int\frac{d\Omega_{e\gamma'}}{4\pi}\,\Big(\rho^{(1)}_{e\gamma\gamma'}(E_e,\vec{k}_e,
\omega, \vec{n}_{\vec{k}}, q_0, \vec{n}_{\vec{q}}) +
\rho^{(2)}_{e\gamma\gamma'}(E_e,\vec{k}_e, \omega, \vec{n}_{\vec{k}},
q_0, \vec{n}_{\vec{q}}) + \rho^{(2)}_{e\gamma\gamma'}(E_e,\vec{k}_e,
q_0, \vec{n}_{\vec{q}}, \omega, \vec{n}_{\vec{k}})\Big).
\end{eqnarray}
For the numerical calculation we use the following definitions

\begin{eqnarray}\label{eq:E.19}
\hspace{-0.3in}&&\int\frac{d\Omega_{e\gamma}}{4\pi}\,\ldots =
\frac{1}{4 \pi}\int^{\pi}_0d\vartheta_{e\gamma}\,
\sin\vartheta_{e\gamma} \int^{2\pi}_0d\varphi_{e\gamma}\,\ldots,\nonumber\\
\hspace{-0.3in}&&\int\frac{d\Omega_{e\gamma'}}{4\pi}\,\ldots =
\frac{1}{4 \pi}\int^{\pi}_0d\vartheta_{e\gamma'}\,
\sin\vartheta_{e\gamma'}
\int^{2\pi}_0d\varphi_{e\gamma'}\,\ldots,\nonumber\\
\hspace{-0.3in}&&\vec{k}_e \cdot \vec{n}_{\vec{k}} = k_e\,
\cos\vartheta_{e\gamma},\nonumber\\
\hspace{-0.3in}&&\vec{k}_e \cdot \vec{n}_{\vec{q}} = k_e\,
\cos\vartheta_{e\gamma'},\nonumber\\
\hspace{-0.3in}&&\vec{n}_{\vec{k}}\cdot \vec{n}_{\vec{q}} =
\cos\vartheta_{e\gamma}\,\cos\vartheta_{e\gamma'} +
\sin\vartheta_{e\gamma}\,\sin\vartheta_{e\gamma'}\,\cos(\varphi_{e\gamma}
- \varphi_{e\gamma'}).
\end{eqnarray}
The integrals in Eq.(\ref{eq:E.18}) we calculate for three photon
energy regions $15\,{\rm keV} \le \omega \le 340\,{\rm keV}$,
$14\,{\rm keV} \le \omega \le 782\,{\rm keV}$ and $0.4\,{\rm keV} \le
\omega \le 14\,{\rm keV}$, respectively.

\end{document}